\pdfoutput=1
\documentclass[]{aa}  

\usepackage{graphicx}
\usepackage{txfonts}
\usepackage{amssymb}
\usepackage{natbib}
\usepackage{fixltx2e}
\usepackage{rotating}
\bibpunct{(}{)}{;}{a}{}{,}
\usepackage{definitions}

\begin{document}
   \title{APEX-CHAMP$^+$ high-$J$ CO observations of \\low--mass young stellar objects}
   \subtitle{IV. Mechanical and radiative feedback}

  \authorrunning{Y{\i}ld{\i}z et al.}
  \titlerunning{Mechanical and radiative feedback}

\author{
           U.~A.~Y{\i}ld{\i}z\inst{1,2}
      \and L.~E.~Kristensen\inst{3}
      \and E.~F.~van~Dishoeck\inst{1,4}
      \and M.~R.~Hogerheijde\inst{1}
	  \and A.~Karska\inst{4}
      \and A.~Belloche\inst{5}
      \and A.~Endo\inst{6}
      \and W.~Frieswijk\inst{7,8}
      \and R.~G{\"u}sten\inst{5}
      \and T.~A.~van~Kempen\inst{1}
      \and S.~Leurini\inst{5}
      \and Z.~Nagy\inst{9}
      \and J.~P.~P{\'e}rez-Beaupuits\inst{5}
      \and C.~Risacher\inst{5,7}
      \and N.~van~der~Marel\inst{1}
      \and R.~J.~van~Weeren\inst{3}
	  \and F.~Wyrowski\inst{5}        
}

\institute{
Leiden Observatory, Leiden University, PO Box 9513, 2300 RA Leiden, The Netherlands\label{inst1} 
\and
Jet Propulsion Laboratory, California Institute of Technology, 4800 Oak Grove Drive, Pasadena CA, 91109, USA\label{inst2}, \\ \email{Umut.Yildiz@jpl.nasa.gov}
\and
Harvard-Smithsonian Center for Astrophysics, 60 Garden Street, Cambridge, MA 02138, USA\label{inst3}
\and
Max-Planck-Institut f\"{u}r Extraterrestrische Physik (MPE), Giessenbachstrasse 1, 85748 Garching, Germany\label{inst4}
\and
Max-Planck-Institut f\"{u}r Radioastronomie, Auf dem H\"{u}gel 69, D-53121, Bonn, Germany\label{inst5}
\and
Kavli Institute of Nanoscience, Delft University of Technology, Lorentzweg 1, 2628 CJ, Delft, the Netherlands\label{inst6}
\and
Kapteyn Institute, University of Groningen, Landleven 12, 9747 AD, Groningen, the Netherlands\label{inst7}
\and
ASTRON, Oude Hoogeveensedijk 4, 7991 PD, Dwingeloo, the Netherlands\label{inst8}
\and
I. Physikalisches Institut der Universit\"{a}t zu K\"{o}ln, Z\"{u}lpicher Strasse 77, 50937 K\"{o}ln, Germany\label{inst9}
}

   \date{Accepted: 2015/01/07}

 
  \abstract
  {During the embedded stage of star formation, bipolar molecular
    outflows and UV radiation from the protostar are important
    feedback processes. Both processes reflect the accretion 
    onto the forming star and affect subsequent collapse or
    fragmentation of the cloud.}
  {Our aim is to quantify the feedback, mechanical and radiative, for
    a large sample of low-mass sources in a consistent manner. The outflow
    activity is compared to radiative feedback in the form of UV
    heating by the accreting protostar to search for correlations and
    evolutionary trends.}
  {Large-scale maps of 26 young stellar objects, which are part of the
    {\it Herschel} WISH key program are obtained using the CHAMP$^+$
    instrument on the Atacama Pathfinder EXperiment (\twco\ and \thco\
    6--5; $E_{\rm up}$$\sim$100 K), and the HARP-B
    instrument on the James Clerk Maxwell Telescope (\twco\ and \thco\
    3--2; $E_{\rm up}$$\sim$30 K). The maps have high spatial
    resolution, particularly the CO~6--5 maps taken with a 9$\arcsec$
    beam, resolving the morphology of the outflows. The maps are used
    to determine outflow parameters and the results are compared with
    higher-$J$ CO lines obtained with {\it Herschel}. Envelope models
    are used to quantify the amount of UV-heated gas and its
    temperature from \thco\ 6--5 observations.}
{All sources in our sample show outflow activity, with the 
spatial extent decreasing from the Class 0 to the Class I stage. 
Consistent with previous
studies, the outflow force, \FCO, is larger for Class~0 sources 
than for Class~I sources, even if their luminosities are comparable.
The outflowing gas
typically extends to much greater distances than the power-law
envelope and therefore influences the surrounding cloud material directly.
Comparison of the CO~6--5 results with HIFI \hho\ 
and PACS high-$J$ CO lines, both tracing currently shocked
gas, shows that the two components are linked, even though the
transitions do not probe the same gas.  The link does not extend down
to CO~3--2. The conclusion is that CO~6--5 depends on the shock
characteristics (density and velocity), whereas CO~3--2 is more
sensitive to conditions in the surrounding environment (density). The
radiative feedback is responsible for increasing the gas temperature
by a factor of two, up to 30--50 K, on scales of a few thousand AU,
particularly along the direction of the outflow. The mass of the UV
heated gas exceeds the mass contained in the entrained outflow in the inner
$\sim$3000 AU and is therefore at least as important on small scales.
}
{}

\maketitle


\section{Introduction}

During the early phases of star-formation, material surrounding the
newly forming star accretes onto the protostar. At the same time,
winds or jets are launched at supersonic speeds from the star-disk
system, which sweep up surrounding envelope material in large bipolar
outflows. The material is accelerated and pushed to distances of
several tens of thousands of AU, and these outflows play a pivotal
role in the physics and chemistry of the star-forming cores
\citep{Snell80, Goldsmith84, Lada87,Greene94,Bachiller99,
  ArceSargent06,Tafalla13}. The youngest protostars have highly
collimated outflows driven by jets, whereas at later stages wide-angle
winds drive less collimated outflows. However, there is still not a
general consensus to explain the launching mechanisms and nature of
these outflows \citep{Arce07,Frank14}.

The goal of this paper is to investigate how the outflow activity
varies with evolution and how this compares with other measures of the
accretion processes for low-mass sources. The outflows reflect the
integrated activity over the entire lifetime of the protostar, which
could be the result of multiple accretion and ejection events.  It is
important to distinguish this probe from the current accretion rate,
as reflected for example in the luminosity of the source, in order to
understand the accretion history. The well-known ``luminosity
problem'' in low-mass star-formation indicates that protostars are
underluminous compared to theoretical models
\citep{Kenyon90,Evans09c2d,Enoch09,Dunham10,Dunham13}. One of the
possible resolutions to this problem is that of ``episodic
accretion'', in which the star builds up through short bursts of rapid
accretion over long periods of time rather than continuous
steady-state accretion. An accurate and consistent quantification of
outflow properties, such as the outflow force and mass, is essential 
for addressing this problem.

Outflows have been observed in CO emission in the last few decades 
towards many sources, but those observations were mainly done via 
lower-$J$ CO rotational transitions (\Ju$\leq$3), which probe colder 
swept-up or entrained gas ($T$$\sim$50--100~K) \citep[e.g.,][and many
others]{Bachiller90,Blake95,Bontemps96,Tafalla00,Curtis10_2outflows}.
One of the most important parameters that is used for the evolutionary 
studies of star formation is the ``outflow force'', which is known as the
strength of an outflow and defined similar to any $r^{-2}$-type force.
These studies conclude that the outflow force correlates well with
bolometric luminosity, \Lbol, a correlation which holds over several
orders of magnitude. Furthermore, the outflow force from Class 0
sources is stronger than for Class I sources, indicating an
evolutionary trend. The correlations, however, often show some degree
of scatter, typically more than an order of magnitude in \FCO\ for any
value of \Lbol. Some of the uncertainties in these studies include the
opacity in the line wings, the adopted inclination angle and cloud
contamination at low outflow velocities \citep[e.g.,][]{vanderMarel13}. 
Comparison with other outflow tracers such as water recently observed 
with the {\it Herschel} Space Observatory is further complicated because 
the various studies use different analysis methods to derive outflow 
parameters from low-$J$ CO maps. One of the goals of this paper is to 
provide a consistent set of outflow parameters determined by the same 
method using data from the same telescopes for comparison with the 
{\it Herschel} lines.

Recently, the importance of radiative feedback from low-mass protostars 
on all scales of star formation has been acknowledged. 
On cloud scales ($>$10$^4$~AU) the feedback sets the
efficiency at which cores fragment from the cloud and form stars
\citep{Offner09, Offner10, Hansen12} because the Jeans length scales
as $T^{0.5}$. Simulations including radiative feedback and radiative
transfer reproduce the observed initial mass function (IMF) better
than models without these effects included \citep{Offner09}. On the
scales of individual cores ($<$3000 AU), the radiative feedback
suppresses the fragmentation into multiple systems and serves to
stabilize the protostellar disk \citep{Offner10}. Thus, quantifying
observationally the temperature changes as a function of position from
the protostar are important steps toward more accurate models of star
formation. The first observational evidence of heating of the gas
around low-mass protostars on scales of $\sim$1000 AU by UV radiation
escaping through the outflow cavities dates back to \citet{Spaans95}
based on strong narrow \thco\ 6--5 lines, and has since been
demonstrated and quantified for a few more sources by
\citet{vanKempen09champ,Yildiz12,Visser12}. We note that this UV-heated gas
is warm gas with temperatures higher than that of the dust, and is
thus in excess of warm material in the envelope that has been heated
by the protostellar luminosity, where the gas temperature is equal to
the dust temperature. Although UV heating toward photo-dissociation regions 
(PDRs) is readily traced by emission from polycyclic aromatic hydrocarbons 
(PAHs), the PAH abundance toward embedded protostars is too low for them to 
be used as a tool in this context \citep{geers09}. Here we investigate 
the importance of radiative feedback for a much larger sample of low-mass 
sources and compare the gas temperatures and involved mass with that 
of the outflows.

Tracing warm gas ($T$$\gtrsim$30~K) in the envelope or in the
surroundings requires observations of higher-$J$ transitions of CO,
e.g., \Ju$\geq$5, for which ground-based telescopes demand excellent
weather conditions on dry observing sites. The CHAMP$^+$ instrument,
mounted on the Atacama Pathfinder EXperiment (APEX) telescope 
is ideally suited to observe \mbox{higher-$J$} CO transitions and
efficiently map extended sources. The broad line wings of CO\,6--5
(\Eupk=115\,K) suffer less from opacity effects than CO\,3--2
(\Eupk=33\,K) \citep{vanKempen09champ2,Yildiz12}. Moreover, the
ambient cloud contribution is smaller for these \mbox{higher-$J$}
transitions, except close to the source position, where the dense
protostellar envelope may still contribute. Even higher-$J$ CO lines
up to \Ju$\sim$50 were routinely observed with the \herschel\
\citep{Pilbratt10} and provide information on the shocked gas in the
{\it Herschel} beam
\citep{Herczeg12,Goicoechea12,Benedettini12,Manoj13,Green13,Nisini13,Karska13}.
This currently shocked gas is different from that observed in \mbox{low-$J$}
CO transitions, as is evident from their different spatial
distributions \citep{Tafalla13,Santangelo13}.

In this paper, we present an APEX-CHAMP$^{+}$ survey of 26 low-mass
young stellar objects (YSOs), which were mapped in \mbox{CO~$J$=6--5}
and isotopologues in order to trace their outflow activity, following
\citet{vanKempen09champ2,vanKempen09champ} and \citet{Yildiz12},
papers I, II and III in this series, on individual or more limited
samples of sources.
These data complement our earlier surveys at lower frequency of CO and
other molecules with the James Clerk Maxwell Telescope (JCMT) and APEX
\citep[e.g.,][]{Jorgensen02,Jorgensen04,vanKempen09_southc+}. The same
sources are covered in the {\it Herschel} key project, ``Water in
star-forming regions with \herschel'' \citep[WISH;][]{vanDishoeck11},
which has observed H$_2$O and selected high-$J$ CO lines with HIFI and
PACS instruments. Many of the sources are also included in the ``Dust, 
Ice and Gas in Time'' program \citep[DIGIT; PI: N.\,Evans;][]{Green13}, 
which has obtained full PACS spectral scans.
The results obtained from the \twco\ maps are complemented by \thco\
6--5 data of the same sources, with the narrower \thco\ 6--5 lines
probing the UV photon-heated gas.

The YSOs in our sample cover both the deeply embedded Class 0 stage as
well as the less embedded Class~I stage \citep{Andre00,Robitaille06}.
Physical models of the dust temperature and density structure of the
envelopes have been developed for all sources by \citet{Kristensen12}
through spherically symmetric radiative transfer models of the
continuum emission. The full data set covering many sources, together
with the envelope models, allows us to address important
characteristics of YSOs through the evolution from Class~0 to Class~I
in a more consistent manner.  These characteristics can be inferred
from their different morphologies, outflow forces, envelope masses,
etc.\ and eventually be compared with evolutionary models.  The study
presented here is also complementary to that of \citet{Yildiz13hifi},
where only the source position was studied with spectrally resolved CO
line profiles from $J$=2--1 to 10--9 (\Eup$\sim$300 K), and trends
with evolution were examined.

The outline of the paper is as follows. In
Section~\ref{sec:observations}, the observations and the telescopes
where the data have been obtained are described. In Section
\ref{5:sec:results}, physical parameters obtained from molecular
outflows are given and the UV heated gas component is identified. In
Sect.~\ref{5:sec:discussion}, these results are discussed, and
conclusions from this work are presented in
Sect.~\ref{5:sec:conclusion}.

\begin{figure}[!t]
    \centering
    \includegraphics[scale=0.67]{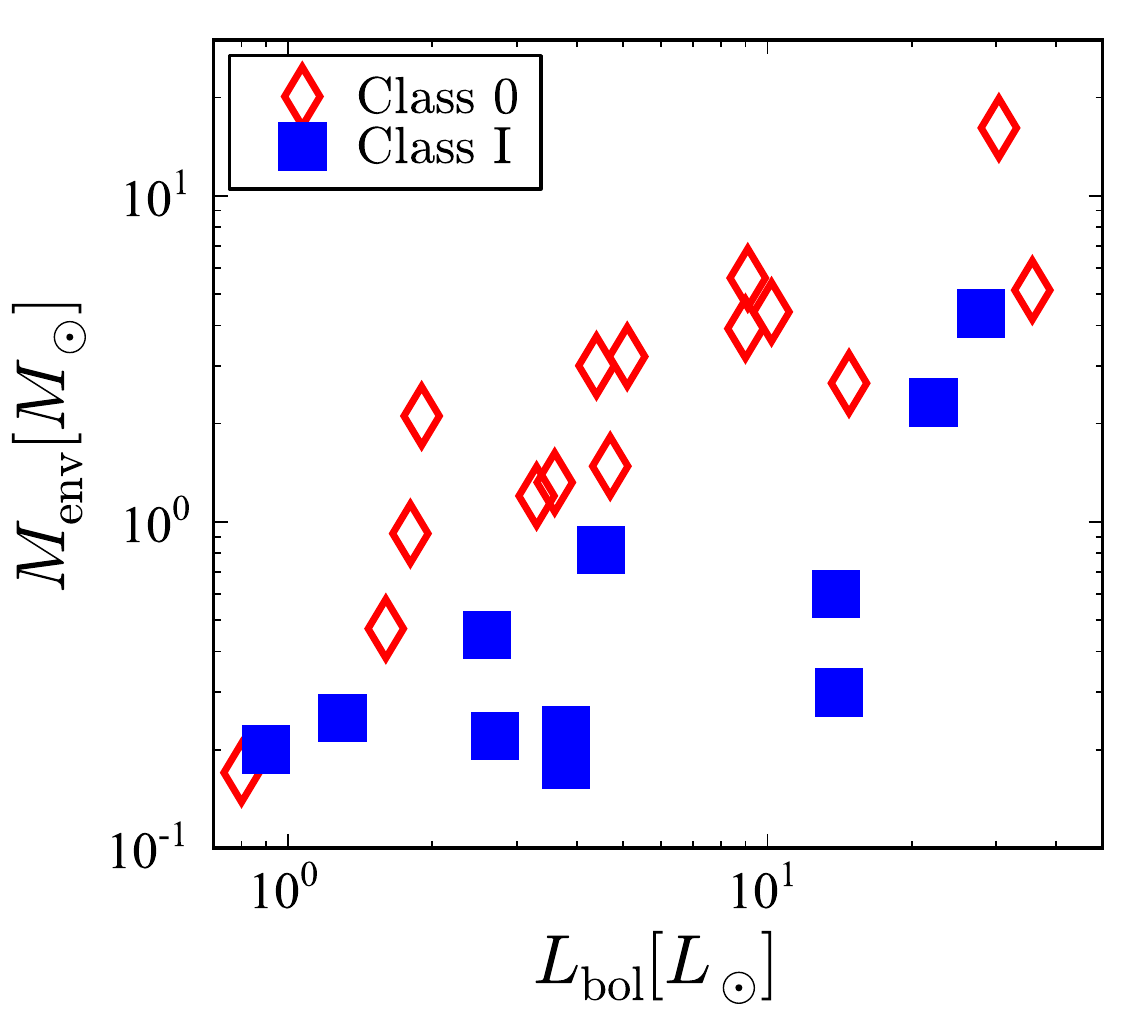}
    \caption{Envelope mass,\Menv, vs bolometric luminosity, \Lbol,
      for the surveyed sources. Red diamonds and blue squares
      indicate Class~0 and Class~I sources, respectively.}
    \label{fig:corr_Lbol_Menv}
\end{figure}

\section{Sample and observations}
\label{sec:observations}
\begin{figure*}[!ht]
\centering
\begin{minipage}{17cm}
    \includegraphics[scale=0.59]{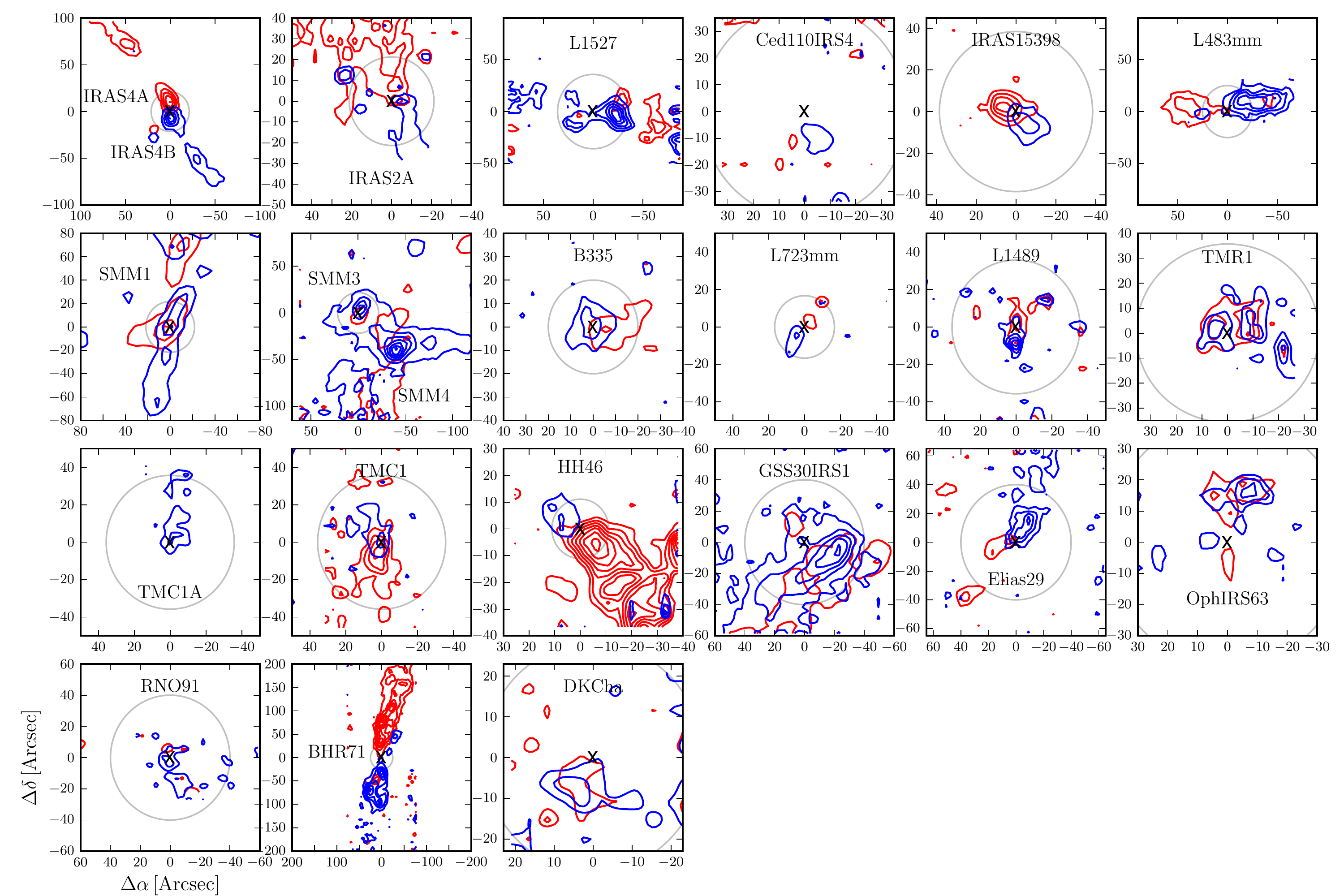}
\end{minipage}
\caption{\small Overview of the outflows traced by the \twco\ 6--5 observations 
with the APEX-CHAMP$^+$ instrument. Contour levels are given in 
Table~\ref{tbl:contourlevels} and the source is located at (0,0) in each map, with 
the exception of the maps of NGC~1333-IRAS~4A and IRAS~4B, and Ser-SMM3 and Ser-SMM4, 
which are located in the same maps and centered on NGC~1333-IRAS~4A and Ser-SMM3, respectively. 
The circle in each plot corresponds to a region of 5000 AU radius at the distance 
of each source. Velocity ranges over which the integration was done are 
provided in Table \ref{tbl:contourlevels}.}
\label{fig:outflows12co65}
\end{figure*}

\begin{figure*}[!ht]
\centering
\begin{minipage}{17cm}
    \includegraphics[scale=0.59]{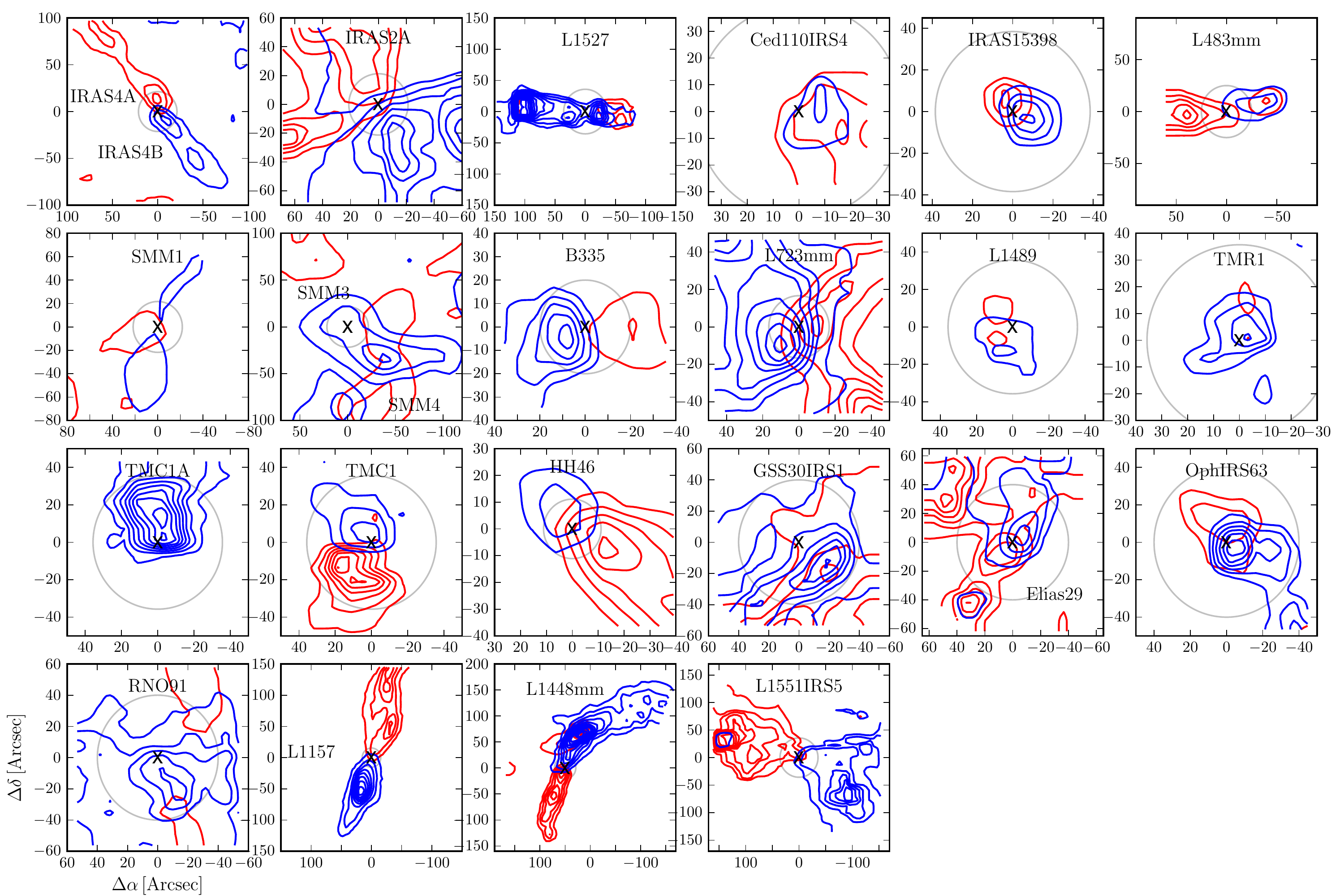}
\end{minipage}
\caption{\small Overview of the entire set of outflows traced by the \twco\ 
3--2 observations with the JCMT and APEX. Contour levels are given in 
Table~\ref{tbl:contourlevels} and the source is located at (0,0) in each map, 
with the exception of the maps of NGC~1333-IRAS~4A and NGC~1333-IRAS~4B, and Ser-SMM3 
and Ser-SMM4, which are located in the same maps and centered on NGC~1333-IRAS~4A and Ser-SMM3, 
respectively. The circle in each plot corresponds to a region of 5000 AU radius 
at the distance of each source. Velocity ranges over which the integration 
was done are provided in Table \ref{tbl:contourlevels}.}
\label{fig:outflows12co32}
\end{figure*}

\subsection{Sample}

The sample selection criteria with the coordinates and other basic
information of the source list are presented in \citet{vanDishoeck11}
with updates in \citet{Kristensen12}, and is the same as the sample
presented in \citet{Yildiz13hifi}. It consists of 15 Class~0 and 11
Class~I embedded protostellar sources located in the Perseus,
Ophiuchus, Taurus, Chamaeleon, and Serpens molecular clouds. The average 
distance is 200\,pc, with a maximum distance of 450\,pc.

Figure \ref{fig:corr_Lbol_Menv} presents the envelope mass (\Menv) as
a function of bolometric luminosity (\Lbol) for all sources.  The
parameters are taken from the continuum radiative transfer modeling by
\citet{Kristensen12} based on fits of the spectral energy
distributions (SEDs) including new {\it Herschel}-PACS fluxes, as well
as the spatial extent of the envelopes observed at submillimeter
wavelengths.  The envelope mass is measured either at the \Tdust =10~K
radius or at the $n$=10$^4$ \cmthree\ radius, depending on which is
smaller. Class~0 and Class~I sources are well separated in the
diagram, with the Class~0 sources having higher envelope masses.  This
type of correlation diagram has been put forward by \citet{Saraceno96}
and subsequently used as an evolutionary diagram for embedded YSOs
with lower envelope masses representing later stages
\citep[e.g.,][]{Bontemps96,Hogerheijde98,Hatchell07}. In our sample,
envelope masses range from 0.04 \Msun\ (Elias 29) to 16 \Msun\ (Ser-SMM1)
and the luminosities range from 0.8~\Lsun\ (Ced110\,IRS4) to 35.7 \Lsun\
(NGC~1333-IRAS~2A). The large range of masses and luminosities makes the
sample well suited for studying trends with various source
parameters. The range of luminosities studied is similar to that of
\citet{Bontemps96}, $\sim$0.5 to 15~\Lsun, but our sample is more
weighted toward higher luminosities and earlier stages.

\subsection{Observations}

Molecular line observations of CO in the $J$=6--5 transitions were
done with the \mbox{12-m} submillimeter Atacama Pathfinder EXperiment,
APEX\footnote{This publication is based on data acquired with the
  Atacama Pathfinder Experiment (APEX). APEX is a collaboration
  between the Max-Planck-Institut f\"{u}r Radioastronomie, the
  European Southern Observatory, and the Onsala Space Observatory.}
\citep{Guesten08} at Llano de Chajnantor in Chile, whereas the
$J$=3--2 transition was primarily observed at the 15-m James Clerk
Maxwell Telescope, JCMT\footnote{The JCMT is operated by The Joint
  Astronomy Centre on behalf of the Science and Technology Facilities
  Council of the United Kingdom, the National Research Council of
  Canada, and (until 31 March 2013) the Netherlands Organisation for
  Scientific Research.} at Mauna Kea, Hawaii.

{\it APEX:} \twco\ and \thco\ 6--5 maps of the survey were obtained
with the CHAMP$^{+}$ instrument on APEX between June 2007 and
September 2012.  The CHAMP$^{+}$ instrument consists of two heterodyne
receiver arrays, each with seven pixel detector elements for
simultaneous operations in the \mbox{620--720 GHz} and \mbox{780--950}
GHz frequency ranges \citep{Kasemann06, Guesten08}. The observational
procedures are explained in detail in
\citet{vanKempen09champ2,vanKempen09champ,vanKempen09_southc+} and
\citet{Yildiz12}.  Simultaneous observations were done with the
following settings of the lower and higher frequency bands:
\twco\ 6--5 with \mbox{\twco\ 7--6}; \thco\ 6--5 with [\ion{C}{i}]~2--1.
\twco\ maps cover the entire outflow extent with a few exceptions (L1527, 
Ced110\,IRS4, and L1551-IRS5), whereas \thco\ maps cover only a 
$\sim$100$''$$\times$100$''$ region around the central source position. 
L1157 is part of the WISH survey, but because it is not accessible from 
APEX (dec = $+$68\degr), no CO 6--5 data are presented.

The APEX beam size is $\sim$9$''$ ($\sim$1800~AU for a source at 200~pc) 
at 691~GHz.  The observations were done using position-switching toward 
an emission-free reference position. The CHAMP$^{+}$ instrument uses the
Fast Fourier Transform Spectrometer (FFTS) backend \citep{Klein06} for
all seven pixels with a resolution of 0.183~MHz (0.079~\kms\ at
691~GHz). The \rms\ at the source position is listed in \citet{Yildiz13hifi} 
for the \co\ 6--5 and \thco\ 6--5 observations and is typically 0.3--0.5 K for 
the former and 0.1--0.3 K for the latter, both in 0.2 \kms\ channels. The 
rms increases near the map edges where the effective integration time per 
beam was significantly smaller than in the central parts; near the edges 
the rms may be twice as high.
Apart from the high-$J$~CO
observations, some of the 3--2 line observations were also conducted
with APEX for a few southern sources, e.g., DK~Cha, Ced110\,IRS4, and
HH\,46 \citep{vanKempen09_southc+}.

{\it JCMT:} Fully sampled jiggle maps of \twco\ and \mbox{\thco~3--2}
were obtained using the HARP-B instrument mounted on the JCMT. HARP-B
consists of 16 SIS detectors with 4$\times$4 pixel elements of 15$''$
each at 30$''$ separation. Most of the maps were obtained through our own
dedicated proposals, with a subset obtained from the JCMT
public archive\footnote{This research used the facilities of the
Canadian Astronomy Data Centre operated by the National Research
Council of Canada with the support of the Canadian Space Agency.}.

The data were acquired on the $T_{\rm A}^{*}$ antenna temperature
scale and were converted to main-beam brightness temperatures $T_{\rm
  MB}$=$T_{\rm A}^{*}/ \eta_{\rm MB}$ using the beam efficiencies
($\eta_{\rm MB}$). The CHAMP$^+$ beam efficiencies were taken from the
CHAMP$^+$
website\footnote{\url{http://www3.mpifr-bonn.mpg.de/div/submmtech/heterodyne/champplus/champ_efficiencies.29-11-13.html}} 
and forward efficiencies are
0.95 in all observations. The various beam efficiencies are all given
in \citet[][their Appendix C]{Yildiz13hifi} and are typically $\sim$0.5. 
The JCMT beam efficiencies were taken from the JCMT Efficiencies
Database\footnote{\url{http://www.jach.hawaii.edu/JCMT/spectral_line/Standards/eff_web.html}},
and 0.63 is used for all HARP-B observations. Calibration errors are
estimated to be $\sim$20\% for both telescopes. Typical \rms\ noise
levels of the $J$=3--2 data are from 0.05~K to 0.1~K in 0.2 \kms\ channels.

For the data reduction and analysis, the ``Continuum and Line Analysis
Single Dish Software'', \verb1CLASS1 program, which is part of the
GILDAS software\footnote{\url{http://www.iram.fr/IRAMFR/GILDAS}}, is used. 
In particular, linear baselines were subtracted from all spectra.
\twco\ and \thco\ 6--5 and 3--2 line profiles of the central source
positions of all the sources in the sample are presented in \citet{Yildiz13hifi}.

\subsection{$^{12}$CO maps}
\label{5:sec:12comaps}

All spectra are binned to a 0.5 \kms\ velocity resolution for
analyzing the outflows. The intensities of the blue and red outflow
lobes are calculated by integrating the blue and red emission in each
of the spectra separately, where the integration limits are carefully
selected for each source by using the 0.2~\kms\ resolution CO~3--2 or 6--5 
spectra if the former is not available (see Fig.~\ref{fig:AllBCR}).
First, the inner velocity limit, $V_{\rm in}$, closest to the source
velocity is determined by selecting a spatial region not associated
with the outflow. The \twco spectra in this region are averaged to determine
the narrow line emission coming from the envelope and surrounding
cloud, and $V_{\rm in}$ is estimated from the width of the quiescent
emission (see Fig. \ref{fig:AllBCR} in the Appendix). Second, the outer 
velocity limits $V_{\rm out}$ are determined from the highest $S/N$ spectrum 
inside each of the blue and red outflow lobes. The outer velocity limits are 
selected as the velocity where the emission in the spectrum goes down to the
1$\sigma$ limit for the first time. It therefore excludes extremely
high velocity or `bullet' emission which is seen for a few
sources. The blue- and red-shifted integrated intensity is measured by
integrating over these velocity limits across the entire map, 
but excluding any extremely high velocity (EHV) or ``bullet'' emission.

\subsubsection{Outflow velocity}
The maximum outflow velocity, \vmax\ is defined as $\lvert$$V_{\rm
  out}$--\vlsr$\lvert$, the total velocity extent measured relative to
the source velocity.  In order to estimate \vmax, representative
spectra from the blue and red outflow lobes observed in CO~3--2 are
selected separately, and $V_{\rm out}$ is measured as described above.
 The differences between the velocity, where
the emission reaches 1$\sigma$ level (\Vout) with \Vlsr\ are taken as 
the global \vmax\ values for the corresponding blue and red-shifted lobes
\citep{CabritBertout92}.

Two issues arise when determining \vmax\ \citep[e.g.,][]{vanderMarel13, 
Dunham14}: First, \vmax\ is a function
of the \rms\ noise level and generally decreases with increasing
\rms. For noisy data, \vmax\ may be underestimated compared to its
true value.  For this reason, the 3--2 lines are chosen to determine
\vmax\ because of their higher $S/N$ than the 6--5 lines.  Second, if
the outflow lobes are inclined, \vmax\ suffers from projection
effects. Both effects will increase the value of \vmax\ if properly
taken into account. 

Concerning the second issue, the inclination is 
difficult to estimate from these data alone; proper-motion studies along with 
radial velocities are required to obtain an accurate estimate of the inclination. 
Alternatively, the velocity structure may be modeled assuming some distribution 
of material, e.g., a wind-driven shell with a Hubble-like flow \citep{lee00}, 
where the inclination then enters as a free parameter.
It is defined as the angle between the outflow direction and
the line of sight \citep[][$i$=0\degr\ is pole on]{CabritBertout90}.
Small radial velocities are expected for outflows which lie in the
plane of the sky. Therefore a correction factor for inclination
$c_{i}$ is applied in the calculations. In Table~\ref{tbl:inclcorr},
the correction factors from \citet{DownesCabrit07} are tabulated; these correction 
factors come from detailed outflow modeling and synthetic observations of the model 
results. Moreover, we note that these correction factors include correction for 
missing mass within $\pm$ 2 \kms\ from the source velocity. The
correction factors have been applied to the outflow rate, force and
luminosity as listed in Tables \ref{tbl:outflowparam_C0} and
\ref{tbl:outflowparam_CI}. The velocity, as a measured parameter, is not corrected 
for inclination. The inclination angles are estimated from the outflow maps as follows:
if the outflow lobes are overlapping, the outflow is likely very
inclined. If the outflow shows low-velocity line wings but a large
extent on the sky, the inclination is very likely low. In this way
each outflow is classified individually, and divided into
inclination bins at 10$\degr$, 30$\degr$, 50$\degr$, and 70$\degr$.
Our estimates are
listed in Tables~\ref{tbl:outflowparam_C0} and
\ref{tbl:outflowparam_CI}, and are consistent with the literature
where available
\citep{CabritBertout92,Gueth96,Bourke97,Hogerheijde97,Micono98,Brown99,Lommen08,Tobin08,vanKempen09champ},
except for IRAS~15398 for which we find a larger inclination than
\citet{vanKempen09_southc+}. Our inclination of IRAS~15398 is consistent with 
newer values from \citep{oya14}. Although the method for determining the outflow 
inclinations is subjective, the inclinations agree with literature values where 
available, which lends some credibility to the method, and we estimate that the 
uncertainty is 30\degr. That is, the correction introduces a potential 
systematic error of up to a factor of 2 in the outflow parameters.

\begin{table}[!t]
\caption{Inclination correction factors.}
\begin{center}
\begin{tabular}{l l l l l l}
\hline
\hline
 $i$($\degr$) & 10 & 30 & 50 & 70 \\
\hline  
$c_{i}$       & 1.2 & 2.8 & 4.4 & 7.1 \\
\hline
\end{tabular}
\end{center}
\tablefoot{Line-of-sight inclinations, where $i$=0\degr\ indicates pole-on 
\citep{DownesCabrit07}.}
\label{tbl:inclcorr}
\end{table}

The resulting maps of all sources are presented in
Figs.~\ref{fig:outflows12co65} and~\ref{fig:outflows12co32} for
$^{12}$CO 6--5 and 3--2, respectively, where blue and red contours
show the blue- and red-shifted outflow lobes, respectively. The
velocity limits are summarized in Table~\ref{tbl:contourlevels} in the
Appendix. A few maps cover only the central
$\sim$2\arcmin$\times$2\arcmin, specifically the three Class 0 sources
NGC~1333-IRAS~2A, L723mm, L1527, and the two Class I sources Elias~29 and
L1551-IRS5. Source-by-source outflow and intensity maps obtained from
the CO 6--5 and 3--2 data are presented in
Figs.~\ref{fig:AllIntensityCO_1}--\ref{fig:AllIntensityCO_3} in the
On-line Appendix.

\subsection{$^{13}$CO maps}
\label{5:sec:13comaps}

The \thco\ 6--5 and 3--2 transitions were mapped around the central 
$\sim$1\arcmin$\times$1\arcmin~region, corresponding to typically 
$\sim$10$^4$~AU~$\times$~10$^4$~AU.
The total integrated intensity is measured for all the sources and
presented in Table C.1-26 of \citet{Yildiz13hifi} for the source positions.
All maps are presented as contour maps in Figs.~\ref{fig:All13COs1}
and \ref{fig:All13COs2} and as spectral maps in
Figs.~\ref{fig:specmap13CO65_1}-\ref{fig:specmap13CO65_6} in the
Appendix.


\section{Results}
\label{5:sec:results}

\subsection{Outflow morphology}
\label{5:sec:morp}

All sources show strong outflow activity in both CO transitions,
$J$=6--5 and 3--2, as is evident from both the maps and spectra
(Figs. \ref{fig:outflows12co65}, \ref{fig:outflows12co32}, and
Figs. \ref{fig:AllBCR}--\ref{fig:AllIntensityCO_3}). The advantage of
the CO~6--5 maps is that they have higher spatial resolution by a
factor of 2 than the CO~3--2 maps. On the other hand, the CO~3--2
maps have the advantage of higher \SN\ than the CO~6--5 maps by
typically a factor of 4 in main beam temperature.

Most sources show a clear blue-red bipolar structure. In a few cases
only one lobe is observed. Specific examples are TMC1A, which shows
no red-shifted outflow lobe, and HH\,46, which has only a very small
blue-shifted outflow lobe. One explanation is that these sources are at the
edge of the cloud and that there is no cloud material to run into
\citep{vanKempen09champ}. For L723mm, NGC~1333-IRAS~2A and BHR71, two outflows are
driven by two independent protostars \citep{Lee02, Parise06,Codella14} and 
both outflows are detected in our CO 3--2 maps. In CO 6--5, only one 
outflow shows up toward L723mm and NGC~1333-IRAS~2A, whereas both outflows 
are seen toward BHR71.

\begin{figure}[!t]
    \centering
    \includegraphics[scale=0.45]{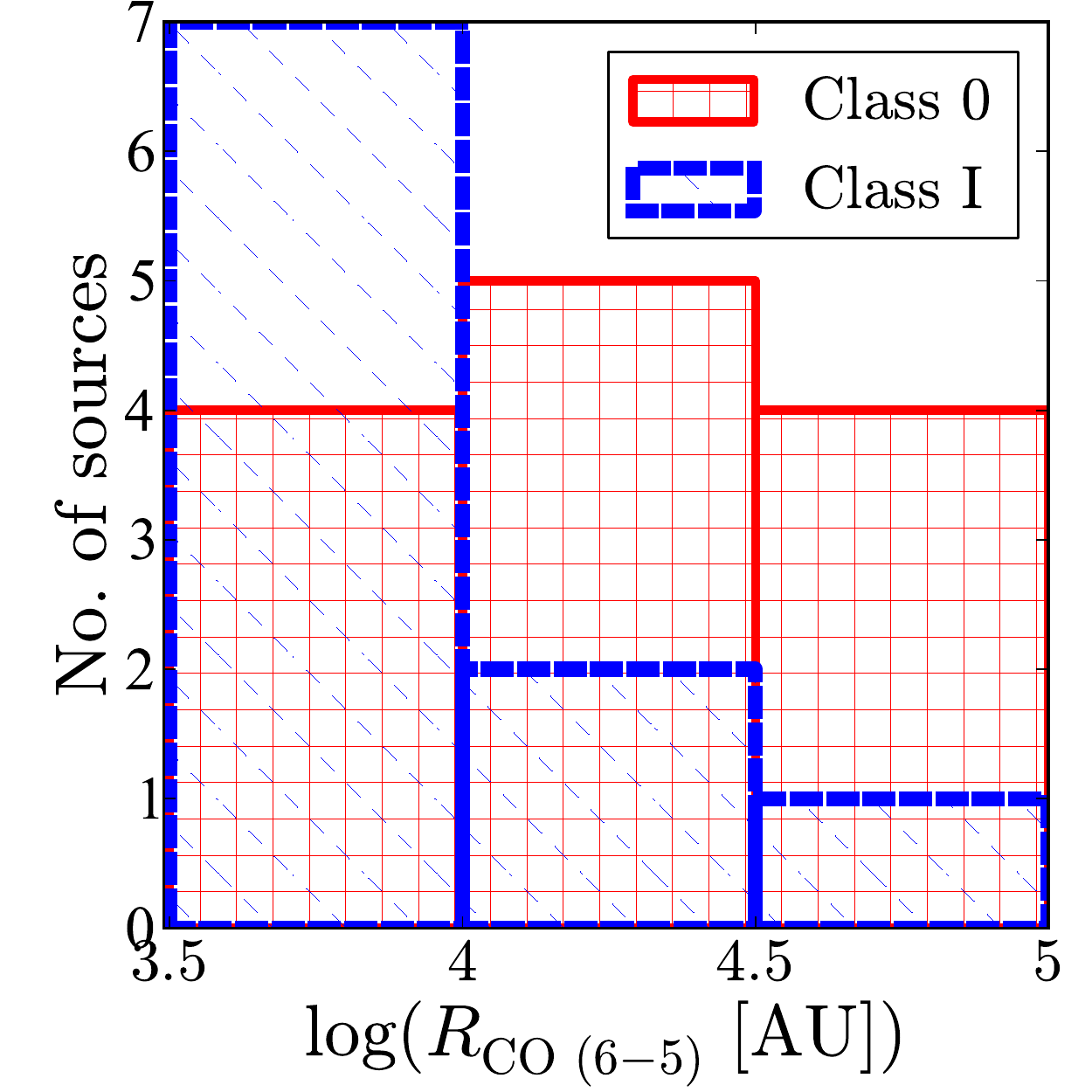}
    \caption{\small Histogram of total \RCO\ (blue- and red-shifted outflows 
    combined) is shown for Class 0 (red) and Class I (blue) sources. 
    (\RCO\ is not corrected for inclination.)}
    \label{fig:HistRCO}
\end{figure}

\begin{figure}[!t]
    \centering
    \includegraphics[scale=0.55]{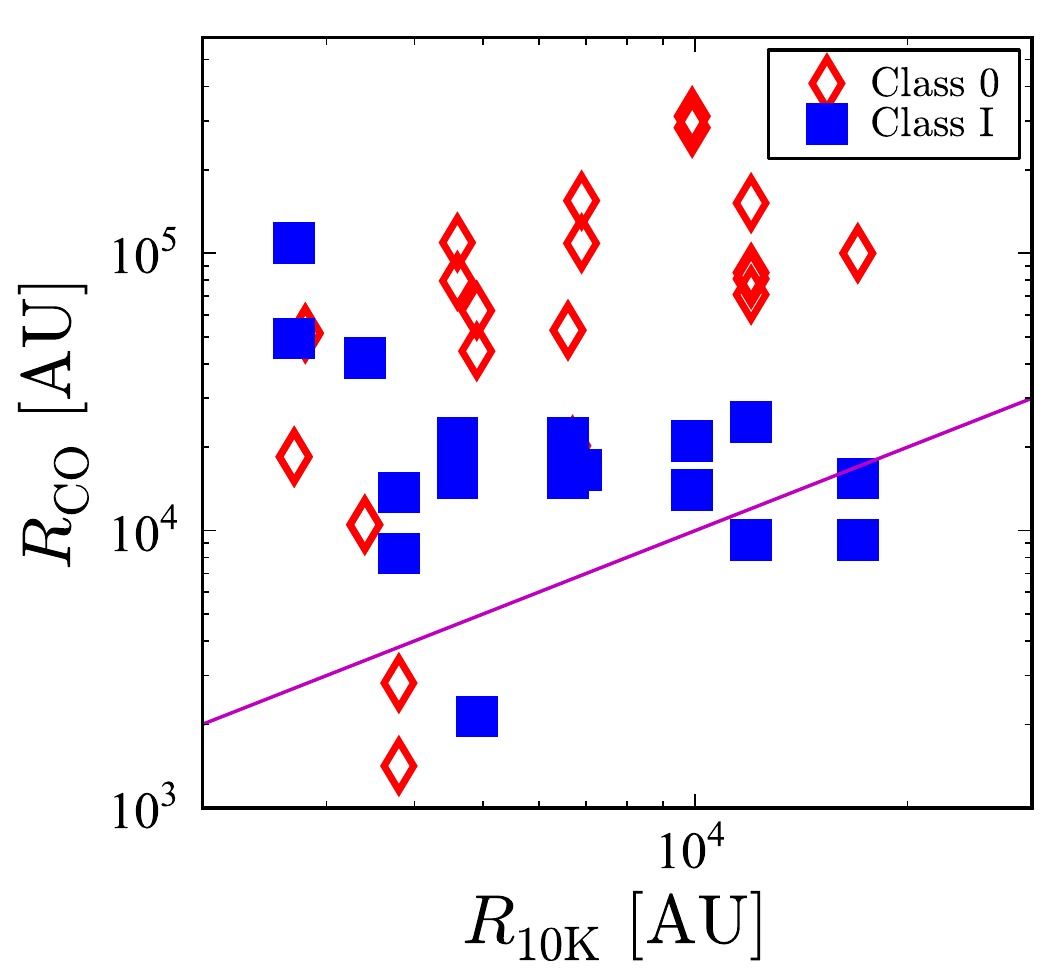}
    \caption{\small \RCO\ is plotted against \R10K, the radius of the modeled 
    envelope within 10~K radius. The black line is for \RCO\ = \R10K, showing 
    that almost all sources follow \RCO\ $>$ \R10K and that \RCO\ is larger 
    for Class 0 than Class I sources.}
    \label{fig:PlotRCOR10K}
\end{figure}

Visual inspection shows that the Class~0 outflows are more collimated
than their Class~I counterparts as expected \citep[e.g.,][]{Arce07}.
The length of the outflows can be quantified for most of the sources.
\RCO\ is defined as the total outflow extent assuming that the
outflows are fully covered in the map. \RCO\ is measured separately
for the blue and red outflow lobes as the projected size, with
sometimes significantly different values. \RCO\ as measured from CO 6--5 is 
applied to CO 3--2 in the cases where the CO 6--5 maps are larger than 
their 3--2 counterparts. Toward some sources, e.g.,
DK~Cha and NGC~1333-IRAS~4B, the blue and red outflow lobes overlap, likely because 
the outflows are observed nearly pole on. In other cases the outflow lobes
cannot be properly isolated from nearby neighboring outflow lobes. Such a 
confusion is most pronounced in Ophiuchus (e.g., GSS30-IRS1). In those 
cases, \RCO\ could not be properly estimated and the estimated value is a 
lower limit. Figure~\ref{fig:HistRCO} shows a histogram of total \RCO\ for
Class 0 and I sources. Class 0 sources show a nearly flat distribution
across the measured range of extents, whereas few Class I sources show
large outflows (L1551 is a notable exception). In
Fig.~\ref{fig:PlotRCOR10K}, \RCO\ is plotted against \R10K, the radius
of the modeled envelope within a 10~K radius.  The outflowing gas
typically extends to much greater distances than the surrounding
envelope and thus influences the surrounding cloud material directly.

\begin{table*}[!th]
\caption{Outflow properties of the red and blue outflow lobes of Class~0 sources.}
\small  
\begin{center}
\begin{tabular}{@{}l c c c c c c c c c c}
\hline \hline
Source & Trans. & Inclination & Lobe & $R_{\rm CO}$\tablefootmark{a} & $t_{\rm{dyn}}$\tablefootmark{a,b} & $M_{\rm outflow}$\tablefootmark{a,c} & $\dot{M}$\tablefootmark{d,e} & $F_{\rm CO}$\tablefootmark{d,f} & $L_{\rm{kin}}$\tablefootmark{d,g}  \\ 
& & [$\degr$] & & [AU] & [10$^3$ yr] & [M$_\odot$] & [M$_\odot$ yr$^{-1}$]  &[M$_\odot$ yr$^{-1}$km s$^{-1}$] & [L$_\odot$]  \\  
\hline
\object{L1448MM} &  CO 3--2  & 50 &  Blue  & 5.9$\times$10$^{4}$ & 5.5 & 9.0$\times$10$^{-2}$ & 7.2$\times$10$^{-5}$ & 2.0$\times$10$^{-3}$ & 2.8$\times$10$^{0}$  \\  & & &  Red  & 5.9$\times$10$^{4}$ & 9.7 & 6.2$\times$10$^{-2}$ & 2.8$\times$10$^{-5}$ & 1.7$\times$10$^{-3}$ & 2.3$\times$10$^{0}$  \\ 
\hline
\object{NGC1333-IRAS~2A} &  CO 6--5  & 70 &  Blue  & 1.4$\times$10$^{4}$ & 2.9 & 7.9$\times$10$^{-3}$ & 2.0$\times$10$^{-5}$ & 3.4$\times$10$^{-4}$ & 1.7$\times$10$^{-1}$  \\  & & &  Red  & 1.4$\times$10$^{4}$ & 3.9 & 2.2$\times$10$^{-2}$ & 4.0$\times$10$^{-5}$ & 2.0$\times$10$^{-3}$ & 1.7$\times$10$^{0}$  \\ 
  &  CO 3--2  & 70 &  Blue  & 2.4$\times$10$^{4}$ & 4.8 & 8.5$\times$10$^{-2}$ & 1.3$\times$10$^{-4}$ & 2.6$\times$10$^{-3}$ & 1.2$\times$10$^{0}$  \\  & & &  Red  & 2.4$\times$10$^{4}$ & 6.4 & 6.9$\times$10$^{-2}$ & 7.7$\times$10$^{-5}$ & 4.8$\times$10$^{-3}$ & 5.4$\times$10$^{0}$  \\ 
\hline
\object{NGC1333-IRAS~4A} &  CO 6--5  & 50 &  Blue  & 2.5$\times$10$^{4}$ & 5.3 & 8.1$\times$10$^{-3}$ & 6.7$\times$10$^{-6}$ & 1.5$\times$10$^{-4}$ & 5.6$\times$10$^{-2}$  \\  & & &  Red  & 3.5$\times$10$^{4}$ & 8.4 & 1.9$\times$10$^{-2}$ & 9.9$\times$10$^{-6}$ & 5.4$\times$10$^{-4}$ & 5.4$\times$10$^{-1}$  \\ 
  &  CO 3--2  & 50 &  Blue  & 2.8$\times$10$^{4}$ & 6.1 & 2.1$\times$10$^{-2}$ & 1.5$\times$10$^{-5}$ & 3.5$\times$10$^{-4}$ & 1.6$\times$10$^{-1}$  \\  & & &  Red  & 3.9$\times$10$^{4}$ & 9.3 & 2.5$\times$10$^{-2}$ & 1.2$\times$10$^{-5}$ & 1.7$\times$10$^{-3}$ & 1.8$\times$10$^{0}$  \\ 
\hline
\object{NGC1333-IRAS~4B} &  CO 6--5  & 10 &  Blue  & 2.4$\times$10$^{3}$ & 0.6 & 8.2$\times$10$^{-4}$ & 1.6$\times$10$^{-6}$ & 3.2$\times$10$^{-5}$ & 7.4$\times$10$^{-3}$  \\  & & &  Red  & 1.2$\times$10$^{3}$ & 0.4 & 7.3$\times$10$^{-4}$ & 2.2$\times$10$^{-6}$ & 1.6$\times$10$^{-4}$ & 1.8$\times$10$^{-1}$  \\ 
  &  CO 3--2  & 10 &  Blue  & 3.5$\times$10$^{3}$ & 0.8 & 8.3$\times$10$^{-4}$ & 1.3$\times$10$^{-6}$ & 2.9$\times$10$^{-5}$ & 1.1$\times$10$^{-2}$  \\  & & &  Red  & 2.4$\times$10$^{3}$ & 0.9 & 2.7$\times$10$^{-3}$ & 3.6$\times$10$^{-6}$ & 1.9$\times$10$^{-4}$ & 1.8$\times$10$^{-1}$  \\ 
\hline
\object{L1527} &  CO 6--5  & 70 &  Blue  & 1.5$\times$10$^{4}$ & 9.1 & 2.3$\times$10$^{-3}$ & 1.8$\times$10$^{-6}$ & 3.1$\times$10$^{-5}$ & 9.9$\times$10$^{-3}$  \\  & & &  Red  & 1.1$\times$10$^{4}$ & 6.5 & 2.5$\times$10$^{-3}$ & 2.7$\times$10$^{-6}$ & 1.1$\times$10$^{-4}$ & 7.5$\times$10$^{-2}$  \\ 
  &  CO 3--2  & 70 &  Blue  & 3.2$\times$10$^{4}$ & 20.6 & 1.0$\times$10$^{-2}$ & 3.5$\times$10$^{-6}$ & 6.1$\times$10$^{-5}$ & 2.0$\times$10$^{-2}$  \\  & & &  Red  & 1.1$\times$10$^{4}$ & 6.5 & 9.0$\times$10$^{-3}$ & 9.8$\times$10$^{-6}$ & 3.8$\times$10$^{-4}$ & 2.6$\times$10$^{-1}$  \\ 
\hline
\object{Ced110-IRS4} &  CO 6--5  & 30 &  Blue  & 3.8$\times$10$^{3}$ & 4.2 & 2.7$\times$10$^{-4}$ & 1.8$\times$10$^{-7}$ & 2.1$\times$10$^{-6}$ & 4.6$\times$10$^{-4}$  \\  & & &  Red  & 3.8$\times$10$^{3}$ & 4.7 & 2.5$\times$10$^{-4}$ & 1.5$\times$10$^{-7}$ & 4.1$\times$10$^{-6}$ & 2.0$\times$10$^{-3}$  \\ 
\hline
\object{BHR71} &  CO 6--5  & 70 &  Blue  & 4.4$\times$10$^{4}$ & 13.4 & 3.4$\times$10$^{-2}$ & 1.8$\times$10$^{-5}$ & 7.7$\times$10$^{-4}$ & 6.2$\times$10$^{-1}$  \\  & & &  Red  & 4.0$\times$10$^{4}$ & 8.5 & 6.9$\times$10$^{-2}$ & 5.8$\times$10$^{-5}$ & 7.7$\times$10$^{-4}$ & 3.3$\times$10$^{-1}$  \\ 
\hline
\object{IRAS~15398} &  CO 6--5  & 30 &  Blue  & 2.6$\times$10$^{3}$ & 1.4 & 3.4$\times$10$^{-4}$ & 1.7$\times$10$^{-6}$ & 6.4$\times$10$^{-6}$ & 1.4$\times$10$^{-3}$  \\  & & &  Red  & 2.6$\times$10$^{3}$ & 1.2 & 2.7$\times$10$^{-4}$ & 1.5$\times$10$^{-6}$ & 2.6$\times$10$^{-5}$ & 2.0$\times$10$^{-2}$  \\ 
  &  CO 3--2  & 30 &  Blue  & 3.2$\times$10$^{3}$ & 1.8 & 4.4$\times$10$^{-4}$ & 1.8$\times$10$^{-6}$ & 9.2$\times$10$^{-6}$ & 2.5$\times$10$^{-3}$  \\  & & &  Red  & 2.0$\times$10$^{3}$ & 0.9 & 2.5$\times$10$^{-4}$ & 1.9$\times$10$^{-6}$ & 2.8$\times$10$^{-5}$ & 2.0$\times$10$^{-2}$  \\ 
\hline
\object{L483MM} &  CO 6--5  & 70 &  Blue  & 1.2$\times$10$^{4}$ & 5.2 & 4.2$\times$10$^{-3}$ & 5.7$\times$10$^{-6}$ & 6.7$\times$10$^{-5}$ & 1.6$\times$10$^{-2}$  \\  & & &  Red  & 1.0$\times$10$^{4}$ & 4.4 & 3.4$\times$10$^{-3}$ & 5.4$\times$10$^{-6}$ & 2.1$\times$10$^{-4}$ & 1.5$\times$10$^{-1}$  \\ 
  &  CO 3--2  & 70 &  Blue  & 1.4$\times$10$^{4}$ & 6.2 & 7.0$\times$10$^{-3}$ & 8.0$\times$10$^{-6}$ & 7.7$\times$10$^{-5}$ & 1.6$\times$10$^{-2}$  \\  & & &  Red  & 1.0$\times$10$^{4}$ & 4.4 & 8.5$\times$10$^{-3}$ & 1.4$\times$10$^{-5}$ & 5.1$\times$10$^{-4}$ & 3.4$\times$10$^{-1}$  \\ 
\hline
\object{Ser-SMM1} &  CO 6--5  & 50 &  Blue  & 3.4$\times$10$^{4}$ & 8.4 & 1.6$\times$10$^{-2}$ & 8.2$\times$10$^{-6}$ & 1.5$\times$10$^{-4}$ & 5.7$\times$10$^{-2}$  \\  & & &  Red  & 1.8$\times$10$^{4}$ & 3.9 & 1.2$\times$10$^{-2}$ & 1.4$\times$10$^{-5}$ & 8.7$\times$10$^{-4}$ & 9.8$\times$10$^{-1}$  \\ 
  &  CO 3--2  & 50 &  Blue  & 3.4$\times$10$^{4}$ & 8.6 & 6.4$\times$10$^{-2}$ & 3.3$\times$10$^{-5}$ & 6.7$\times$10$^{-4}$ & 2.8$\times$10$^{-1}$  \\  & & &  Red  & 1.8$\times$10$^{4}$ & 3.9 & 3.3$\times$10$^{-2}$ & 3.7$\times$10$^{-5}$ & 2.3$\times$10$^{-3}$ & 2.7$\times$10$^{0}$  \\ 
\hline
\object{Ser-SMM4} &  CO 6--5  & 30 &  Blue  & 1.8$\times$10$^{4}$ & 4.6 & 2.4$\times$10$^{-2}$ & 1.5$\times$10$^{-5}$ & 2.5$\times$10$^{-4}$ & 9.3$\times$10$^{-2}$  \\  & & &  Red  & 1.8$\times$10$^{4}$ & 7.3 & 2.8$\times$10$^{-2}$ & 1.1$\times$10$^{-5}$ & 5.9$\times$10$^{-4}$ & 5.8$\times$10$^{-1}$  \\ 
  &  CO 3--2  & 30 &  Blue  & 1.8$\times$10$^{4}$ & 4.6 & 1.6$\times$10$^{-1}$ & 9.9$\times$10$^{-5}$ & 2.0$\times$10$^{-3}$ & 8.3$\times$10$^{-1}$  \\  & & &  Red  & 1.8$\times$10$^{4}$ & 7.6 & 1.3$\times$10$^{-1}$ & 4.7$\times$10$^{-5}$ & 2.8$\times$10$^{-3}$ & 2.9$\times$10$^{0}$  \\ 
\hline
\object{Ser-SMM3} &  CO 6--5  & 50 &  Blue  & 4.6$\times$10$^{3}$ & 1.0 & 6.9$\times$10$^{-3}$ & 3.1$\times$10$^{-5}$ & 6.0$\times$10$^{-4}$ & 2.6$\times$10$^{-1}$  \\  & & &  Red  & 4.6$\times$10$^{3}$ & 1.6 & 3.0$\times$10$^{-3}$ & 8.1$\times$10$^{-6}$ & 4.9$\times$10$^{-4}$ & 5.4$\times$10$^{-1}$  \\ 
  &  CO 3--2  & 50 &  Blue  & 4.6$\times$10$^{3}$ & 1.0 & 2.7$\times$10$^{-2}$ & 1.2$\times$10$^{-4}$ & 2.4$\times$10$^{-3}$ & 1.0$\times$10$^{0}$  \\  & & &  Red  & 4.6$\times$10$^{3}$ & 1.6 & 1.1$\times$10$^{-2}$ & 3.0$\times$10$^{-5}$ & 1.8$\times$10$^{-3}$ & 1.9$\times$10$^{0}$  \\ 
\hline
\object{B335} &  CO 6--5  & 70 &  Blue  & 6.2$\times$10$^{3}$ & 3.4 & 4.7$\times$10$^{-4}$ & 9.9$\times$10$^{-7}$ & 2.3$\times$10$^{-5}$ & 9.4$\times$10$^{-3}$  \\  & & &  Red  & 8.8$\times$10$^{3}$ & 4.8 & 1.3$\times$10$^{-3}$ & 1.9$\times$10$^{-6}$ & 9.0$\times$10$^{-5}$ & 7.6$\times$10$^{-2}$  \\ 
  &  CO 3--2  & 70 &  Blue  & 1.0$\times$10$^{4}$ & 5.3 & 3.7$\times$10$^{-3}$ & 4.9$\times$10$^{-6}$ & 1.3$\times$10$^{-4}$ & 6.2$\times$10$^{-2}$  \\  & & &  Red  & 7.5$\times$10$^{3}$ & 4.1 & 5.4$\times$10$^{-3}$ & 9.3$\times$10$^{-6}$ & 4.7$\times$10$^{-4}$ & 4.3$\times$10$^{-1}$  \\ 
\hline
\object{L723MM} &  CO 6--5  & 50 &  Blue  & 1.2$\times$10$^{4}$ & 4.1 & 6.0$\times$10$^{-3}$ & 6.6$\times$10$^{-6}$ & 1.8$\times$10$^{-4}$ & 1.0$\times$10$^{-1}$  \\  & & &  Red  & 1.2$\times$10$^{4}$ & 3.8 & 7.5$\times$10$^{-3}$ & 8.5$\times$10$^{-6}$ & 5.5$\times$10$^{-4}$ & 6.3$\times$10$^{-1}$  \\ 
  &  CO 3--2  & 50 &  Blue  & 1.8$\times$10$^{4}$ & 6.0 & 3.0$\times$10$^{-2}$ & 2.2$\times$10$^{-5}$ & 6.5$\times$10$^{-4}$ & 3.7$\times$10$^{-1}$  \\  & & &  Red  & 1.8$\times$10$^{4}$ & 5.8 & 4.2$\times$10$^{-2}$ & 3.2$\times$10$^{-5}$ & 2.2$\times$10$^{-3}$ & 2.8$\times$10$^{0}$  \\ 
\hline
\object{L1157} &  CO 3--2  & 70 &  Blue  & 4.4$\times$10$^{4}$ & 16.8 & 1.2$\times$10$^{-1}$ & 4.9$\times$10$^{-5}$ & 5.0$\times$10$^{-4}$ & 1.9$\times$10$^{-1}$  \\  & & &  Red  & 5.2$\times$10$^{4}$ & 14.1 & 1.5$\times$10$^{-1}$ & 7.3$\times$10$^{-5}$ & 3.2$\times$10$^{-3}$ & 3.1$\times$10$^{0}$  \\ 
\hline
\hline
\end{tabular}\\
\end{center}
\tablefoot{
\tablefoottext{a}{Outflow extents and outflow masses are not corrected for inclination.} 
\tablefoottext{b}{Dynamical timescale. }
\tablefoottext{c}{Constant temperature of 75~K is assumed for both 
CO~6--5 and CO~3--2 calculations. }\tablefoottext{d}{Corrected for inclination 
as explained in Sect. \ref{5:sec:outflowforces}.}
\tablefoottext{e}{Mass outflow rate}
\tablefoottext{f}{Outflow force}
\tablefoottext{g}{Kinetic luminosity.}
}
\label{tbl:outflowparam_C0}
\end{table*}

\begin{table*}[!th]
\caption{Outflow properties of the red and blue outflow lobes of Class~I sources.}
\small  
\begin{center}
\begin{tabular}{@{}l c c c c c c c c c c c}
\hline \hline
Source & Trans. & Inclination & Lobe & $R_{\rm CO}$\tablefootmark{a} & $t_{\rm{dyn}}$\tablefootmark{a,b} & $M_{\rm outflow}$\tablefootmark{a,c} & $\dot{M}$\tablefootmark{d,e} & $F_{\rm CO}$\tablefootmark{d,f} & $L_{\rm{kin}}$\tablefootmark{d,g}  \\ 
& & [$\degr$] & & [AU] & [10$^3$ yr] & [M$_\odot$] & [M$_\odot$ yr$^{-1}$]  &[M$_\odot$ yr$^{-1}$km s$^{-1}$] & [L$_\odot$]  \\ %
\hline
\object{L1489} &  CO 6--5  & 50 &  Blue  & 3.5$\times$10$^{3}$ & 1.2 & 4.0$\times$10$^{-5}$ & 1.5$\times$10$^{-7}$ & 1.8$\times$10$^{-6}$ & 3.2$\times$10$^{-4}$  \\  & & &  Red  & 2.1$\times$10$^{3}$ & 1.3 & 1.3$\times$10$^{-4}$ & 4.6$\times$10$^{-7}$ & 2.3$\times$10$^{-5}$ & 2.0$\times$10$^{-2}$  \\ 
  &  CO 3--2  & 50 &  Blue  & 3.5$\times$10$^{3}$ & 1.2 & 6.9$\times$10$^{-4}$ & 2.5$\times$10$^{-6}$ & 3.9$\times$10$^{-5}$ & 1.3$\times$10$^{-2}$  \\  & & &  Red  & 2.1$\times$10$^{3}$ & 1.3 & 7.1$\times$10$^{-4}$ & 2.5$\times$10$^{-6}$ & 1.2$\times$10$^{-4}$ & 1.1$\times$10$^{-1}$  \\ 
\hline
\object{L1551-IRS5}  &  CO~3--2  &  70  &  Blue  &  1.7$\times$10$^{4}$  &  8.2  &  7.4$\times$10$^{-3}$  &  6.4$\times$10$^{-6}$  &  9.3$\times$10$^{-5}$  &  2.7$\times$10$^{-2}$  \\ 	 & & & Red & 1.7$\times$10$^{4}$ & 6.7 & 9.6$\times$10$^{-3}$ & 1.0$\times$10$^{-5}$ & 4.2$\times$10$^{-4}$ & 3.1$\times$10$^{-1}$ \\   
\hline
\object{TMR1} &  CO 6--5  & 50 &  Blue  & 4.9$\times$10$^{3}$ & 2.9 & 1.2$\times$10$^{-4}$ & 1.8$\times$10$^{-7}$ & 2.0$\times$10$^{-6}$ & 4.9$\times$10$^{-4}$  \\  & & &  Red  & 3.5$\times$10$^{3}$ & 4.5 & 2.4$\times$10$^{-4}$ & 2.3$\times$10$^{-7}$ & 8.0$\times$10$^{-6}$ & 4.8$\times$10$^{-3}$  \\ 
  &  CO 3--2  & 50 &  Blue  & 4.9$\times$10$^{3}$ & 3.0 & 2.6$\times$10$^{-4}$ & 3.8$\times$10$^{-7}$ & 5.1$\times$10$^{-6}$ & 1.4$\times$10$^{-3}$  \\  & & &  Red  & 3.5$\times$10$^{3}$ & 4.5 & 5.8$\times$10$^{-4}$ & 5.7$\times$10$^{-7}$ & 2.0$\times$10$^{-5}$ & 1.3$\times$10$^{-2}$  \\ 
\hline
\object{TMC1A} &  CO 6--5  & 50 &  Blue  & 5.6$\times$10$^{3}$ & 1.4 & 2.3$\times$10$^{-4}$ & 7.2$\times$10$^{-7}$ & 1.3$\times$10$^{-5}$ & 7.6$\times$10$^{-3}$  \\  & & &  Red  & 2.1$\times$10$^{3}$ & 1.8 & 4.0$\times$10$^{-6}$ & 9.5$\times$10$^{-9}$ & 3.7$\times$10$^{-7}$ & 2.4$\times$10$^{-4}$  \\ 
  &  CO 3--2  & 50 &  Blue  & 5.6$\times$10$^{3}$ & 1.6 & 2.8$\times$10$^{-3}$ & 7.8$\times$10$^{-6}$ & 1.1$\times$10$^{-4}$ & 3.7$\times$10$^{-2}$  \\  & & &  Red  & 1.7$\times$10$^{3}$ & 1.5 & 1.4$\times$10$^{-4}$ & 4.1$\times$10$^{-7}$ & 1.8$\times$10$^{-5}$ & 1.4$\times$10$^{-2}$  \\ 
\hline
\object{TMC1} &  CO 6--5  & 50 &  Blue  & 3.5$\times$10$^{3}$ & 1.2 & 1.3$\times$10$^{-4}$ & 4.7$\times$10$^{-7}$ & 3.2$\times$10$^{-6}$ & 4.0$\times$10$^{-4}$  \\  & & &  Red  & 4.9$\times$10$^{3}$ & 1.6 & 3.8$\times$10$^{-4}$ & 1.1$\times$10$^{-6}$ & 5.0$\times$10$^{-5}$ & 4.4$\times$10$^{-2}$  \\ 
  &  CO 3--2  & 50 &  Blue  & 3.5$\times$10$^{3}$ & 1.2 & 5.6$\times$10$^{-4}$ & 2.0$\times$10$^{-6}$ & 2.7$\times$10$^{-5}$ & 7.9$\times$10$^{-3}$  \\  & & &  Red  & 2.1$\times$10$^{3}$ & 0.7 & 1.4$\times$10$^{-3}$ & 8.9$\times$10$^{-6}$ & 4.2$\times$10$^{-4}$ & 3.7$\times$10$^{-1}$  \\ 
\hline
\object{HH46-IRS} &  CO 6--5  & 50 &  Blue  & 1.1$\times$10$^{4}$ & 9.9 & 2.6$\times$10$^{-3}$ & 1.2$\times$10$^{-6}$ & 1.2$\times$10$^{-5}$ & 2.9$\times$10$^{-3}$  \\  & & &  Red  & 2.5$\times$10$^{4}$ & 7.9 & 3.2$\times$10$^{-2}$ & 1.8$\times$10$^{-5}$ & 7.7$\times$10$^{-4}$ & 6.5$\times$10$^{-1}$  \\ 
  &  CO 3--2  & 50 &  Blue  & 1.6$\times$10$^{4}$ & 13.6 & 2.2$\times$10$^{-2}$ & 7.2$\times$10$^{-6}$ & 2.5$\times$10$^{-4}$ & 1.7$\times$10$^{-1}$  \\  & & &  Red  & 2.5$\times$10$^{4}$ & 7.9 & 2.2$\times$10$^{-2}$ & 1.2$\times$10$^{-5}$ & 8.1$\times$10$^{-4}$ & 9.3$\times$10$^{-1}$  \\ 
\hline
\object{DK~Cha} &  CO 6--5  & 10 &  Blue  & 1.8$\times$10$^{3}$ & 1.6 & 1.8$\times$10$^{-4}$ & 1.3$\times$10$^{-7}$ & 6.6$\times$10$^{-7}$ & 7.3$\times$10$^{-5}$  \\  & & &  Red  & 1.8$\times$10$^{3}$ & 0.9 & 1.1$\times$10$^{-4}$ & 1.4$\times$10$^{-7}$ & 2.5$\times$10$^{-6}$ & 4.4$\times$10$^{-4}$  \\ 
\hline
\object{GSS30-IRS1} &  CO 6--5  & 30 &  Blue  & 1.5$\times$10$^{4}$ & 5.5 & 1.5$\times$10$^{-2}$ & 7.9$\times$10$^{-6}$ & 5.1$\times$10$^{-5}$ & 1.1$\times$10$^{-2}$  \\  & & &  Red  & 1.5$\times$10$^{4}$ & 4.9 & 8.9$\times$10$^{-3}$ & 5.1$\times$10$^{-6}$ & 2.0$\times$10$^{-4}$ & 1.6$\times$10$^{-1}$  \\ 
  &  CO 3--2  & 30 &  Blue  & 1.5$\times$10$^{4}$ & 5.5 & 2.1$\times$10$^{-2}$ & 1.1$\times$10$^{-5}$ & 6.0$\times$10$^{-5}$ & 8.8$\times$10$^{-3}$  \\  & & &  Red  & 1.5$\times$10$^{4}$ & 4.9 & 2.4$\times$10$^{-2}$ & 1.4$\times$10$^{-5}$ & 4.6$\times$10$^{-4}$ & 3.0$\times$10$^{-1}$  \\ 
\hline 
\object{Elias~29} &  CO 6--5  & 30 &  Blue  & 7.5$\times$10$^{3}$ & 3.1 & 6.4$\times$10$^{-4}$ & 5.7$\times$10$^{-7}$ & 4.4$\times$10$^{-6}$ & 1.2$\times$10$^{-3}$  \\  & & &  Red  & 5.0$\times$10$^{3}$ & 1.7 & 6.3$\times$10$^{-4}$ & 1.0$\times$10$^{-6}$ & 3.9$\times$10$^{-5}$ & 2.7$\times$10$^{-2}$  \\ 
  &  CO 3--2  & 30 &  Blue  & 7.5$\times$10$^{3}$ & 3.6 & 1.4$\times$10$^{-3}$ & 1.1$\times$10$^{-6}$ & 6.6$\times$10$^{-6}$ & 1.0$\times$10$^{-3}$  \\  & & &  Red  & 7.5$\times$10$^{3}$ & 3.3 & 1.8$\times$10$^{-3}$ & 1.5$\times$10$^{-6}$ & 5.7$\times$10$^{-5}$ & 3.9$\times$10$^{-2}$  \\ 
\hline
\object{Oph-IRS63} &  CO 6--5  & 50 &  Blue  & 3.8$\times$10$^{3}$ & 1.6 & 1.0$\times$10$^{-4}$ & 2.8$\times$10$^{-7}$ & 4.3$\times$10$^{-6}$ & 2.1$\times$10$^{-3}$  \\  & & &  Red  & 3.8$\times$10$^{3}$ & 4.2 & 8.6$\times$10$^{-5}$ & 9.0$\times$10$^{-8}$ & 2.1$\times$10$^{-6}$ & 8.3$\times$10$^{-4}$  \\ 
  &  CO 3--2  & 50 &  Blue  & 8.8$\times$10$^{3}$ & 3.7 & 7.0$\times$10$^{-4}$ & 8.4$\times$10$^{-7}$ & 5.6$\times$10$^{-6}$ & 1.5$\times$10$^{-3}$  \\  & & &  Red  & 5.0$\times$10$^{3}$ & 7.4 & 5.0$\times$10$^{-4}$ & 3.0$\times$10$^{-7}$ & 5.2$\times$10$^{-6}$ & 2.2$\times$10$^{-3}$  \\ 
\hline
\object{RNO91} &  CO 6--5  & 50 &  Blue  & 3.1$\times$10$^{3}$ & 1.0 & 2.5$\times$10$^{-4}$ & 1.1$\times$10$^{-6}$ & 2.9$\times$10$^{-5}$ & 2.1$\times$10$^{-2}$  \\  & & &  Red  & 1.9$\times$10$^{3}$ & 2.5 & 7.3$\times$10$^{-5}$ & 1.3$\times$10$^{-7}$ & 1.1$\times$10$^{-6}$ & 1.8$\times$10$^{-4}$  \\ 
  &  CO 3--2  & 50 &  Blue  & 6.2$\times$10$^{3}$ & 2.0 & 3.5$\times$10$^{-3}$ & 7.7$\times$10$^{-6}$ & 1.0$\times$10$^{-4}$ & 4.1$\times$10$^{-2}$  \\  & & &  Red  & 1.9$\times$10$^{3}$ & 2.5 & 2.3$\times$10$^{-4}$ & 4.0$\times$10$^{-7}$ & 2.3$\times$10$^{-6}$ & 2.6$\times$10$^{-4}$  \\ 
\hline
\hline
\end{tabular}\\
\end{center}
\tablefoot{
\tablefoottext{a}{Outflow extents and outflow masses are not corrected for inclination.} 
\tablefoottext{b}{Dynamical timescale. }
\tablefoottext{c}{Constant temperature of 75~K is assumed for both 
CO~6--5 and CO~3--2 calculations.}\tablefoottext{d}{Corrected for 
inclination as explained in Sect. \ref{5:sec:outflowforces}.}
\tablefoottext{e}{Mass outflow rate}
\tablefoottext{f}{Outflow force} 
\tablefoottext{g}{Kinetic luminosity.}
}
\label{tbl:outflowparam_CI}
\end{table*}

\subsection{Outflow parameters}
\label{5:sec:outflowforces}

In the following, different outflow parameters, including mass, force
and luminosity, are measured. These parameters have previously been
determined from lower-$J$ lines for several young stellar objects
\citep[e.g.,][]{CabritBertout92, Bontemps96, Hogerheijde98,
Hatchell07,Curtis10_2outflows, vanderMarel13, Dunham14} and more recently 
from CO 6--5 by \citet{vanKempen09champ} and \citet{Yildiz12} for a small 
subset of the sources presented here. All results are listed in Tables
\ref{tbl:outflowparam_C0} and \ref{tbl:outflowparam_CI}.
Uncertainties in the methods are discussed extensively in
\citet{vanderMarel13}.

\subsubsection{Outflow mass}

One of the most basic outflow parameters is the mass. The inferred
mass depends on three assumptions: the line opacity, the
distribution of level populations, and the CO abundance with respect
to H$_2$. In the following, we assume that the line wings are optically
thin, as has been demonstrated observationally for CO 6--5 for a few
sources with massive outflows \citep[e.g.,
NGC~1333-IRAS~4A,][]{Yildiz12}. CO 3--2 emission is also assumed
optically thin in the following, although that assumption may not be fully 
valid (see discussion below). The level populations are assumed to
follow a Boltzmann distribution with a single temperature, \Tex.
Finally, the abundance ratio is taken as
[H$_{2}$/$^{12}$CO]=1.2$\times$10$^{4}$, as in \citet{Yildiz12}.

The upper level column density per statistical weight in a single pixel 
(4\farcs5$\times$4\farcs5 for \co\ 6--5, 7\farcs5$\times$7\farcs5 for \co\ 3--2) is calculated as
\begin{equation}
\frac{N_{\rm u}}{g_{\rm u}} = \frac{\beta \nu ^{2} \int T_{\rm mb} {\rm d}V}{A_{\rm ul}\, g_{\rm u}}\ .
\end{equation}
The constant $\beta$ is $8\pi
k/hc^{3}$=1937~cm$^{-2}$ (GHz$^2$ K km)$^{-1}$. The remaining parameters are for the
specific transition, where $\nu$ is the frequency, $A_{\rm
  ul}$ is the Einstein $A$ coefficient and $g_{\rm
  u}$=2$J$+1.

The total CO column density in a pixel, $N_{\rm total}$, is
\begin{equation}
N_{\rm total} = \frac{N_{\rm u}}{g_{\rm u}} Q(T)  e^{E_{\rm u}/kT_{\rm ex}}\ ;
\end{equation}
$Q(T)$ is the partition function corresponding to a specific
excitation temperature, $T_{\rm ex}$, which is assumed to be 75~K for
both CO~3--2 and CO~6--5 observations
\citep{vanKempen09champ,Yildiz12, Yildiz13hifi}.  Changing $T_{\rm
  ex}$ by $\pm 30$~K changes the inferred column densities by only
10--20\%.
  
The mass is calculated as
\begin{equation}
M_{\rm outflow} = \mu_{\rm H_{2}} \,m_{\rm H}\, A \, \left[ \frac{\rm H_2}{^{12}{\rm CO}} \right] \sum_j N_{{\rm total},j}
\end{equation}
where the factor $\mu_{\mathrm{H_{2}}}$=2.8 includes the contribution
of helium \citep{Kauffmann08} and $m_{\rm H}$ is the mass of the
hydrogen atom.  $A$ is the surface area of one pixel $j$.  The sum is
over all outflow pixels.

The mass may be underestimated if the \twco\ line emission is optically thick. \thco\ data exist toward most outflows (see above) but the $S/N$ of these data is typically too low to properly measure the opacity in the line wings, except for at the source position where the signal naturally is the strongest. The \thco\ line wings do not extend beyond the inner velocity limits. NGC1333-IRAS~4A is one of the few sources where line wings are detected in \thco\ at the outflow positions \citep[Fig. 11 in ][]{Yildiz12}, and it is clear that at the velocity ranges considered here, the line emission is optically thin ($\tau<1$); the same is true for the outflows studied by \citet{vanderMarel13} in CO 3--2 emission in Ophiuchus (their Fig. 4), where deep pointed observations of \thco\ 3--2 were required to measure the opacity. That study concluded that the opacity does not play a significant role when determining the outflow parameters. Similarly, \citet{Dunham14} conclude that CO 3--2 may be optically thick at velocities less than 2 \kms\ offset from the source velocity, velocities which are  excluded from our analysis because of the risk of cloud contamination. Potentially more problematic is the missing mass at low velocities. The missing mass is moving close to the systemic velocity and it is not possible to disentangle this mass from the surrounding cloud material, an effect which may introduce a typical uncertainty of a factor of 2--3 \citep{DownesCabrit07}. However, the correction factors derived by the same authors and implemented here account for that missing mass. \twco\ 6--5 emission will be less affected by this confusion than the \twco\ 3--2 emission, simply because of the different excitation conditions required.

\subsubsection{Outflow force}

One of the most important outflow parameters is the outflow force, \FCO. 
The best method for computing the outflow force is still debated and the
results suffer from ill-constrained observational parameters, such as
inclination, $i$. \citet{vanderMarel13} compare seven different
methods proposed in the literature to calculate outflow forces.  The
``separation method'' (see below) in their paper is found to be the
preferred method, which is less affected by the observational
biases. The method can also be applied to low spatial resolution
observations or incomplete maps. Uncertainties are estimated to be a
factor of 2--3.

In the following, the outflow force is calculated separately for the
blue-- and red--shifted lobes, only including emission above the 3$\sigma$ 
level.  The mass is calculated for each channel separately and multiplied 
by the central velocity of that particular channel. The integral runs over 
velocities from \Vin\ to \Vout. They are then summed and the sum is over all 
pixels $j$ in the map with outflow emission. The outflow force is calculated 
for the red- and blue-shifted outflow lobes separately. This method is formulated as:
\begin{equation}
F_{\rm CO}=c_{i} \frac{ V_{\rm max} \sum_{j}^{} \left[ \int M(V')V' {\rm d}V' \right]_{j}}{R_{\rm CO}},
\label{5:eq:fco}
\end{equation}
where $c_{i}$ is the inclination correction (Table \ref{tbl:inclcorr}), and $R_{\rm CO}$ 
is the projected size of the red- or blue-shifted outflow lobe.  
The outflow force is computed
separately from the CO~3--2 and 6--5 maps of the same source (see
Tables~\ref{tbl:outflowparam_C0} and \ref{tbl:outflowparam_CI}).

The difference in outflow force between the red and blue outflow lobes
ranges from $\sim$1 up to a factor of 10. For sources with a low
outflow force such as Oph IRS63 ($<$10$^{-5}$ M$_\odot$ yr$^{-1}$ km
s$^{-1}$) this is a result of differences in the inferred outflow mass
per lobe, which, in these specific cases, is primarily a result of low
\SN. In these cases, the overall uncertainty on the outflow force is
high, up to a factor of 10. In other cases, such as HH\,46 as
mentioned above, there is a real asymmetry between the different lobes
which is caused by a difference in the surrounding environment. In the
following, only the sum of the outflow forces of both lobes as
measured from each outflow lobe will be used.

\begin{figure}[!t]
    \centering
    \includegraphics[width=\columnwidth]{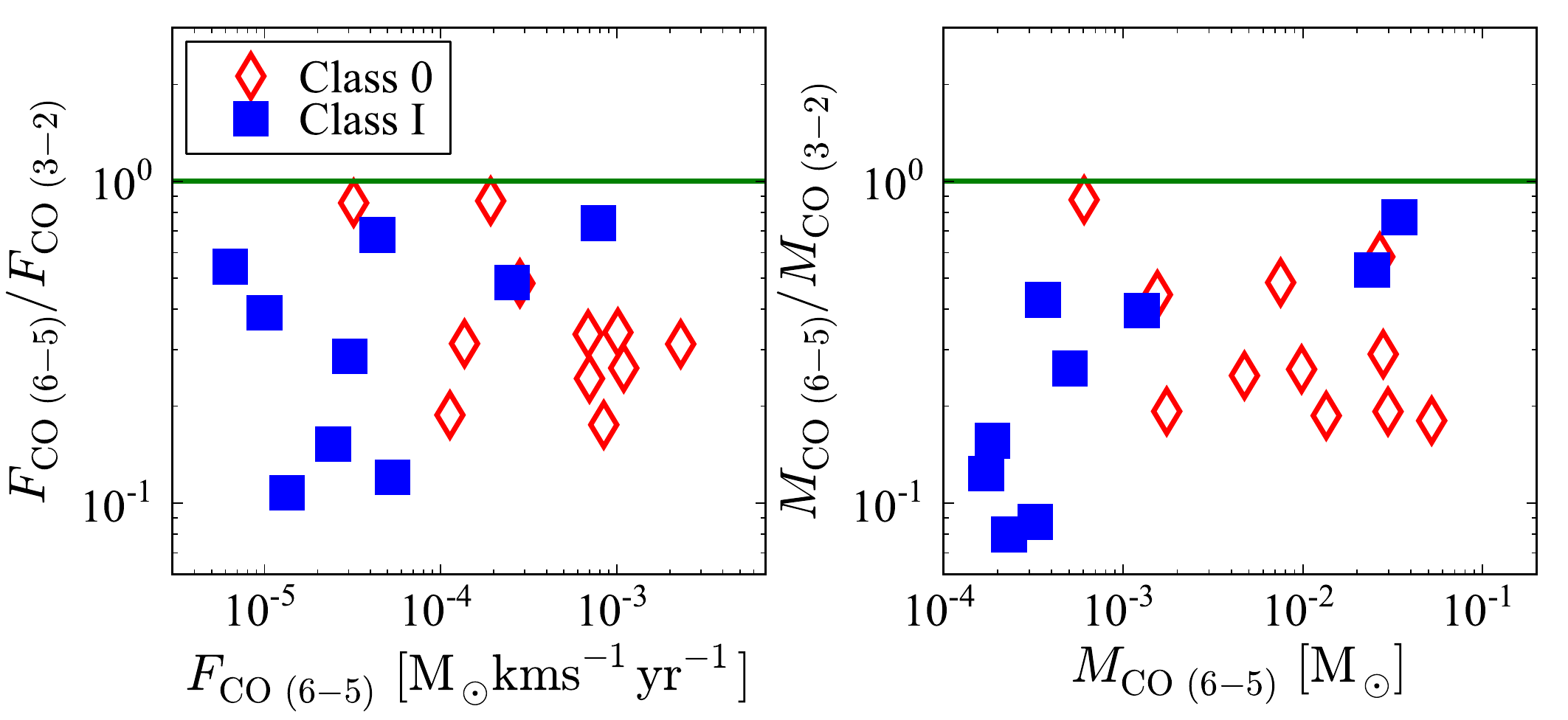}
    \caption{\small Outflow forces ({\it left}) and outflow masses ({\it right}), 
    calculated from CO 6--5 and 3--2 emission are compared for Class 0 and I sources. 
    Green lines are for a ratio of 1.}
    \label{fig:ratioFCO3265}
\end{figure}

\begin{figure}[!t]
    \centering
    \includegraphics[scale=0.4]{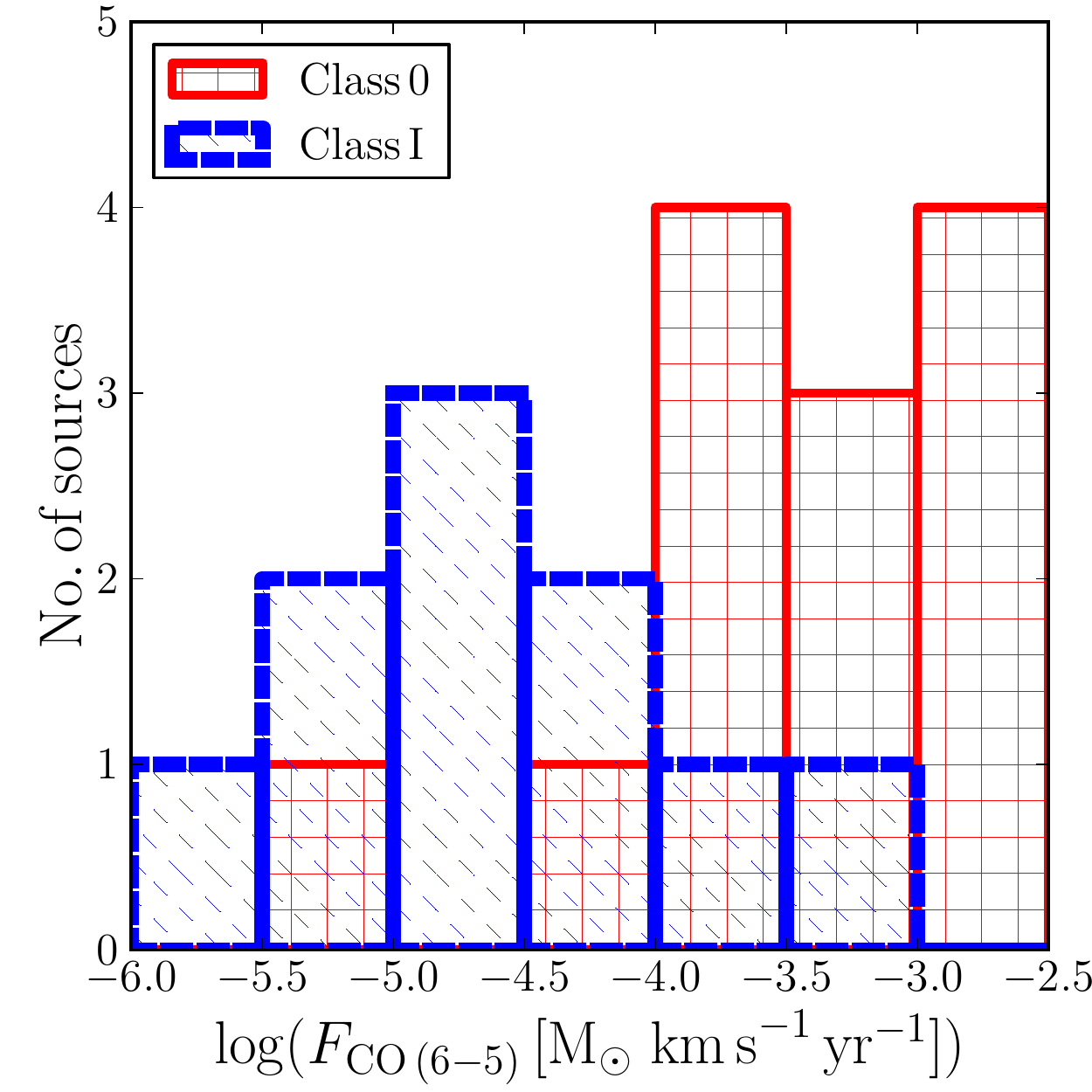}
    \caption{\small Histograms of calculated total outflow force \FCO\ are shown 
    for Class~0 (red) and Class~I (blue) sources.}
    \label{fig:histFCO}
\end{figure}

Figure~\ref{fig:ratioFCO3265} shows how the outflow forces and outflow
masses calculated from CO~3--2 and 6--5 differ. For strong outflows,
there is a factor of a few difference in the two calculations, with
differences up to an order of magnitude for the weaker outflow
sources.  Although the CO 6--5 emission suffers less from opacity
effects and so recovers more emission / mass at lower velocities, this
effect is overwhelmed by the lower \SN\ of the CO 6--5 emission.  The
fact that the masses and outflow forces derived from the 6--5 data are
systematically lower than those from the 3--2 data is likely due to
the same effect \citep{vanderMarel13}. Moreover, if CO~6--5 traces slightly 
warmer gas than CO 3--2 \citep{Yildiz13hifi} then the mass traced by this 
line will be lower than that traced by CO 3--2. Both effects work to 
systematically lower the CO 6--5 masses, which in turn leads to lower 
outflow forces. 

Figure~\ref{fig:histFCO} displays \FCO\ from CO 6--5 for Class~0 and
Class~I sources separately. Generally, Class~0 sources have higher
outflow forces and are thus more powerful than their Class~I
counterparts \citep{Bontemps96}. The Class I source with an exceptionally
high outflow force is HH46.

\subsection{Other outflow parameters}
Other outflow parameters that characterize the outflow activity are the 
dynamical age, \tdyn, mass outflow  rate, $\Mdot_{\rm outflow}$, and 
kinetic luminosity, \Lkin. 

Assuming that the outflow moves with a constant velocity over the extent 
of the outflow, the dynamical age is determined as
\begin{equation}
t_{\rm dyn} = \frac{R_{\rm CO}}{V_{\rm max}}\ .
\end{equation}
This age is a lower limit on the age of the protostar
\citep{Curtis10_2outflows} if the outflowing material is decelerated, e.g., 
through interactions with the ambient surrounding material. On the other hand, 
the outflow may be significantly younger since the velocities of the central jet
that drives the molecular outflow are typically higher than 100 km
s$^{-1}$ and what is observed in these colder low-$J$ CO lines may just be the 
outer shell which is currently undergoing acceleration, not deceleration. See, 
e.g., \citet{DownesCabrit07} for a more complete discussion.
The outflow mass loss rate is computed according to
\begin{equation}
\Mdot_{\rm outflow} = \frac{M_{\rm outflow}}{t_{\rm dyn}}\ .
\end{equation}
The kinetic luminosity is given by
\begin{equation}
L_{\rm kin} = \frac{1}{2} F_{\rm CO} V_{\rm max}\ .
\label{5:eq:lkin }
\end{equation}

Outflow parameters of \FCO, $\Mdot$, and \Lkin\ with inclination
corrections are presented in Tables \ref{tbl:outflowparam_C0} and
\ref{tbl:outflowparam_CI}. However, \Moutflow,  \RCO, \tdyn, and \Vmax\ are not 
corrected for inclination, since they are measured quantities. 
The median values of the results are given in Table~\ref{tbl:smallstat}.

\begin{table}[!t]
\small
\caption{Median values of the outflow parameters.}
\begin{center}
\begin{tabular}{l l l l l l l l l}
\hline
\hline
     & $M_{\rm outflow}$ & $\dot{M}$ & $F_{\rm CO}$ & $L_{\rm{kin}}$ \\
     & [$M_\odot$] & [$M_\odot$\,yr$^{-1}$] & [$M_\odot$\,km$^{s}$\,yr$^{-1}$] & [$L_\odot$] \\
\hline  
CO~6--5 \\
\hline
Class~0 & 9.8$\times$10$^{-3}$ & 1.5$\times$10$^{-5}$ & 6.9$\times$10$^{-4}$ & 6.0$\times$10$^{-1}$  \\ 
Class~I & 3.4$\times$10$^{-4}$ & 8.4$\times$10$^{-6}$ & 2.8$\times$10$^{-5}$ & 1.2$\times$10$^{-1}$  \\ 
Total & 2.2$\times$10$^{-3}$ & 1.0$\times$10$^{-5}$ & 1.4$\times$10$^{-4}$ & 1.9$\times$10$^{-1}$  \\ 
\hline
CO~3--2 \\
\hline
Class~0 & 7.2$\times$10$^{-2}$ & 5.4$\times$10$^{-5}$ & 2.9$\times$10$^{-3}$ & 2.9$\times$10$^{0}$  \\ 
Class~I & 3.1$\times$10$^{-3}$ & 3.0$\times$10$^{-5}$ & 1.4$\times$10$^{-4}$ & 1.4$\times$10$^{0}$  \\ 
Total & 1.7$\times$10$^{-2}$ & 3.3$\times$10$^{-5}$ & 5.2$\times$10$^{-4}$ & 1.9$\times$10$^{0}$  \\ 
\hline
\end{tabular}
\end{center}
\label{tbl:smallstat}
\end{table}

\begin{figure*}[!t]
    \centering
    \includegraphics[scale=0.6]{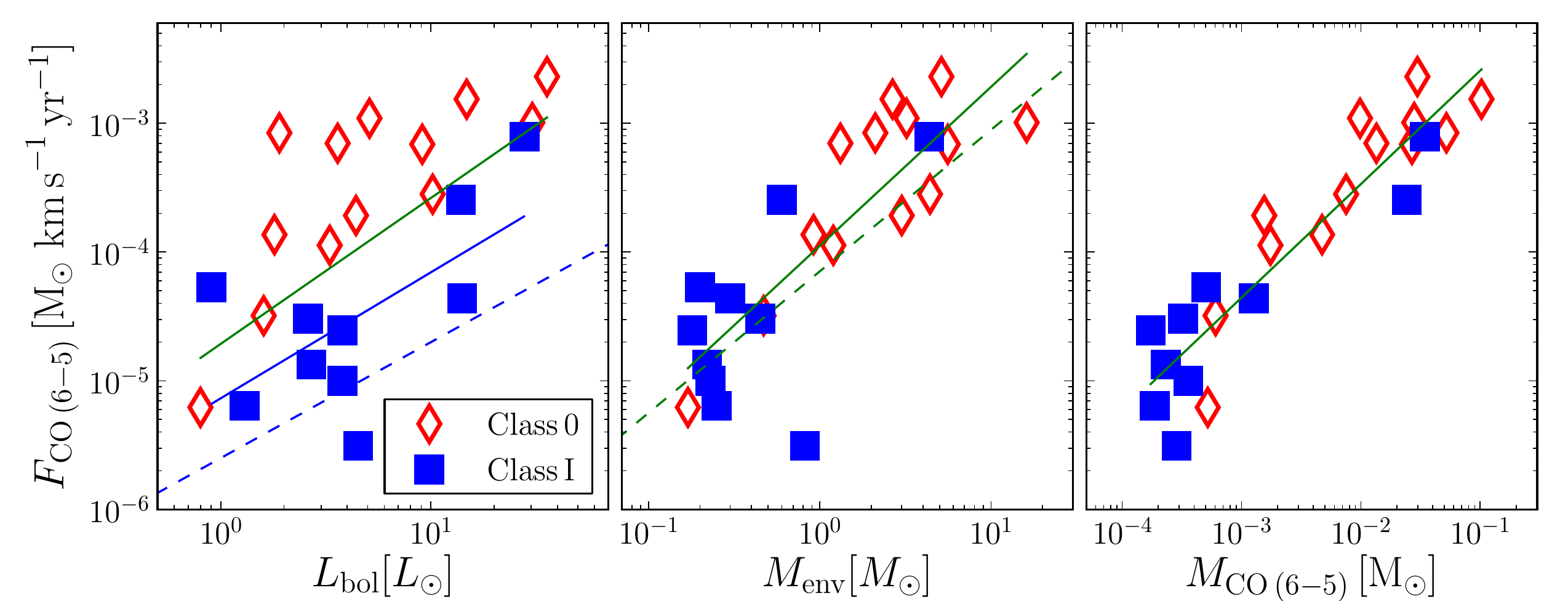}
    \caption{\small Correlations of \FCO\ with \Lbol, \Menv, and \Moutflow, where 
    \FCO\ is determined from the CO 6--5 data. Red and blue symbols indicate Class~0 
    and Class~I sources, respectively. The green solid line is the fit to all values 
    and the blue solid line is the fit to the Class~I sources alone. Blue and green 
    dashed lines are the best fits from \citet{Bontemps96}.}
    \label{fig:corrFCO_ML}
\end{figure*}

\subsection{Correlations}

Most previous studies of the outflow force were done using CO 1--0,
2--1, or 3--2 \citep[e.g.,][]{CabritBertout92, Bontemps96,
Hogerheijde98, Hatchell07, vanKempen09_southc+, Dunham14}. The opacity 
decreases with excitation, as suggested by, e.g., the observations reported 
in \citet{Dunham14}, but without targeted, deep surveys of \thco, it is 
difficult to quantify how much the CO column density is underestimated. 
Furthermore, cloud or envelope emission may contribute to the emission at 
the lowest outflow velocities at which the bulk of the mass is flowing. 
With our CO 6--5 observations, some of the above-mentioned issues can be 
avoided, or their effects can be lessened. Thus, it is important to revisit the
correlations of outflow force with bolometric luminosity and envelope
mass using these new measurements.

In Fig.~\ref{fig:corrFCO_ML}, \FCO\ is plotted against \Lbol, \Menv,
and \Moutflow, where the \FCO\ and \Moutflow\ values are taken from the CO~6--5
data.  The best fit between \FCO\ and \Lbol\ is shown
with the green line corresponding to
\begin{equation}
{\rm log}(F_{\rm CO}) = -(4.71\pm0.02)+(1.13\pm0.37)\,{\rm log}(L_{\rm bol})\ . 
\end{equation}
Outflows from Class~0 and Class~I sources are well-separated; Class~0 sources 
show more powerful outflows compared to Class~I sources of similar luminosity.
The Pearson correlation
coefficients are $r$=0.62, 0.83, and 0.64 for all sources, Class~0, and
Class~I sources, corresponding to confidences 
of 2.9, 2.9, and 1.9$\sigma$, respectively.

The best fit between \FCO\ and \Menv\ is described as
\begin{equation}
{\rm log}(F_{\rm CO}) = -(3.95\pm0.37)+(1.24\pm0.21)\,{\rm log}(M_{\rm env}) 
\end{equation}
and Pearson correlation coefficients are $r$=0.81, 0.82, and 0.56 (3.8, 2.8 and 
1.7$\sigma$) for all sources, Class 0, and Class~I, respectively. Since early
Class~0 sources have higher accretion rates their outflow force is much
higher than for the Class~I sources \citep[see, e.g.,][for a full discussion]{Bontemps96}.
Finally, as expected, a strong correlation is found between \FCO\ and
\Moutflow\ with a Pearson correlation coefficient of $r$=0.92 for all
sources (4.3$\sigma$), not surprisingly since \FCO\ is nearly proportional 
to \Moutflow. The best fit is described as
\begin{equation}
{\rm log}(F_{\rm CO}) = -(1.71\pm0.02)+(0.88\pm0.62)\,{\rm log}(M_{\rm CO})\ .
\end{equation}

Previously, \citet{Bontemps96} surveyed 45 sources using CO~2--1
observations with small-scale maps. In Fig.~\ref{fig:corrFCO_ML}, the
blue and green dashed lines of \FCO\ vs. \Lbol\ and \Menv\ show the
fit results from their Figs. 5 and 6 \citep[][]{Bontemps96}. Since
their number of Class~I sources is higher than Class~0 sources, the
fit was only done for Class~I sources in \FCO\ vs. \Lbol. In
Fig.~\ref{fig:corrFCO_ML}, the blue solid line only shows the fit for
Class~I sources and the correlation is described by,
\begin{equation}
{\rm log}(F_{\rm CO}) = -(5.14\pm0.29)+(0.98\pm0.55)\,{\rm log}(L_{\rm bol})\ . 
\end{equation}
In the \FCO\ vs. \Menv\ plot, the fits are shown as green lines for
the entire sample.  The \citet{Bontemps96} sample is weighted toward
lower luminosities ($<10$ L$_{\rm bol}$), where our \FCO\ measurements
from the CO 6--5 data follow their relation for Class I sources
obtained from 2--1 data, but with a shift to a factor of a few higher
values of $F_{\rm CO}$. However, given the scatter in the results for
low \Lbol\ sources, this difference is hardly significant.

Examining the same outflow parameters measured using the CO 3--2
transition, and their correlation with the same outflow parameters, a
similar picture arises (Fig.~\ref{fig:corrFCO32}). However, for the sources
in our sample, the correlations follow the same trend but they are somewhat
weaker. In particular, the correlation with $L_{\rm bol}$ is at the
$\sim$2.7$\sigma$ level, whereas the correlation with $M_{\rm env}$ is
3.1$\sigma$. Although the measured values of, e.g., $F_{\rm CO}$, fill
out the same parameter space as when the measurements are done with CO
6--5, the scatter is larger. The scatter remains on the order of one
order of magnitude, which is similar to the scatter reported in the
literature \citep[e.g.,][]{Bontemps96}, but because of the limited
source sample (20 sources with $F_{\rm CO}$ measurements) it is
difficult to compare these 3--2 measurements with what is presented in
the literature.

\subsection{Radiative feedback from UV heating}

The quiescent gas is traced by the narrow (FWHM $\lesssim$1~\kms)
\thco\ 6--5 emission, which has been mapped over a 1$'$ region around
the source position. As the contour maps in Figs.~\ref{fig:All13COs1} and
\ref{fig:All13COs2} show, the emission is strongly centrally
concentrated and does not extend beyond the mapped region except for
special cases like NGC~1333~IRAS~4A \citep{Yildiz12}. The
observed emisison has two contributions: ({\it i}) the dense envelope heated
`passively' by the luminosity of the protostar, i.e., the dust in the
envelope absorbs all the protostellar luminosity and is heated by it,
and this temperature is then transferred to the gas through gas-dust
collisions; ({\it ii}) the gas heated by UV photons created by protostellar
accretion or by shocks in the outflow, and escaping from the immediate
protostellar surroundings, for example through outflow cavities, to
larger distances. Here the temperature of the gas is higher than that
of the dust.

The first component has been modeled by \citet{Kristensen12} for all
our sources and dust temperatures in excess of 10~K are typically
found out to between 2.5$\times$10$^{3}$ up to 1.5$\times$10$^{4}$~AU
from the sources. There is evidence that the dust may be further
heated on large scales by the UV photons generated by the accreting
protostar \citep{Hatchell13, Sicilia-Aguilar13}. We here quantify the
second component, which is the gas with temperatures higher than that
of the dust, {\it in excess of the passively heated envelope}. This
second mechanism operates on larger scales and is most relevant to
preventing further collapse or fragmentation of the core
\citep{Offner09, Offner10}.

To isolate this second component, the method outlined in
\citet{Yildiz12} is used. The \thco\ 6--5 envelope emission (component
({\it i})) is modeled using the temperature and density profiles from
\citet{Kristensen12} together with the \ceio\ constant abundance
results provided in \citet[][Table~5]{Yildiz13hifi}. For the three NGC
1333 sources, drop abundance profiles are used in which CO is frozen
out in some part of the envelope; for NGC~1333-IRAS~2A the results from
\citet{Yildiz10} are taken, whereas for NGC~1333-IRAS~4A and NGC~1333-IRAS~4B the models
from \citet{Yildiz12} are adopted.  These \ceio\ abundances are then
multiplied by the [$^{13}$C]/[$^{18}$O] abundance ratio of 8.5
\citep{LangerPenzias90} and the \thco\ emission is computed using the
non-LTE excitation line radiative transfer code RATRAN
\citep{Hogerheijde00}.  The turbulent width for
all the model \thco\ spectra is taken as 0.9~\kms, except for 
NGC~1333-IRAS~4A and NGC~1333-IRAS~4B where the values from \citet{Yildiz12} are used.
The resulting emission map is convolved with the relevant observing
beam.

\begin{figure}
\includegraphics[trim = 0mm 0mm 30mm 30mm, clip, scale=0.177]{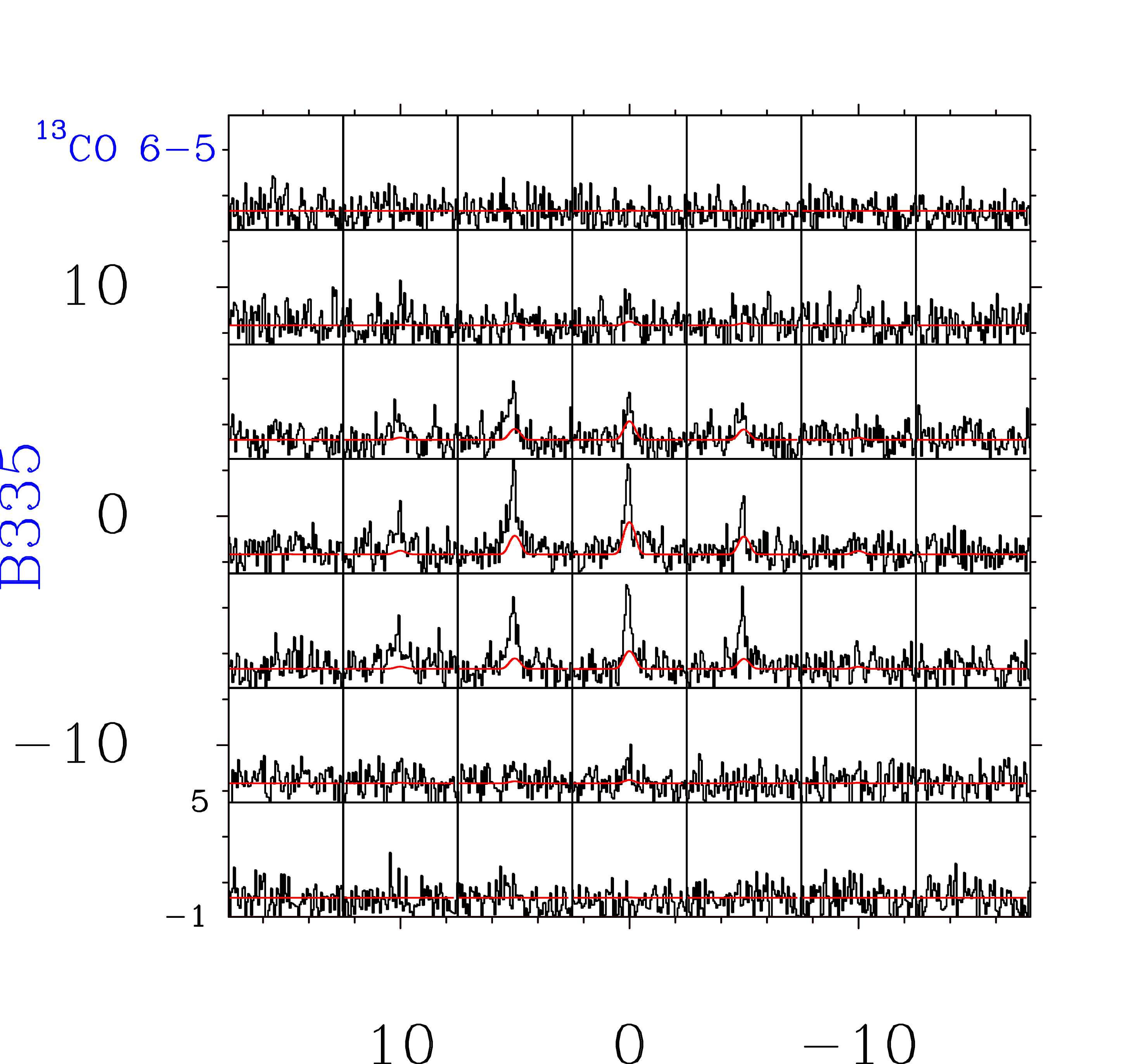}
\includegraphics[trim = 50mm 0mm 0mm 0mm, clip, scale=0.177]{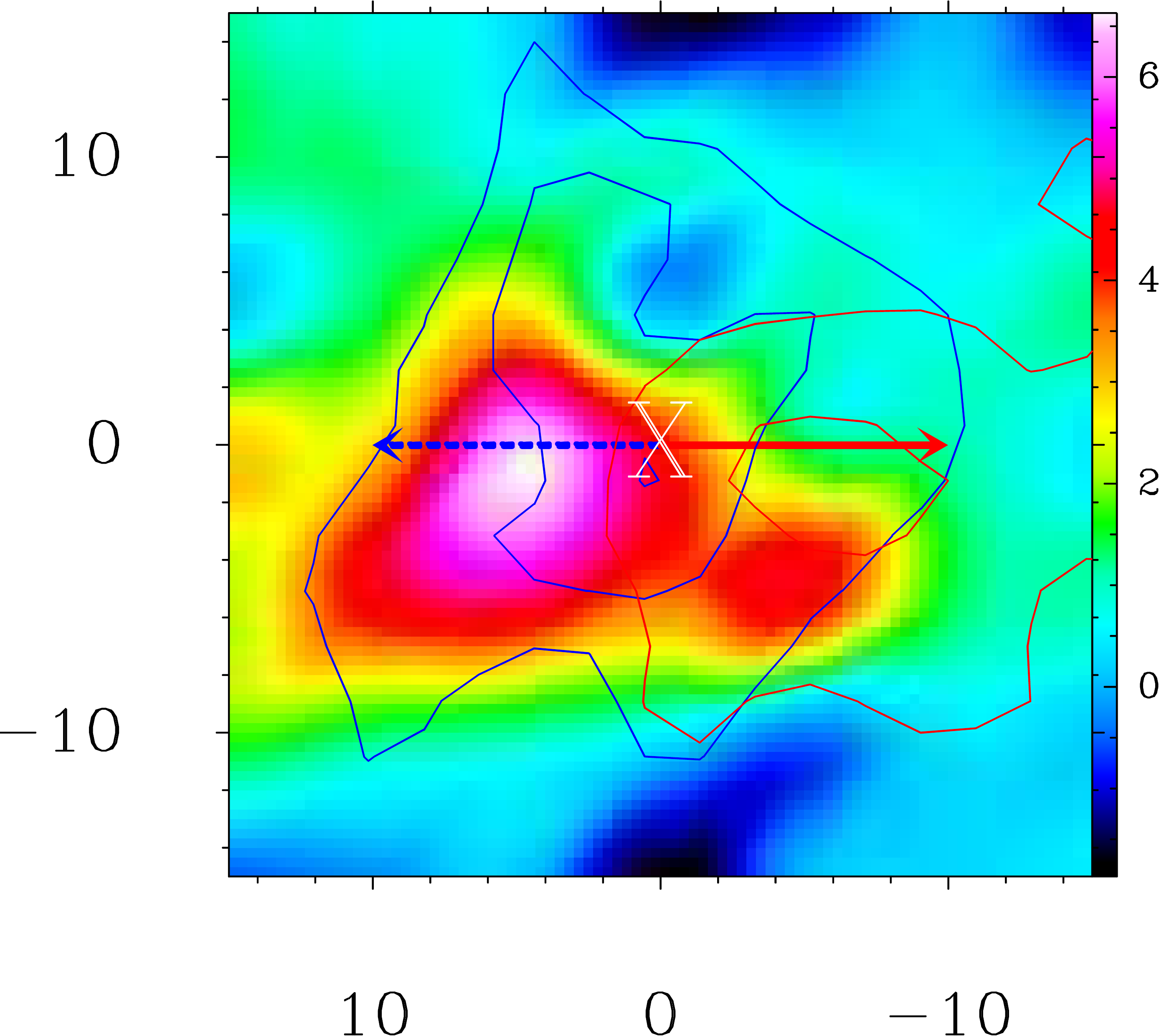}
\includegraphics[trim = 0mm 0mm 30mm 30mm, clip, scale=0.177]{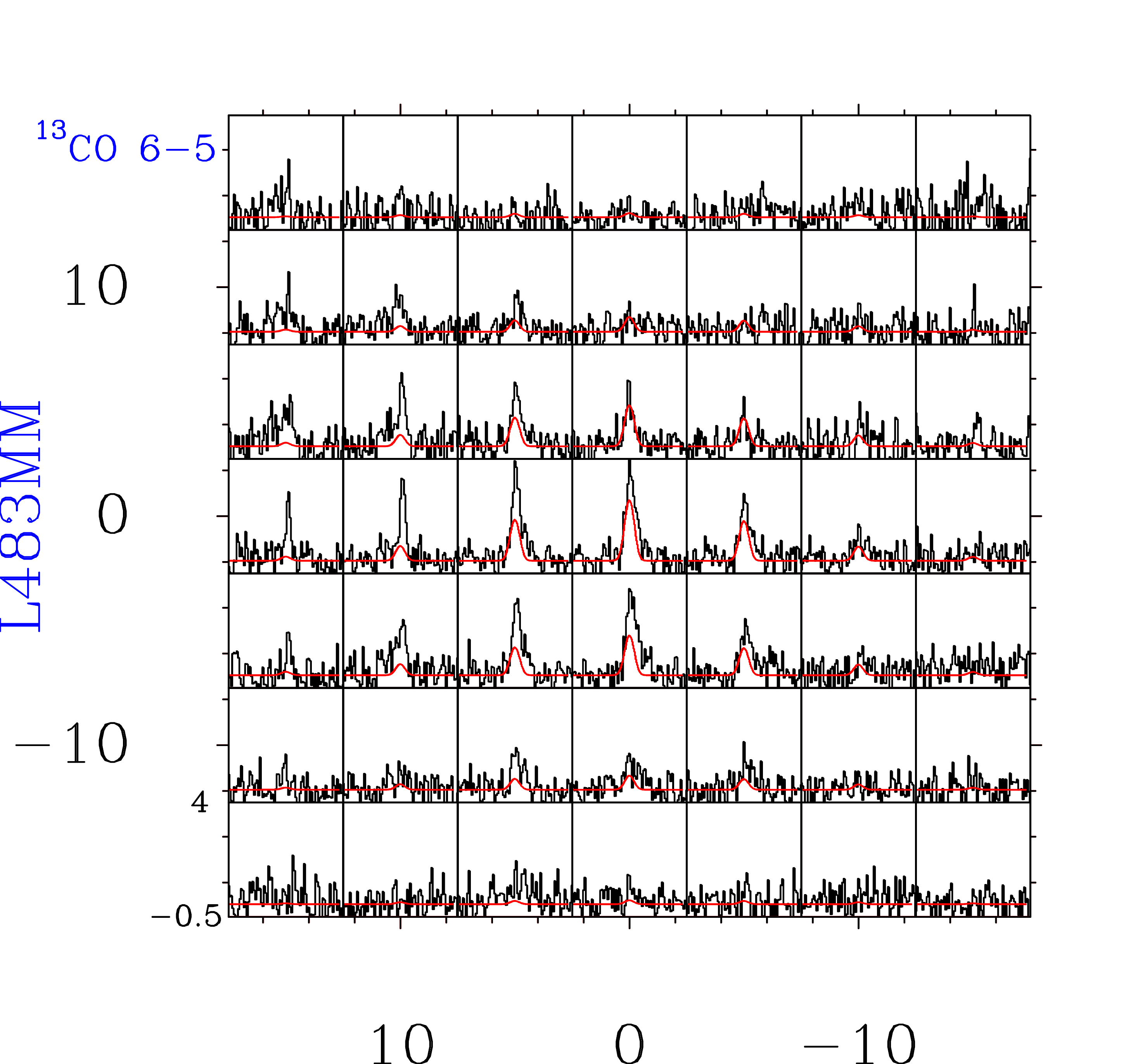}
\includegraphics[trim = 50mm 0mm 0mm 0mm, clip, scale=0.177]{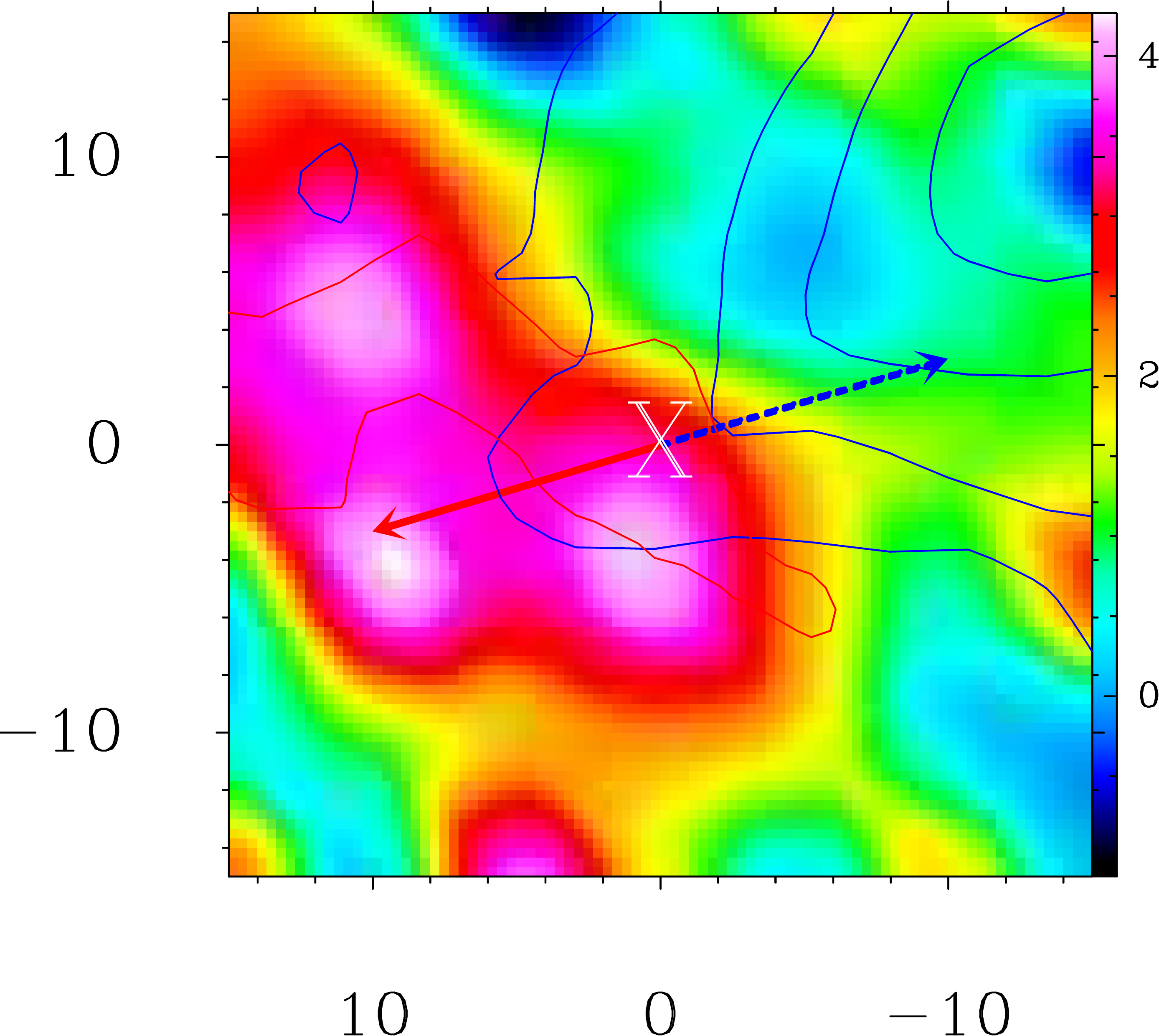}
\caption{\thco\ spectral maps in black overlaid with the model envelope spectra 
in red shown on the left panels. On the right panels, color maps of the UV heated 
gas distribution are shown. These are obtained by subtracting the model envelope 
emission from the observed spectra on a pixel-by-pixel basis. The sources are 
B335 ({\it top}) and L483mm ({\it bottom}). The axes show the offsets 
($\Delta$$\alpha$, $\Delta$$\delta$) in arcsec. The color scale is in units 
of K km s$^{-1}$.}
\label{fig:13CO65contourillus}
\end{figure}

Figures~\ref{fig:specmap13CO65_1}-\ref{fig:specmap13CO65_6} present
7$\times$7 pixel maps ($\sim$30\arcsec\ $\times$ 30\arcsec; 1 pixel =
4.5\arcsec) around the central position of each source in the \twco\
and \thco\ 6--5 transitions. The modeled envelope emission (component
({\it i})) is shown as red lines overplotted on the \thco\ maps. The
right-most panels present the difference between the model envelope
emission and the observed emission, which is the UV-heated gas. Two
illustrative maps are shown for B335 and L483mm in
Fig.~\ref{fig:13CO65contourillus}.

Most sources show some excess \thco\ emission on scales of
5\arcsec--10\arcsec\ or 1000--2000 AU at the average distance of 200
pc. The only exceptions are L1527 and Oph\,IRS63.  The emission is
almost always aligned with one (12/24 sources) or both (4/24 sources)
outflow lobes. A few sources show widely distributed \thco\ emission
(6/24 sources). More Class 0 sources show excess emission along the
direction of the outflow (11/13 sources) than Class I's do (5/11) but
this may be a $S/N$ effect.

The typical \thco\ 6--5 line width is $\lesssim$1 \kms, and so the
emission is not part of the swept-up outflow gas as illustrated in
more detail for the case of NGC~1333-IRAS~4A by \citet{Yildiz12}. The only
known mechanism to create this excess narrow emission is by UV~photons
generated from the protostellar accretion process and subsequently
escaping through the outflow cavities \citep{Spaans95}.

To estimate the effects of the UV radiation on these scales, it is
first important to estimate the temperature of the gas compared with
that of the dust. 
Figure~\ref{fig:radex13coratio} shows model
$^{13}$CO 3--2/6--5 line ratios for a grid of kinetic temperatures and
densities, with the observed values for each source overplotted at the
$7.5''$ radius density.  The inferred temperatures are in the range of 
30--80~K, consistent with the model predictions from \citet{Visser12}
on spatial scales of a few 1000~AU. For comparison, the typical dust
envelope temperature at this distance is $\sim$15--25~K and thus the
gas is heated to higher temperatures by more than a factor of 2.

The mass of the UV-heated gas (component {\it (ii)}) is calculated on
the basis of the residual after subtracting the \thco\ model envelope
emission (component ({\it i})) from the observed \thco\ emission. The
mass is then calculated via the residual emission by assuming
\Tex=50~K and CO/\hh=1.2$\times$10$^{-4}$, where the value of \Tex\ is
chosen because it is the median value for $^{13}$CO as reported in
\citet{Yildiz13hifi} based on transitions from 2--1 up to 10--9.
In order to compare UV-heated gas mass to the total outflow gas mass,
the outflow mass is recalculated from \twco~6--5 over the same
  ($\sim$30\arcsec\ $\times$ 30\arcsec) area, using \Tex=75~K to be
  consistent with all other \twco\ mass calculations.  In
Table~\ref{tbl:masscomparison}, the masses calculated for the
envelope, UV-heated gas and outflow gas are tabulated.  

\begin{figure*}[tb]
    \centering
    \includegraphics[scale=0.55]{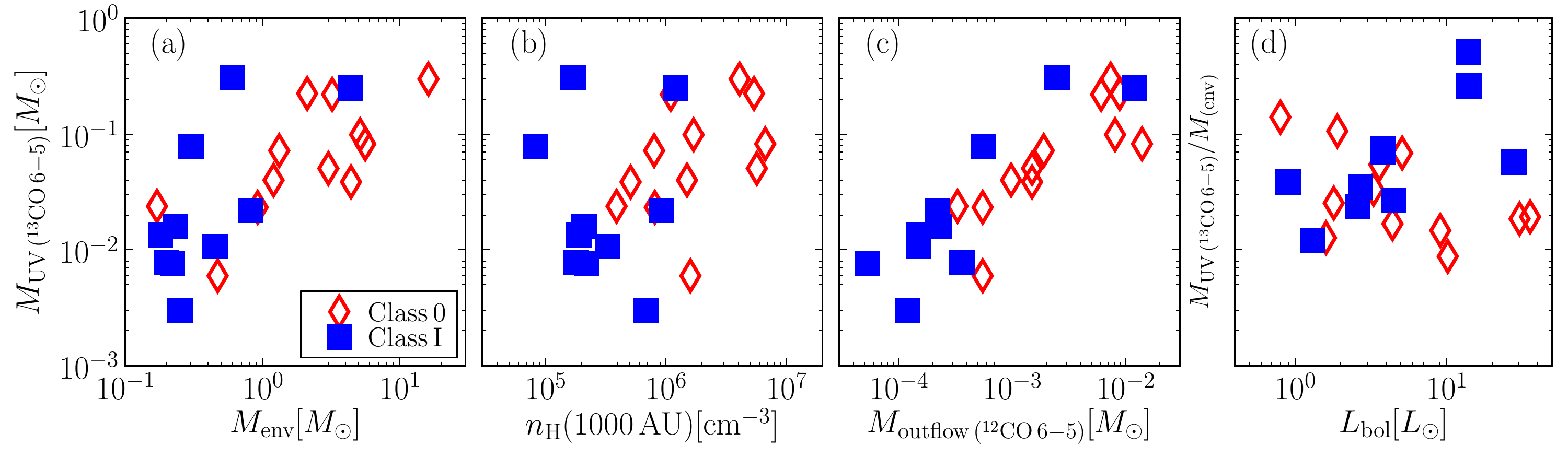}
    \caption{\small (a) UV heated gas mass is shown as a function of envelope 
    mass (\Menv), (b) density at 1000~AU (\n1000AU), and (c) the outflow mass 
    calculated from the \twco~6--5 lines for the same region 
    (\Mobs$_{(^{12}{\rm CO}\,6-5)}$). (d) The fraction of the UV heated gas 
    mass over envelope mass as a function of bolometric luminosity (\Lbol). 
    Figure \ref{fig:13comassCorr}(d) has the same y-axis values as (a--c).}
    \label{fig:13comassCorr}
\end{figure*}

The mass of the UV-heated gas is typically a factor of 10 to 100 times
lower than the total envelope mass (Fig. \ref{fig:13comassCorr}a) and
a factor of just a few up to 50 compared to the envelope mass within
the 30$\arcsec$$\times$30$\arcsec$ region. There is no correlation
with evolution; i.e., the fraction of UV-heated gas compared to the
total envelope mass does not change from Class~0 to Class~I. Similarly
there is no correlation between the mass of the UV-heated gas and the
density at 1000~AU (Fig.~\ref{fig:13comassCorr}b), which may
  suggest that the emission is independent of density and thus the
emission is thermalized.

Compared with the outflow masses, the UV-heated gas masses (component
({\it ii})) are typically a few times higher, as also found for
NGC~1333-IRAS~4A in \citet{Yildiz12}. They follow a remarkably tight
correlation with a Pearson correlation coefficient of 0.86
(3.1$\sigma$; Fig.~\ref{fig:13comassCorr}c).  Furthermore, the
fraction of UV-heated to envelope gas mass is constant as a function
of bolometric luminosity at a median value of $\sim$0.03
(Fig.~\ref{fig:13comassCorr}d). The two outstanding high \MUV/\Menv\
Class~I sources are DK\,Cha and GSS30\,IRS1.

Many protostellar envelopes show varying degrees of asymmetry and are
not spherical; most striking is the flattened envelope surrounding
L1157 \citep{Tobin10}. This asymmetry naturally introduces systematic
uncertainties in the envelope modeling which is then propagated
through to the determination of the mass of the UV-heated
gas. However, most envelopes are elongated perpendicular to the
direction of the outflow (e.g., L1157) whereas the residual \thco\
6--5 emission is typically elongated along the outflow
direction. Therefore we do not think that the use of spherical envelope 
models changes any of the conclusions regarding the effects of the 
UV-heated gas.

\begin{figure}[!t]
    \centering
    \includegraphics[width=\columnwidth]{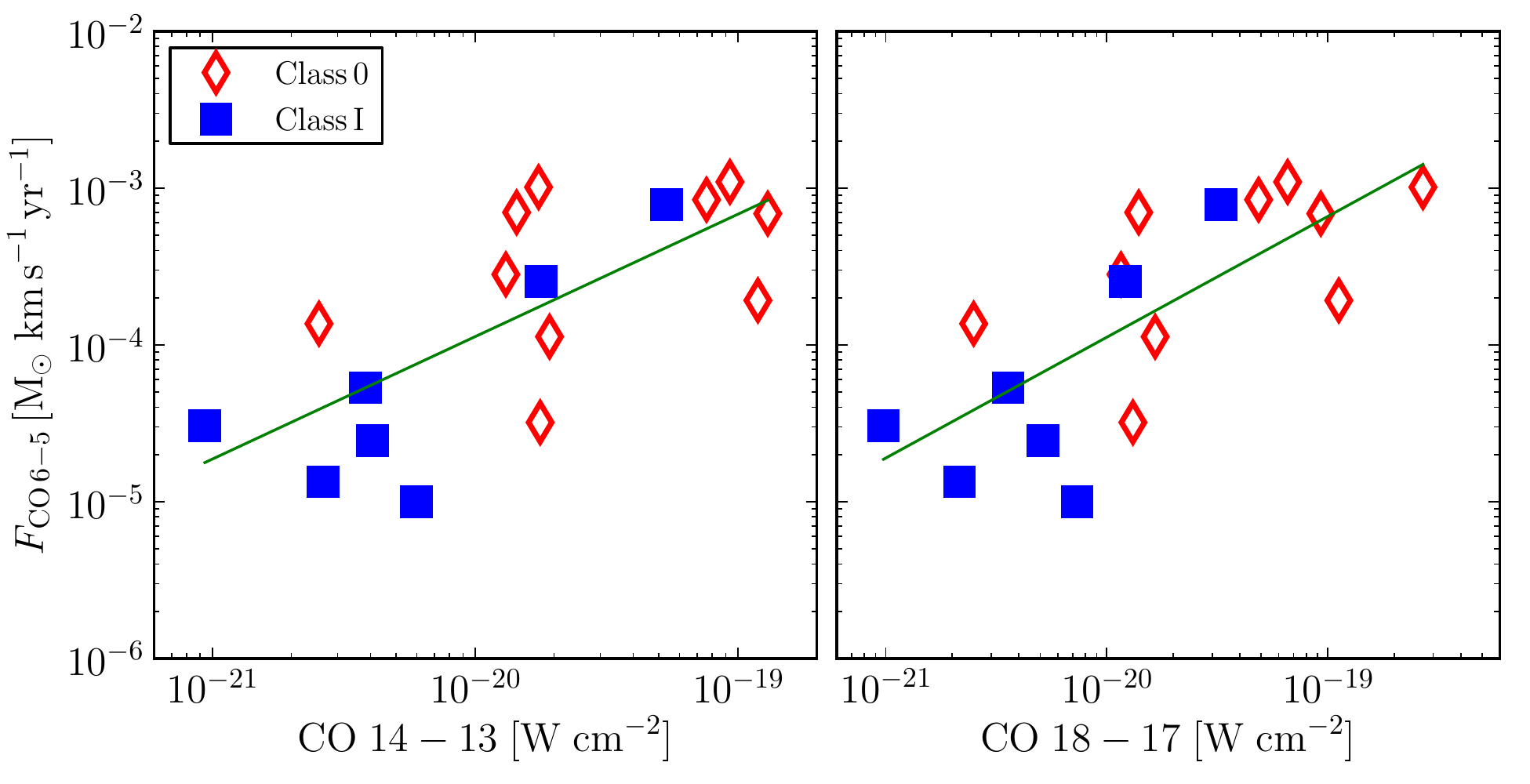}
    \caption{\small Correlation between \FCO\ and the CO~14--13 and 18--17 
    fluxes obtained from {\it Herschel}-PACS. The integrated intensities are 
    scaled to a common distance of 200~pc. The green lines shows the best-fit 
    power-laws to the data and are simple least-squares fits.}
    \label{fig:corrPACSCO}
\end{figure}

\begin{table}[!t]
\small
\caption{\small Comparison of envelope, photon-heated and outflow masses over 
the 30$\times$30$''$ area surrounding the central protostar.}
\begin{center}
\begin{tabular}{l l l c c }
\hline \hline
Source & $M_{\rm Envelope}$\tablefootmark{a} & $M_{\rm Envelope}$\tablefootmark{b} & $M_{\rm UV}$\tablefootmark{c} & $M_{\rm outflow}$\tablefootmark{d} \\
       & Total & $\leq 15''$ & $^{13}$CO~6--5    & $^{12}$CO~6--5\\ 
\hline
\object{L1448mm}         & 9.0   & 1.69  &  4.0$\times$10$^{-2}$  & \dots \\
\object{NGC1333-IRAS~2A}  & 5.13  & 0.67  &  9.9$\times$10$^{-2}$  & 8.1$\times$10$^{-3}$ \\ 
\object{NGC1333-IRAS~4A}  & 5.59  & 2.56  &  8.2$\times$10$^{-2}$  & 1.4$\times$10$^{-2}$ \\ 
\object{NGC1333-IRAS~4B}  & 3.01  & 2.60  &  5.1$\times$10$^{-2}$  & 2.2$\times$10$^{-3}$ \\ 
\object{L1527}           & 0.92  & 0.18  &  2.3$\times$10$^{-2}$  & $<$5.5$\times$10$^{-4}$ \\ 
\object{Ced110-IRS4}     & 0.17  & 0.07  &  2.4$\times$10$^{-2}$  & 3.3$\times$10$^{-4}$ \\ 
\object{IRAS~15398}       & 0.47  & 0.28  &  6.0$\times$10$^{-3}$  & 5.5$\times$10$^{-4}$ \\ 
\object{L483mm}          & 4.4   & 0.25  &  3.9$\times$10$^{-2}$  & 1.5$\times$10$^{-3}$ \\ 
\object{Ser-SMM1}        & 16.13 & 1.99  &  3.0$\times$10$^{-1}$  & 7.4$\times$10$^{-3}$ \\ 
\object{Ser-SMM4}        & 2.11  & 3.17  &  2.2$\times$10$^{-1}$  & 8.9$\times$10$^{-3}$ \\ 
\object{Ser-SMM3}        & 3.21  & 0.74  &  2.2$\times$10$^{-1}$  & 6.1$\times$10$^{-3}$ \\ 
\object{L723mm}          & 1.32  & 0.67  &  7.2$\times$10$^{-2}$  & 1.9$\times$10$^{-3}$ \\ 
\object{B335}            & 1.2   & 0.79  &  4.0$\times$10$^{-2}$  & 9.8$\times$10$^{-4}$ \\ 
\hline  
\object{L1489}           & 0.18  & 0.04  &  1.4$\times$10$^{-2}$  & 1.5$\times$10$^{-4}$ \\ 
\object{TMR1}            & 0.23  & 0.04  &  1.6$\times$10$^{-2}$  & 2.3$\times$10$^{-4}$ \\ 
\object{TMC1A}           & 0.22  & 0.04  &  7.6$\times$10$^{-3}$  & 5.3$\times$10$^{-5}$ \\ 
\object{TMC1}            & 0.2   & 0.04  &  7.7$\times$10$^{-3}$  & 3.6$\times$10$^{-4}$ \\ 
\object{HH46-IRS}        & 4.36  & 1.21  &  2.5$\times$10$^{-1}$  & 1.2$\times$10$^{-2}$ \\ 
\object{DK~Cha}          & 0.82  & 0.26  &  2.2$\times$10$^{-2}$  & 2.2$\times$10$^{-4}$ \\ 
\object{GSS30-IRS1}      & 0.6   & 0.03  &  3.1$\times$10$^{-1}$  & 2.5$\times$10$^{-3}$ \\ 
\object{Elias~29}        & 0.3   & 0.01  &  7.8$\times$10$^{-2}$  & 5.6$\times$10$^{-4}$ \\ 
\object{Oph-IRS63}       & 0.25  & 0.12  &  3.0$\times$10$^{-3}$  & $<$1.2$\times$10$^{-4}$ \\ 
\object{RNO91}           & 0.45  & 0.05  &  1.1$\times$10$^{-2}$  & 1.5$\times$10$^{-4}$ \\
\object{L1551-IRS5}      & 22.2  & 0.23  &  2.0$\times$10$^{-2}$  & \dots \\
\hline 
\end{tabular}
\end{center}
\tablefoot{
  All masses are given in M$_{\odot}$.
  \tablefoottext{a}{Total mass of the spherical envelope inferred from
  the continuum radiative-transfer modeling \citep{Kristensen12}.}
  \tablefoottext{b}{As $a$, but limited to the mass within the 15$\arcsec$ 
  radius over which the UV-heated component is determined.}
  \tablefoottext{c}{UV photon-heated gas mass (component {\it (ii))}
  as calculated
  from the \thco~6--5 spectra over the mapped area after subtracting the modeled 
  envelope emission.}
  \tablefoottext{d}{Outflow mass calculated from the \twco~6--5 map for the 
  same area as the $^{13}$CO maps.}
  All masses except the total envelope mass ({\it a}) are obtained over a 
  $\sim$30$\arcsec$$\times$30$\arcsec$ (Fig.~\ref{fig:specmap13CO65_1}-\ref{fig:specmap13CO65_6})
}
\label{tbl:masscomparison}
\end{table}

\section{Discussion}
\label{5:sec:discussion}

\subsection{Mechanical feedback}

Our results show that the outflow parameters inferred from the \co\
6--5 data show the same trends with \Lbol\ and evolutionary stage as
found previously in the literature, but with stronger correlations
than for the 3--2 data. Even though the same telescope and methods are
used for all sources and the spatial resolution is high, there remains
a scatter of at least an order of magnitude in the correlation between
\FCO\ and \Lbol.  This could point to the importance of ``episodic
accretion'' as a resolution to the ``luminosity problem''
\citep{Evans09c2d,Dunham10,Dunham13}. Some Class~0 sources are very
luminous, which is likely due to a current rapid burst in accretion
which may happen every 10$^3$--10$^4$ years \citep{Dunham10}.
However, their location in the high state is not constant and would
drop in the course of time, on timescales as fast as 10$^2$ years
\citep{Johnstone13}. The envelope mass, on the other hand, is
independent of the current luminosity, and the stronger correlation
of \FCO\ with \Menv\ may simply reflect that more mass is swept up.

Since the outflow force gives the integrated activity over the entire
lifetime of a YSO, it is also interesting to compare this parameter with
the currently shocked gas probed by the \herschel-PACS high-$J$
CO observations (\Jup$>$14). In Fig.~\ref{fig:corrPACSCO}, \FCO\
is plotted against CO~14--13 and CO~18--17 fluxes ($E_{\rm up}$ $\sim$580 
and 940~K) obtained from \citet{Karska13, Goicoechea12, Herczeg12,Green13}
and \citet{vanKempen10dkcha}.
There is a strong correlation with the \mbox{CO~14--13} and \mbox{CO~18--17} 
fluxes with \FCO\ ($r$=$\sim$0.76 
$\sim$3.1$\sigma$; Fig.~\ref{fig:corrPACSCO}). This
correlation illustrates that although \mbox{CO~18--17} likely traces a
different outflow component than \mbox{CO~6--5},
 a component closer to the
shock front \citep{Santangelo12,Nisini13,Tafalla13}, the underlying 
driving mechanism is the same. Furthermore, CO 18--17 emission is often 
extended along the outflow direction \citep{Karska13} and clearly traces, 
spatially, a component related to that traced by CO 6--5. Although the 
excitation of CO 18--17 requires higher densities and temperatures
(\ncrit$\sim$10$^6$ \cmthree; \Eup$\sim$940 K) than \mbox{CO~6--5}
(\ncrit$\sim$10$^5$ \cmthree; \Eup$\sim$120 K), CO~6--5 likely follows
in the wake of the shocks traced by the higher-$J$ lines and therefore
the excitation of both lines ultimately depend on the actual shock
conditions. Testing this scenario requires velocity-resolved line
profiles of high-$J$ lines such as CO~16--15 (Kristensen et al. in
prep.).

 \begin{figure}[!t]
     \centering
     \includegraphics[width=\columnwidth]{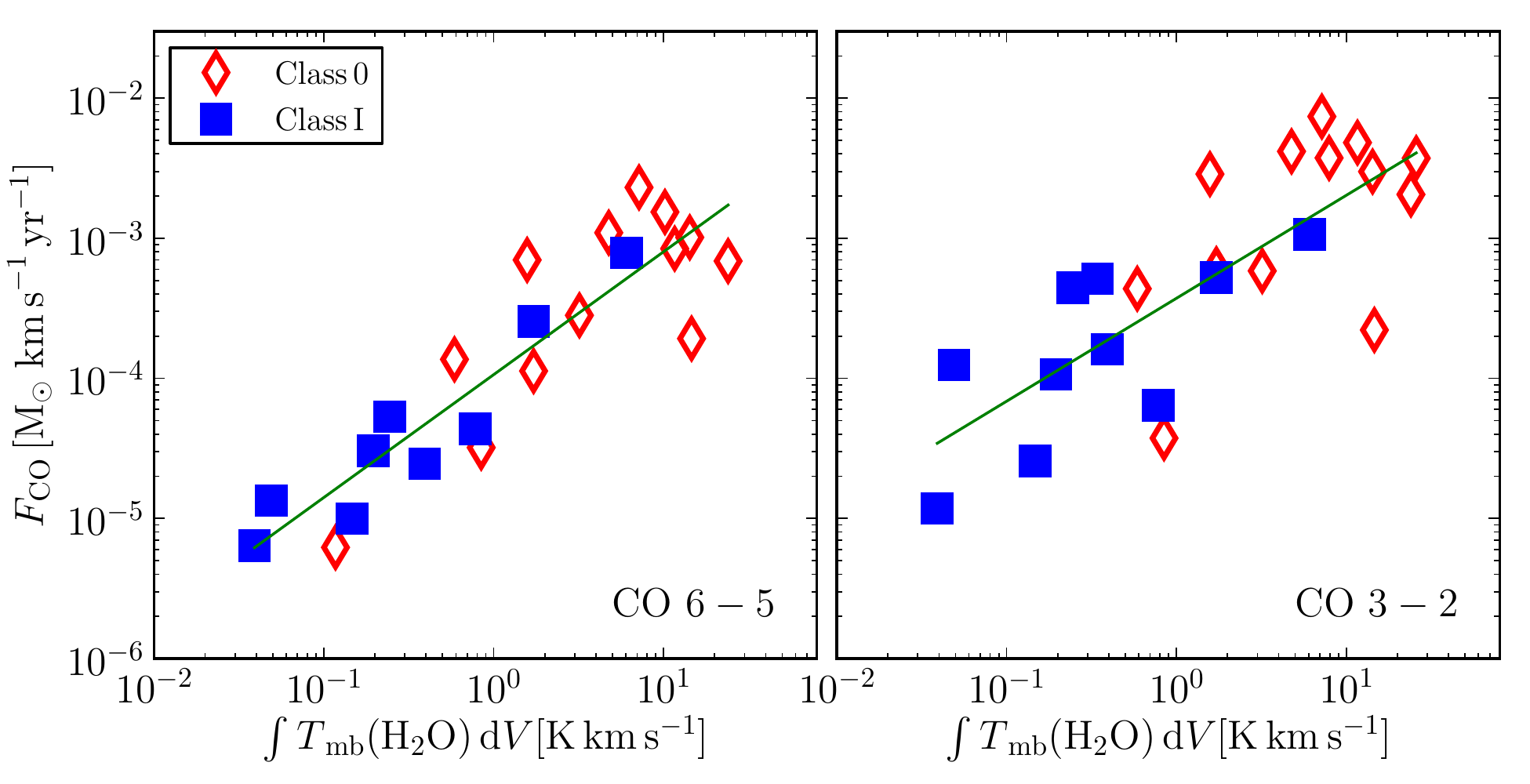}
     \caption{\small Correlation between \FCO\ measured from the CO
       6--5 ({\it left}) and CO 3--2 ({\it right}) data and the integrated 
       intensity of the ground-state H$_2$O 1$_{10}$--1$_{01}$ transition at
       557~GHz. The integrated intensities are scaled to a common
       distance of 200~pc \citep{Kristensen12}. The correlation is
       strong for CO 6--5, $r$=0.90 with 4.1$\sigma$. The green lines shows 
       the best-fit power-laws to the data and are simple least-squares fits.}
    \label{fig:corrHIFIH2O}
 \end{figure}

Another indication that the outflow force as measured from CO 6--5 is
more closely linked to the currently shocked gas than that from 3--2 comes 
from comparing H$_2$O and $F_{\rm CO}$. Water is one of the best shock
tracers, as shown most recently by several \textit{Herschel}
observations
\citep{vanKempen10hh46,LeFloch10,Kristensen10hifi,Nisini10,Vasta12,Tafalla13}.
\citet{Kristensen12} compared the integrated intensity of the H$_2$O
1$_{10}$--1$_{01}$ transition at 557 GHz with the outflow forces
presented in the literature.  These observed line intensities are
scaled by the square of the source distance to a common distance of
$d_{\rm avg}$=200~pc. The literature values of the outflow force 
used in that paper were calculated using a
variety of methods and data sets, and provided an inhomogeneous
sample. No correlation of H$_2$O integrated intensity with $F_{\rm
  CO}$ was found. Revisiting this comparison with the newly measured
outflow forces in a consistent way reveals a correlation with
the force measured from \mbox{CO~3--2} data ($r$=0.78; 3.6$\sigma$) and a 
stronger correlation with the force measured from the \mbox{CO~6--5} data
($r$=0.90; 4.1$\sigma$, Fig. \ref{fig:corrHIFIH2O}) \citep[see
  also][]{Bjerkeli12}.  Thus, $F_{\rm CO}$ as deduced from 6--5 can
be used as a measure of the outflow force of the shocked gas, rather
than just the entrained, swept-up gas.

\subsection{Radiative feedback}

The observational data demonstrate that \thco\ 6--5 traces UV-heated
gas, and that the UV-heated gas is predominantly found along the same 
direction as the outflow. The $A_{V}$ is lower inside the outflow 
cavity, because the density is lower, and so UV radiation from the 
accretion can escape more easily along this direction
\citep{Spaans95,Bruderer09,Visser12}. If there are also external UV
sources, the UV-heated gas could have a more isotropic component as
well but this is not traced by our \mbox{\thco\ 6--5} data at the current
\SN\ level except for the case of two sources in Ophiuchus, Elias 29 
and GSS30\,IRS1.  Narrow \twco\ lines may be used instead at positions well
away from the outflow cone, as illustrated by previous observations
\citep{vanKempen09champ}.

The estimated gas temperature of the UV-heated gas of 30--50 K is
likely a lower limit to the maximum temperature achieved by this
process. Model calculations by \citet{Spaans95} and \citet{Visser12} show 
that the gas temperatures can reach values up to a few hundred K at 1000 AU
radius in a narrow layer along the outflow cavity, depending on source
characteristics. Gas temperatures $>$30 K are maintained out to 10$^4$
AU radius.  Thus, in clustered environments such as NGC~1333 or
Ophiuchus, it is unlikely that the gas temperature ever drops down to
10 K because the protostars heat the gas radiatively.  Gas and dust
temperatures are clearly decoupled, with dust temperatures significantly 
lower than the gas temperature, by about a factor of 2. Thus, estimates 
of the radiative feedback based on dust observations alone \citep{Hatchell13, 
Sicilia-Aguilar13} likely underestimate both the temperature and the extent 
of the feedback. Indeed, the continuum emission, as observed with e.g. SCUBA 
at 450 and 850 $\mu$m, typically does not show extended structure along the 
direction of the outflows.

The tight correlation between the mass of the UV-heated gas and the outflow 
mass, when measured over the same area, is puzzling. Naively, one would expect 
the two properties to be unrelated as they are caused by two different physical 
mechanisms, UV excitation and outflow entrainment. However, the cause of these 
two physical processes is linked, accretion and ejection 
\citep[e.g.][]{Bontemps96, Frank14}. The UV photons are generated in the 
accretion shocks onto the protostar, and during this accretion process part 
of the material is ejected. Thus, higher-luminosity sources at a given envelope 
mass should show higher UV luminosities. It is not possible to verify this 
hypothesis directly as all of these sources are deeply embedded. A second 
component is required to efficiently UV-heat the surrounding gas: an outflow 
cavity needs to be cleared out for the radiation to escape which requires the 
outflow to have been active for at least one dynamic time-scale. Thus, there 
may be good reason to expect a correlation between the masses of the UV-heated 
and outflowing gas, when measured over the same area.

\section{Conclusions}
\label{5:sec:conclusion}
In this paper, we present large-scale maps of 26 YSOs obtained with
the APEX-CHAMP$^+$ instrument (\twco\ and \mbox{\thco\ 6--5}), together with
the JCMT-HARP-B instrument (\twco\ and \mbox{\thco\ 3--2}). Our sample
consists of deeply embedded Class~0 sources as well as less deeply
embedded Class~I sources. With these high spatial and spectral
resolution maps, we have studied the outflow activity of these two
different evolutionary stages of YSOs in a consistent manner.  All
embedded sources show large-scale outflow activity that can be traced
by the CO line wings, however their activity is reduced over the
course of evolution to the later evolutionary stages as indicated by
the decrease of several outflow parameters, including the spatial
extent of the outflow as seen in the \twco\ 6--5 maps.

One of the crucial parameters, the outflow force, \FCO\ is
quantified and correlations with other physical parameters are
sought. In agreement with previous studies, Class 0 sources have
higher outflow forces than Class I sources.
\FCO\ is directly proportional to \Menv\ and \Moutflow, showing that
higher outflow forces are associated with higher envelope mass and outflow
mass, as present in Class~0 sources. Comparing the outflow force as
measured from CO 6--5 data to H$_2$O observed with
\textit{Herschel}-HIFI and high-$J$ CO observed with
\textit{Herschel}-PACS reveals a correlation, suggesting that the
outflow force from 6--5 is related to current shock
activity. This is in contrast with the outflow force measured from CO
3--2, where the correlation with water and the
high-$J$ CO fluxes is weaker.

The quiescent gas is traced by narrow (FWHM $\sim$1~\kms)
\thco\ 6--5 emission. 
For this purpose, maps are obtained in \thco\ 6--5 transition for the 
sources $\sim$1$'$ region around the source position. 
Envelope emission is 
modeled via radiative transfer models and is subtracted from the observed 
\thco\ 6--5 emission. It is shown that an excess emission exists in most 
sources on scales of a 1000--2000 AU and this emission is caused by 
UV~photons generated from the protostellar accretion process and subsequently
escaping through the outflow cavities. The fraction of the UV-heated gas 
compared to the total envelope mass does not change from Class~0 to 
Class~I and there are no clear signs of evolutionary trends.

UV heating is prominent along the outflow direction and this is a
general observable trend. This directional preference suggests that
the UV feedback on large scales is most important in the same regions
as the outflows. The UV heating observed in $^{13}$CO 6-5 is important
on scales of $<$10$^{4}$~AU, i.e., not on cluster scales. Future
models of core and disk fragmentation should take these effects into
account.


\begin{acknowledgements}
The authors would like to thank the anonymous referee for suggestions 
  and comments, which improved this paper.
  We are grateful to the APEX and JCMT staff for support with these
  observations. Astrochemistry in Leiden is supported by the
  Netherlands Research School for Astronomy (NOVA), by a Spinoza grant
  and grant 614.001.008 from the Netherlands Organisation for
  Scientific Research (NWO), and by the European Community's Seventh
  Framework Programme FP7/2007-2013 under grant agreement 238258
  (LASSIE). This work was carried out in part at the Jet Propulsion 
  Laboratory, which is operated by the California Institute of Technology 
  under contract with NASA. Construction of CHAMP$^+$ is a collaboration 
  between the Max-Planck-Institut fur Radioastronomie Bonn, Germany; SRON
  Netherlands Institute for Space Research, Groningen, the
  Netherlands; the Netherlands Research School for Astronomy (NOVA);
  and the Kavli Institute of Nanoscience at Delft University of
  Technology, the Netherlands; with support from the Netherlands
  Organization for Scientific Research (NWO) grant 600.063.310.10. 
  The APEX data was obtained via Max Planck Institute observing time.
\end{acknowledgements}

\bibliographystyle{aa}
\bibliography{bibdata}

\begin{thebibliography}{85}
\expandafter\ifx\csname natexlab\endcsname\relax\def\natexlab#1{#1}\fi

\bibitem[{{Andr{\'e}} {et~al.}(2000){Andr{\'e}}, {Ward-Thompson}, \&
  {Barsony}}]{Andre00}
{Andr{\'e}}, P., {Ward-Thompson}, D., \& {Barsony}, M. 2000, Protostars \&
  Planets IV, 59

\bibitem[{{Arce} \& {Sargent}(2006)}]{ArceSargent06}
{Arce}, H.~G. \& {Sargent}, A.~I. 2006, \apj, 646, 1070

\bibitem[{{Arce} {et~al.}(2007){Arce}, {Shepherd}, {Gueth}, {Lee}, {Bachiller},
  {Rosen}, \& {Beuther}}]{Arce07}
{Arce}, H.~G., {Shepherd}, D., {Gueth}, F., {et~al.} 2007, Protostars and
  Planets V, 245

\bibitem[{{Bachiller} {et~al.}(1990){Bachiller}, {Martin-Pintado}, {Tafalla},
  {Cernicharo}, \& {Lazareff}}]{Bachiller90}
{Bachiller}, R., {Martin-Pintado}, J., {Tafalla}, M., {Cernicharo}, J., \&
  {Lazareff}, B. 1990, \aap, 231, 174

\bibitem[{{Bachiller} \& {Tafalla}(1999)}]{Bachiller99}
{Bachiller}, R. \& {Tafalla}, M. 1999, in NATO ASIC Proc. 540: The Origin of
  Stars and Planetary Systems, ed. {C.~J.~Lada \& N.~D.~Kylafis}, 227

\bibitem[{{Benedettini} {et~al.}(2012){Benedettini}, {Busquet}, {Lefloch},
  {Codella}, {Cabrit}, {Ceccarelli}, {Giannini}, {Nisini}, {Vasta},
  {Cernicharo}, {Lorenzani}, \& {di Giorgio}}]{Benedettini12}
{Benedettini}, M., {Busquet}, G., {Lefloch}, B., {et~al.} 2012, \aap, 539, L3

\bibitem[{{Bjerkeli} {et~al.}(2012){Bjerkeli}, {Liseau}, {Larsson}, {Rydbeck},
  {Nisini}, {Tafalla}, {Antoniucci}, {Benedettini}, {Bergman}, {Cabrit},
  {Giannini}, {Melnick}, {Neufeld}, {Santangelo}, \& {van
  Dishoeck}}]{Bjerkeli12}
{Bjerkeli}, P., {Liseau}, R., {Larsson}, B., {et~al.} 2012, \aap, 546, A29

\bibitem[{{Blake} {et~al.}(1995){Blake}, {Sandell}, {van Dishoeck},
  {Groesbeck}, {Mundy}, \& {Aspin}}]{Blake95}
{Blake}, G.~A., {Sandell}, G., {van Dishoeck}, E.~F., {et~al.} 1995, \apj, 441,
  689

\bibitem[{{Bontemps} {et~al.}(1996){Bontemps}, {Andr\'e}, {Terebey}, \&
  {Cabrit}}]{Bontemps96}
{Bontemps}, S., {Andr\'e}, P., {Terebey}, S., \& {Cabrit}, S. 1996, \aap, 311,
  858

\bibitem[{{Bourke} {et~al.}(1997){Bourke}, {Garay}, {Lehtinen}, {Koehnenkamp},
  {Launhardt}, {Nyman}, {May}, {Robinson}, \& {Hyland}}]{Bourke97}
{Bourke}, T.~L., {Garay}, G., {Lehtinen}, K.~K., {et~al.} 1997, \apj, 476, 781

\bibitem[{{Brown} \& {Chandler}(1999)}]{Brown99}
{Brown}, D.~W. \& {Chandler}, C.~J. 1999, \mnras, 303, 855

\bibitem[{{Bruderer} {et~al.}(2009){Bruderer}, {Benz}, {Doty}, {van Dishoeck},
  \& {Bourke}}]{Bruderer09}
{Bruderer}, S., {Benz}, A.~O., {Doty}, S.~D., {van Dishoeck}, E.~F., \&
  {Bourke}, T.~L. 2009, \apj, 700, 872

\bibitem[{{Cabrit} \& {Bertout}(1990)}]{CabritBertout90}
{Cabrit}, S. \& {Bertout}, C. 1990, \apj, 348, 530

\bibitem[{{Cabrit} \& {Bertout}(1992)}]{CabritBertout92}
{Cabrit}, S. \& {Bertout}, C. 1992, \aap, 261, 274

\bibitem[{{Codella} {et~al.}(2014){Codella}, {Maury}, {Gueth}, {Maret},
  {Belloche}, {Cabrit}, \& {Andr{\'e}}}]{Codella14}
{Codella}, C., {Maury}, A.~J., {Gueth}, F., {et~al.} 2014, \aap, 563, L3

\bibitem[{{Curtis} {et~al.}(2010){Curtis}, {Richer}, {Swift}, \&
  {Williams}}]{Curtis10_2outflows}
{Curtis}, E.~I., {Richer}, J.~S., {Swift}, J.~J., \& {Williams}, J.~P. 2010,
  \mnras, 408, 1516

\bibitem[{{Downes} \& {Cabrit}(2007)}]{DownesCabrit07}
{Downes}, T.~P. \& {Cabrit}, S. 2007, \aap, 471, 873

\bibitem[{{Dunham} {et~al.}(2013){Dunham}, {Arce}, {Allen}, {Evans},
  {Broekhoven-Fiene}, {Chapman}, {Cieza}, {Gutermuth}, {Harvey}, {Hatchell},
  {Huard}, {Kirk}, {Matthews}, {Mer{\'{\i}}n}, {Miller}, {Peterson}, \&
  {Spezzi}}]{Dunham13}
{Dunham}, M.~M., {Arce}, H.~G., {Allen}, L.~E., {et~al.} 2013, \aj, 145, 94

\bibitem[{{Dunham} {et~al.}(2014){Dunham}, {Arce}, {Mardones}, {Lee},
  {Matthews}, {Stutz}, \& {Williams}}]{Dunham14}
{Dunham}, M.~M., {Arce}, H.~G., {Mardones}, D., {et~al.} 2014, \apj, 783, 29

\bibitem[{{Dunham} {et~al.}(2010){Dunham}, {Evans}, {Terebey}, {Dullemond}, \&
  {Young}}]{Dunham10}
{Dunham}, M.~M., {Evans}, II, N.~J., {Terebey}, S., {Dullemond}, C.~P., \&
  {Young}, C.~H. 2010, \apj, 710, 470

\bibitem[{{Enoch} {et~al.}(2009){Enoch}, {Evans}, {Sargent}, \&
  {Glenn}}]{Enoch09}
{Enoch}, M.~L., {Evans}, II, N.~J., {Sargent}, A.~I., \& {Glenn}, J. 2009,
  \apj, 692, 973

\bibitem[{{Evans} {et~al.}(2009){Evans}, {Dunham}, {J{\o}rgensen}, {Enoch},
  {Mer{\'{\i}}n}, {van Dishoeck}, {Alcal{\'a}}, {Myers}, {Stapelfeldt},
  {Huard}, {Allen}, {Harvey}, {van Kempen}, {Blake}, {Koerner}, {Mundy},
  {Padgett}, \& {Sargent}}]{Evans09c2d}
{Evans}, N.~J., {Dunham}, M.~M., {J{\o}rgensen}, J.~K., {et~al.} 2009, \apjs,
  181, 321

\bibitem[{{Frank} {et~al.}(2014){Frank}, {Ray}, {Cabrit}, {Hartigan}, {Arce},
  {Bacciotti}, {Bally}, {Benisty}, {Eisl{\"o}ffel}, {G{\"u}del}, {Lebedev},
  {Nisini}, \& {Raga}}]{Frank14}
{Frank}, A., {Ray}, T.~P., {Cabrit}, S., {et~al.} 2014, ArXiv: 1402.3553

\bibitem[{{Geers} {et~al.}(2009){Geers}, {van Dishoeck}, {Pontoppidan},
  {Lahuis}, {Crapsi}, {Dullemond}, \& {Blake}}]{geers09}
{Geers}, V.~C., {van Dishoeck}, E.~F., {Pontoppidan}, K.~M., {et~al.} 2009,
  \aap, 495, 837

\bibitem[{{Goicoechea} {et~al.}(2012){Goicoechea}, {Cernicharo}, {Karska},
  {Herczeg}, {Polehampton}, {Wampfler}, {Kristensen}, {van Dishoeck},
  {Etxaluze}, {Bern{\'e}}, \& {Visser}}]{Goicoechea12}
{Goicoechea}, J.~R., {Cernicharo}, J., {Karska}, A., {et~al.} 2012, \aap, 548,
  A77

\bibitem[{{Goldsmith} {et~al.}(1984){Goldsmith}, {Snell}, {Hemeon-Heyer}, \&
  {Langer}}]{Goldsmith84}
{Goldsmith}, P.~F., {Snell}, R.~L., {Hemeon-Heyer}, M., \& {Langer}, W.~D.
  1984, \apj, 286, 599

\bibitem[{{Green} {et~al.}(2013){Green}, {Evans}, {J{\o}rgensen}, {Herczeg},
  {Kristensen}, {Lee}, {Dionatos}, {Yildiz}, {Salyk}, {Meeus}, {Bouwman},
  {Visser}, {Bergin}, {van Dishoeck}, {Rascati}, {Karska}, {van Kempen},
  {Dunham}, {Lindberg}, {Fedele}, \& {DIGIT Team1}}]{Green13}
{Green}, J.~D., {Evans}, II, N.~J., {J{\o}rgensen}, J.~K., {et~al.} 2013, \apj,
  770, 123

\bibitem[{{Greene} {et~al.}(1994){Greene}, {Wilking}, {Andr\'e}, {Young}, \&
  {Lada}}]{Greene94}
{Greene}, T.~P., {Wilking}, B.~A., {Andr\'e}, P., {Young}, E.~T., \& {Lada},
  C.~J. 1994, \apj, 434, 614

\bibitem[{{Gueth} {et~al.}(1996){Gueth}, {Guilloteau}, \&
  {Bachiller}}]{Gueth96}
{Gueth}, F., {Guilloteau}, S., \& {Bachiller}, R. 1996, \aap, 307, 891

\bibitem[{{G{\"u}sten} {et~al.}(2008){G{\"u}sten}, {Baryshev}, {Bell},
  {Belloche}, {Graf}, {Hafok}, {Heyminck}, {Hochg{\"u}rtel}, {Honingh},
  {Jacobs}, {Kasemann}, {Klein}, {Klein}, {Korn}, {Kr{\"a}mer}, {Leinz},
  {Lundgren}, {Menten}, {Meyer}, {Muders}, {Pacek}, {Rabanus}, {Sch{\"a}fer},
  {Schilke}, {Schneider}, {Stutzki}, {Wieching}, {Wunsch}, \&
  {Wyrowski}}]{Guesten08}
{G{\"u}sten}, R., {Baryshev}, A., {Bell}, A., {et~al.} 2008, in SPIE Conference
  Series, Vol. 7020

\bibitem[{{Hansen} {et~al.}(2012){Hansen}, {Klein}, {McKee}, \&
  {Fisher}}]{Hansen12}
{Hansen}, C.~E., {Klein}, R.~I., {McKee}, C.~F., \& {Fisher}, R.~T. 2012, \apj,
  747, 22

\bibitem[{{Hatchell} {et~al.}(2007){Hatchell}, {Fuller}, \&
  {Richer}}]{Hatchell07}
{Hatchell}, J., {Fuller}, G.~A., \& {Richer}, J.~S. 2007, \aap, 472, 187

\bibitem[{{Hatchell} {et~al.}(2013){Hatchell}, {Wilson}, {Drabek}, {Curtis},
  {Richer}, {Nutter}, {Di Francesco}, {Ward-Thompson}, \& {JCMT GBS
  Consortium}}]{Hatchell13}
{Hatchell}, J., {Wilson}, T., {Drabek}, E., {et~al.} 2013, \mnras, 429, L10

\bibitem[{{Herczeg} {et~al.}(2012){Herczeg}, {Karska}, {Bruderer},
  {Kristensen}, {van Dishoeck}, {J{\o}rgensen}, {Visser}, {Wampfler}, {Bergin},
  {Y{\i}ld{\i}z}, {Pontoppidan}, \& {Gracia-Carpio}}]{Herczeg12}
{Herczeg}, G.~J., {Karska}, A., {Bruderer}, S., {et~al.} 2012, \aap, 540, A84

\bibitem[{{Hogerheijde} \& {van der Tak}(2000)}]{Hogerheijde00}
{Hogerheijde}, M.~R. \& {van der Tak}, F.~F.~S. 2000, \aap, 362, 697

\bibitem[{{Hogerheijde} {et~al.}(1997){Hogerheijde}, {van Dishoeck}, {Blake},
  \& {van Langevelde}}]{Hogerheijde97}
{Hogerheijde}, M.~R., {van Dishoeck}, E.~F., {Blake}, G.~A., \& {van
  Langevelde}, H.~J. 1997, \apj, 489, 293

\bibitem[{{Hogerheijde} {et~al.}(1998){Hogerheijde}, {van Dishoeck}, {Blake},
  \& {van Langevelde}}]{Hogerheijde98}
{Hogerheijde}, M.~R., {van Dishoeck}, E.~F., {Blake}, G.~A., \& {van
  Langevelde}, H.~J. 1998, \apj, 502, 315

\bibitem[{{Johnstone} {et~al.}(2013){Johnstone}, {Hendricks}, {Herczeg}, \&
  {Bruderer}}]{Johnstone13}
{Johnstone}, D., {Hendricks}, B., {Herczeg}, G.~J., \& {Bruderer}, S. 2013,
  \apj, 765, 133

\bibitem[{{J{\o}rgensen} {et~al.}(2002){J{\o}rgensen}, {Sch{\"o}ier}, \& {van
  Dishoeck}}]{Jorgensen02}
{J{\o}rgensen}, J.~K., {Sch{\"o}ier}, F.~L., \& {van Dishoeck}, E.~F. 2002,
  \aap, 389, 908

\bibitem[{{J{\o}rgensen} {et~al.}(2004){J{\o}rgensen}, {Sch{\"o}ier}, \& {van
  Dishoeck}}]{Jorgensen04}
{J{\o}rgensen}, J.~K., {Sch{\"o}ier}, F.~L., \& {van Dishoeck}, E.~F. 2004,
  \aap, 416, 603

\bibitem[{{Karska} {et~al.}(2013){Karska}, {Herczeg}, {van Dishoeck},
  {Wampfler}, {Kristensen}, {Goicoechea}, {Visser}, {Nisini}, {San
  Jos{\'e}-Garc{\'{\i}}a}, {Bruderer}, {{\'S}niady}, {Doty}, {Fedele},
  {Y{\i}ld{\i}z}, {Benz}, {Bergin}, {Caselli}, {Herpin}, {Hogerheijde},
  {Johnstone}, {J{\o}rgensen}, {Liseau}, {Tafalla}, {van der Tak}, \&
  {Wyrowski}}]{Karska13}
{Karska}, A., {Herczeg}, G.~J., {van Dishoeck}, E.~F., {et~al.} 2013, \aap,
  552, A141

\bibitem[{{Kasemann} {et~al.}(2006){Kasemann}, {G{\"u}sten}, {Heyminck},
  {Klein}, {Klein}, {Philipp}, {Korn}, {Schneider}, {Henseler}, {Baryshev}, \&
  {Klapwijk}}]{Kasemann06}
{Kasemann}, C., {G{\"u}sten}, R., {Heyminck}, S., {et~al.} 2006, in SPIE
  Conference Series, Vol. 6275

\bibitem[{{Kauffmann} {et~al.}(2008){Kauffmann}, {Bertoldi}, {Bourke}, {Evans},
  \& {Lee}}]{Kauffmann08}
{Kauffmann}, J., {Bertoldi}, F., {Bourke}, T.~L., {Evans}, II, N.~J., \& {Lee},
  C.~W. 2008, \aap, 487, 993

\bibitem[{{Kenyon} {et~al.}(1990){Kenyon}, {Hartmann}, {Strom}, \&
  {Strom}}]{Kenyon90}
{Kenyon}, S.~J., {Hartmann}, L.~W., {Strom}, K.~M., \& {Strom}, S.~E. 1990,
  \aj, 99, 869

\bibitem[{{Klein} {et~al.}(2006){Klein}, {Philipp}, {Kr{\"a}mer}, {Kasemann},
  {G{\"u}sten}, \& {Menten}}]{Klein06}
{Klein}, B., {Philipp}, S.~D., {Kr{\"a}mer}, I., {et~al.} 2006, \aap, 454, L29

\bibitem[{{Kristensen} {et~al.}(2012){Kristensen}, {van Dishoeck}, {Bergin},
  {Visser}, {Y{\i}ld{\i}z}, {San Jose-Garcia}, {J{\o}rgensen}, {Herczeg},
  {Johnstone}, {Wampfler}, {Benz}, {Bruderer}, {Cabrit}, {Caselli}, {Doty},
  {Harsono}, {Herpin}, {Hogerheijde}, {Karska}, {van Kempen}, {Liseau},
  {Nisini}, {Tafalla}, {van der Tak}, \& {Wyrowski}}]{Kristensen12}
{Kristensen}, L.~E., {van Dishoeck}, E.~F., {Bergin}, E.~A., {et~al.} 2012,
  \aap, 542, A8

\bibitem[{{Kristensen} {et~al.}(2010){Kristensen}, {Visser}, {van Dishoeck},
  {Y{\i}ld{\i}z}, {Doty}, {Herczeg}, {Liu}, {Parise}, {J{\o}rgensen}, {van
  Kempen}, {Brinch}, {Wampfler}, {Bruderer}, {Benz}, {Hogerheijde}, {Deul},
  {Bachiller}, {Baudry}, {Benedettini}, {Bergin}, {Bjerkeli}, {Blake},
  {Bontemps}, {Braine}, {Caselli}, {Cernicharo}, {Codella}, {Daniel}, {de
  Graauw}, {di Giorgio}, {Dominik}, {Encrenaz}, {Fich}, {Fuente}, {Giannini},
  {Goicoechea}, {Helmich}, {Herpin}, {Jacq}, {Johnstone}, {Kaufman}, {Larsson},
  {Lis}, {Liseau}, {Marseille}, {McCoey}, {Melnick}, {Neufeld}, {Nisini},
  {Olberg}, {Pearson}, {Plume}, {Risacher}, {Santiago-Garc{\'{\i}}a},
  {Saraceno}, {Shipman}, {Tafalla}, {Tielens}, {van der Tak}, {Wyrowski},
  {Beintema}, {de Jonge}, {Dieleman}, {Ossenkopf}, {Roelfsema}, {Stutzki}, \&
  {Whyborn}}]{Kristensen10hifi}
{Kristensen}, L.~E., {Visser}, R., {van Dishoeck}, E.~F., {et~al.} 2010, \aap,
  521, L30

\bibitem[{{Lada}(1987)}]{Lada87}
{Lada}, C.~J. 1987, in IAU Symposium, Vol. 115, Star Forming Regions, ed.
  M.~{Peimbert} \& J.~{Jugaku}, 1--17

\bibitem[{{Langer} \& {Penzias}(1990)}]{LangerPenzias90}
{Langer}, W.~D. \& {Penzias}, A.~A. 1990, \apj, 357, 477

\bibitem[{{Lee} {et~al.}(2000){Lee}, {Mundy}, {Reipurth}, {Ostriker}, \&
  {Stone}}]{lee00}
{Lee}, C.-F., {Mundy}, L.~G., {Reipurth}, B., {Ostriker}, E.~C., \& {Stone},
  J.~M. 2000, \apj, 542, 925

\bibitem[{{Lee} {et~al.}(2002){Lee}, {Mundy}, {Stone}, \& {Ostriker}}]{Lee02}
{Lee}, C.-F., {Mundy}, L.~G., {Stone}, J.~M., \& {Ostriker}, E.~C. 2002, \apj,
  576, 294

\bibitem[{{Lefloch} {et~al.}(2010){Lefloch}, {Cabrit}, {Codella}, {Melnick},
  {Cernicharo}, {Caux}, {Benedettini}, {Boogert}, {Caselli}, {Ceccarelli},
  {Gueth}, {Hily-Blant}, {Lorenzani}, {Neufeld}, {Nisini}, {Pacheco}, {Pagani},
  {Pardo}, {Parise}, {Salez}, {Schuster}, {Viti}, {Bacmann}, {Baudry}, {Bell},
  {Bergin}, {Blake}, {Bottinelli}, {Castets}, {Comito}, {Coutens}, {Crimier},
  {Dominik}, {Demyk}, {Encrenaz}, {Falgarone}, {Fuente}, {Gerin}, {Goldsmith},
  {Helmich}, {Hennebelle}, {Henning}, {Herbst}, {Jacq}, {Kahane}, {Kama},
  {Klotz}, {Langer}, {Lis}, {Lord}, {Maret}, {Pearson}, {Phillips}, {Saraceno},
  {Schilke}, {Tielens}, {van der Tak}, {van der Wiel}, {Vastel}, {Wakelam},
  {Walters}, {Wyrowski}, {Yorke}, {Bachiller}, {Borys}, {de Lange}, {Delorme},
  {Kramer}, {Larsson}, {Lai}, {Maiwald}, {Martin-Pintado}, {Mehdi},
  {Ossenkopf}, {Siegel}, {Stutzki}, \& {Wunsch}}]{LeFloch10}
{Lefloch}, B., {Cabrit}, S., {Codella}, C., {et~al.} 2010, \aap, 518, L113

\bibitem[{{Lommen} {et~al.}(2008){Lommen}, {J{\o}rgensen}, {van Dishoeck}, \&
  {Crapsi}}]{Lommen08}
{Lommen}, D., {J{\o}rgensen}, J.~K., {van Dishoeck}, E.~F., \& {Crapsi}, A.
  2008, \aap, 481, 141

\bibitem[{{Manoj} {et~al.}(2013){Manoj}, {Watson}, {Neufeld}, {Megeath},
  {Vavrek}, {Yu}, {Visser}, {Bergin}, {Fischer}, {Tobin}, {Stutz}, {Ali},
  {Wilson}, {Di Francesco}, {Osorio}, {Maret}, \& {Poteet}}]{Manoj13}
{Manoj}, P., {Watson}, D.~M., {Neufeld}, D.~A., {et~al.} 2013, \apj, 763, 83

\bibitem[{{Micono} {et~al.}(1998){Micono}, {Davis}, {Ray}, {Eisloeffel}, \&
  {Shetrone}}]{Micono98}
{Micono}, M., {Davis}, C.~J., {Ray}, T.~P., {Eisloeffel}, J., \& {Shetrone},
  M.~D. 1998, \apjl, 494, L227

\bibitem[{{Nisini} {et~al.}(2010){Nisini}, {Benedettini}, {Codella},
  {Giannini}, {Liseau}, {Neufeld}, {Tafalla}, {van Dishoeck}, {Bachiller},
  {Baudry}, {Benz}, {Bergin}, {Bjerkeli}, {Blake}, {Bontemps}, {Braine},
  {Bruderer}, {Caselli}, {Cernicharo}, {Daniel}, {Encrenaz}, {di Giorgio},
  {Dominik}, {Doty}, {Fich}, {Fuente}, {Goicoechea}, {de Graauw}, {Helmich},
  {Herczeg}, {Herpin}, {Hogerheijde}, {Jacq}, {Johnstone}, {J{\o}rgensen},
  {Kaufman}, {Kristensen}, {Larsson}, {Lis}, {Marseille}, {McCoey}, {Melnick},
  {Olberg}, {Parise}, {Pearson}, {Plume}, {Risacher}, {Santiago}, {Saraceno},
  {Shipman}, {van Kempen}, {Visser}, {Viti}, {Wampfler}, {Wyrowski}, {van der
  Tak}, {Y{\i}ld{\i}z}, {Delforge}, {Desbat}, {Hatch}, {P{\'e}ron}, {Schieder},
  {Stern}, {Teyssier}, \& {Whyborn}}]{Nisini10}
{Nisini}, B., {Benedettini}, M., {Codella}, C., {et~al.} 2010, \aap, 518, L120

\bibitem[{{Nisini} {et~al.}(2013){Nisini}, {Santangelo}, {Antoniucci},
  {Benedettini}, {Codella}, {Giannini}, {Lorenzani}, {Liseau}, {Tafalla},
  {Bjerkeli}, {Cabrit}, {Caselli}, {Kristensen}, {Neufeld}, {Melnick}, \& {van
  Dishoeck}}]{Nisini13}
{Nisini}, B., {Santangelo}, G., {Antoniucci}, S., {et~al.} 2013, \aap, 549, A16

\bibitem[{{Offner} {et~al.}(2009){Offner}, {Klein}, {McKee}, \&
  {Krumholz}}]{Offner09}
{Offner}, S.~S.~R., {Klein}, R.~I., {McKee}, C.~F., \& {Krumholz}, M.~R. 2009,
  \apj, 703, 131

\bibitem[{{Offner} {et~al.}(2010){Offner}, {Kratter}, {Matzner}, {Krumholz}, \&
  {Klein}}]{Offner10}
{Offner}, S.~S.~R., {Kratter}, K.~M., {Matzner}, C.~D., {Krumholz}, M.~R., \&
  {Klein}, R.~I. 2010, \apj, 725, 1485

\bibitem[{{Oya} {et~al.}(2014){Oya}, {Sakai}, {Sakai}, {Watanabe}, {Hirota},
  {Lindberg}, {Bisschop}, {J{\o}rgensen}, {van Dishoeck}, \&
  {Yamamoto}}]{oya14}
{Oya}, Y., {Sakai}, N., {Sakai}, T., {et~al.} 2014, \apj, 795, 152

\bibitem[{{Parise} {et~al.}(2006){Parise}, {Belloche}, {Leurini}, {Schilke},
  {Wyrowski}, \& {G{\"u}sten}}]{Parise06}
{Parise}, B., {Belloche}, A., {Leurini}, S., {et~al.} 2006, \aap, 454, L79

\bibitem[{{Pilbratt} {et~al.}(2010){Pilbratt}, {Riedinger}, {Passvogel},
  {Crone}, {Doyle}, {Gageur}, {Heras}, {Jewell}, {Metcalfe}, {Ott}, \&
  {Schmidt}}]{Pilbratt10}
{Pilbratt}, G.~L., {Riedinger}, J.~R., {Passvogel}, T., {et~al.} 2010, \aap,
  518, L1

\bibitem[{{Robitaille} {et~al.}(2006){Robitaille}, {Whitney}, {Indebetouw},
  {Wood}, \& {Denzmore}}]{Robitaille06}
{Robitaille}, T.~P., {Whitney}, B.~A., {Indebetouw}, R., {Wood}, K., \&
  {Denzmore}, P. 2006, \apjs, 167, 256

\bibitem[{{Santangelo} {et~al.}(2013){Santangelo}, {Nisini}, {Antoniucci},
  {Codella}, {Cabrit}, {Giannini}, {Herczeg}, {Liseau}, {Tafalla}, \& {van
  Dishoeck}}]{Santangelo13}
{Santangelo}, G., {Nisini}, B., {Antoniucci}, S., {et~al.} 2013, \aap, 557, A22

\bibitem[{{Santangelo} {et~al.}(2012){Santangelo}, {Nisini}, {Giannini},
  {Antoniucci}, {Vasta}, {Codella}, {Lorenzani}, {Tafalla}, {Liseau}, {van
  Dishoeck}, \& {Kristensen}}]{Santangelo12}
{Santangelo}, G., {Nisini}, B., {Giannini}, T., {et~al.} 2012, \aap, 538, A45

\bibitem[{{Saraceno} {et~al.}(1996){Saraceno}, {Andre}, {Ceccarelli},
  {Griffin}, \& {Molinari}}]{Saraceno96}
{Saraceno}, P., {Andre}, P., {Ceccarelli}, C., {Griffin}, M., \& {Molinari}, S.
  1996, \aap, 309, 827

\bibitem[{{Sicilia-Aguilar} {et~al.}(2013){Sicilia-Aguilar}, {Henning}, {Linz},
  {Andr{\'e}}, {Stutz}, {Eiroa}, \& {White}}]{Sicilia-Aguilar13}
{Sicilia-Aguilar}, A., {Henning}, T., {Linz}, H., {et~al.} 2013, \aap, 551, A34

\bibitem[{{Snell} {et~al.}(1980){Snell}, {Loren}, \& {Plambeck}}]{Snell80}
{Snell}, R.~L., {Loren}, R.~B., \& {Plambeck}, R.~L. 1980, \apjl, 239, L17

\bibitem[{{Spaans} {et~al.}(1995){Spaans}, {Hogerheijde}, {Mundy}, \& {van
  Dishoeck}}]{Spaans95}
{Spaans}, M., {Hogerheijde}, M.~R., {Mundy}, L.~G., \& {van Dishoeck}, E.~F.
  1995, \apjl, 455, L167

\bibitem[{{Tafalla} {et~al.}(2013){Tafalla}, {Liseau}, {Nisini}, {Bachiller},
  {Santiago-Garc{\'{\i}}a}, {van Dishoeck}, {Kristensen}, {Herczeg}, \&
  {Y{\i}ld{\i}z}}]{Tafalla13}
{Tafalla}, M., {Liseau}, R., {Nisini}, B., {et~al.} 2013, \aap, 551, A116

\bibitem[{{Tafalla} {et~al.}(2000){Tafalla}, {Myers}, {Mardones}, \&
  {Bachiller}}]{Tafalla00}
{Tafalla}, M., {Myers}, P.~C., {Mardones}, D., \& {Bachiller}, R. 2000, \aap,
  359, 967

\bibitem[{{Tobin} {et~al.}(2008){Tobin}, {Hartmann}, {Calvet}, \&
  {D'Alessio}}]{Tobin08}
{Tobin}, J.~J., {Hartmann}, L., {Calvet}, N., \& {D'Alessio}, P. 2008, \apj,
  679, 1364

\bibitem[{{Tobin} {et~al.}(2010){Tobin}, {Hartmann}, {Looney}, \&
  {Chiang}}]{Tobin10}
{Tobin}, J.~J., {Hartmann}, L., {Looney}, L.~W., \& {Chiang}, H.-F. 2010, \apj,
  712, 1010

\bibitem[{{van der Marel} {et~al.}(2013){van der Marel}, {Kristensen},
  {Visser}, {Mottram}, {Y{\i}ld{\i}z}, \& {van Dishoeck}}]{vanderMarel13}
{van der Marel}, N., {Kristensen}, L.~E., {Visser}, R., {et~al.} 2013, \aap,
  556, A76

\bibitem[{{van Dishoeck} {et~al.}(2011){van Dishoeck}, {Kristensen}, {Benz},
  {Bergin}, {Caselli}, {Cernicharo}, {Herpin}, {Hogerheijde}, {Johnstone},
  {Liseau}, {Nisini}, {Shipman}, {Tafalla}, {van der Tak}, {Wyrowski},
  {Aikawa}, {Bachiller}, {Baudry}, {Benedettini}, {Bjerkeli}, {Blake},
  {Bontemps}, {Braine}, {Brinch}, {Bruderer}, {Chavarr{\'{\i}}a}, {Codella},
  {Daniel}, {de Graauw}, {Deul}, {di Giorgio}, {Dominik}, {Doty}, {Dubernet},
  {Encrenaz}, {Feuchtgruber}, {Fich}, {Frieswijk}, {Fuente}, {Giannini},
  {Goicoechea}, {Helmich}, {Herczeg}, {Jacq}, {J{\o}rgensen}, {Karska},
  {Kaufman}, {Keto}, {Larsson}, {Lefloch}, {Lis}, {Marseille}, {McCoey},
  {Melnick}, {Neufeld}, {Olberg}, {Pagani}, {Pani{\'c}}, {Parise}, {Pearson},
  {Plume}, {Risacher}, {Salter}, {Santiago-Garc{\'{\i}}a}, {Saraceno},
  {St{\"a}uber}, {van Kempen}, {Visser}, {Viti}, {Walmsley}, {Wampfler}, \&
  {Y{\i}ld{\i}z}}]{vanDishoeck11}
{van Dishoeck}, E.~F., {Kristensen}, L.~E., {Benz}, A.~O., {et~al.} 2011,
  \pasp, 123, 138

\bibitem[{{van Kempen} {et~al.}(2010{\natexlab{a}}){van Kempen}, {Green},
  {Evans}, {van Dishoeck}, {Kristensen}, {Herczeg}, {Mer{\'{\i}}n}, {Lee},
  {J{\o}rgensen}, {Bouwman}, {Acke}, {Adamkovics}, {Augereau}, {Bergin},
  {Blake}, {Brown}, {Carr}, {Chen}, {Cieza}, {Dominik}, {Dullemond}, {Dunham},
  {Glassgold}, {G{\"u}del}, {Harvey}, {Henning}, {Hogerheijde}, {Jaffe}, {Kim},
  {Knez}, {Lacy}, {Maret}, {Meeus}, {Meijerink}, {Mulders}, {Mundy}, {Najita},
  {Olofsson}, {Pontoppidan}, {Salyk}, {Sturm}, {Visser}, {Waters}, {Waelkens},
  \& {Y{\i}ld{\i}z}}]{vanKempen10dkcha}
{van Kempen}, T.~A., {Green}, J.~D., {Evans}, N.~J., {et~al.}
  2010{\natexlab{a}}, \aap, 518, L128

\bibitem[{{van Kempen} {et~al.}(2010{\natexlab{b}}){van Kempen}, {Kristensen},
  {Herczeg}, {Visser}, {van Dishoeck}, {Wampfler}, {Bruderer}, {Benz}, {Doty},
  {Brinch}, {Hogerheijde}, {J{\o}rgensen}, {Tafalla}, {Neufeld}, {Bachiller},
  {Baudry}, {Benedettini}, {Bergin}, {Bjerkeli}, {Blake}, {Bontemps}, {Braine},
  {Caselli}, {Cernicharo}, {Codella}, {Daniel}, {di Giorgio}, {Dominik},
  {Encrenaz}, {Fich}, {Fuente}, {Giannini}, {Goicoechea}, {de Graauw},
  {Helmich}, {Herpin}, {Jacq}, {Johnstone}, {Kaufman}, {Larsson}, {Lis},
  {Liseau}, {Marseille}, {McCoey}, {Melnick}, {Nisini}, {Olberg}, {Parise},
  {Pearson}, {Plume}, {Risacher}, {Santiago-Garc{\'{\i}}a}, {Saraceno},
  {Shipman}, {van der Tak}, {Wyrowski}, {Y{\i}ld{\i}z}, {Ciechanowicz},
  {Dubbeldam}, {Glenz}, {Huisman}, {Lin}, {Morris}, {Murphy}, \&
  {Trappe}}]{vanKempen10hh46}
{van Kempen}, T.~A., {Kristensen}, L.~E., {Herczeg}, G.~J., {et~al.}
  2010{\natexlab{b}}, \aap, 518, L121

\bibitem[{{van Kempen} {et~al.}(2009{\natexlab{a}}){van Kempen}, {van
  Dishoeck}, {G{\"u}sten}, {Kristensen}, {Schilke}, {Hogerheijde}, {Boland},
  {Menten}, \& {Wyrowski}}]{vanKempen09champ2}
{van Kempen}, T.~A., {van Dishoeck}, E.~F., {G{\"u}sten}, R., {et~al.}
  2009{\natexlab{a}}, \aap, 507, 1425

\bibitem[{{van Kempen} {et~al.}(2009{\natexlab{b}}){van Kempen}, {van
  Dishoeck}, {G{\"u}sten}, {Kristensen}, {Schilke}, {Hogerheijde}, {Boland},
  {Nefs}, {Menten}, {Baryshev}, \& {Wyrowski}}]{vanKempen09champ}
{van Kempen}, T.~A., {van Dishoeck}, E.~F., {G{\"u}sten}, R., {et~al.}
  2009{\natexlab{b}}, \aap, 501, 633

\bibitem[{{van Kempen} {et~al.}(2009{\natexlab{c}}){van Kempen}, {van
  Dishoeck}, {Hogerheijde}, \& {G{\"u}sten}}]{vanKempen09_southc+}
{van Kempen}, T.~A., {van Dishoeck}, E.~F., {Hogerheijde}, M.~R., \&
  {G{\"u}sten}, R. 2009{\natexlab{c}}, \aap, 508, 259

\bibitem[{{Vasta} {et~al.}(2012){Vasta}, {Codella}, {Lorenzani}, {Santangelo},
  {Nisini}, {Giannini}, {Tafalla}, {Liseau}, {van Dishoeck}, \&
  {Kristensen}}]{Vasta12}
{Vasta}, M., {Codella}, C., {Lorenzani}, A., {et~al.} 2012, \aap, 537, A98

\bibitem[{{Visser} {et~al.}(2012){Visser}, {Kristensen}, {Bruderer}, {van
  Dishoeck}, {Herczeg}, {Brinch}, {Doty}, {Harsono}, \& {Wolfire}}]{Visser12}
{Visser}, R., {Kristensen}, L.~E., {Bruderer}, S., {et~al.} 2012, \aap, 537,
  A55

\bibitem[{{Y{\i}ld{\i}z} {et~al.}(2012){Y{\i}ld{\i}z}, {Kristensen}, {van
  Dishoeck}, {Belloche}, {van Kempen}, {Hogerheijde}, {G{\"u}sten}, \& {van der
  Marel}}]{Yildiz12}
{Y{\i}ld{\i}z}, U.~A., {Kristensen}, L.~E., {van Dishoeck}, E.~F., {et~al.}
  2012, \aap, 542, A86

\bibitem[{{Y{\i}ld{\i}z} {et~al.}(2013){Y{\i}ld{\i}z}, {Kristensen}, {van
  Dishoeck}, {San Jos{\'e}-Garc{\'{\i}}a}, {Karska}, {Harsono}, {Tafalla},
  {Fuente}, {Visser}, {J{\o}rgensen}, \& {Hogerheijde}}]{Yildiz13hifi}
{Y{\i}ld{\i}z}, U.~A., {Kristensen}, L.~E., {van Dishoeck}, E.~F., {et~al.}
  2013, \aap, 556, A89

\bibitem[{{Y{\i}ld{\i}z} {et~al.}(2010){Y{\i}ld{\i}z}, {van Dishoeck},
  {Kristensen}, {Visser}, {J{\o}rgensen}, {Herczeg}, {van Kempen},
  {Hogerheijde}, {Doty}, {Benz}, {Bruderer}, {Wampfler}, {Deul}, {Bachiller},
  {Baudry}, {Benedettini}, {Bergin}, {Bjerkeli}, {Blake}, {Bontemps}, {Braine},
  {Caselli}, {Cernicharo}, {Codella}, {Daniel}, {di Giorgio}, {Dominik},
  {Encrenaz}, {Fich}, {Fuente}, {Giannini}, {Goicoechea}, {de Graauw},
  {Helmich}, {Herpin}, {Jacq}, {Johnstone}, {Larsson}, {Lis}, {Liseau}, {Liu},
  {Marseille}, {McCoey}, {Melnick}, {Neufeld}, {Nisini}, {Olberg}, {Parise},
  {Pearson}, {Plume}, {Risacher}, {Santiago-Garc{\'{\i}}a}, {Saraceno},
  {Shipman}, {Tafalla}, {Tielens}, {van der Tak}, {Wyrowski}, {Dieleman},
  {Jellema}, {Ossenkopf}, {Schieder}, \& {Stutzki}}]{Yildiz10}
{Y{\i}ld{\i}z}, U.~A., {van Dishoeck}, E.~F., {Kristensen}, L.~E., {et~al.}
  2010, \aap, 521, L40

\end{thebibliography}

\appendix

\Online

\section{Additional Material}

\begin{table*}[!t]
\scriptsize
\caption{Integration limits and contour levels.}
\begin{center}
\begin{tabular}{l r r r r r r r r r r r r r r r }
\hline
\hline
   & &  \multicolumn{3}{c}{Blue Lobe\tablefootmark{a}}   &  \multicolumn{3}{c}{Red Lobe\tablefootmark{a}} & \multicolumn{2}{c}{CO~6--5}  &  \multicolumn{2}{c}{CO~3--2} \\ 
\hline
Source & $\varv_{\rm LSR}$ & \vmax & $V_{\rm out,blue}$ & $V_{\rm in,blue}$ & \vmax & $V_{\rm in,red}$ & $V_{\rm out,red}$ & Lowest Cntr\tablefootmark{b} & Step Size\tablefootmark{b} &  Lowest Cntr\tablefootmark{b} & Step Size\tablefootmark{b} \\
              & [\kms] & [\kms] & [\kms] & [\kms] & [\kms] & [\kms] & [\kms] & [\Kkms] & [\Kkms] & [\Kkms] & [\Kkms]\\
\hline  
L1448mm           & $+$5.2 & 50.6 & $-$45.4 & 2.0 & 28.8 & 6.6 & 34.0 & \dots & \dots & 10 & 10\\
NGC1333-IRAS~2A            & $+$7.7 & 23.2 & $-$15.5 & 7.0 & 17.3 & 10.5& 25.0 & 25 & 20 & 15 & 15 \\
NGC1333-IRAS~4A            & $+$7.0 & 22.0 & $-$15.5 & 7.0 & 19.8 & 9.2 & 27.0 & 20 & 20 & 20 & 20 \\
NGC1333-IRAS~4B            & $+$7.1 & 20.0 & $-$12.8 & 4.0 & 12.8 & 9.2 & 20.0 & 20 & 20 & 20 & 20 \\
L1527             & $+$5.9 & 7.4  & $-$1.5  & 4.5 & 8.1  & 7.0 & 14.0 & 7       & 4       & 5        & 3 \\
Ced110-IRS4\tablefootmark{c} & $+$4.2 & 4.2  & 0.0 & 3.5 & 3.8  & 5.5 & 8.0 & 8 & 5 & 5 & 3 \\
BHR71\tablefootmark{c}      & $-$4.4 & 15.5 & $-$20.0 & $-$6.0 & 22.4 & $-$3.9 & 18.0 & 20 & 20 & \dots & \dots \\
IRAS~15398         & $+$5.1 & 8.6  & $-$3.5 & 4.0  & 9.9  & 6.5 & 15.0 & 5 & 5 & 3 & 3 \\
L483MM            & $+$5.2 & 10.7 & $-$5.5 & 3.5  & 10.8 & 6.0 & 16.0 & 8 & 8 & 5 & 8 \\
Ser-SMM1              & $+$8.5 & 19.0 & $-$10.5 & 6.0 & 22.5 &10.5 & 31.0 & 15 & 20 & 30 & 25 \\
Ser-SMM4              & $+$8.0 & 19.0 & $-$10.5 & 6.0 & 11.5 &10.5 & 20.0 & 15 & 20 & 30 & 25 \\
Ser-SMM3              & $+$7.6 & 22.0 & $-$13.5 & 6.0 & 13.5 &10.5 & 22.0 & 15 & 20 & 30 & 25 \\
L723MM            & $+$11.2 & 14.2 & $-$3.0  & 9.0 & 10.8 &12.0 & 26.0 & 15 & 10 & 5 & 5 \\
B335              & $+$8.4 & 8.9  & $-$0.5 & 7.0  & 8.6  & 9.5 & 17.0 & 5 & 5 & \dots & \dots \\
L1157             & $+$2.6 & 12.4 & $-$9.8 & 1.5  & 17.4 & 3.7 & 20.0 & \dots & \dots & 10 & 20 \\
\hline
L1489             & $+$7.2 & 13.7  & $-$6.5 & 5.0   & 7.8 & 8.5 & 15.0 & 3 & 2 & 10 & 10 \\
L1551-IRS5         & $+$6.2 & 9.7   & $-$3.5 & 4.5   &11.8 & 7.5 & 18.0 & \dots & \dots & 10 & 10 \\
TMR1              & $+$6.3 & 7.8   & $-$1.5 & 4.0   & 3.7 & 6.5 & 10.0 & 3 & 3 & 2 & 2 \\
TMC1A             & $+$6.6 & 17.1  & $-$10.5& 5.0   & 5.4 & 6.5 & 10.0 & 5 & 5 & 2 & 5 \\
TMC1              & $+$5.2 & 13.7  & $-$8.5 & 4.0   &14.8 & 6.2 & 20.0 & 4 & 5 & 3 & 3 \\
HH46-IRS              & $+$5.2 & 5.5   & $-$0.3 & 10.0  &14.8 &12.2 & 20.0 & 20 & 20 & 10 & 10 \\
DK~Cha\tablefootmark{c}  & $+$3.1 & 5.3 & $-$2.2 & 1.5 & 8.9 & 4.3 &  12.0 & 5 & 5 & 5 & 5 \\
GSS30-IRS1         & $+$3.5 & 13.0  & $-$9.5 & 1.5   &14.5 & 5.5 & 18.0 & 20 & 30 & 15 & 15 \\
Elias~29           & $+$4.3 & 9.8  & $-$5.5 & 1.5   &10.7 & 7.0 & 15.0 & 15 & 10 & 7.5 & 5.0 \\
Oph-IRS63          & $+$2.8 & 11.3  & $-$8.5&  1.0 & 3.2 & 4.0 & 6.0 & 3 & 1.5 & 2 & 1 \\
RNO91             & $+$0.5 & 15.0  & $-$14.5 &$-$1.0 & 3.5 & 1.0 & 4.0  & 3 & 3 & 3 & 3 \\
\hline
\end{tabular}
\end{center}
\tablefoot{Velocities are not corrected for inclination; $\varv_{\rm LSR}$ values are from \citet{Yildiz13hifi}.
\tablefoottext{a}{Velocity integration limits as shown in Fig.~\ref{fig:AllBCR}.} 
\tablefoottext{b}{Contour levels are given in absolute intensities.}
\tablefoottext{c}{Obtained from \twco\ 6--5.}
}
\label{tbl:contourlevels}
\end{table*}


\begin{figure*}[htb]
\begin{minipage}{18cm}
\begin{center}
    \includegraphics[scale=0.157]{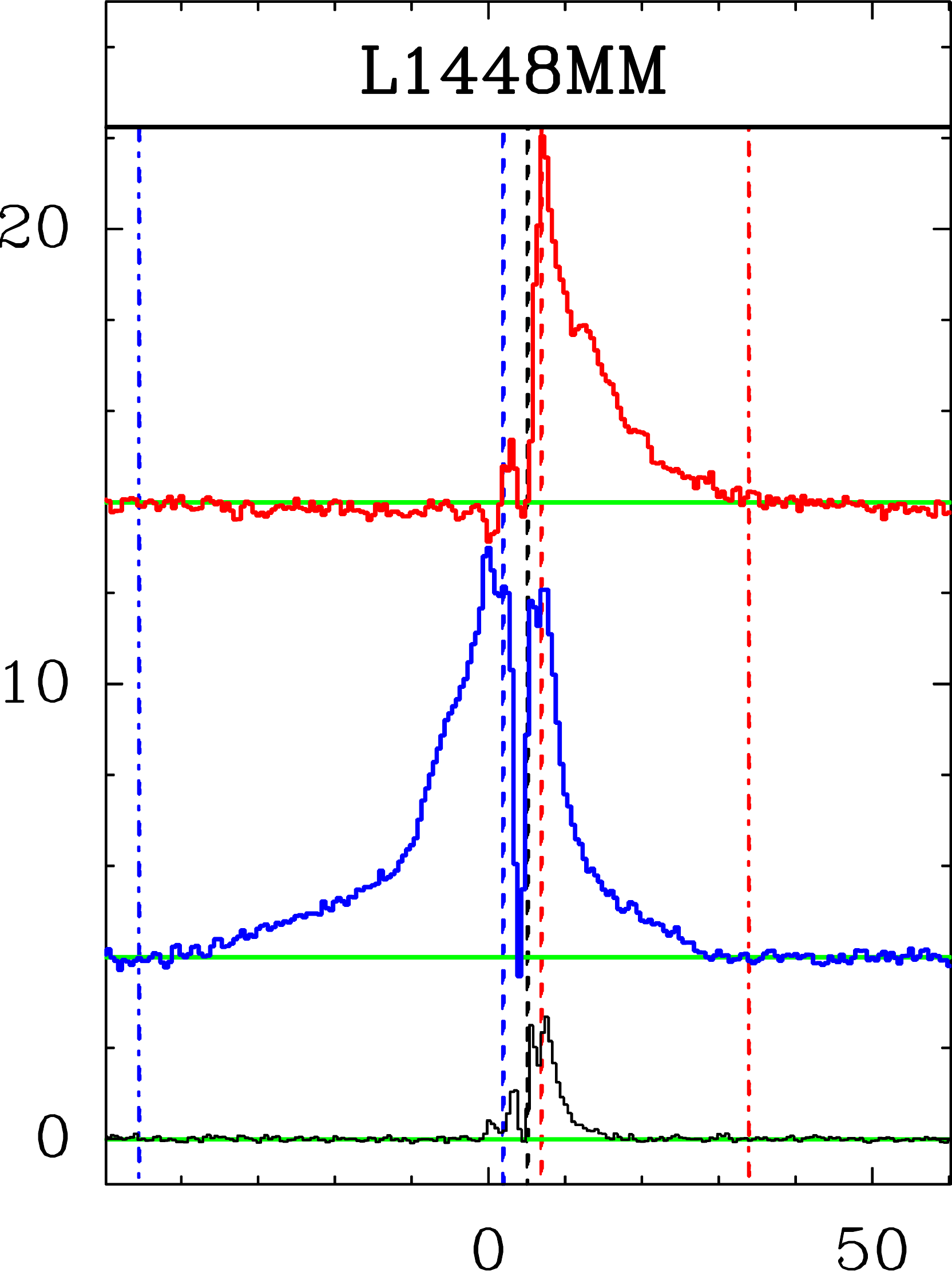}
    \includegraphics[scale=0.157]{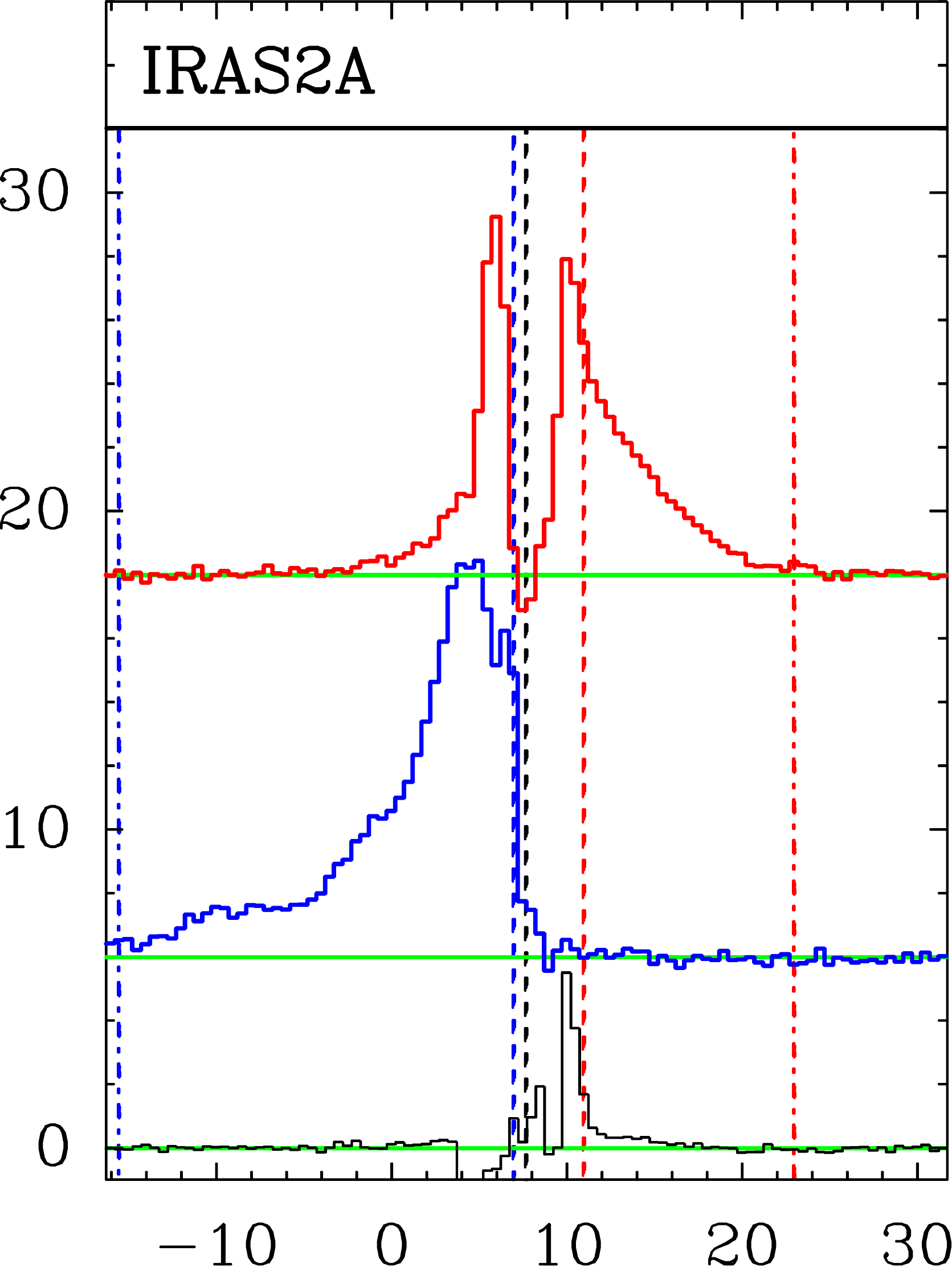}
    \includegraphics[scale=0.157]{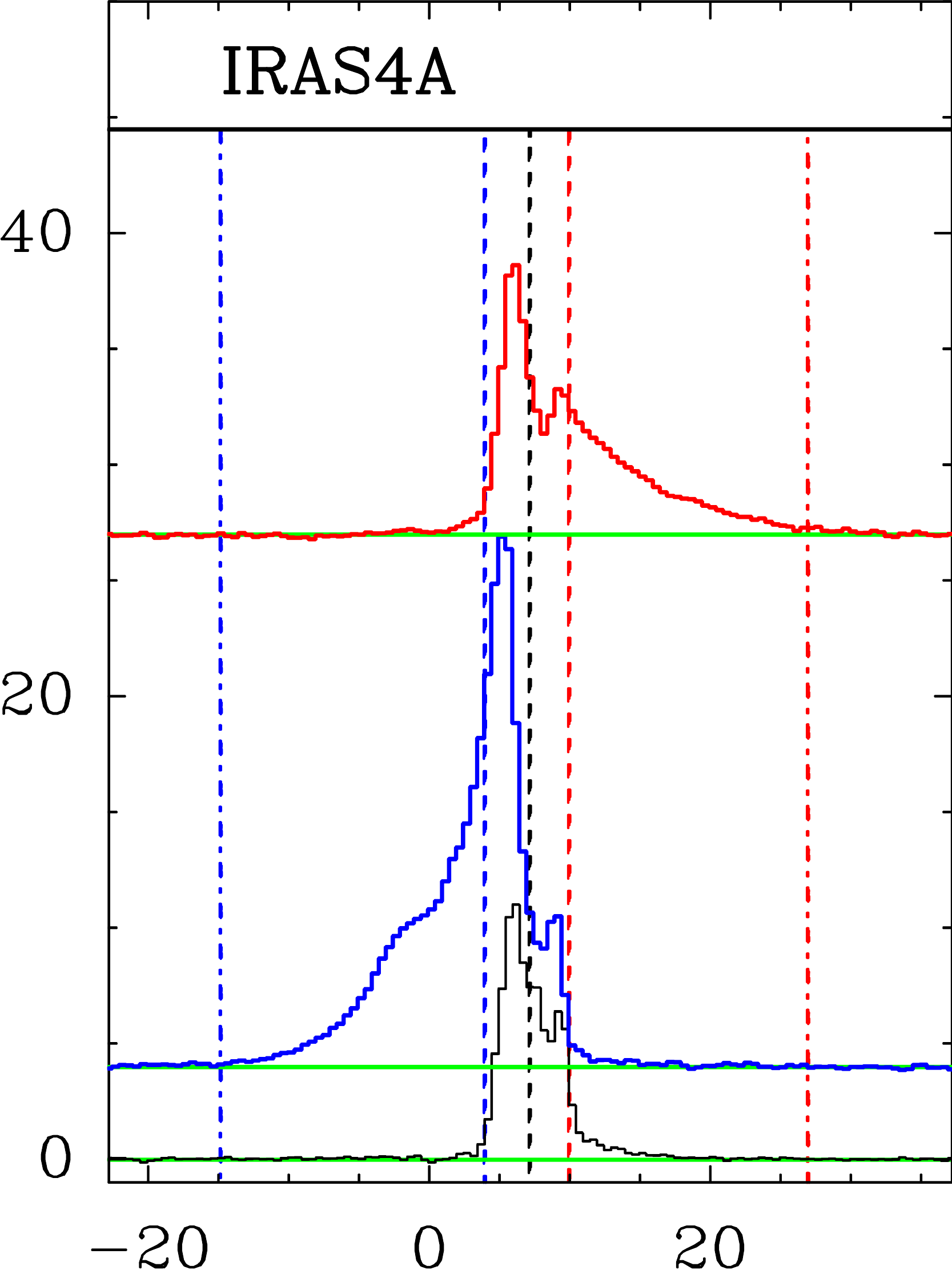}
    \includegraphics[scale=0.157]{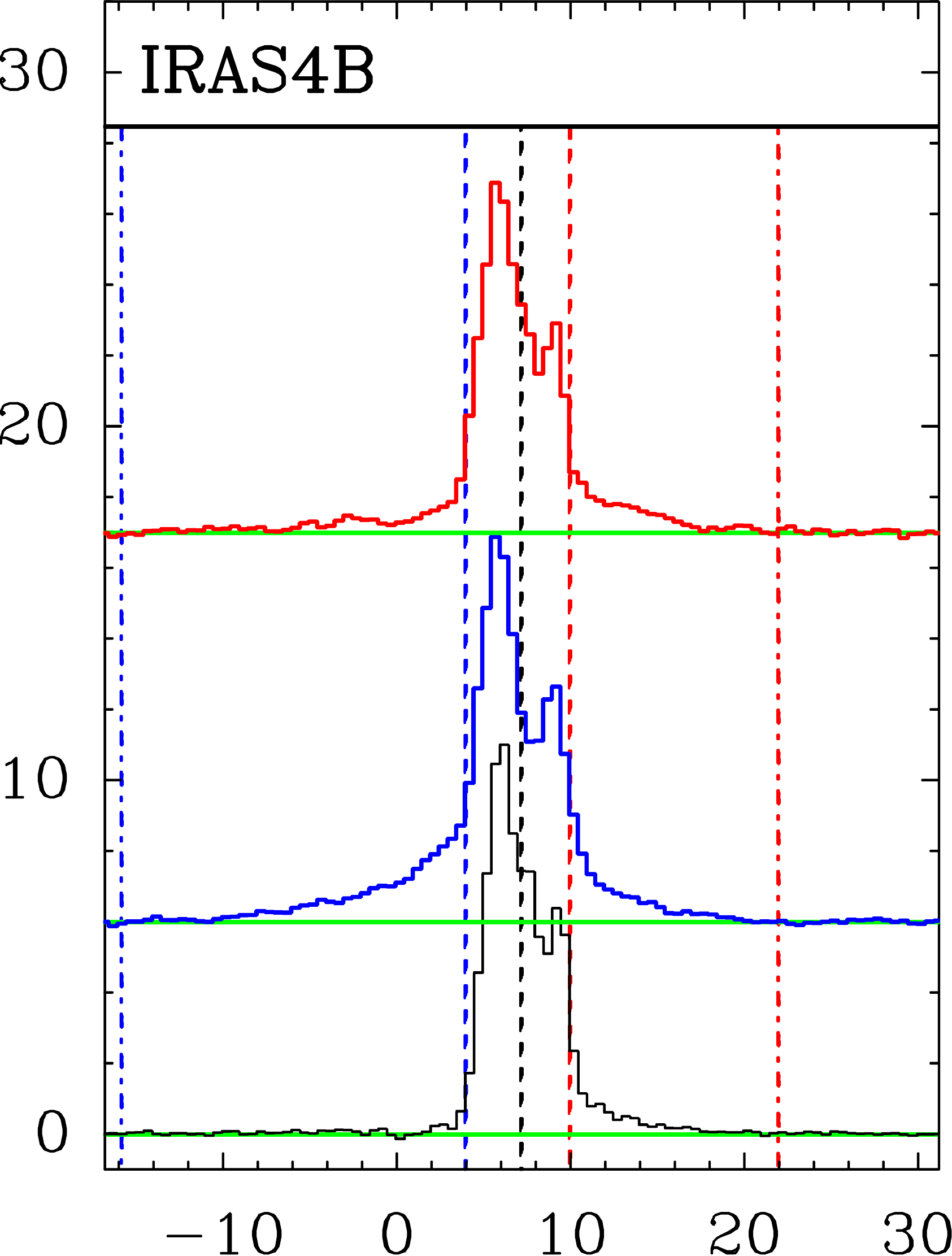}
    \includegraphics[scale=0.157]{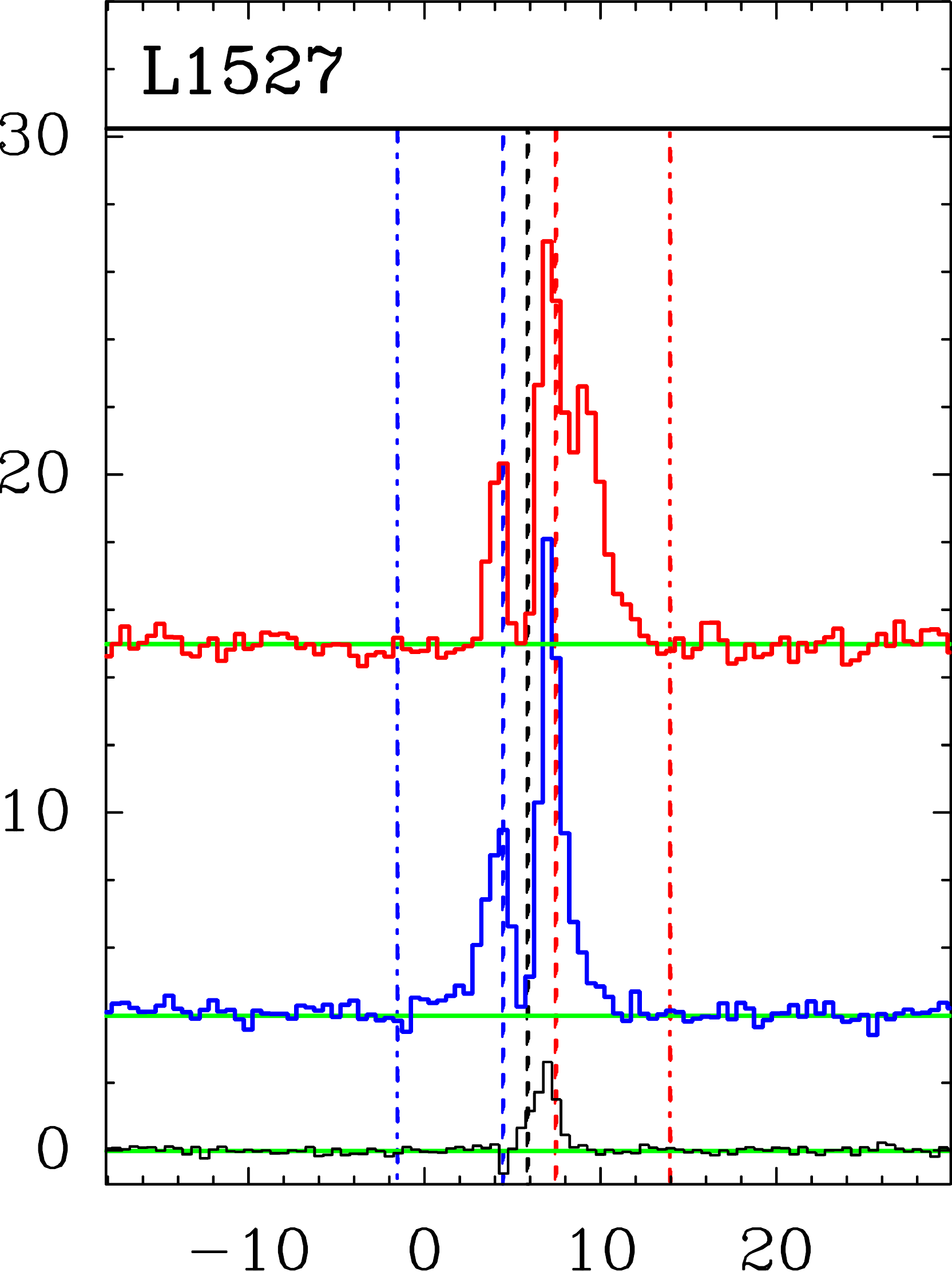}
    \includegraphics[scale=0.157]{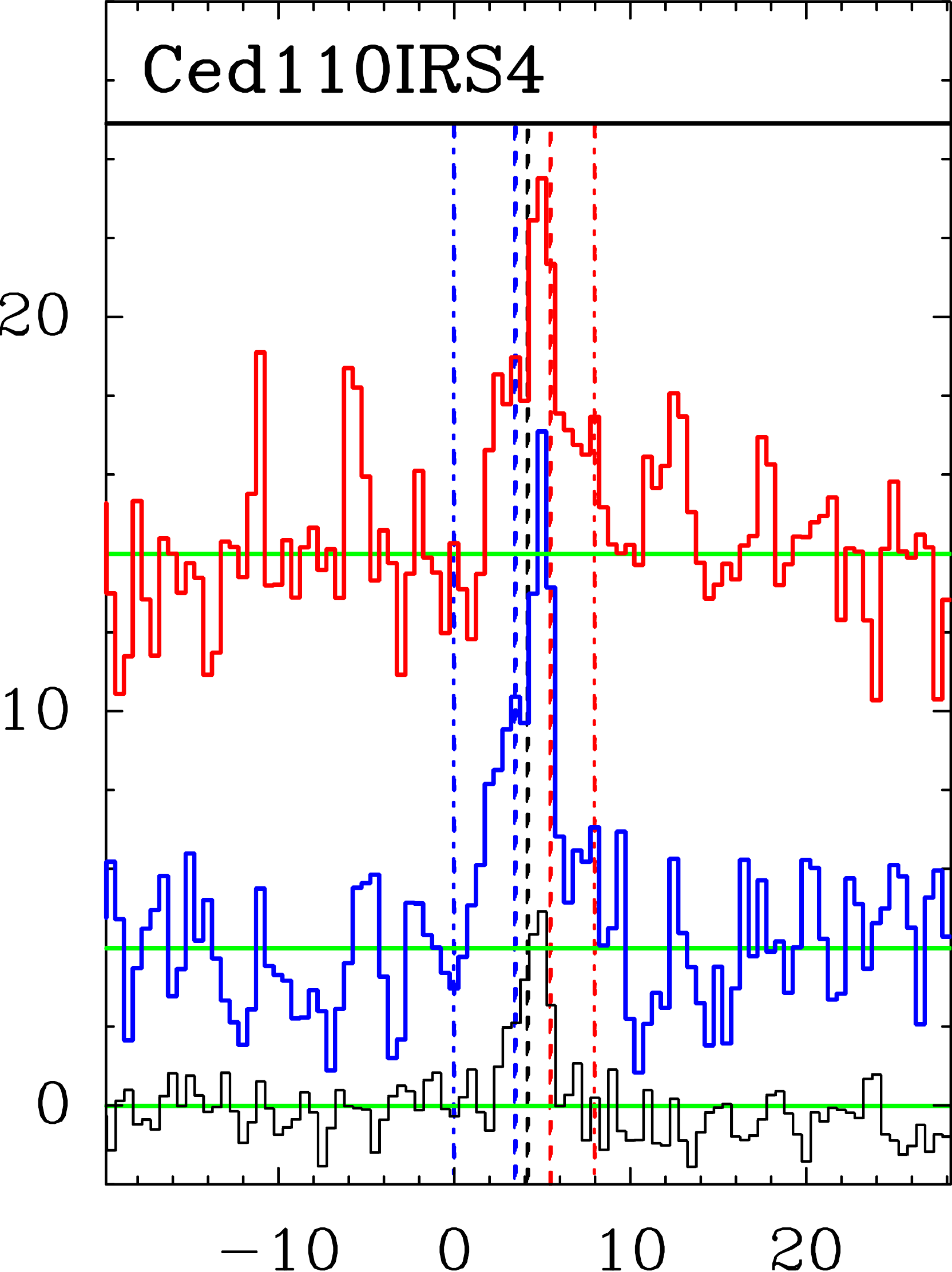}
    \includegraphics[scale=0.157]{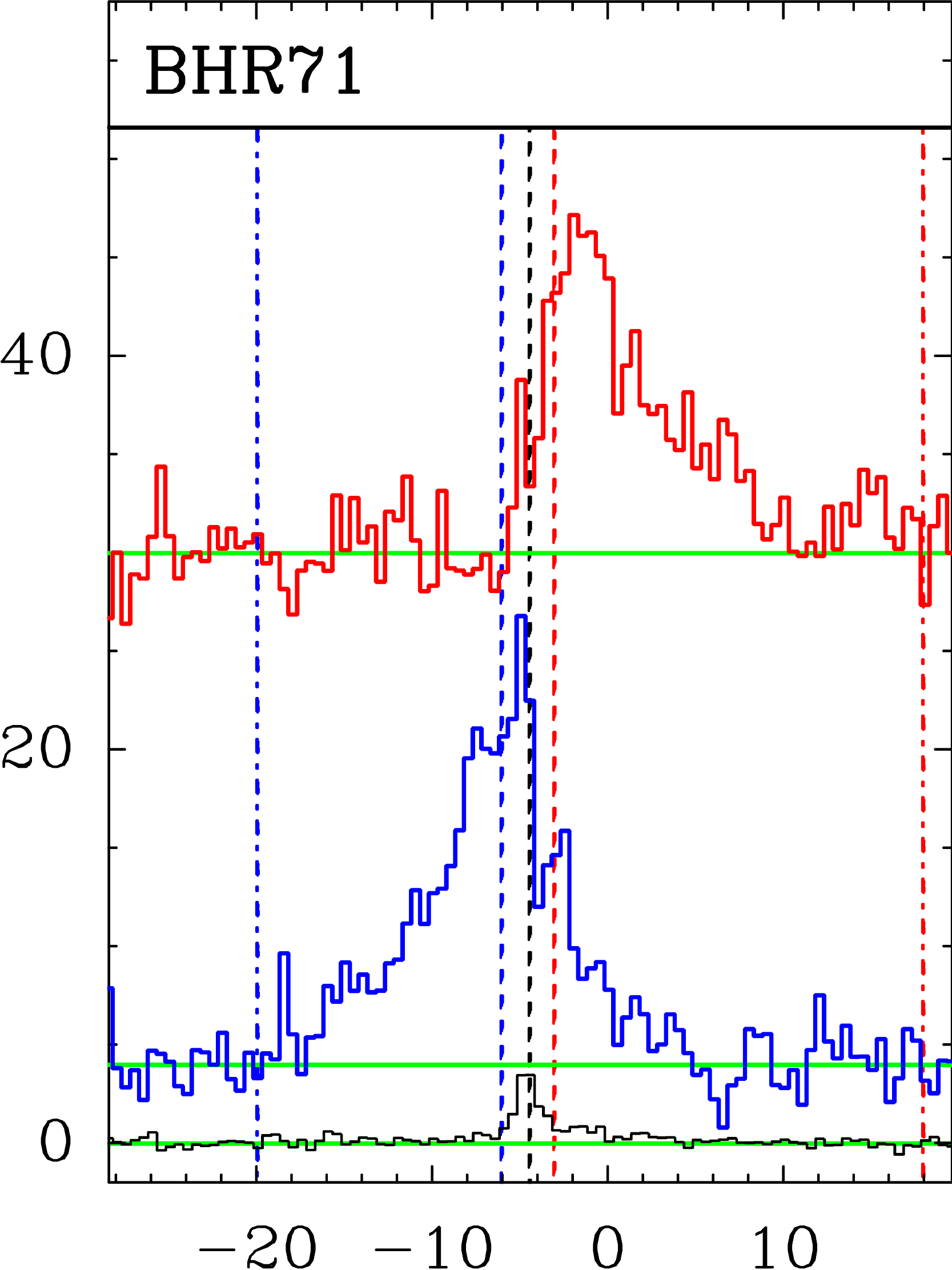}
    \includegraphics[scale=0.157]{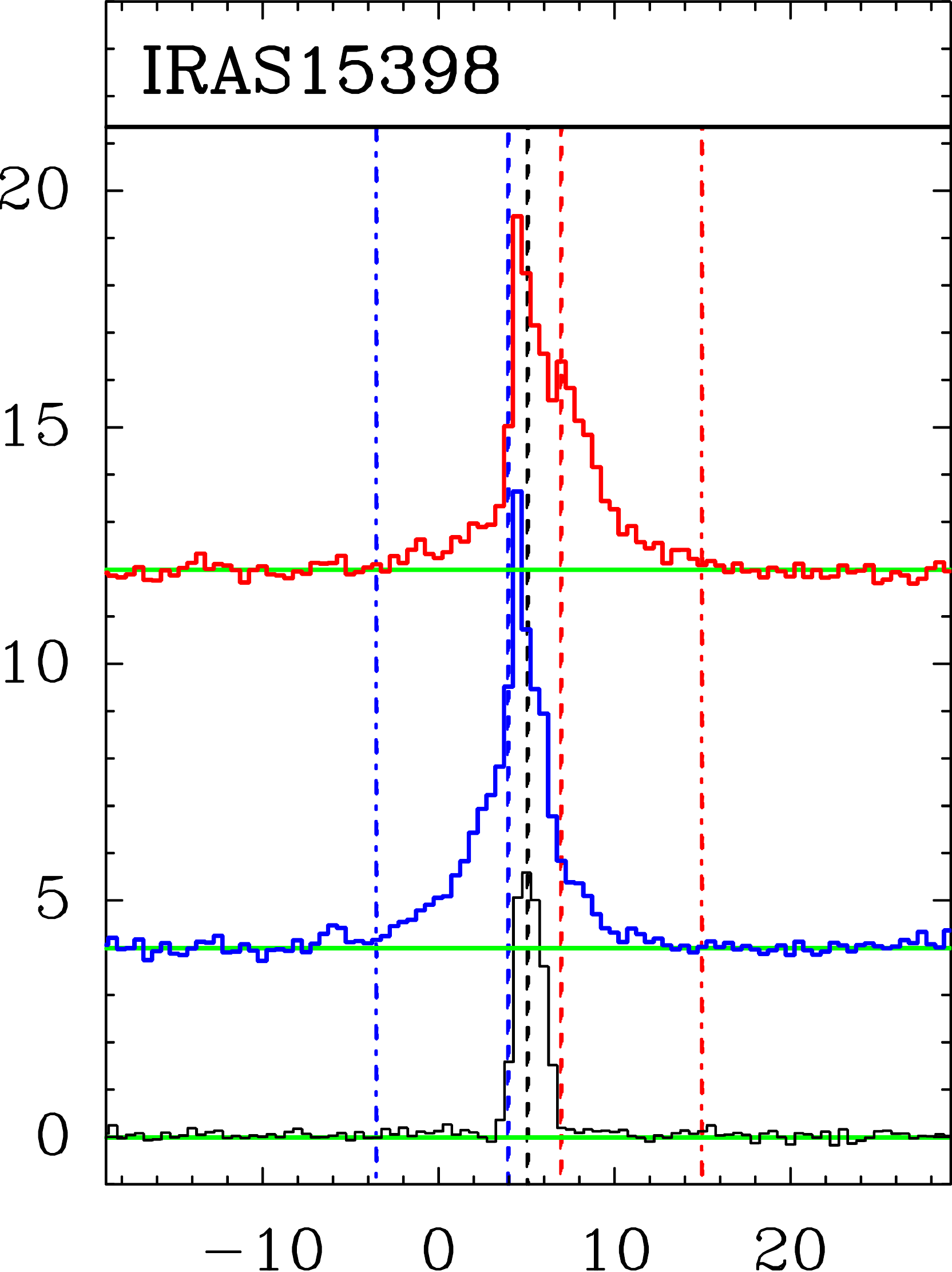}
    \includegraphics[scale=0.157]{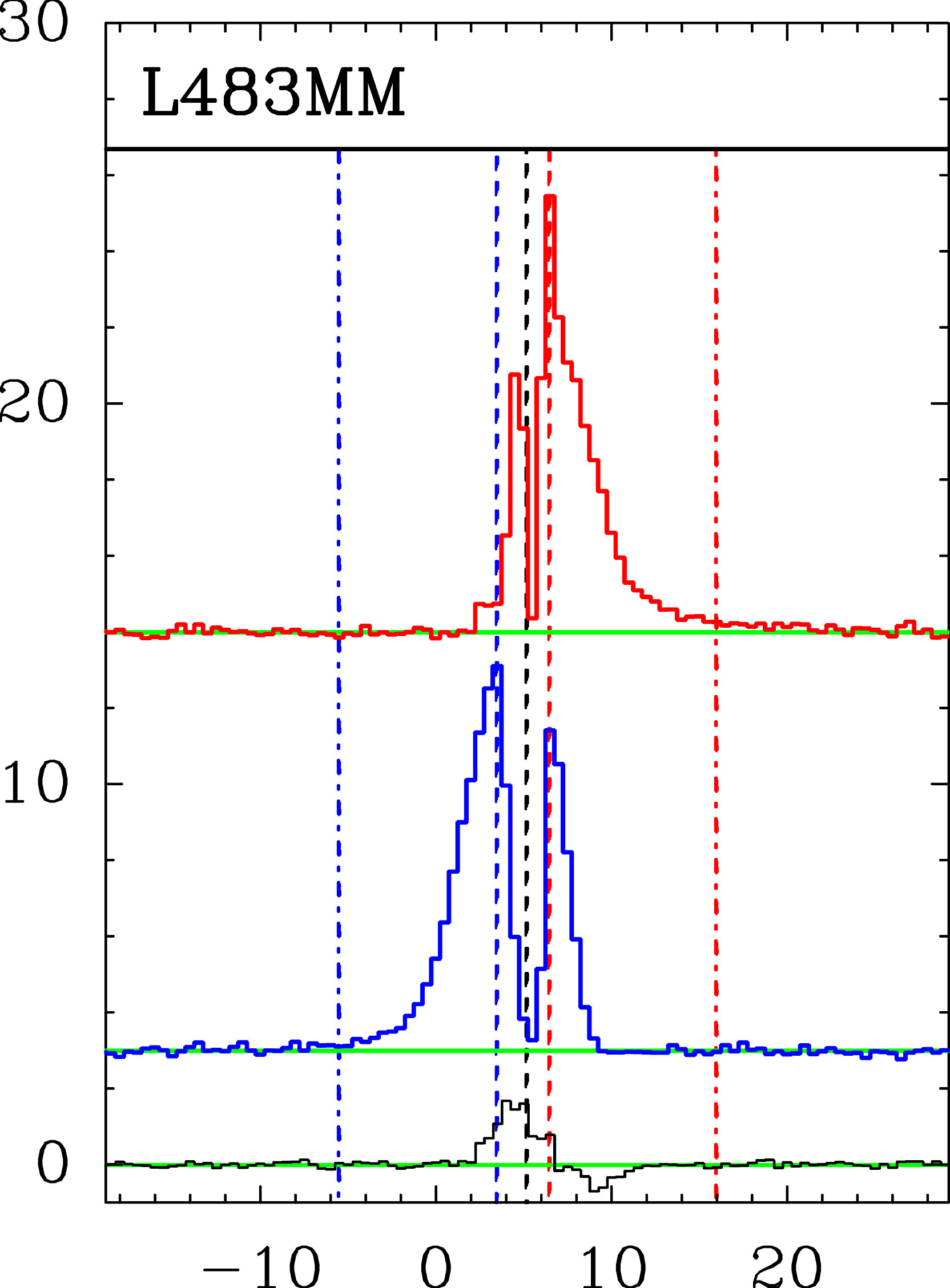}
    \includegraphics[scale=0.157]{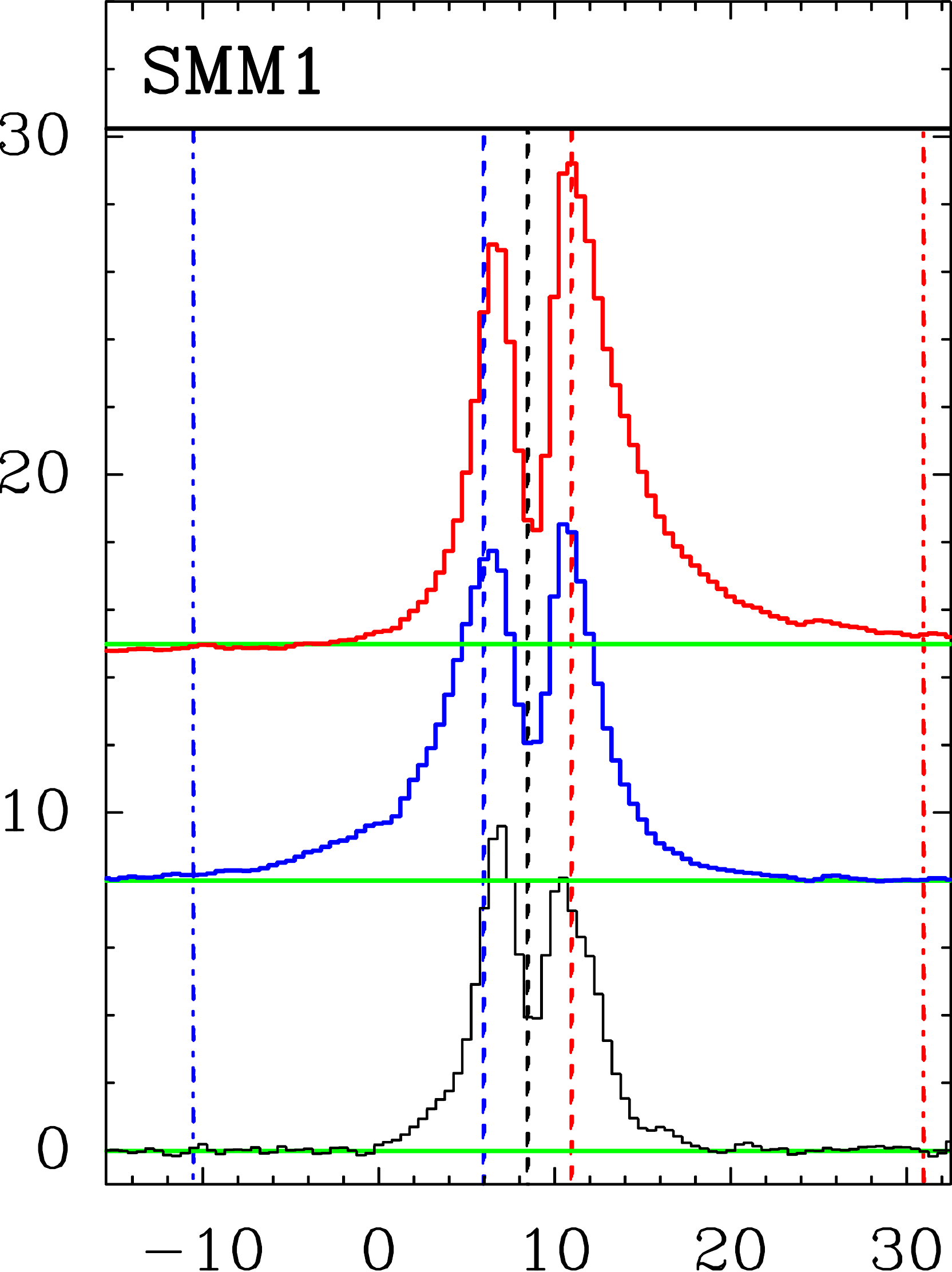}
    \includegraphics[scale=0.157]{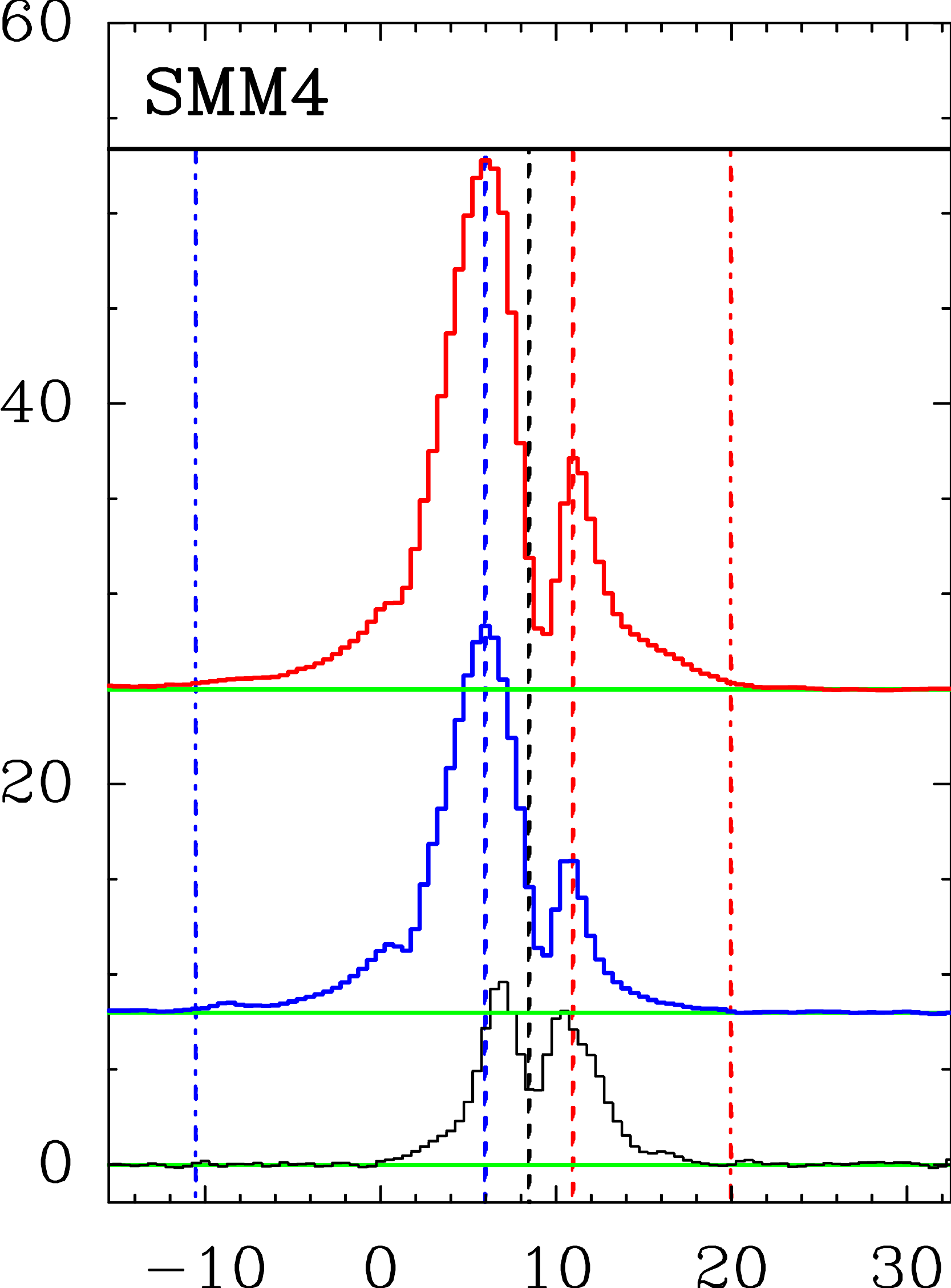}
    \includegraphics[scale=0.157]{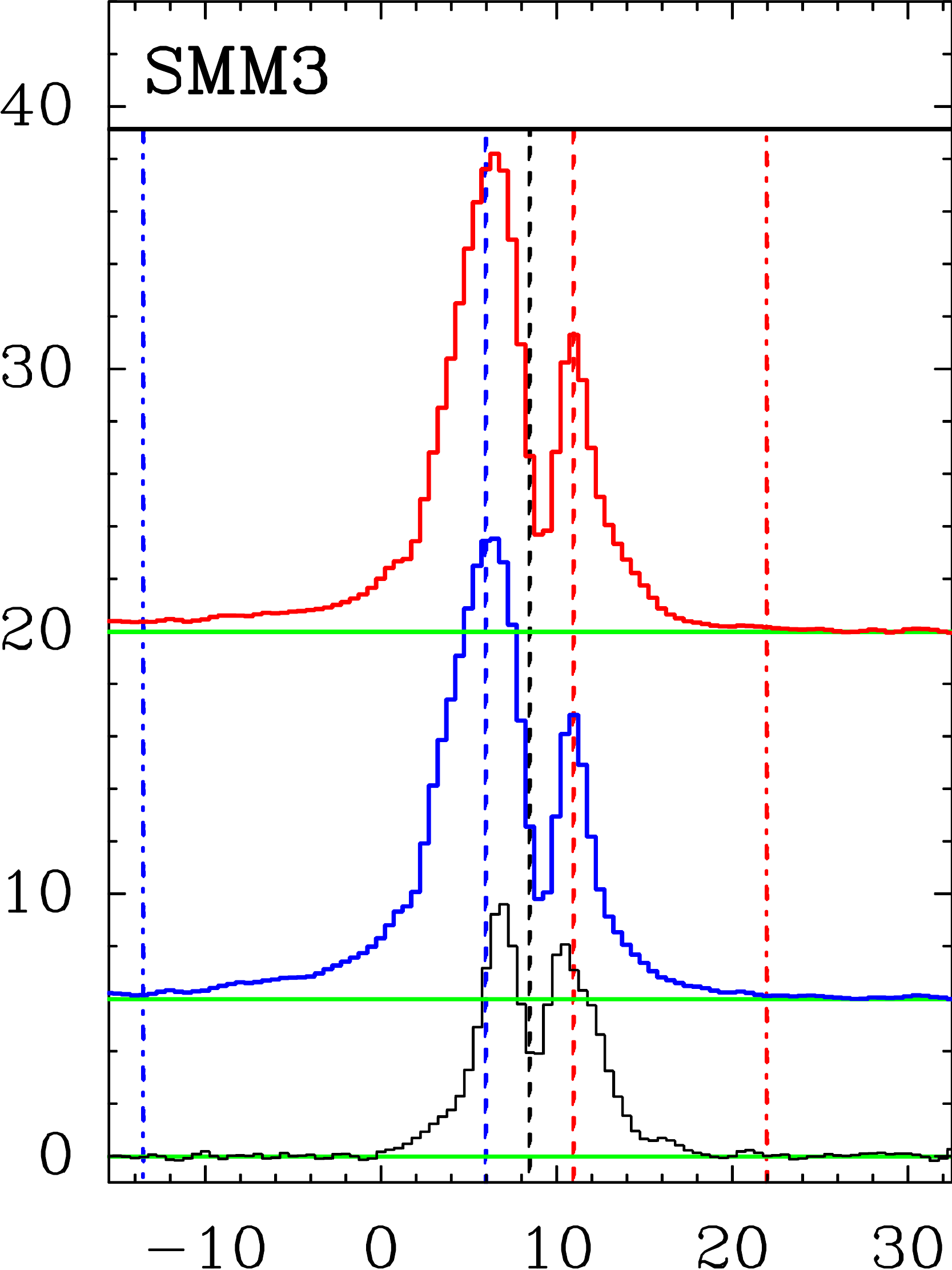}
    \includegraphics[scale=0.157]{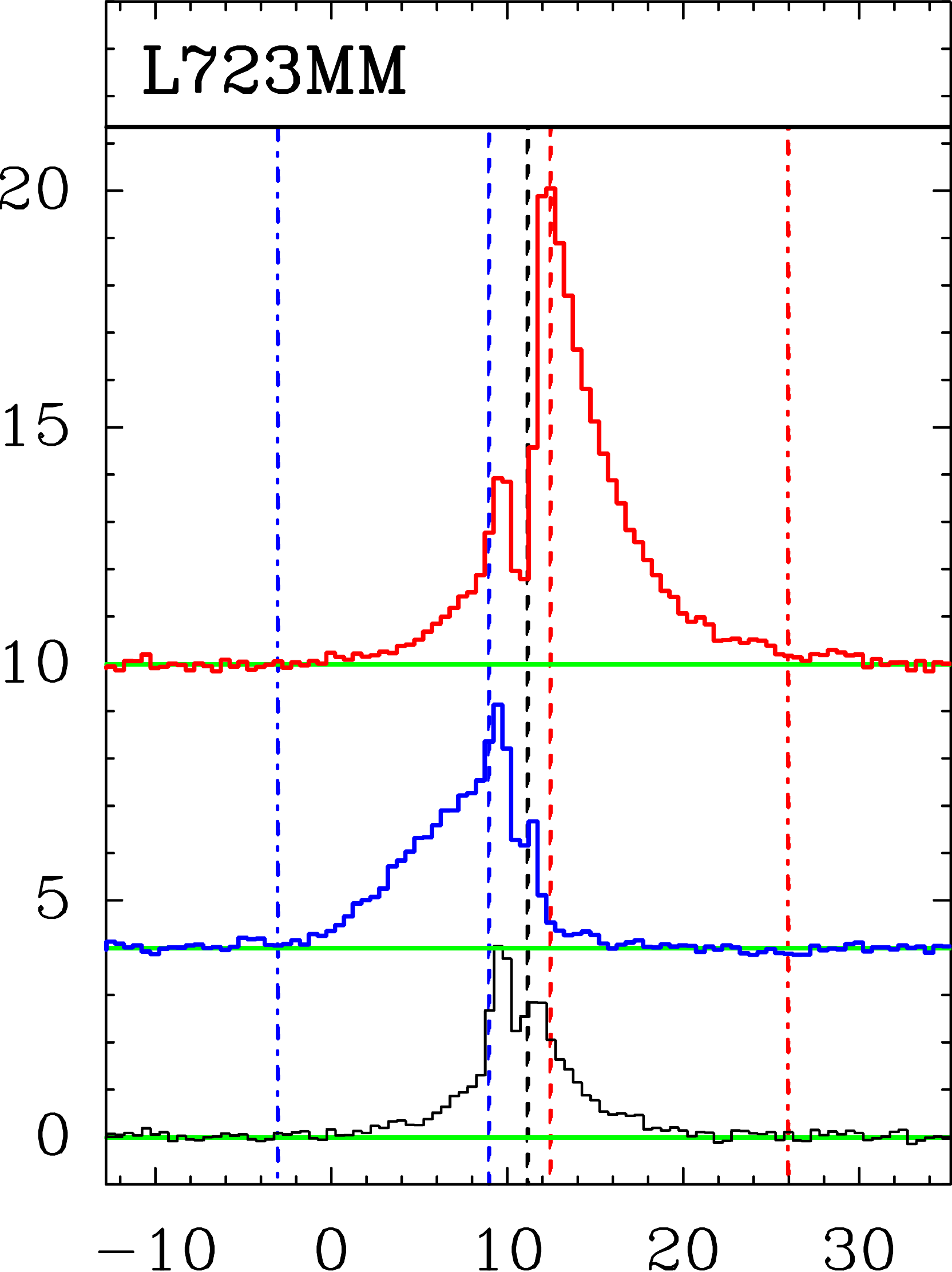}
    \includegraphics[scale=0.157]{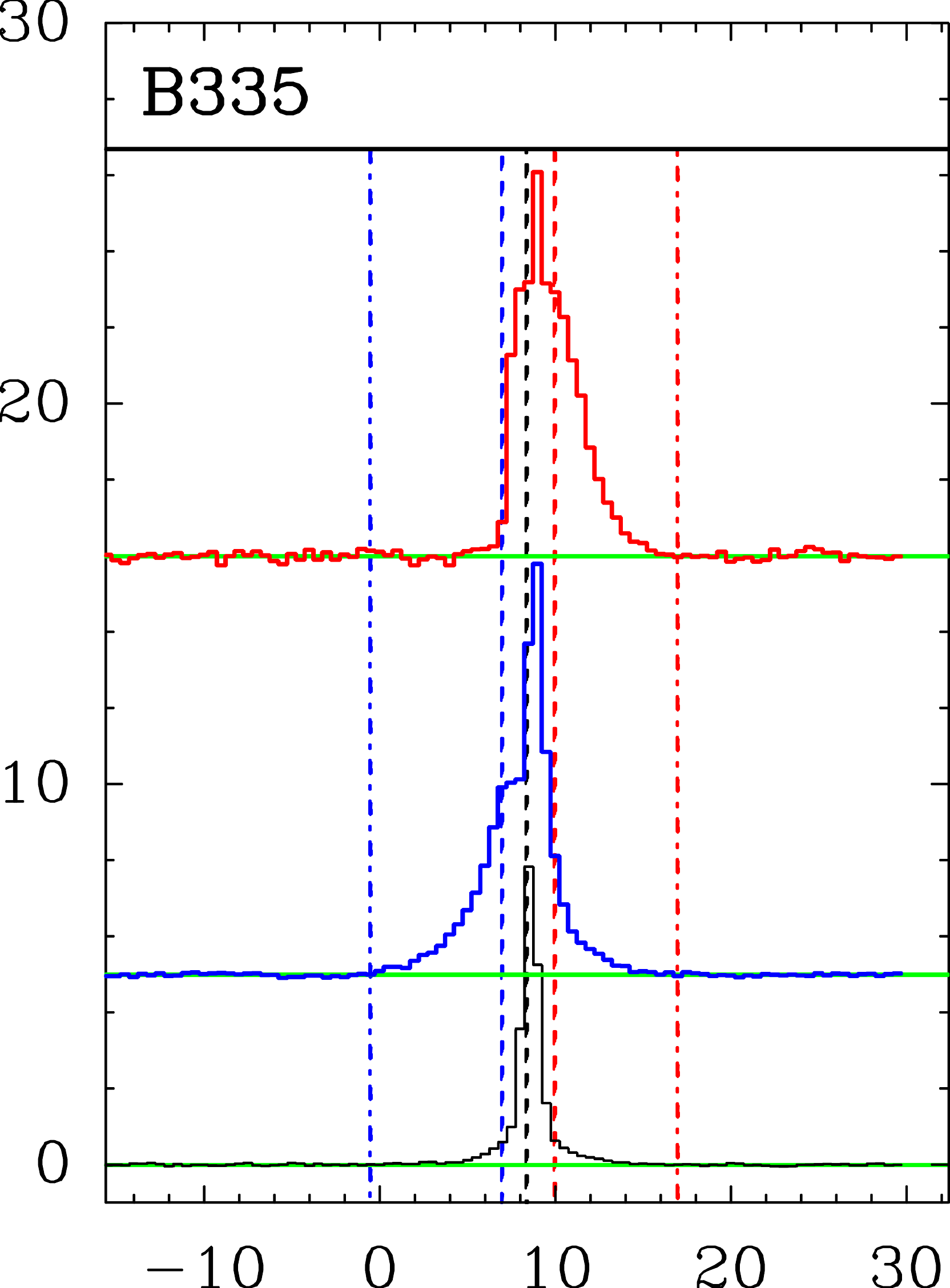}
    \includegraphics[scale=0.157]{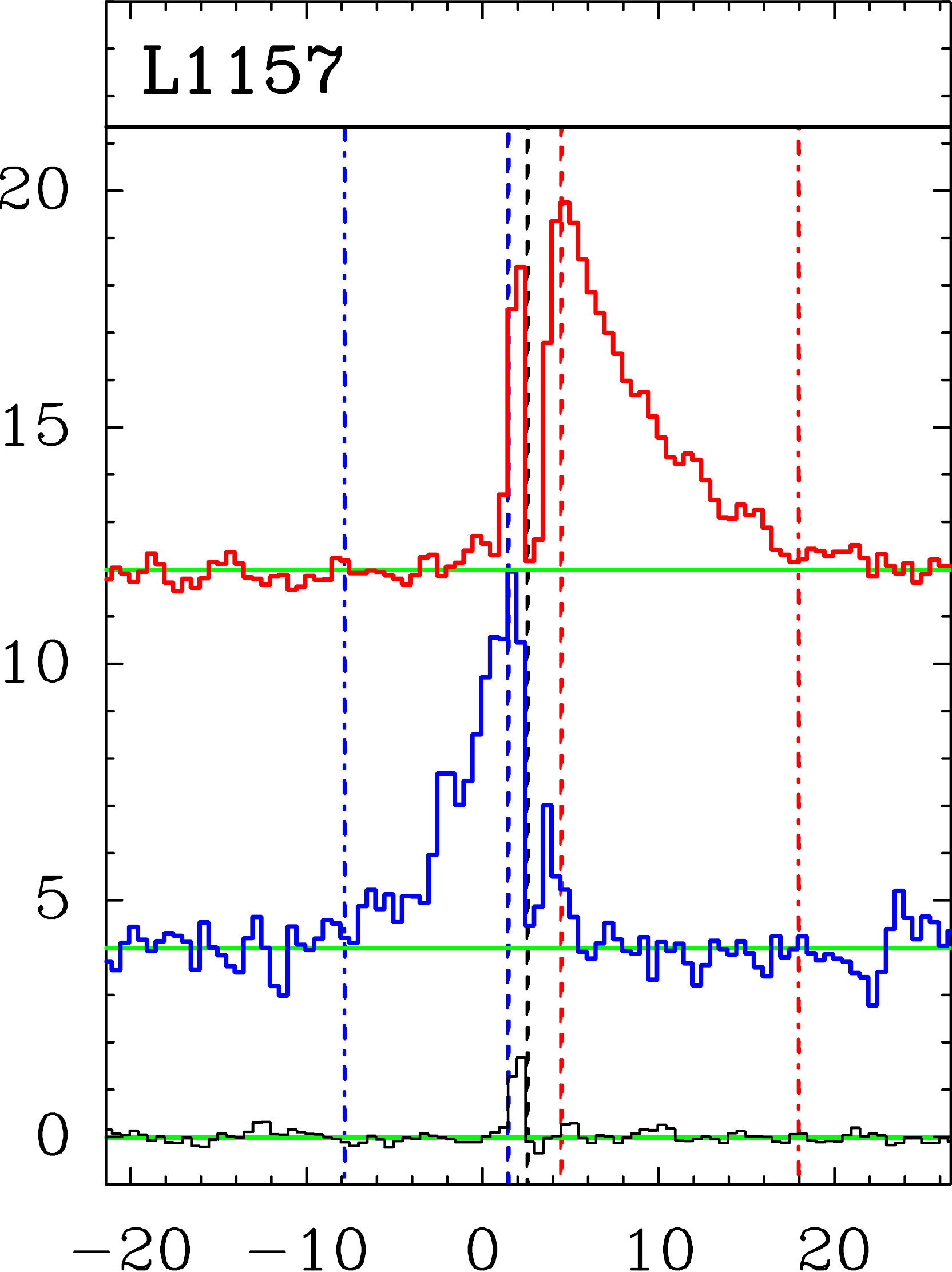}
    \includegraphics[scale=0.157]{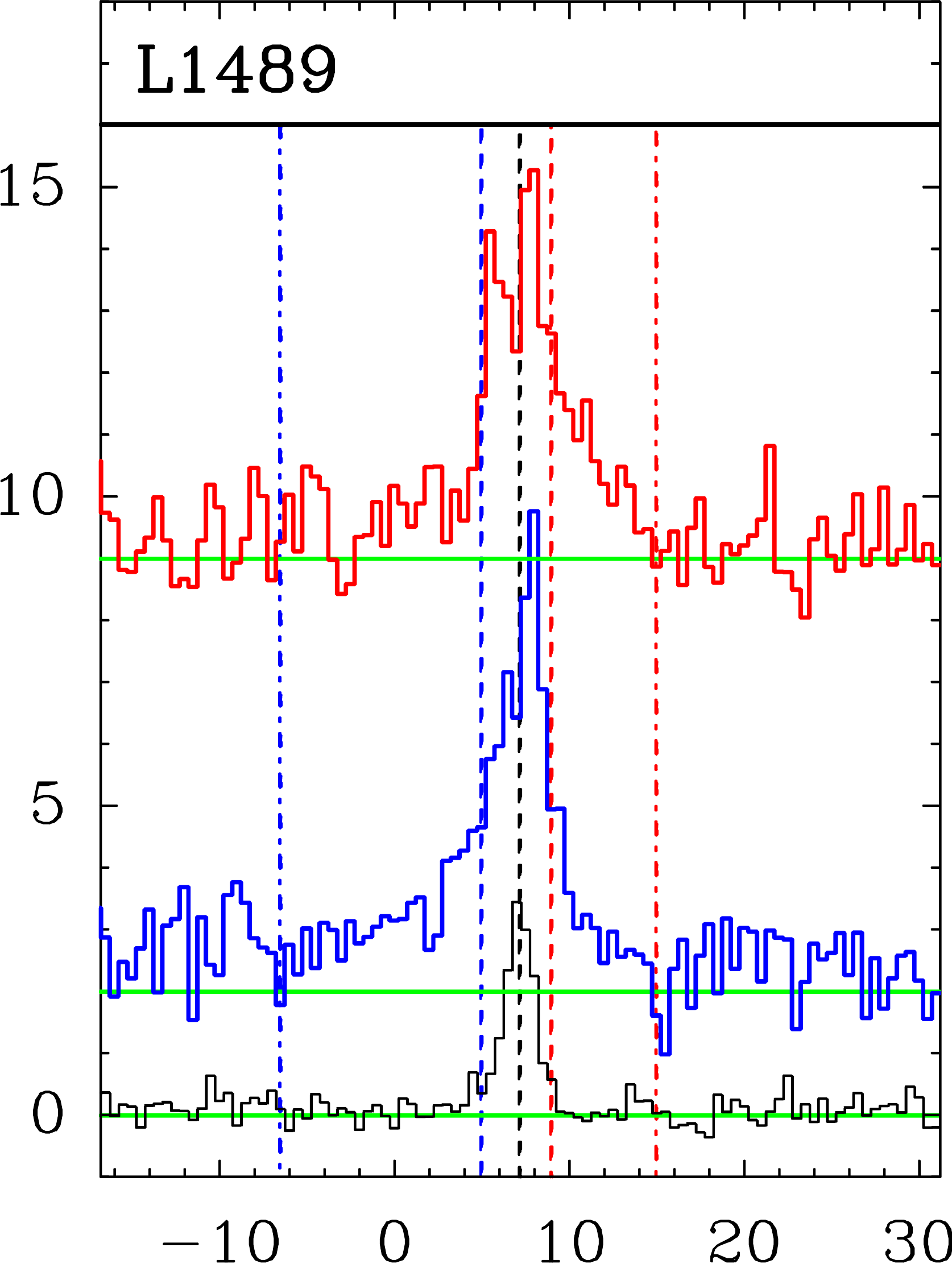}
    \includegraphics[scale=0.157]{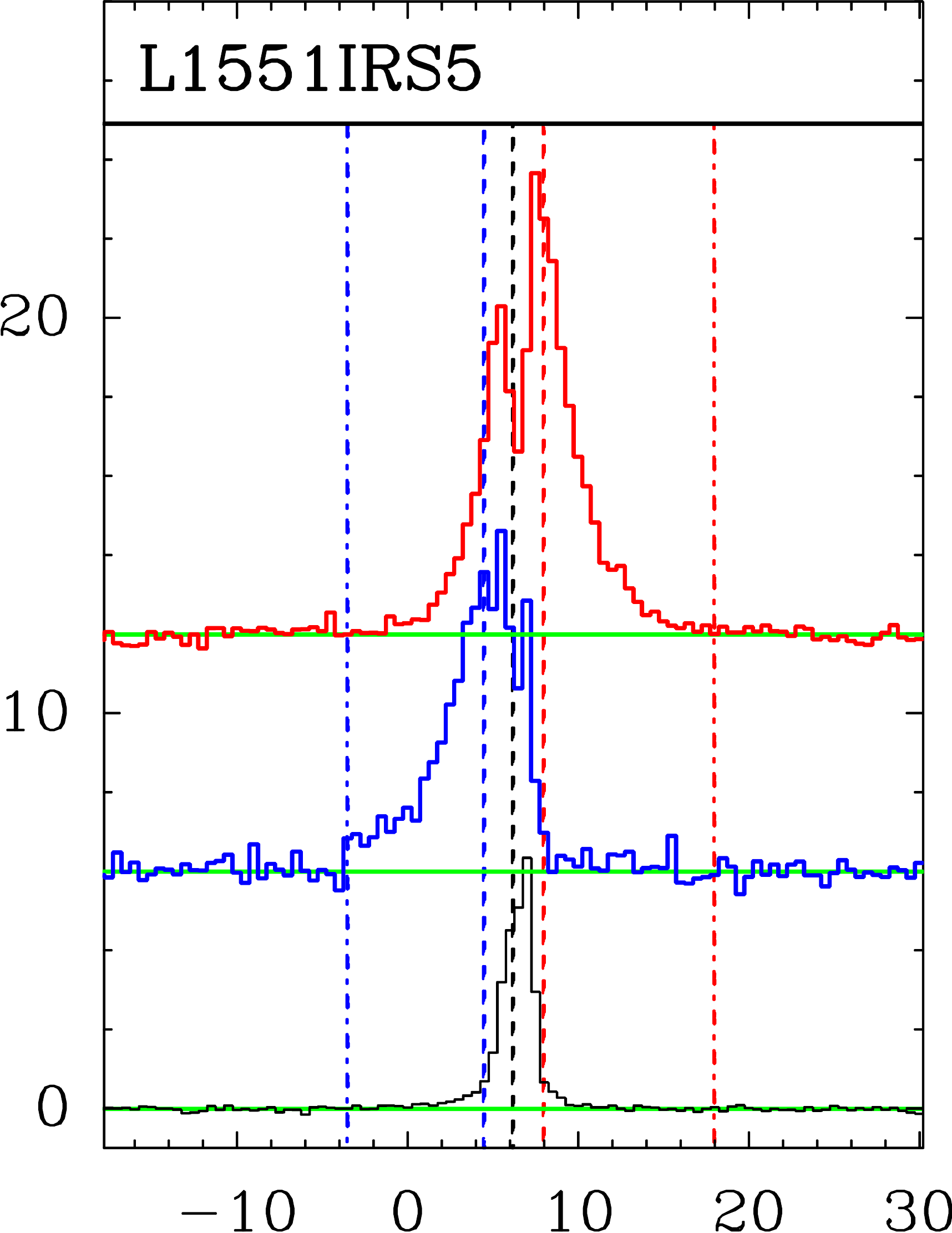}
    \includegraphics[scale=0.157]{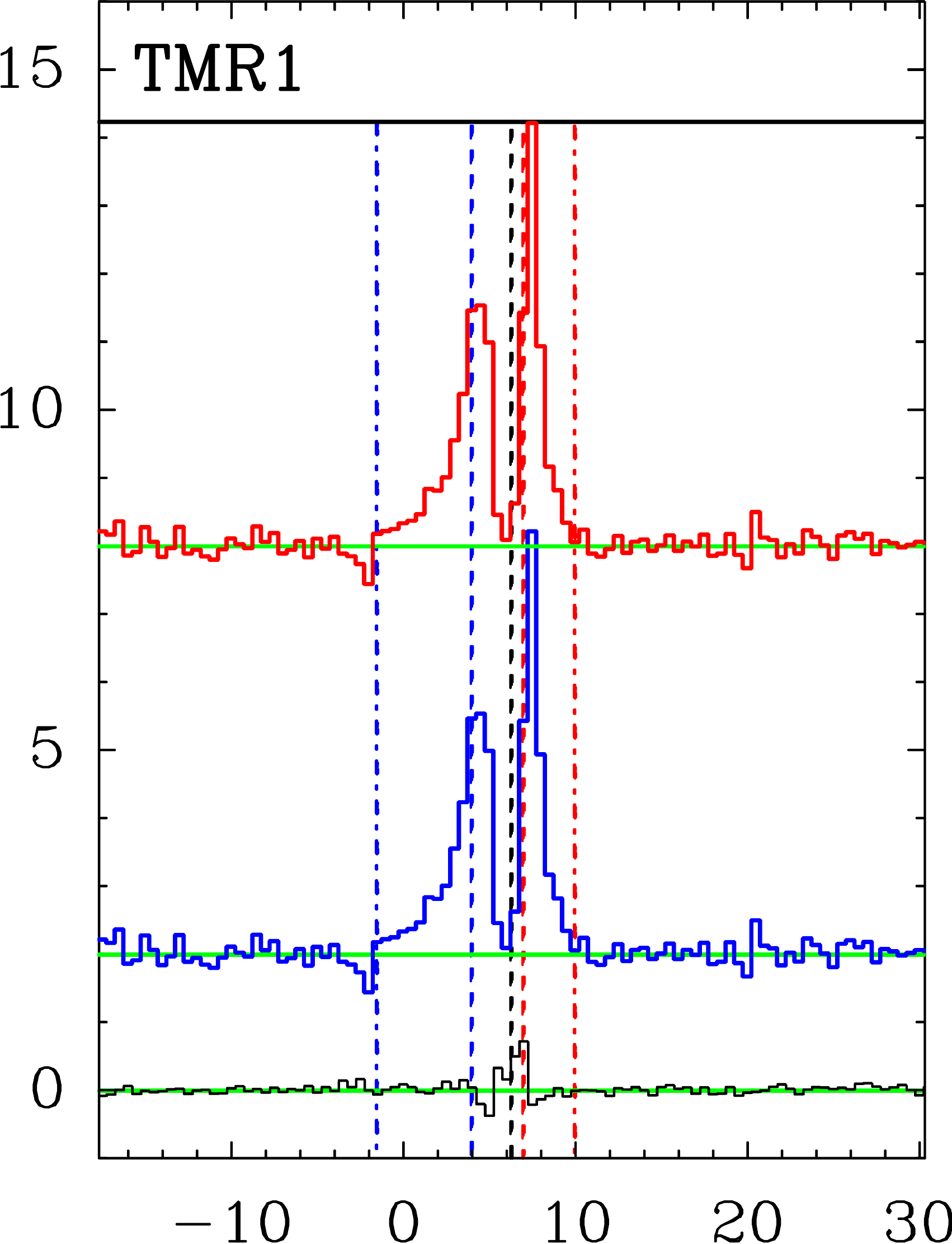}
    \includegraphics[scale=0.157]{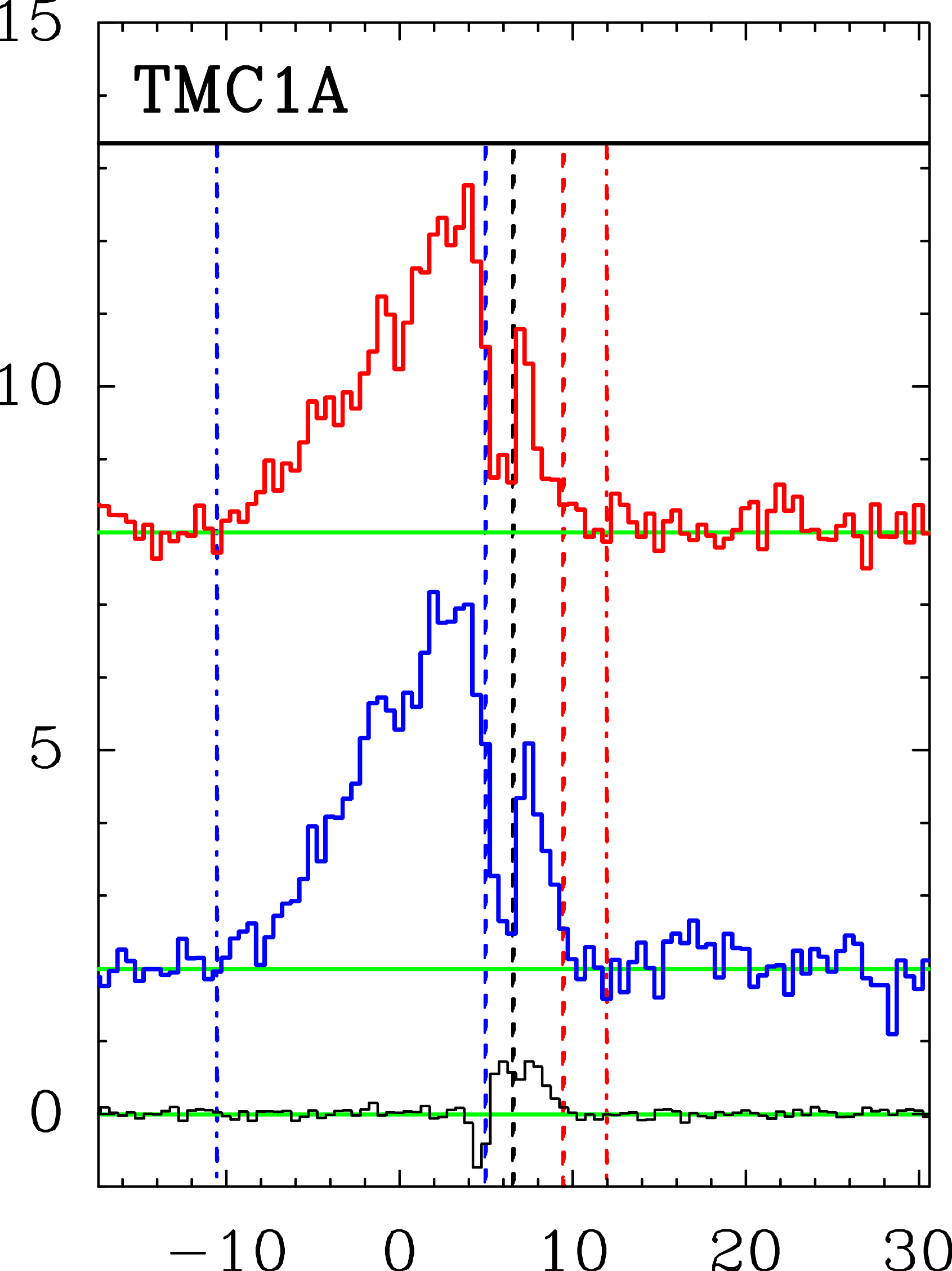}
    \includegraphics[scale=0.157]{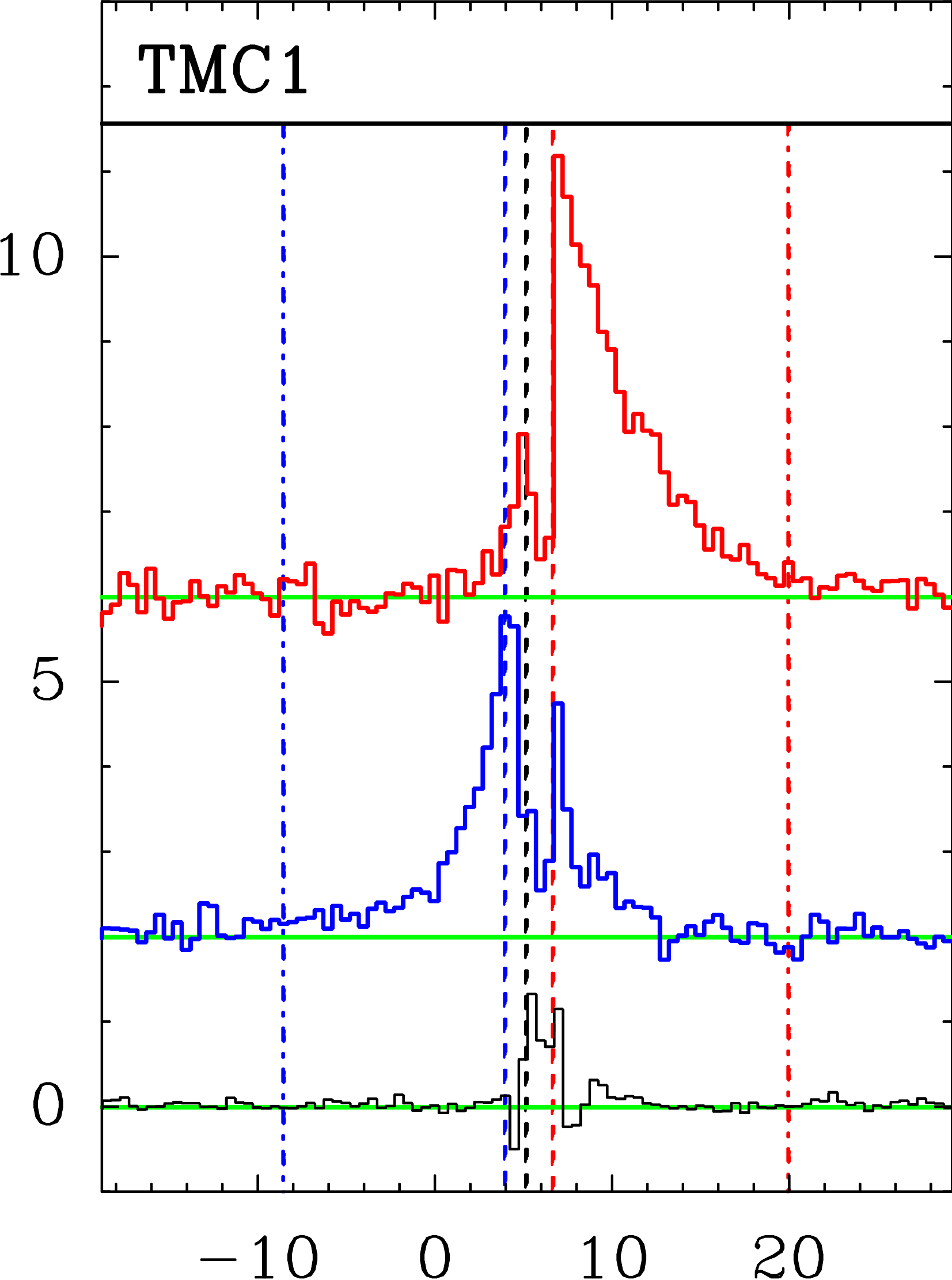}
    \includegraphics[scale=0.157]{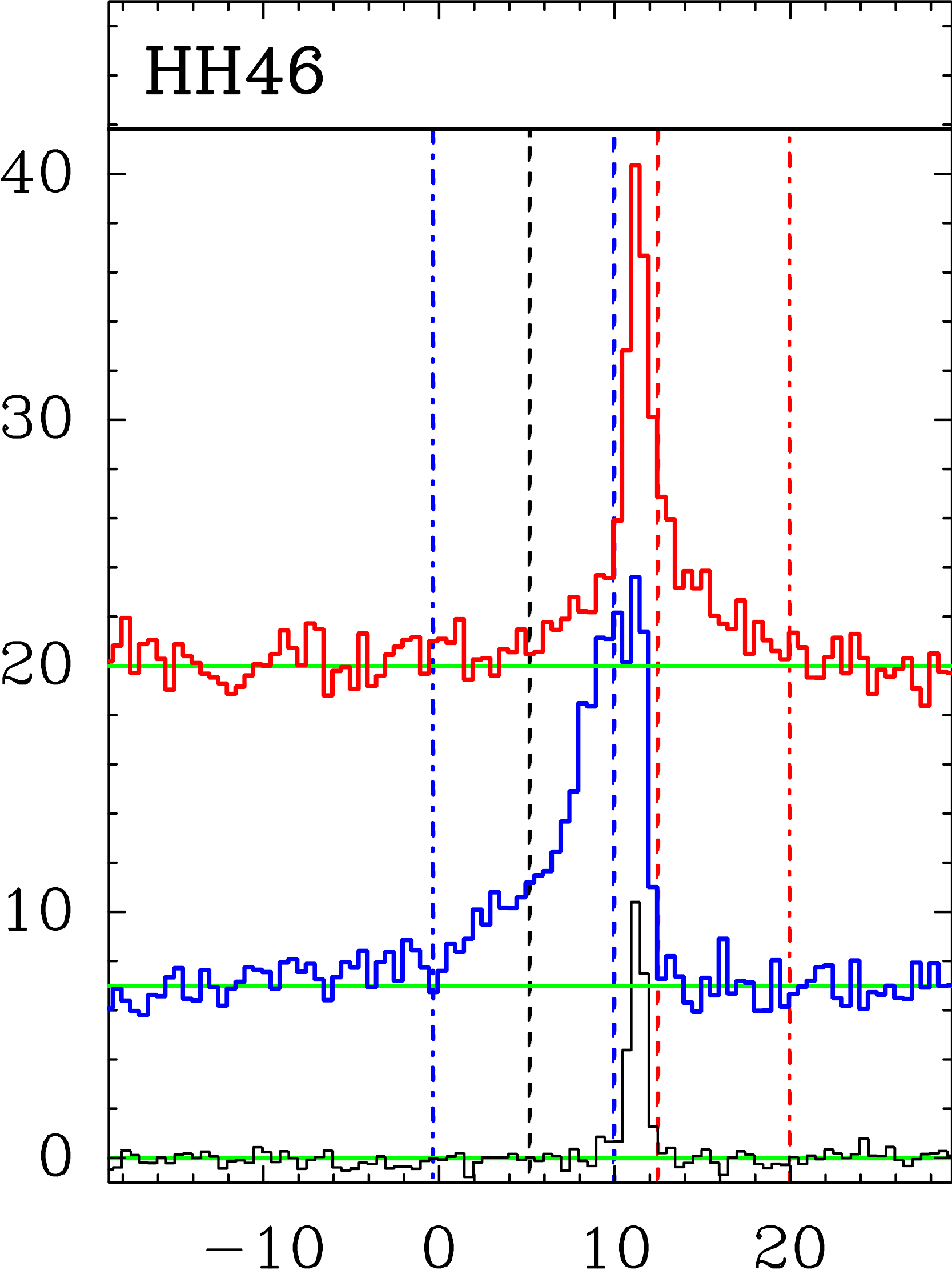}
    \includegraphics[scale=0.157]{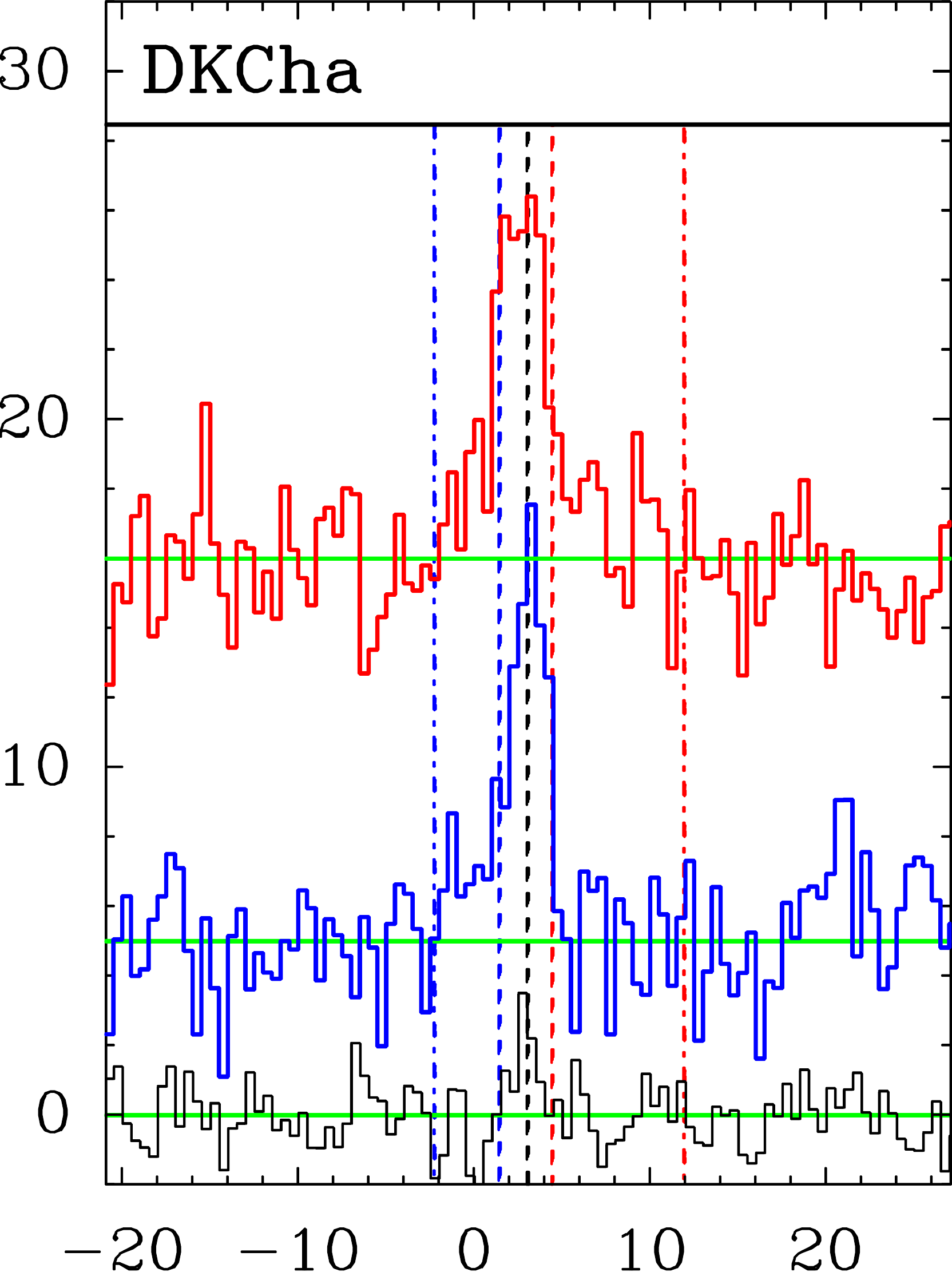}
    \includegraphics[scale=0.157]{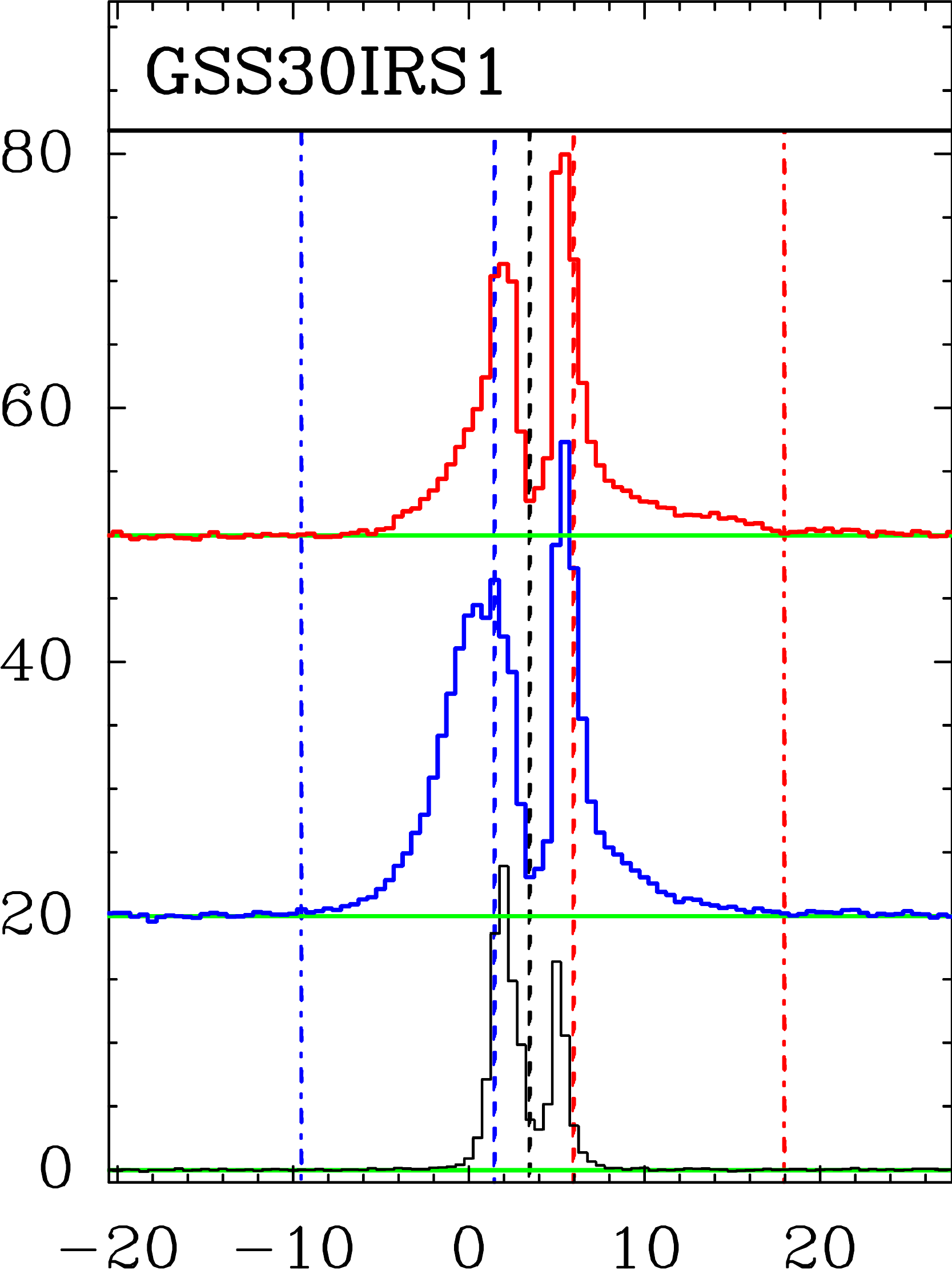}
    \includegraphics[scale=0.157]{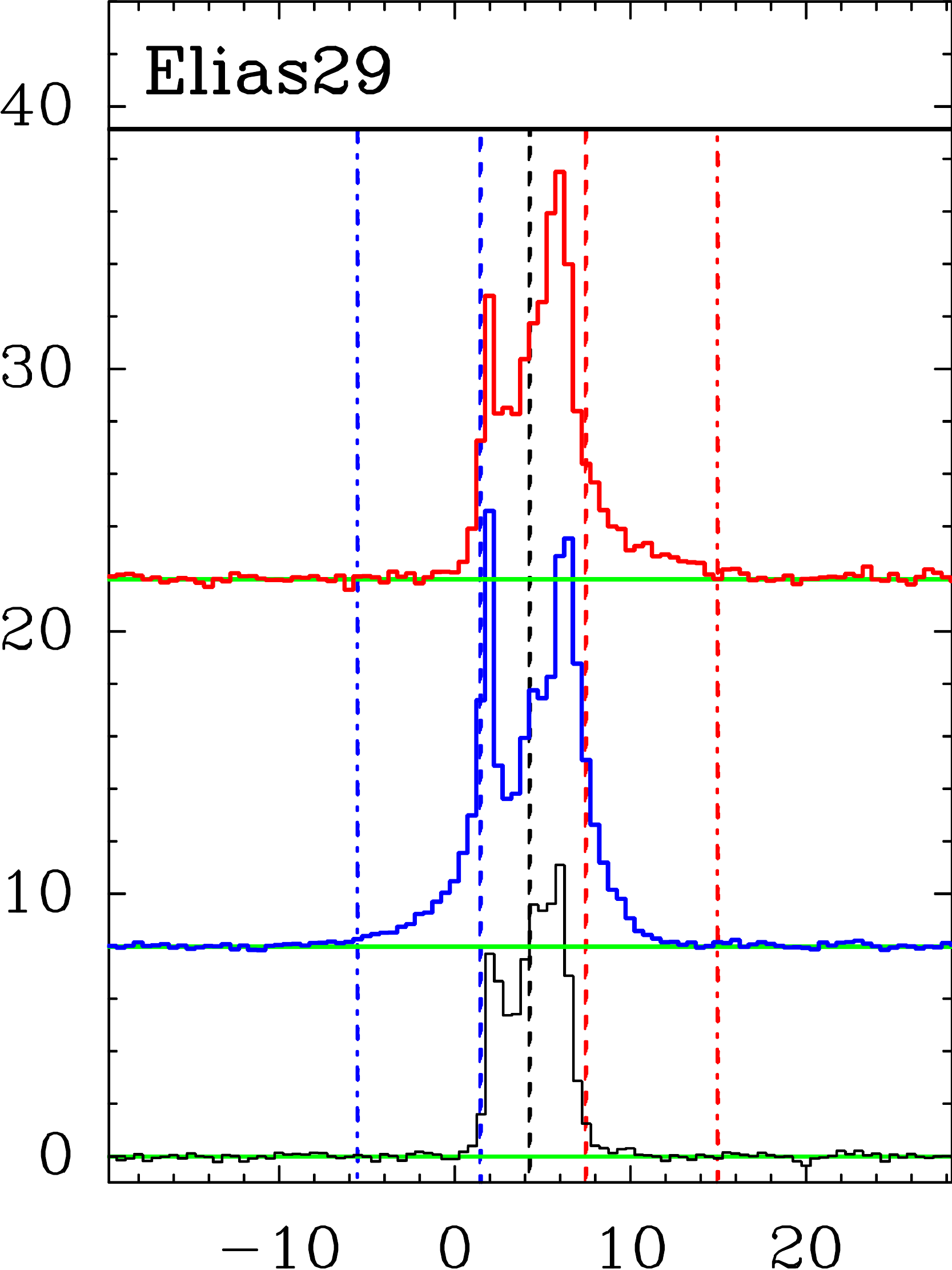}
    \includegraphics[scale=0.157]{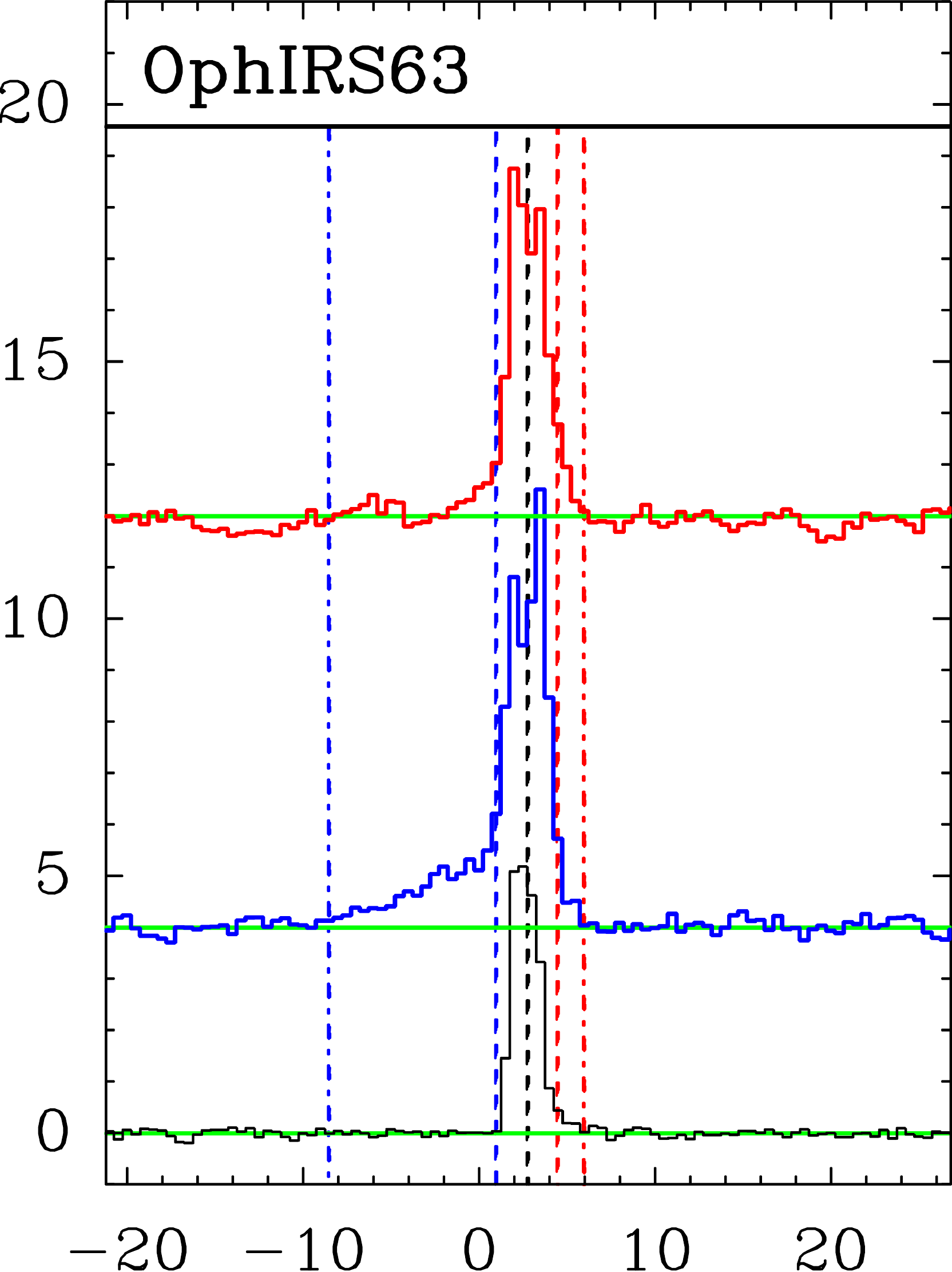}
    \includegraphics[scale=0.157]{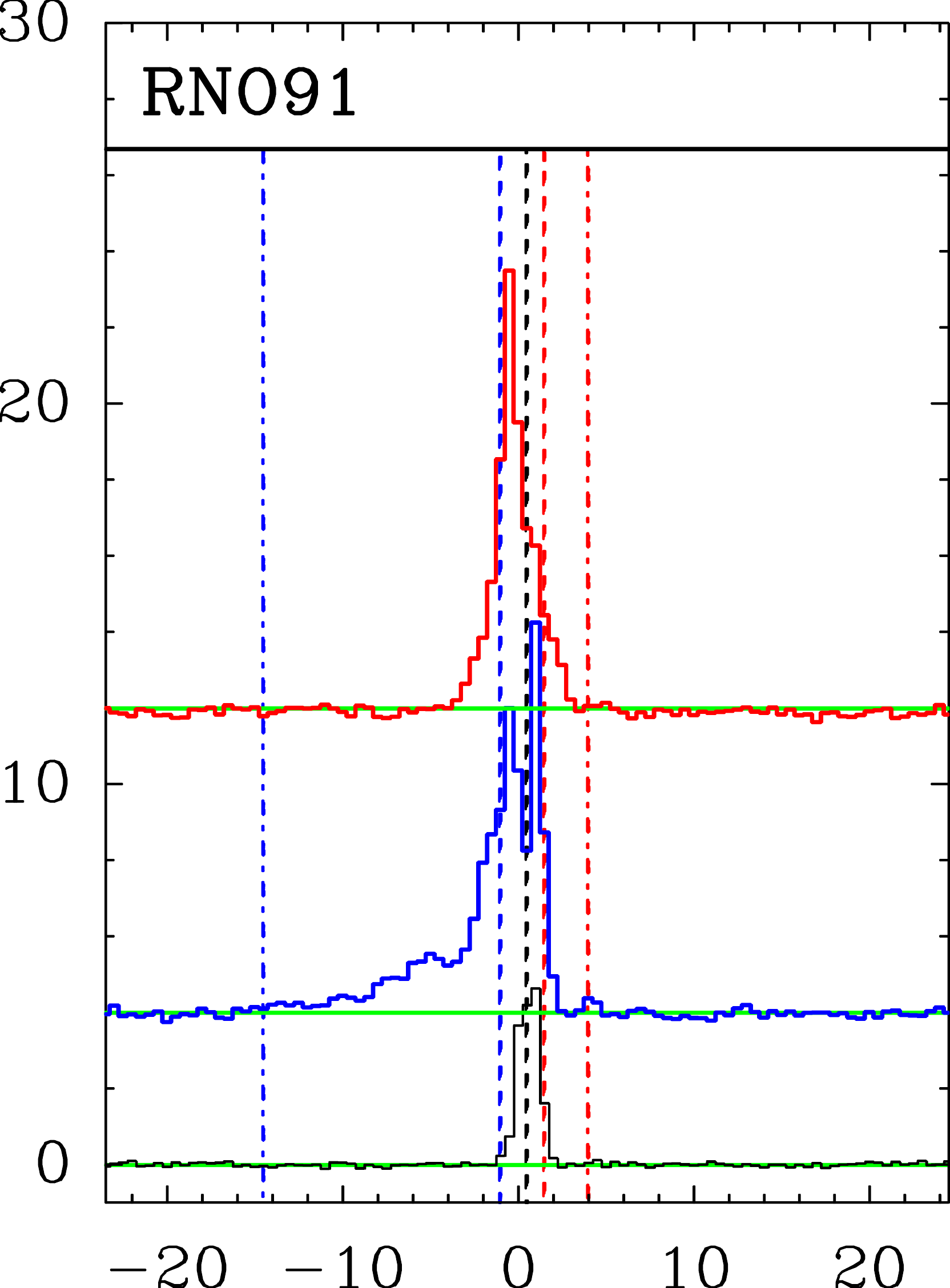}
\end{center}
\end{minipage}
\caption{\footnotesize CO~3--2 spectra with selected integration
  limits indicated, except for Ced110\,IRS4, BHR71, and DK\,Cha where
  CO~6--5 was used. Each panel presents these limits for each
  source. The black spectrum at the {\it bottom} is taken from a clean
  position representative for the envelope emission. The blue spectrum
  at the {\it middle} is the representative spectrum from the blue
  outflow lobe, and red spectrum at the {\it top} is the
  representative spectrum from the red outflow lobe. Each panel shows
  five vertical lines, these are $V_{\rm LSR}$ {\it (black dashed line)},
  $V_{\rm out,blue}$ {\it (dot-dash blue line)}, $V_{\rm in,blue}$
  {\it (dashed blue line)}, $V_{\rm in,red}$ {\it (dashed red line)},
  and $V_{\rm out,red}$ {\it (dot-dash red line)}.}
    \label{fig:AllBCR}
\end{figure*}


\begin{figure*}[htb]
    \centering
    \includegraphics[scale=0.5]{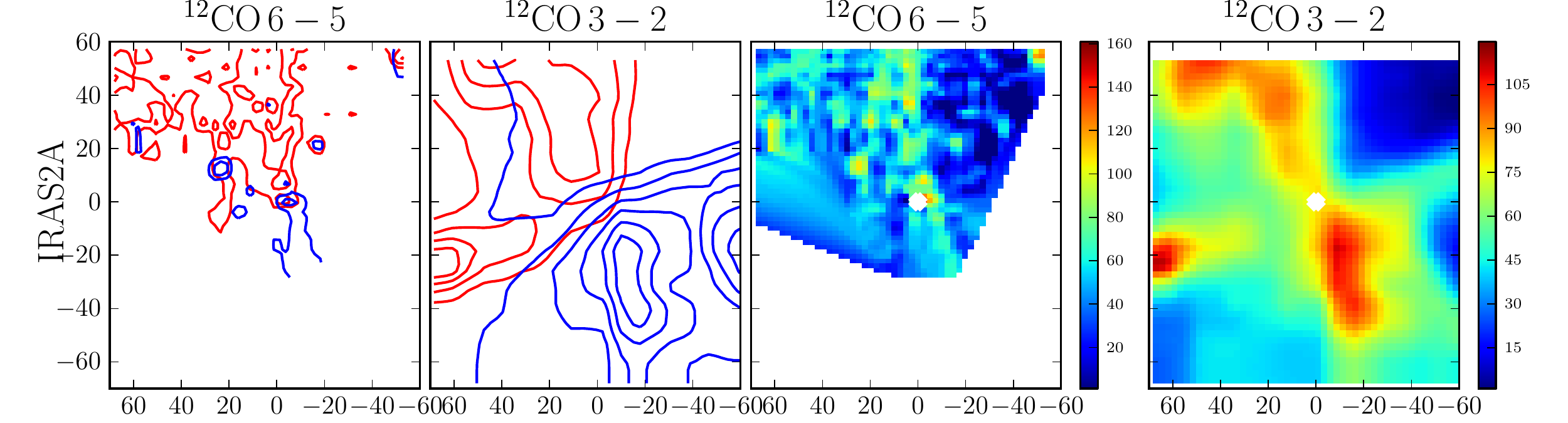} \\
    \includegraphics[scale=0.5]{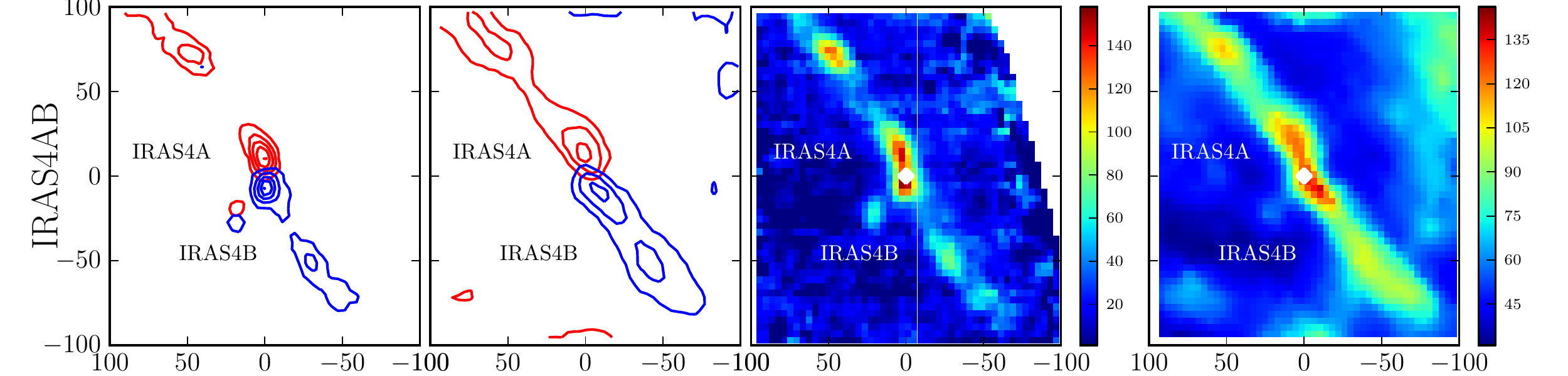} \\ 
    \includegraphics[scale=0.5]{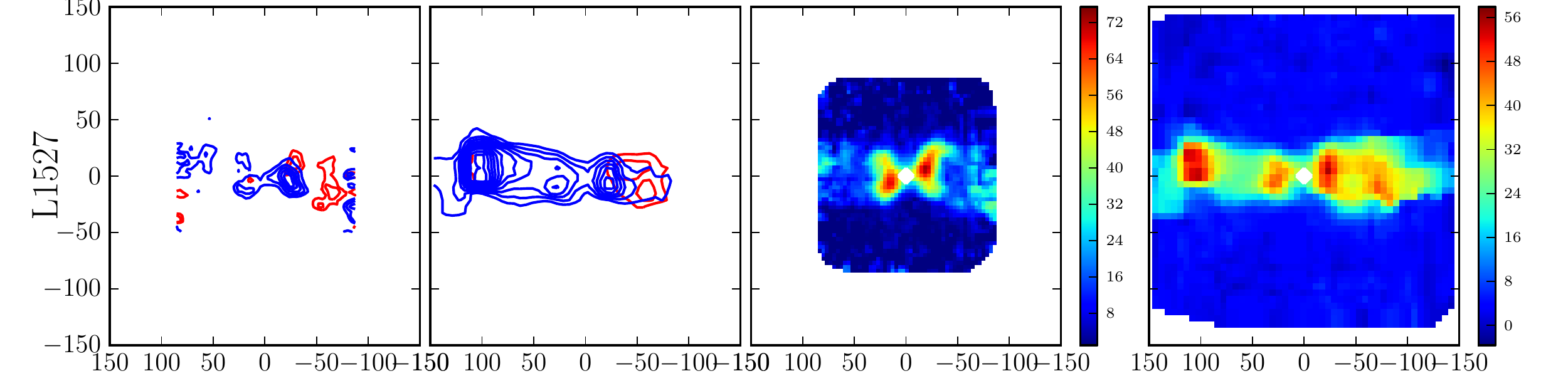} \\
    \includegraphics[scale=0.5]{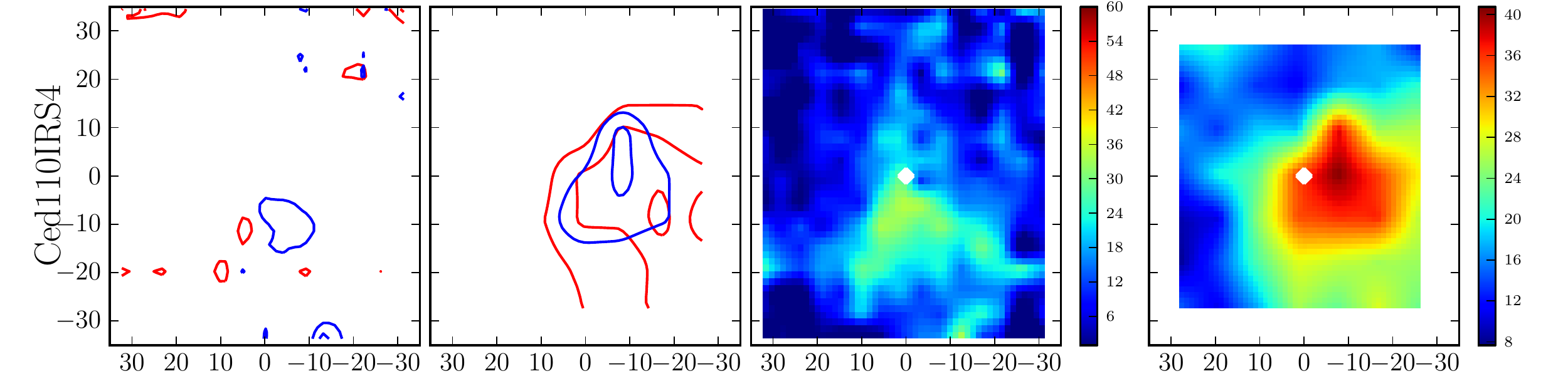} \\
    \includegraphics[scale=0.5]{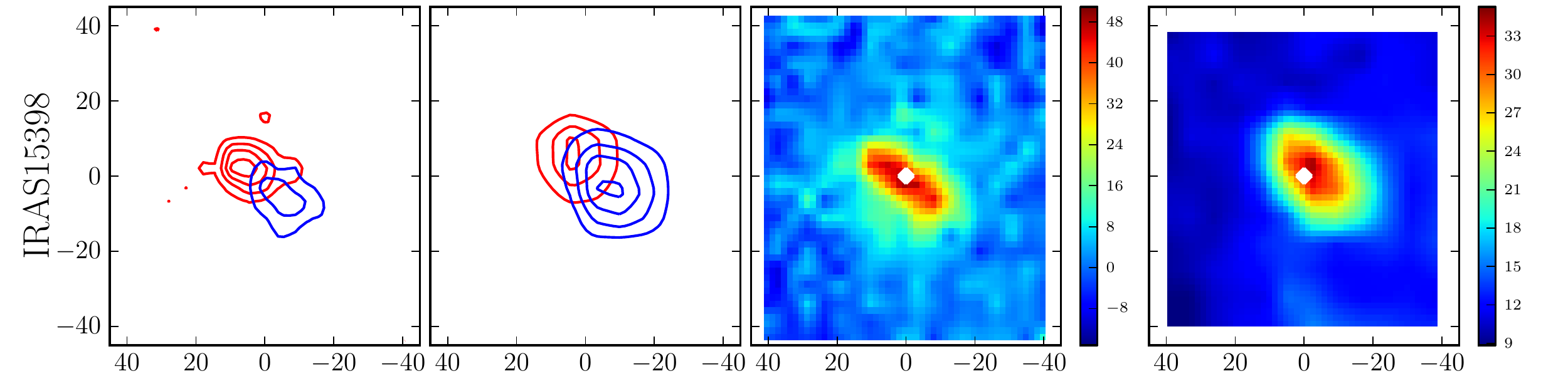} \\
    \includegraphics[scale=0.5]{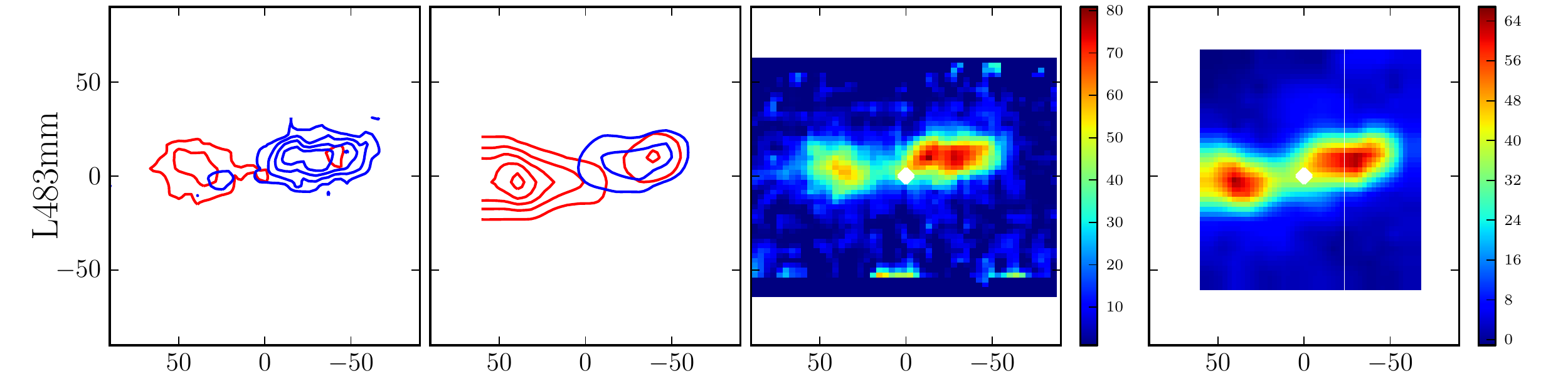} \\
    \includegraphics[scale=0.5]{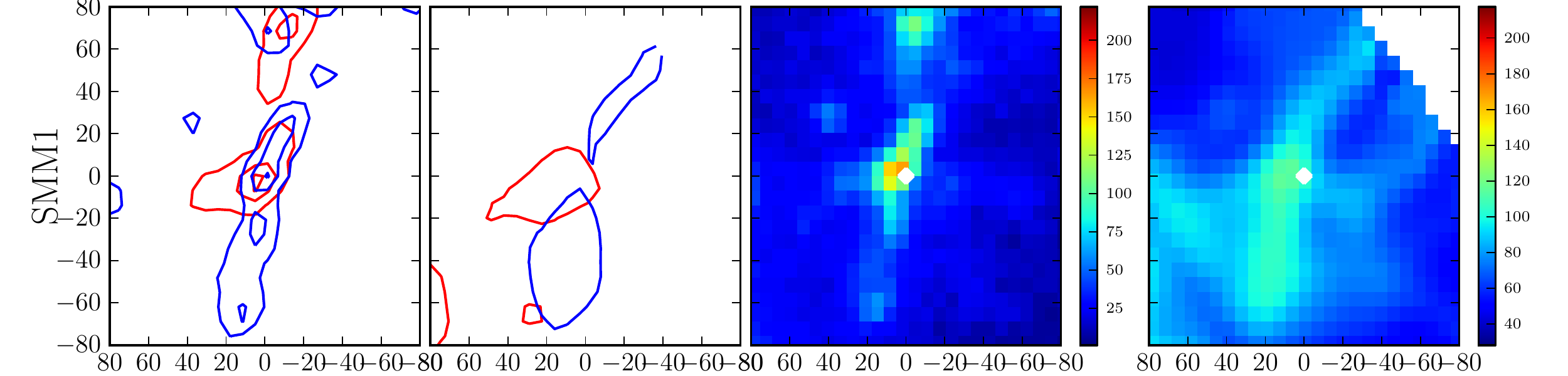} \\
    \includegraphics[scale=0.5]{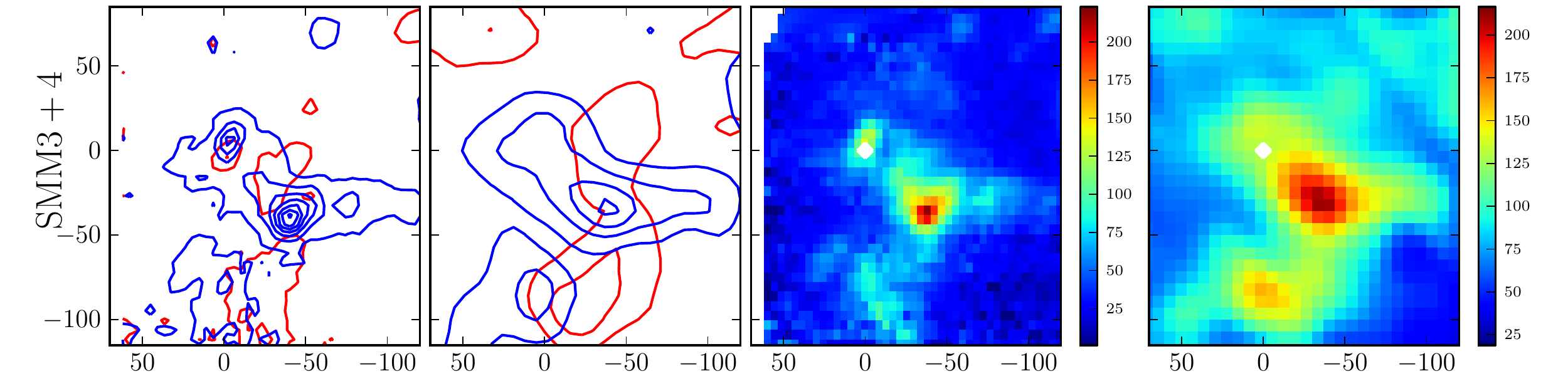}
    \caption{\small Each row contains contour and integrated intensity maps (in K km s$^{-1}$) of 
    sources in \twco\ 6--5 and 3--2. The contour levels and integration limits are 
    given in Table~\ref{tbl:contourlevels} and integration limits shown in Fig.~\ref{fig:AllBCR}. The color images show all emission integrated from $V_{\rm out,red}$ to $V_{\rm out,blue}$, including any minor cloud contribution.}
    \label{fig:AllIntensityCO_1}
\end{figure*}

\begin{figure*}[htb]
    \centering
    \includegraphics[scale=0.5]{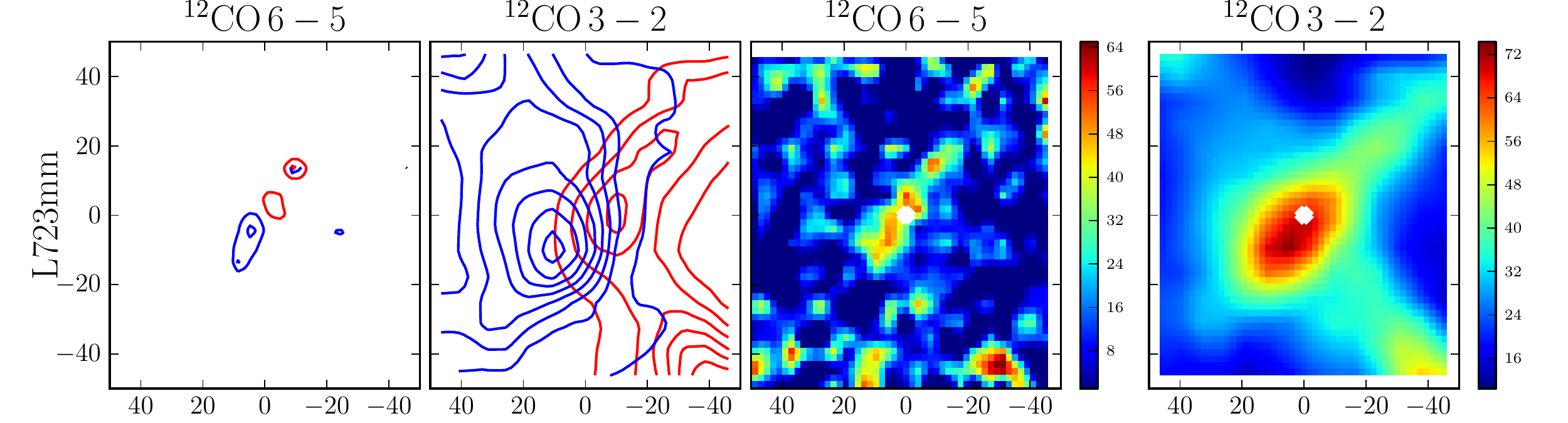} \\
    \includegraphics[scale=0.5]{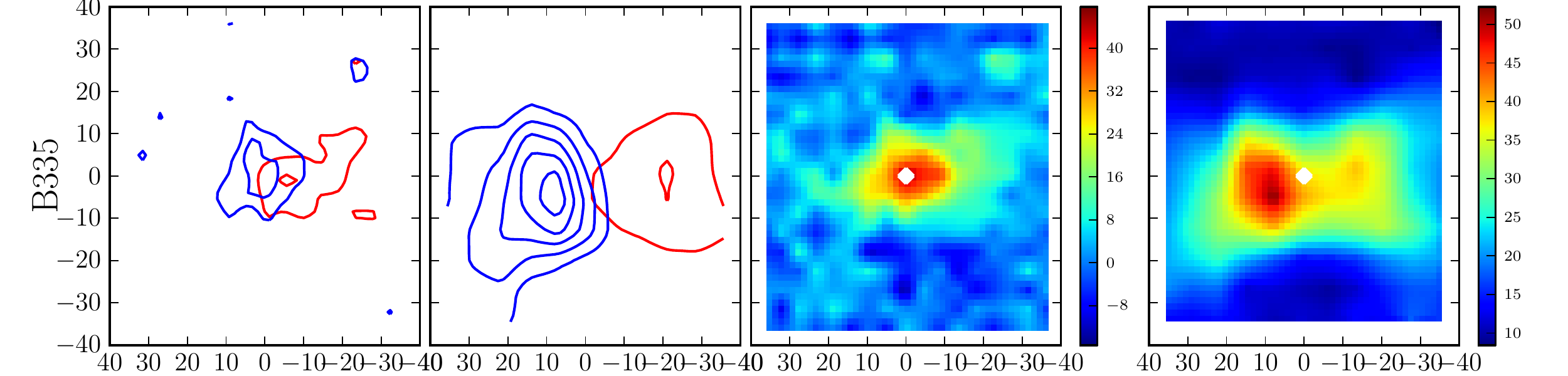} \\
    \includegraphics[scale=0.5]{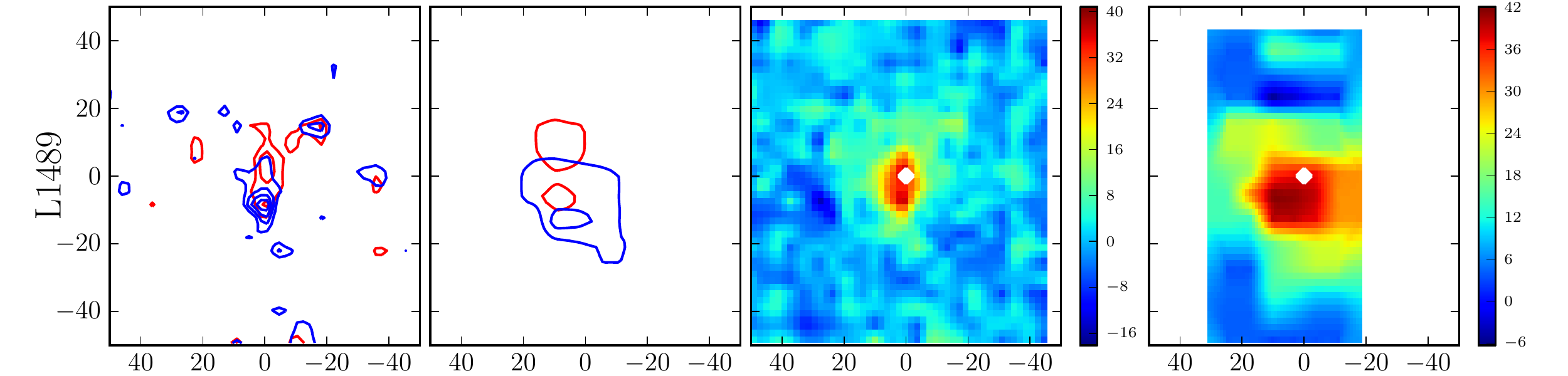} \\
    \includegraphics[scale=0.5]{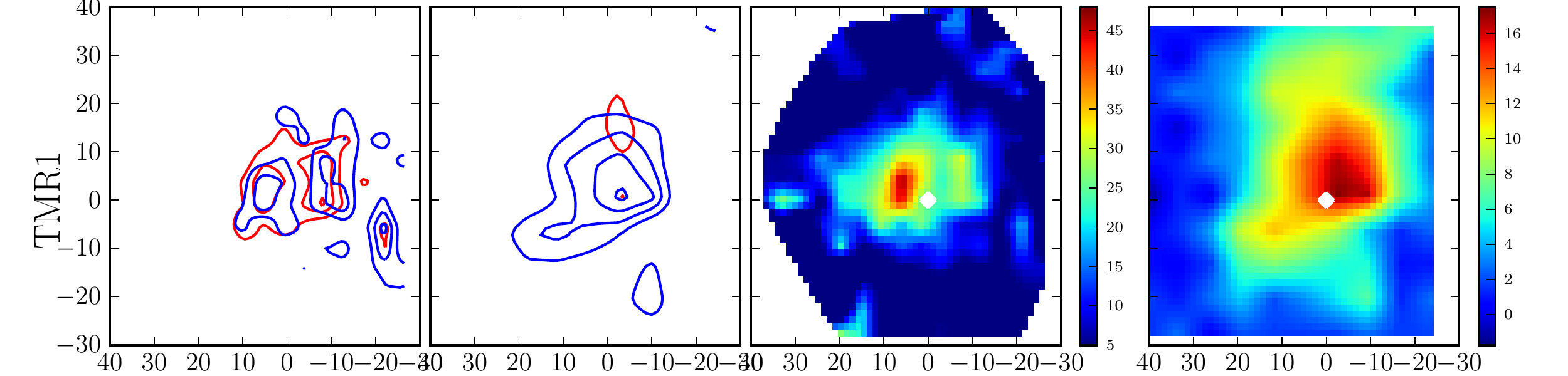} \\
    \includegraphics[scale=0.5]{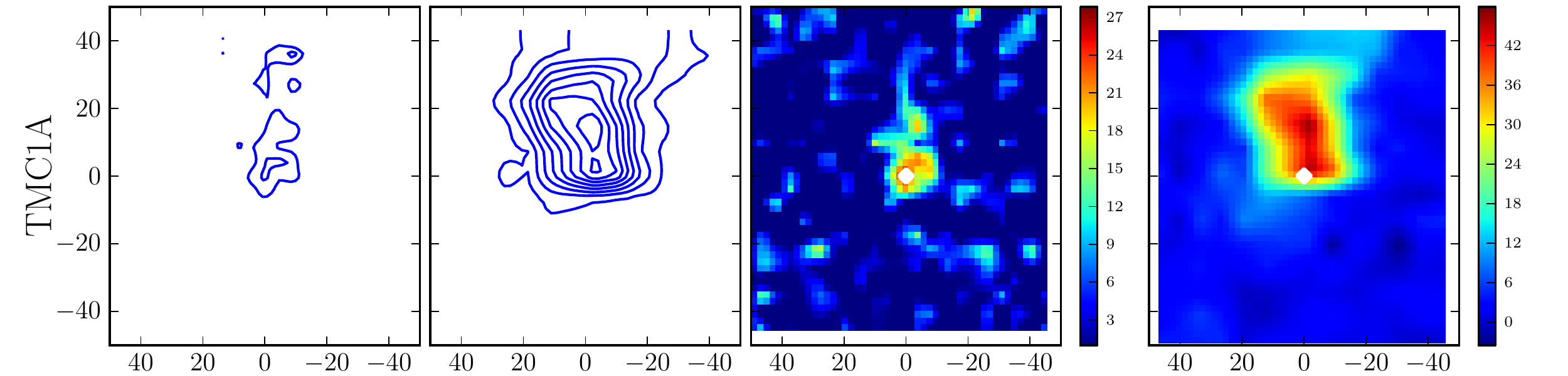} \\
    \includegraphics[scale=0.5]{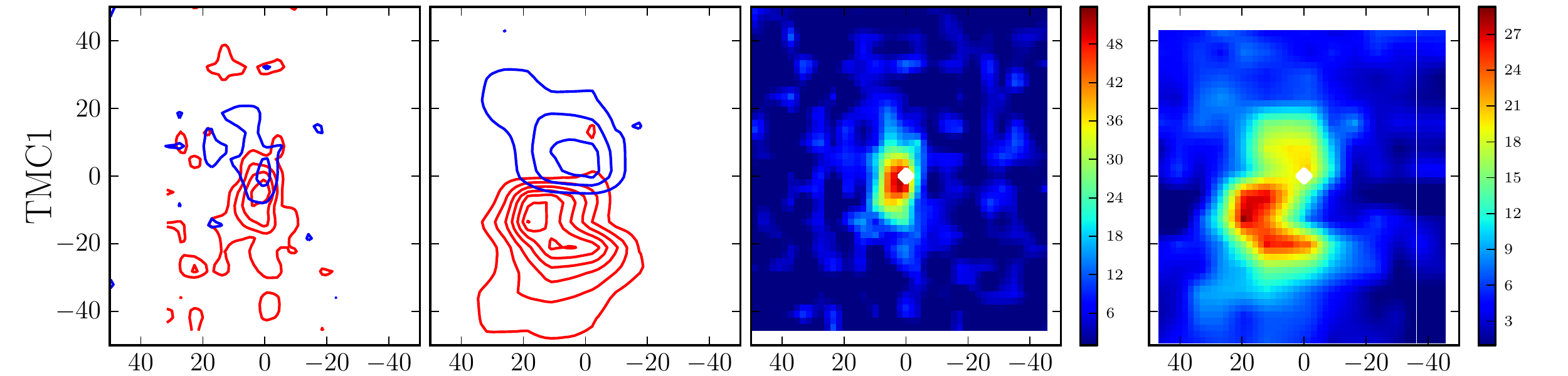} \\
    \includegraphics[scale=0.5]{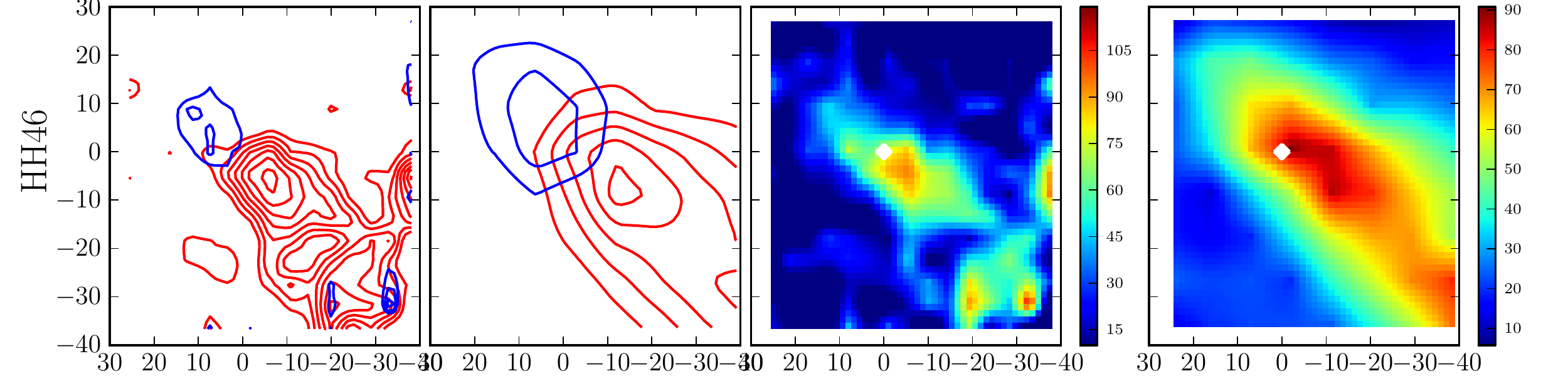} \\
    \includegraphics[scale=0.5]{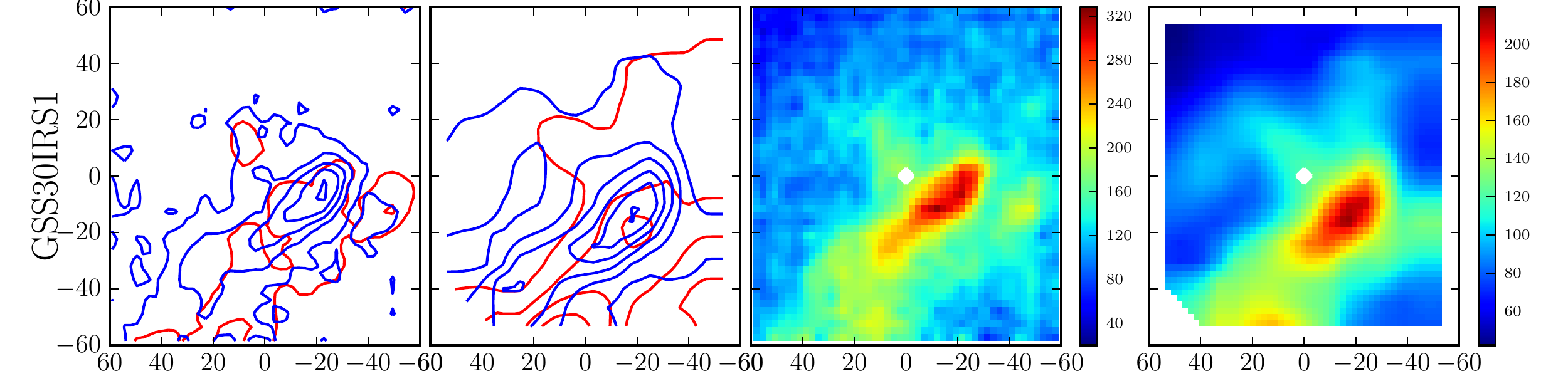}
    \caption{\small Caption is same as for Fig.~\ref{fig:AllIntensityCO_1}.}
    \label{fig:AllIntensityCO_2}
\end{figure*}

\begin{figure*}[htb]
    \centering
    \includegraphics[scale=0.5]{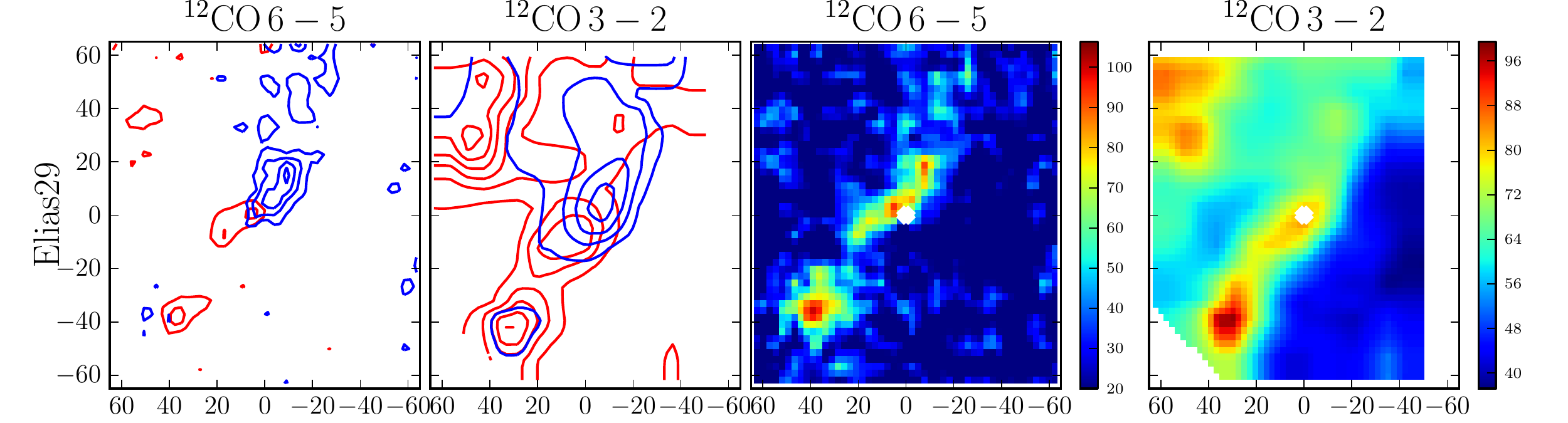} \\
    \includegraphics[scale=0.5]{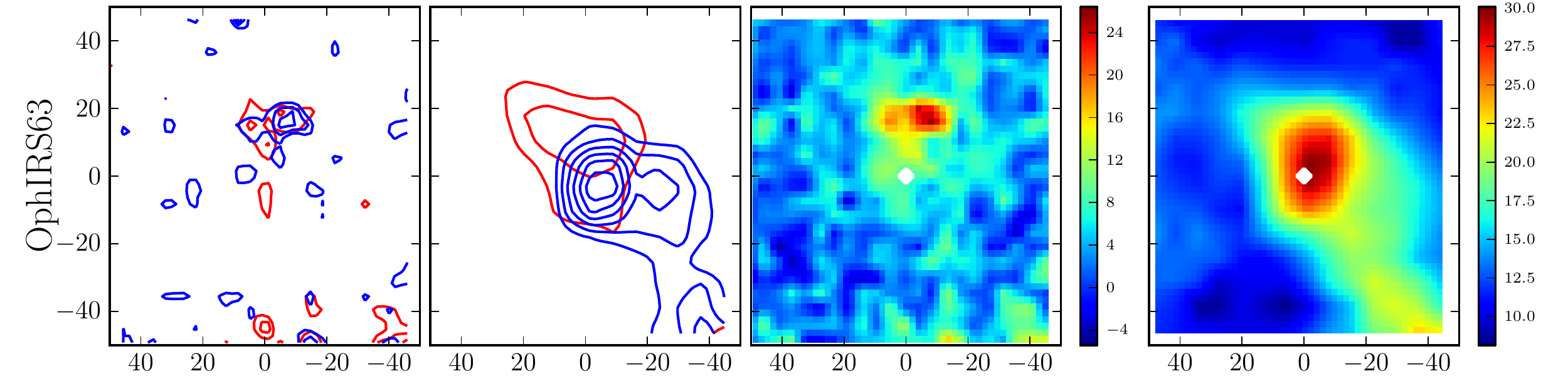} \\
    \includegraphics[scale=0.5]{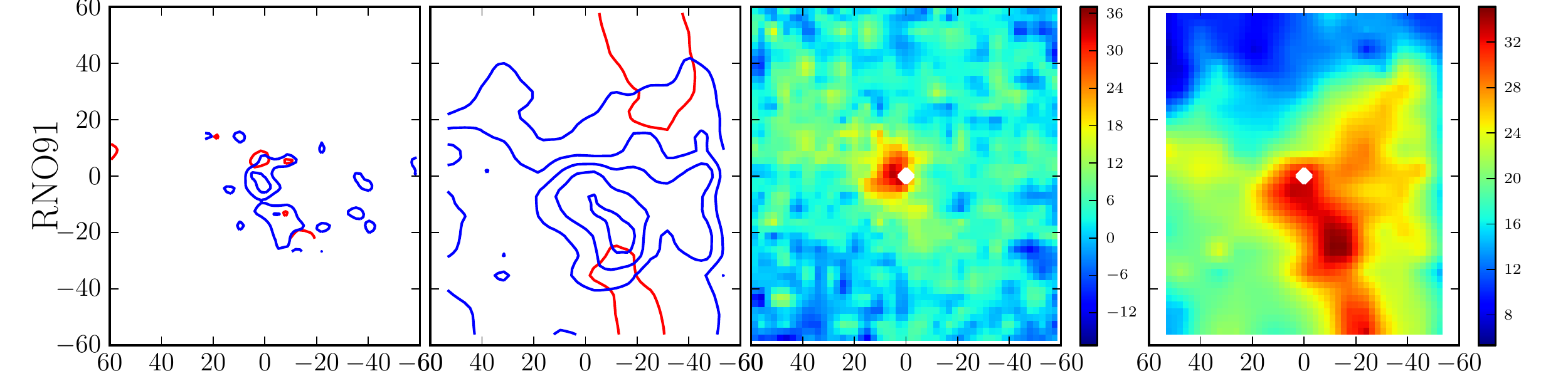} \\
    \includegraphics[scale=0.46]{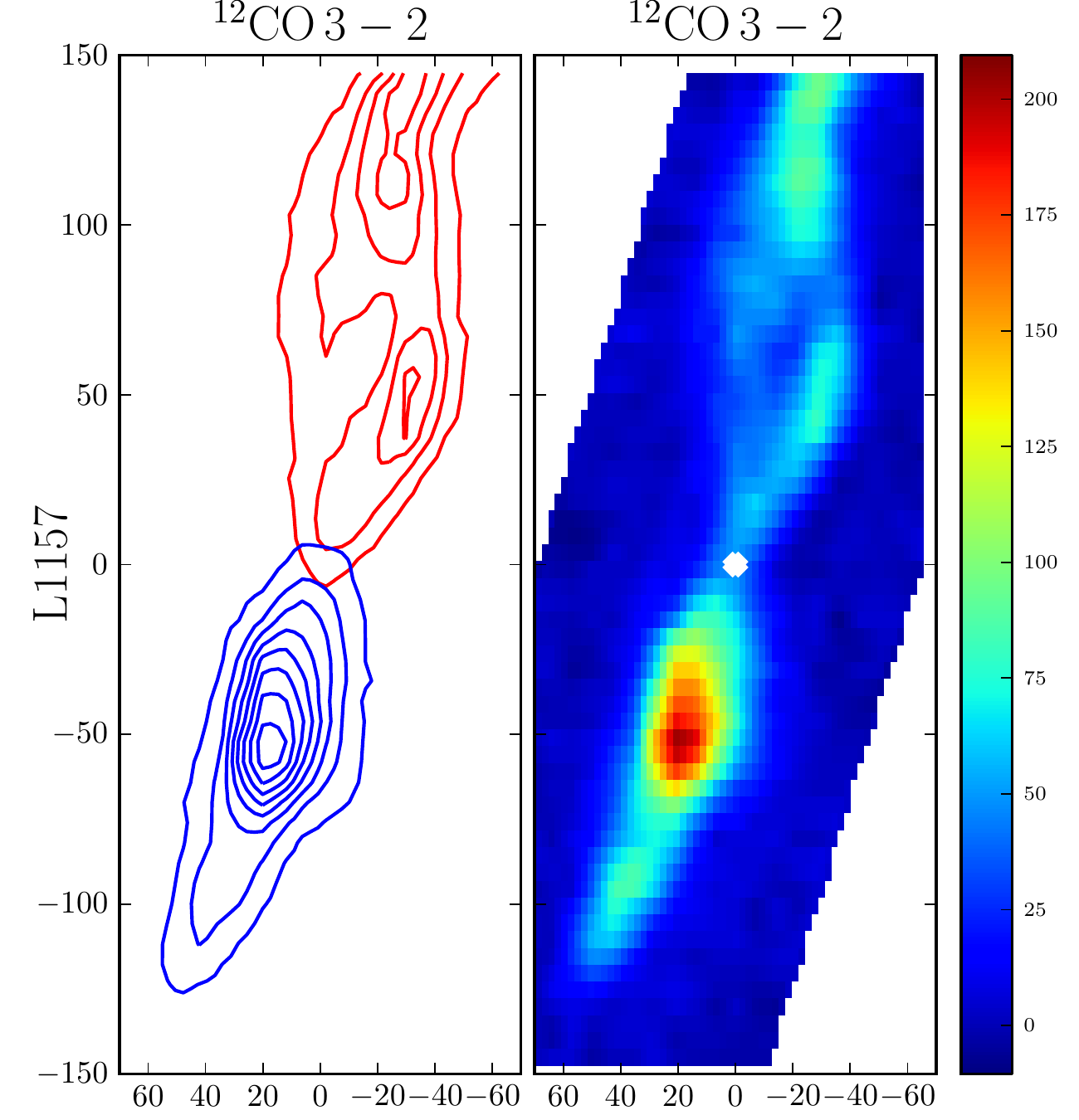}  
    \includegraphics[scale=0.46]{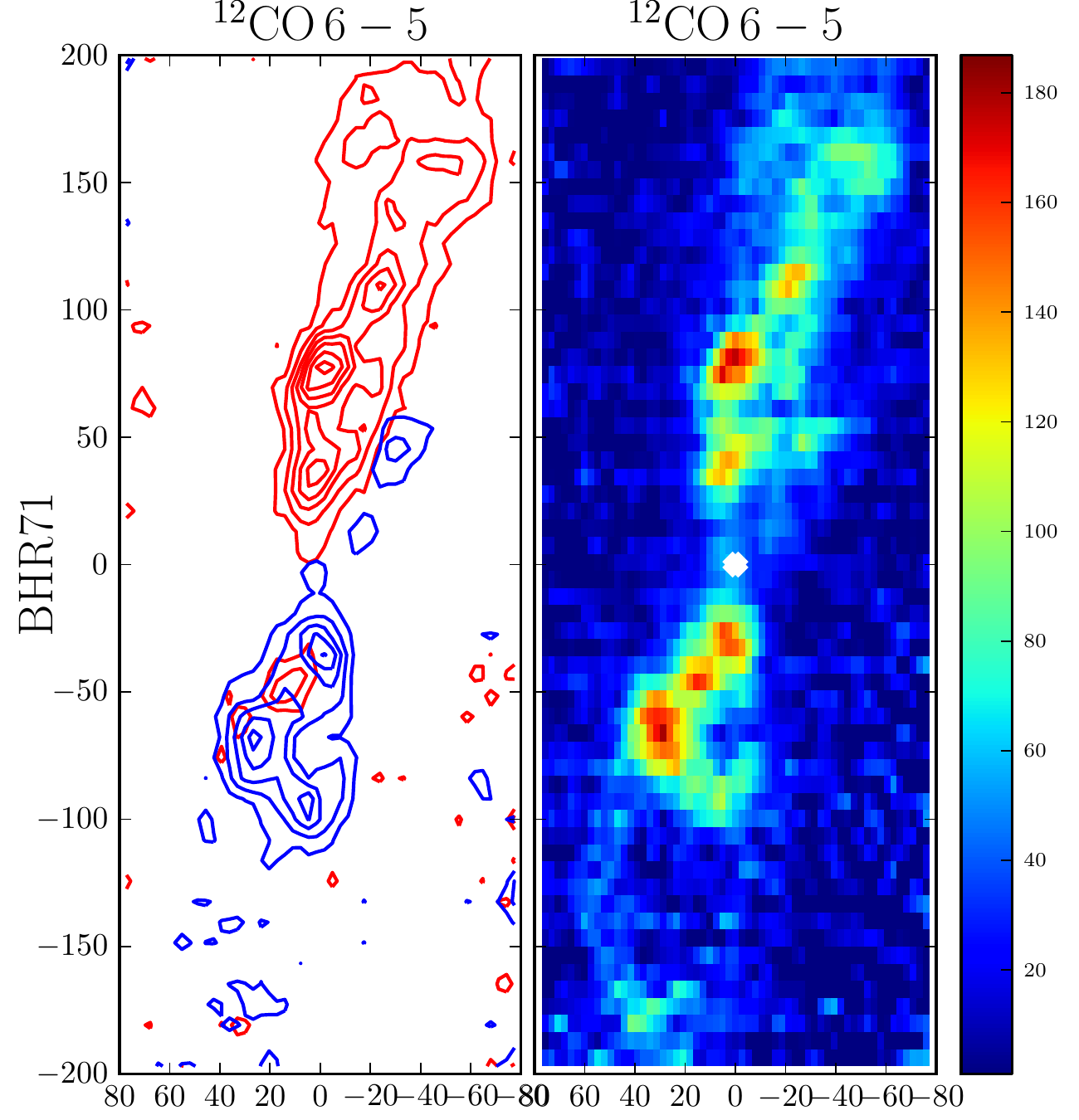} \\
    \includegraphics[scale=0.46]{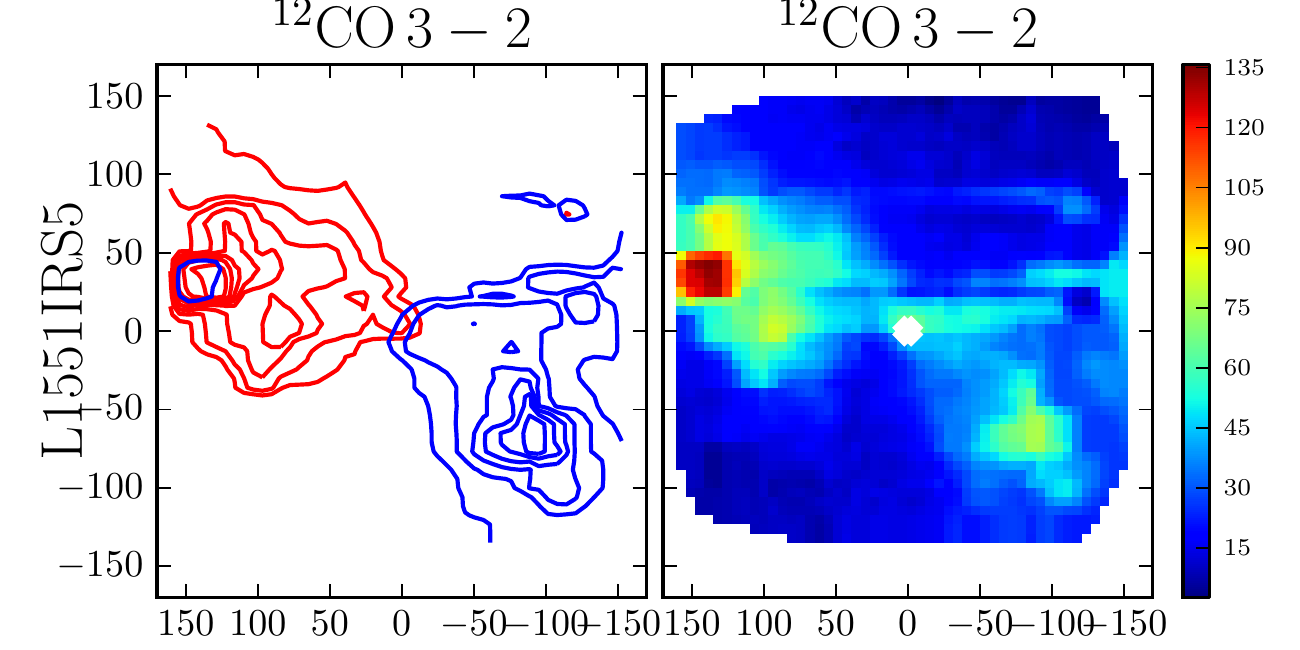} 
    \includegraphics[scale=0.46]{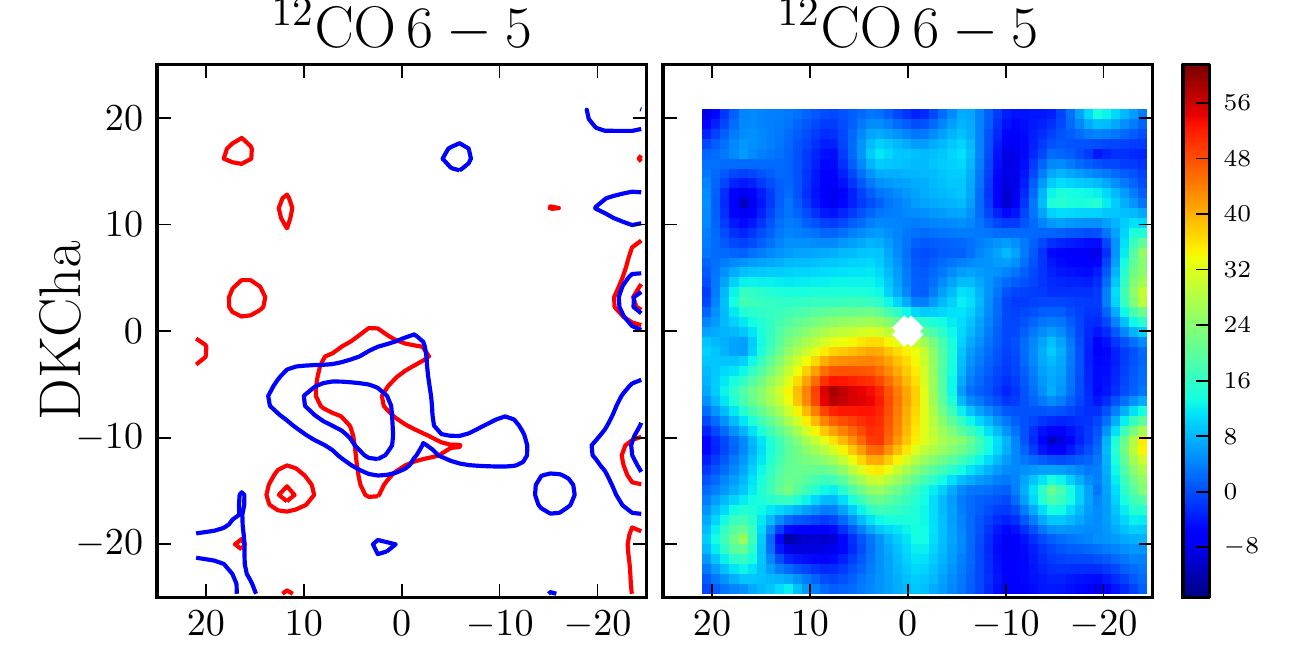} \\
    \includegraphics[scale=0.46]{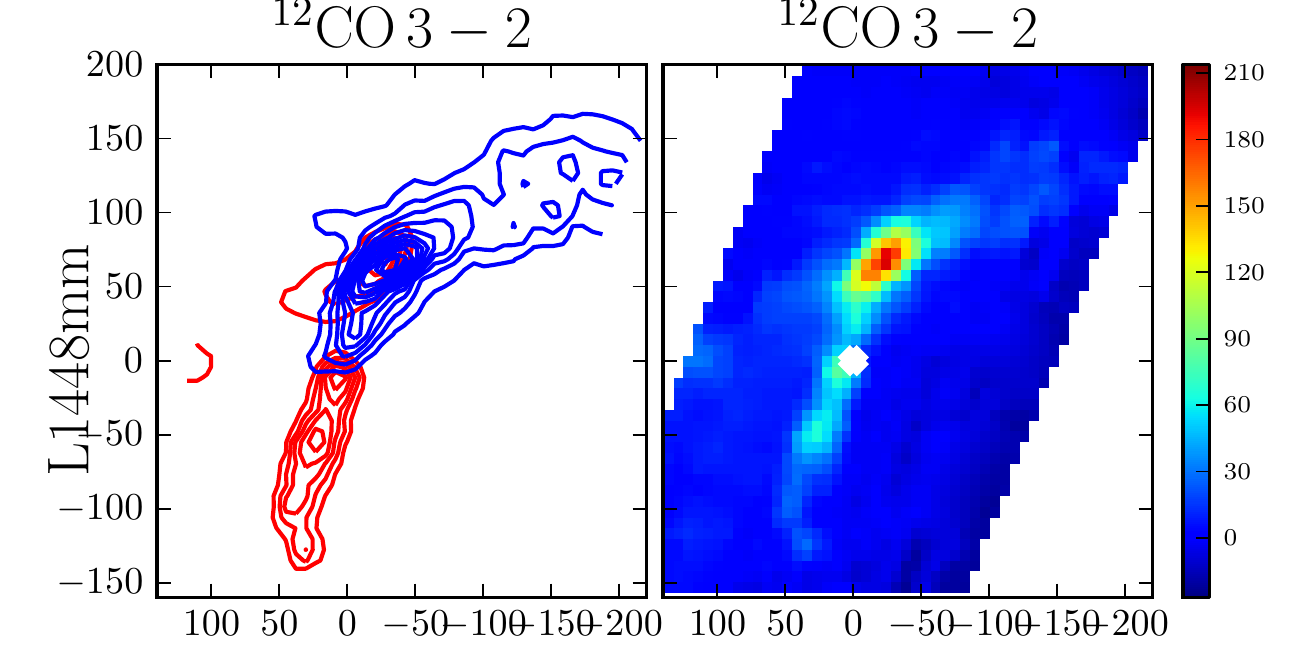}
    \caption{\small Caption is same as for Fig.~\ref{fig:AllIntensityCO_1}.}
    \label{fig:AllIntensityCO_3}
\end{figure*}


\begin{figure*}[htb]
    \centering
    \includegraphics[scale=0.51]{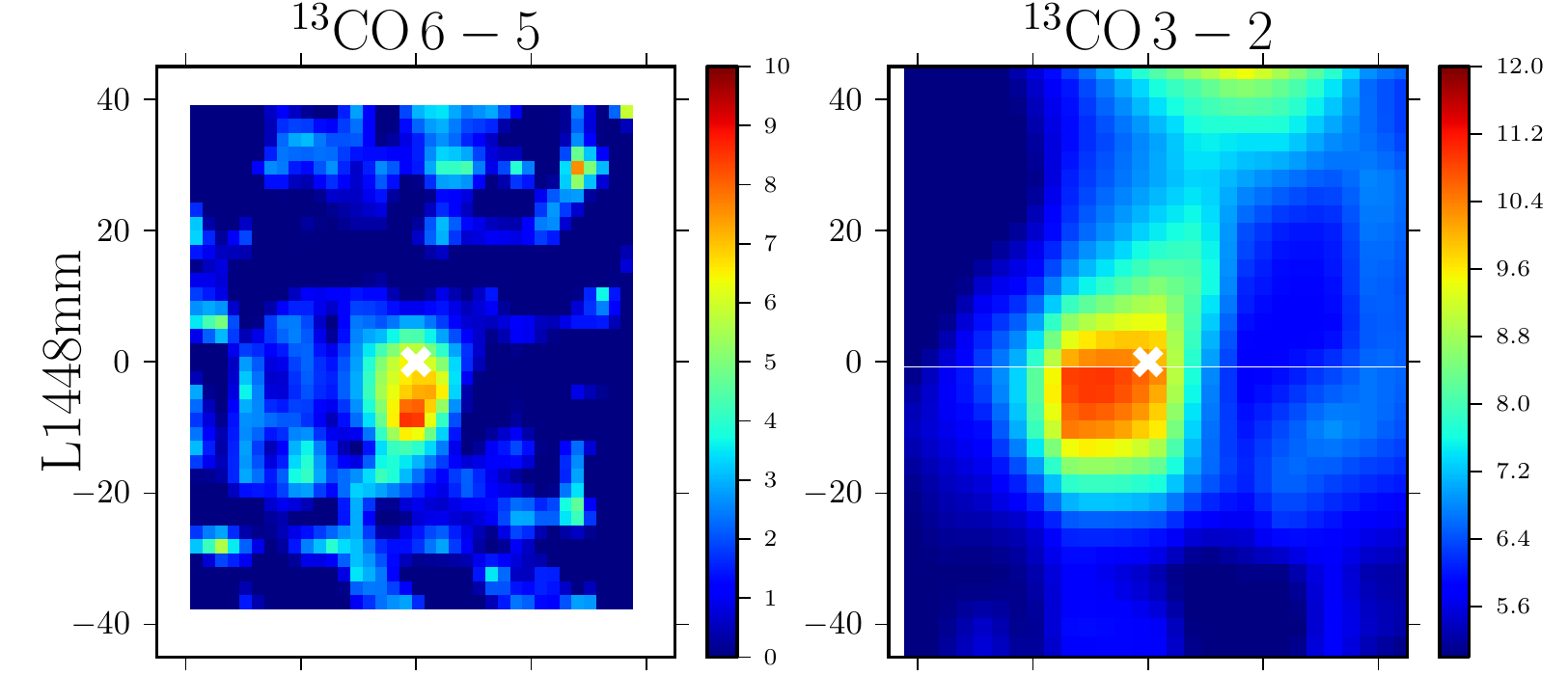}   
    \includegraphics[scale=0.51]{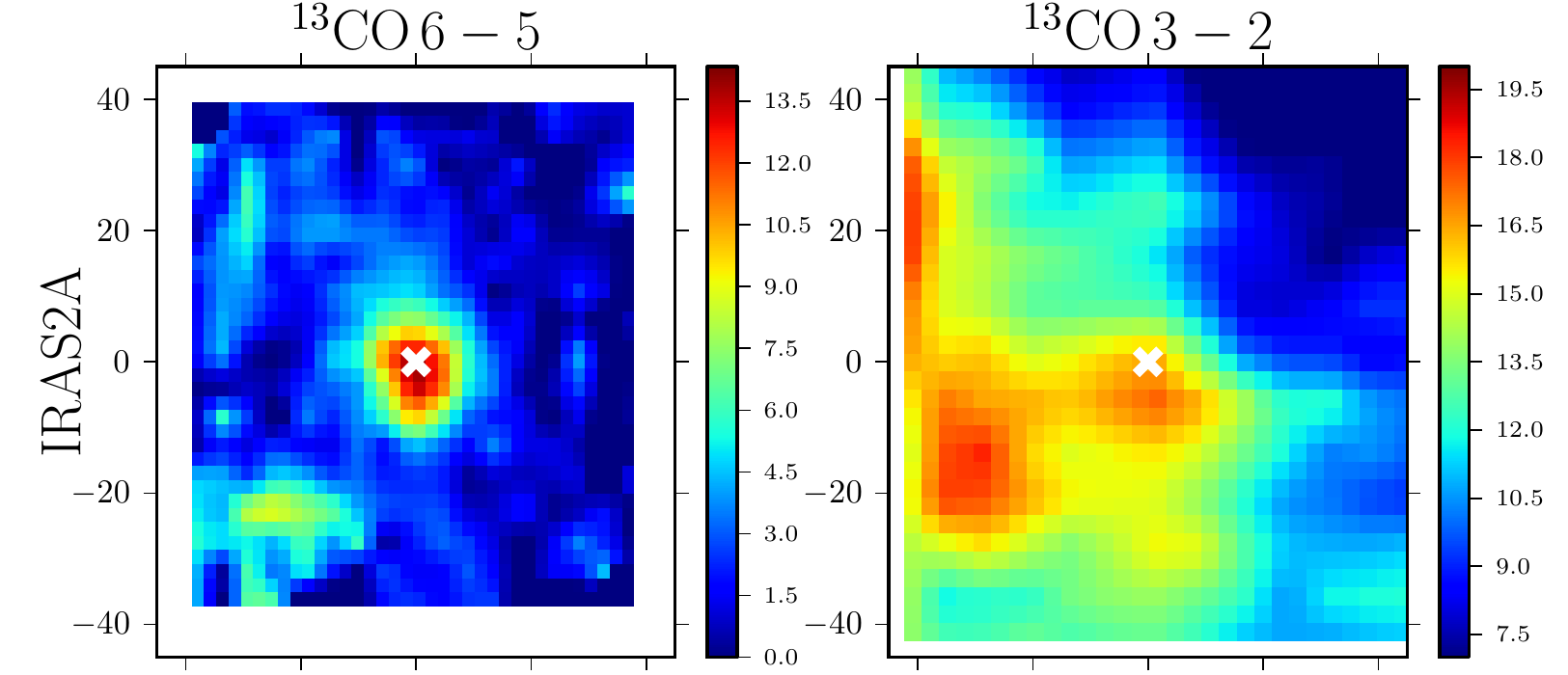} \\
    \includegraphics[scale=0.51]{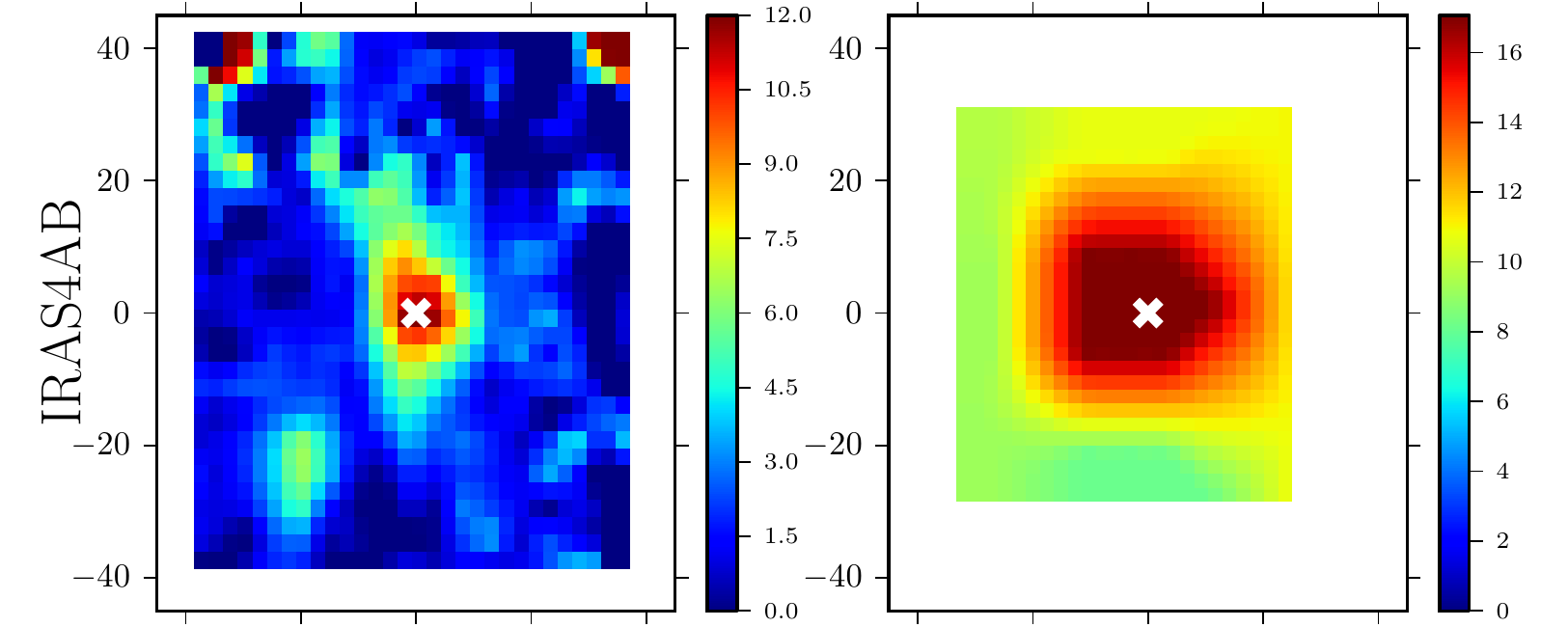}   
    \includegraphics[scale=0.51]{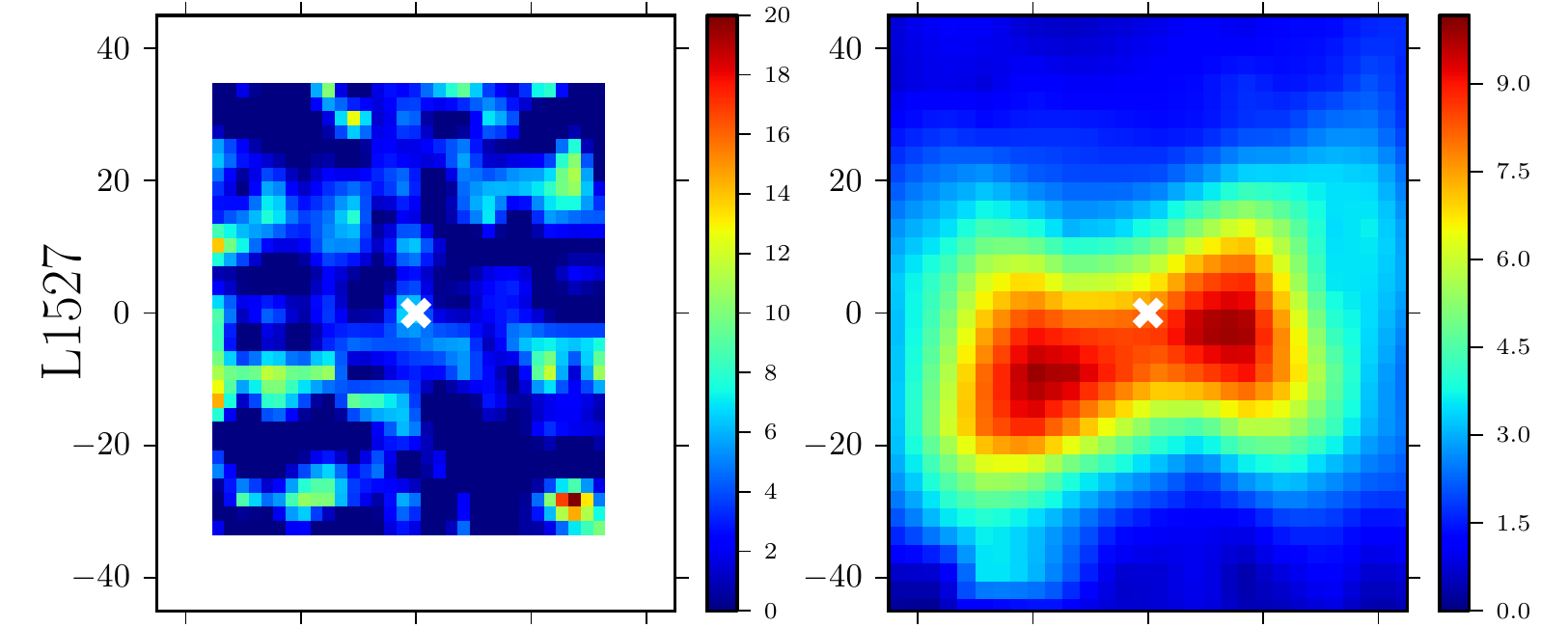} \\
    \includegraphics[scale=0.51]{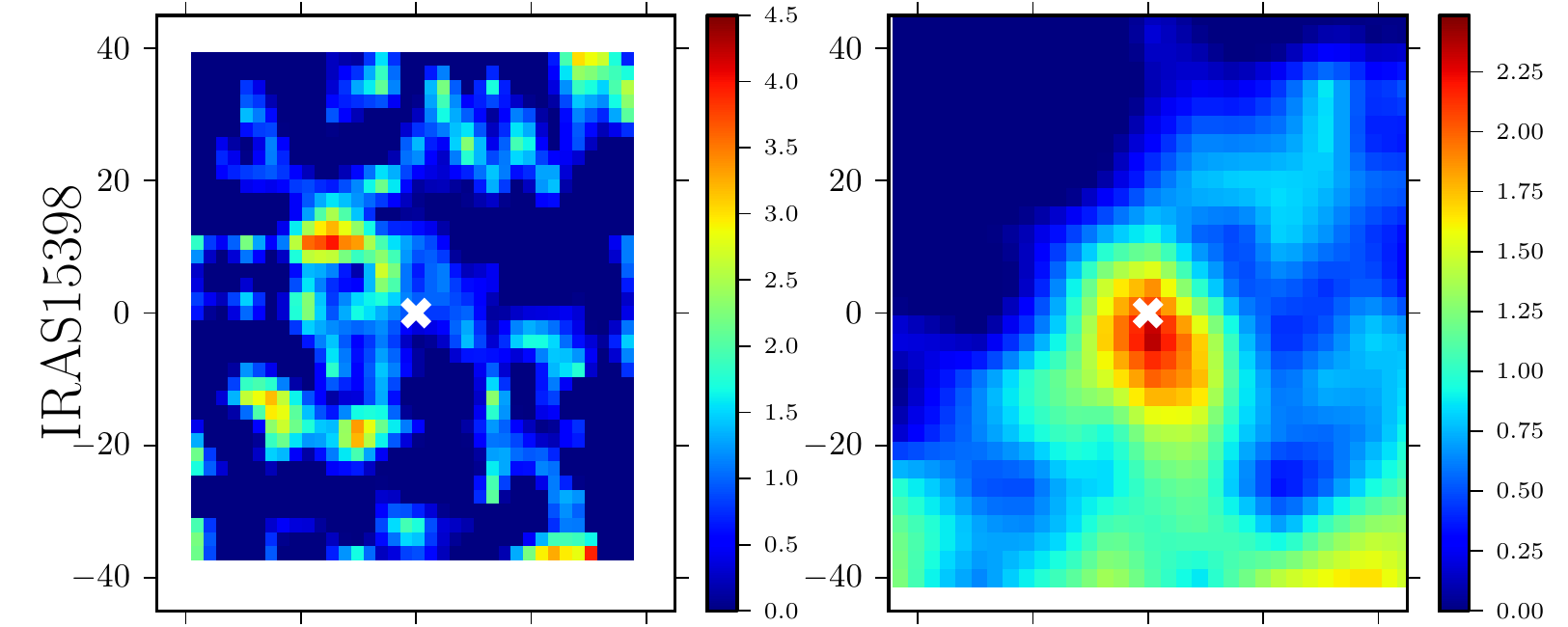}   
    \includegraphics[scale=0.51]{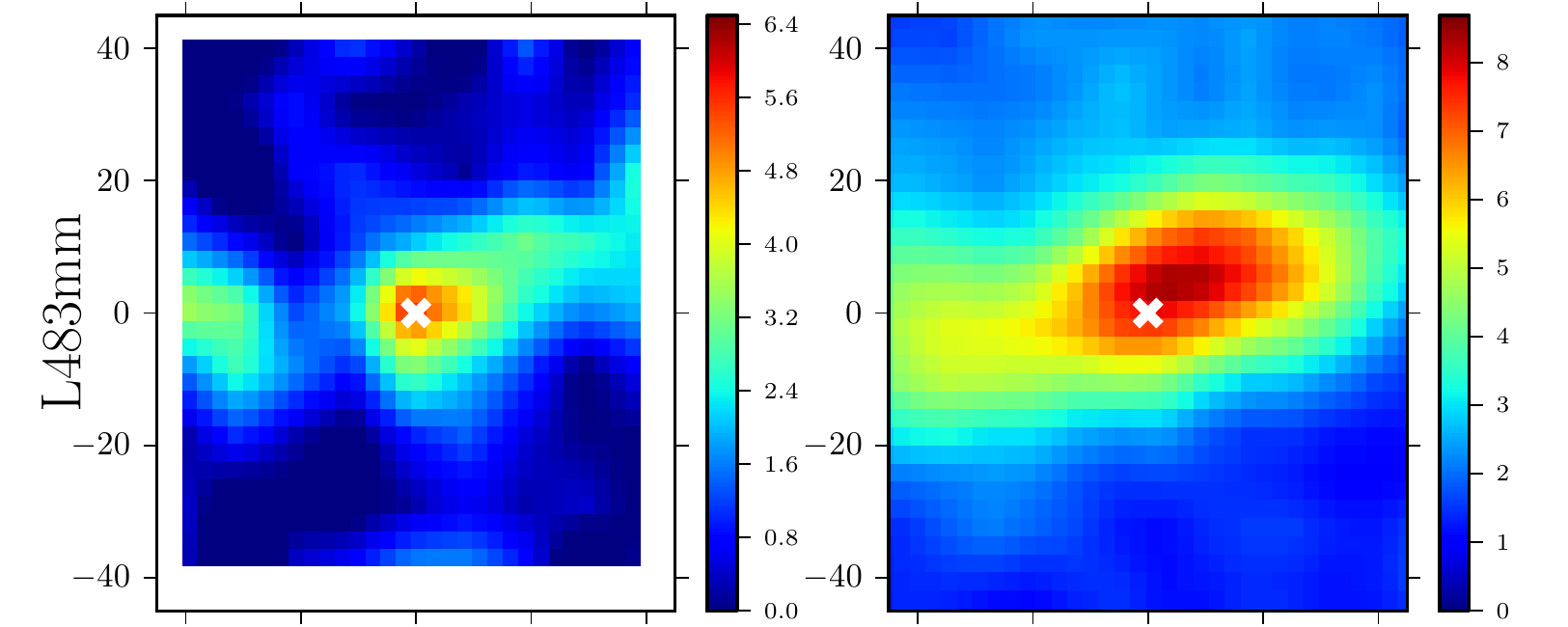} \\
    \includegraphics[scale=0.51]{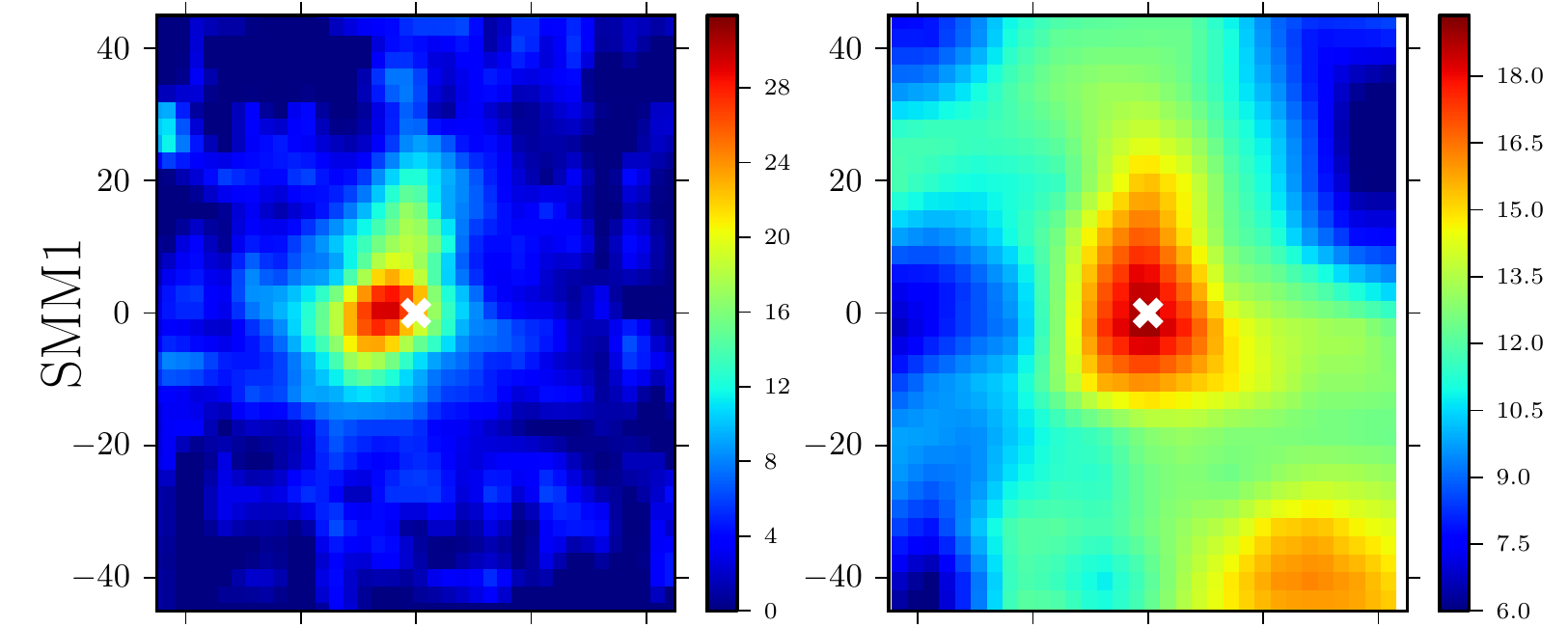}   
    \includegraphics[scale=0.51]{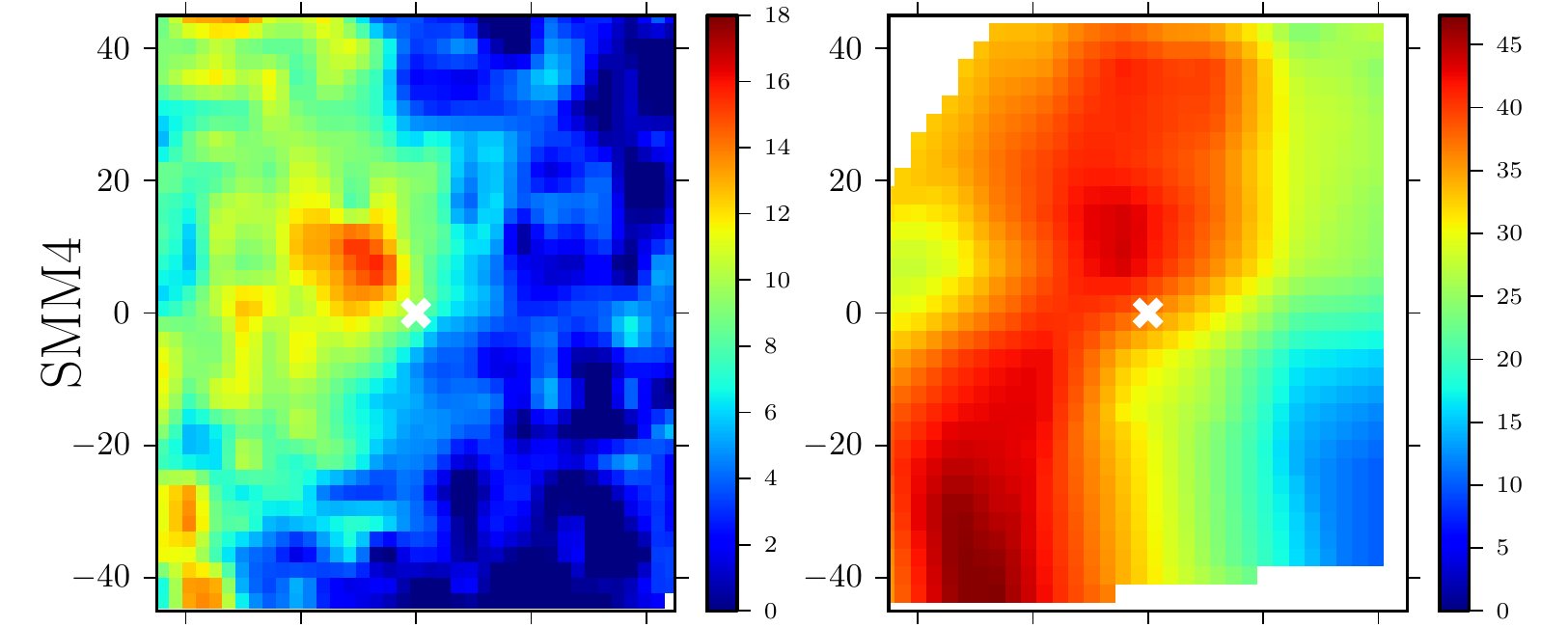} \\
    \includegraphics[scale=0.51]{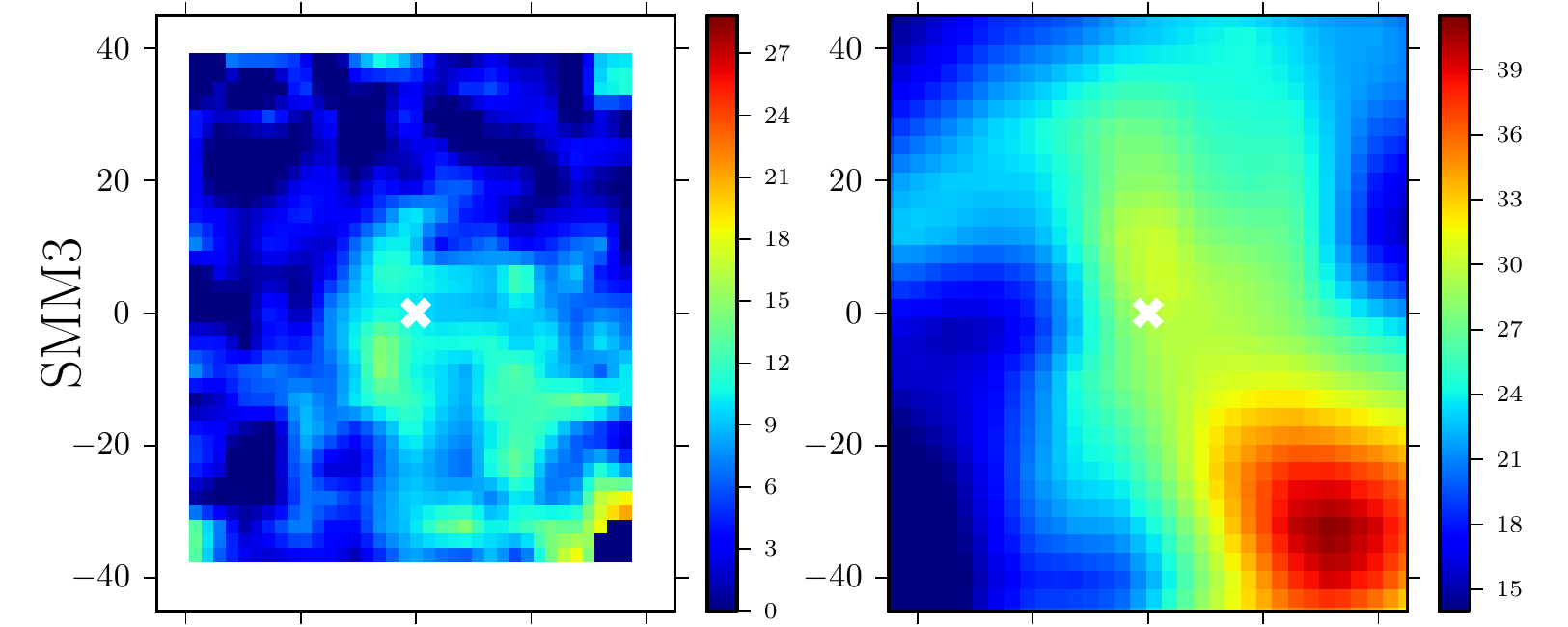}   
    \includegraphics[scale=0.51]{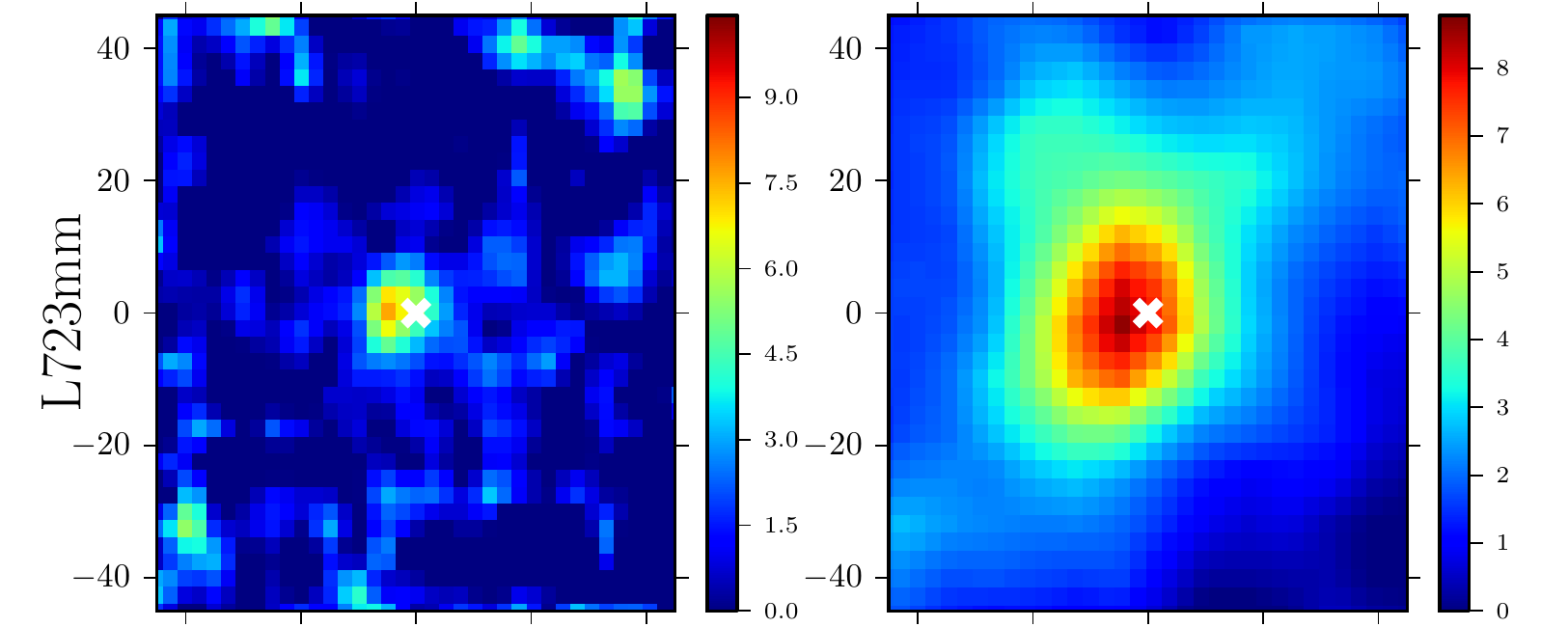} \\
    \includegraphics[scale=0.51]{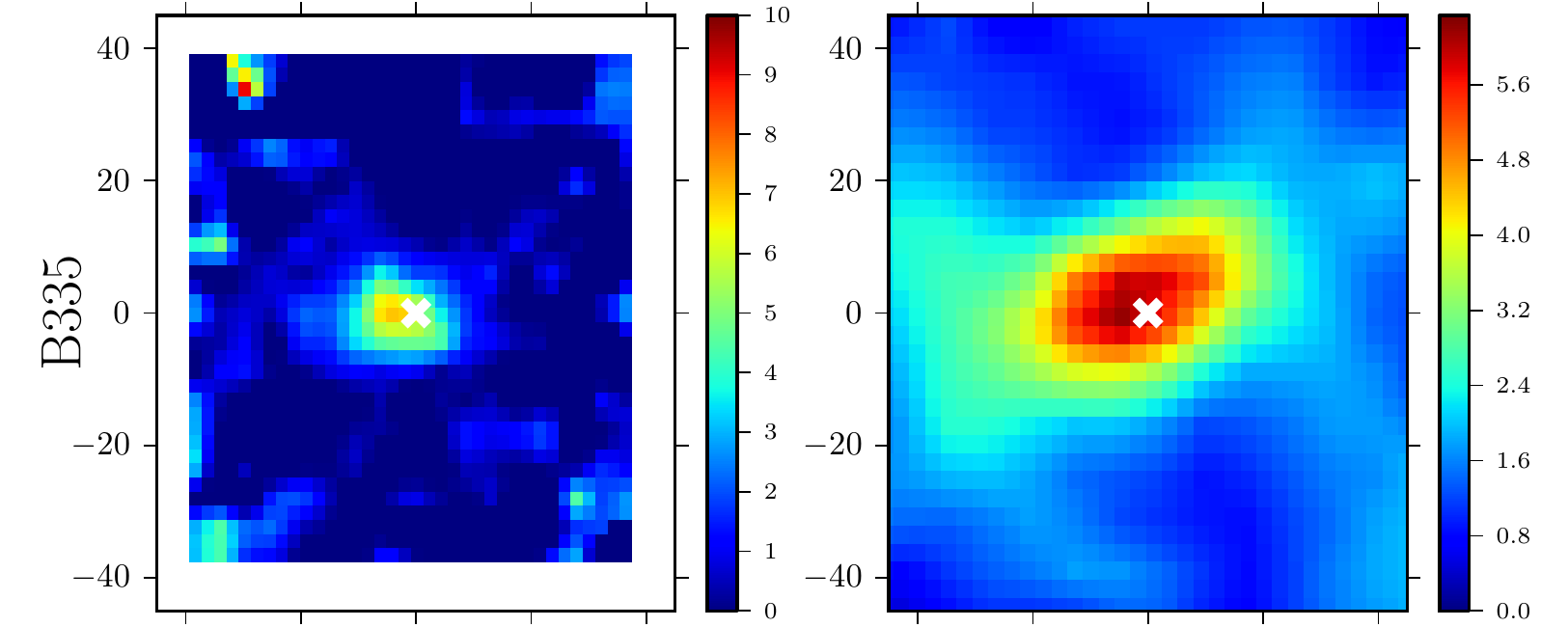}   
    \includegraphics[scale=0.51]{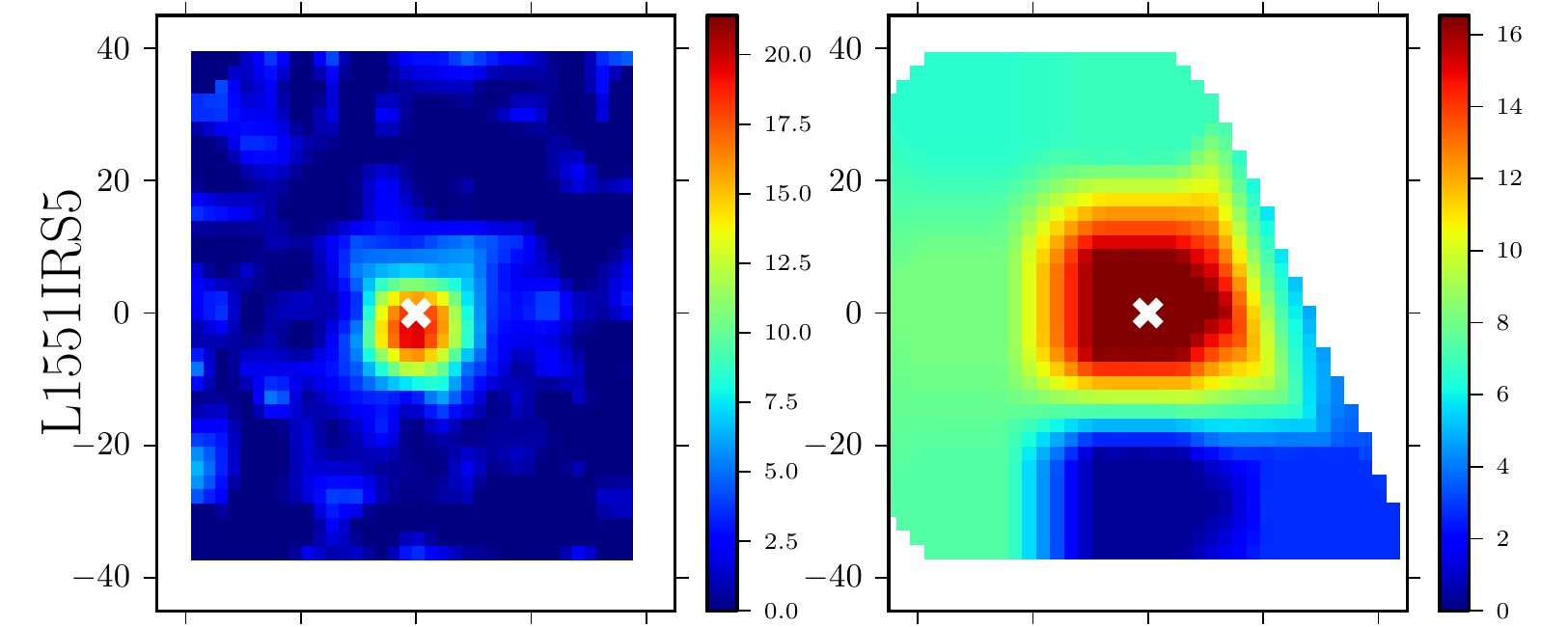} \\
    \includegraphics[scale=0.51]{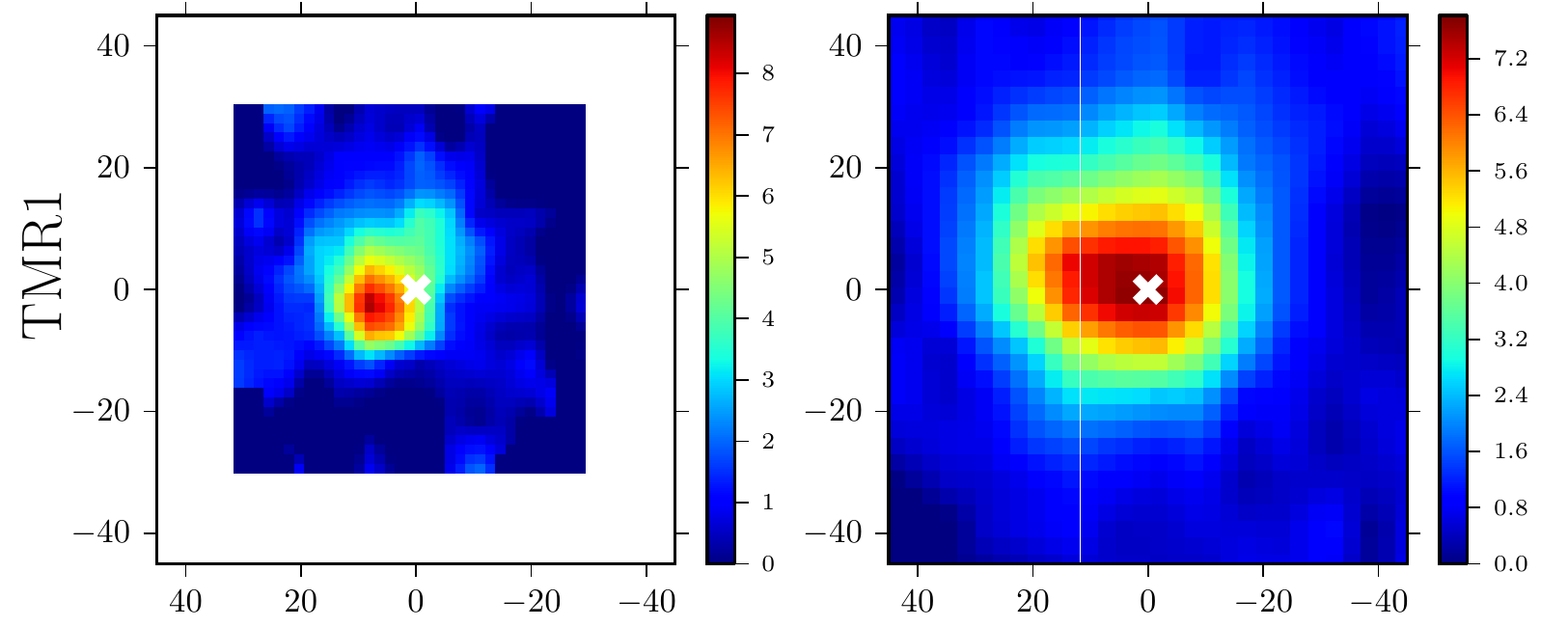}   
    \includegraphics[scale=0.51]{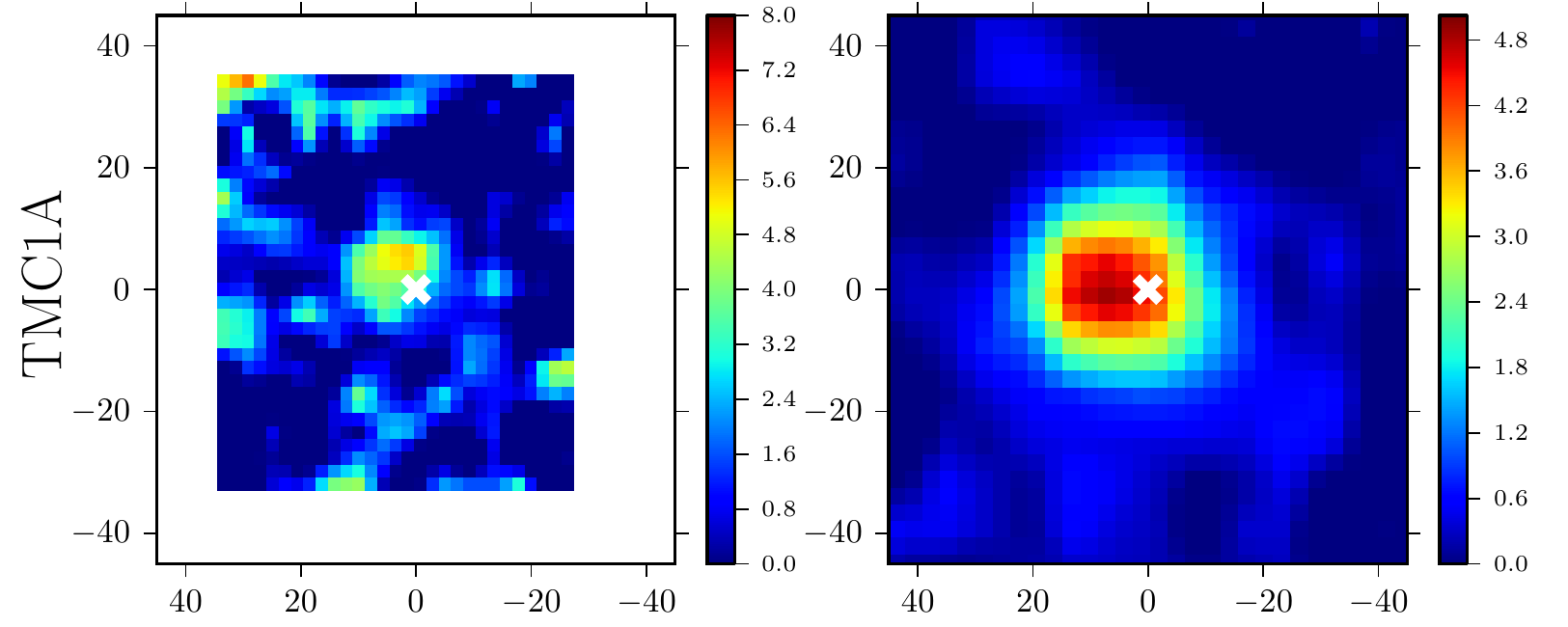}   
    \caption{\small \thco\ 6--5 and 3--2 integrated intensity maps of 
  the sources (in K km s$^{-1}$). }
    \label{fig:All13COs1}
\end{figure*}

\begin{figure*}[htb]
\begin{flushleft}
    \includegraphics[scale=0.51]{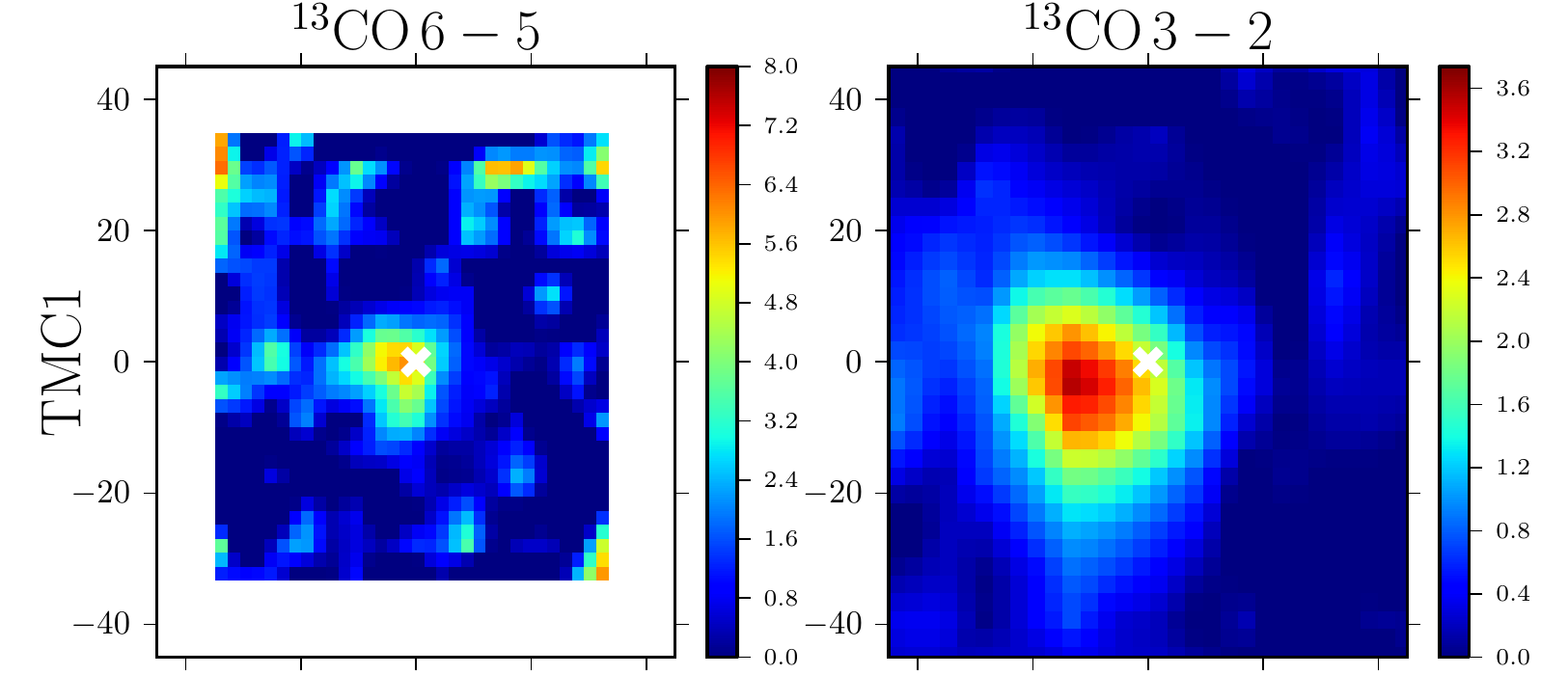}   
    \includegraphics[scale=0.51]{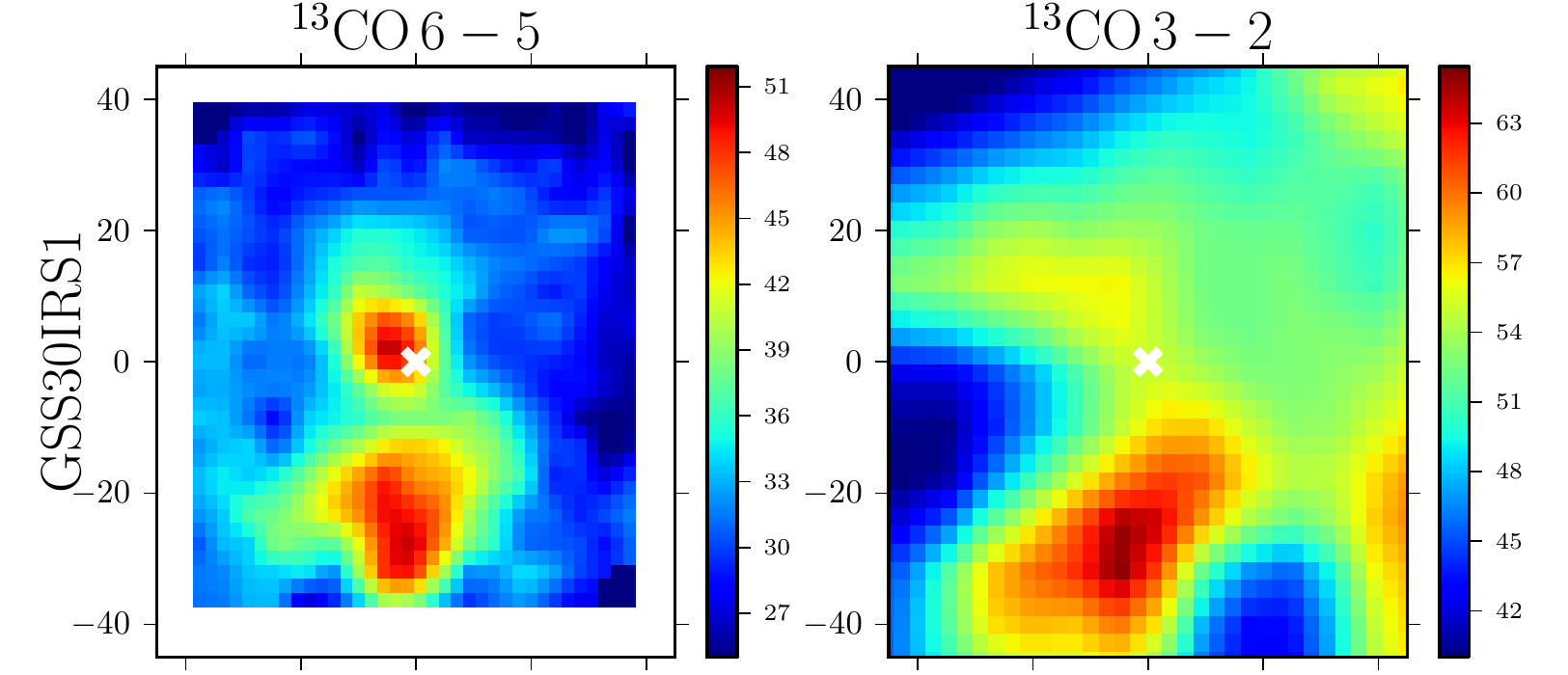} \\
    \includegraphics[scale=0.51]{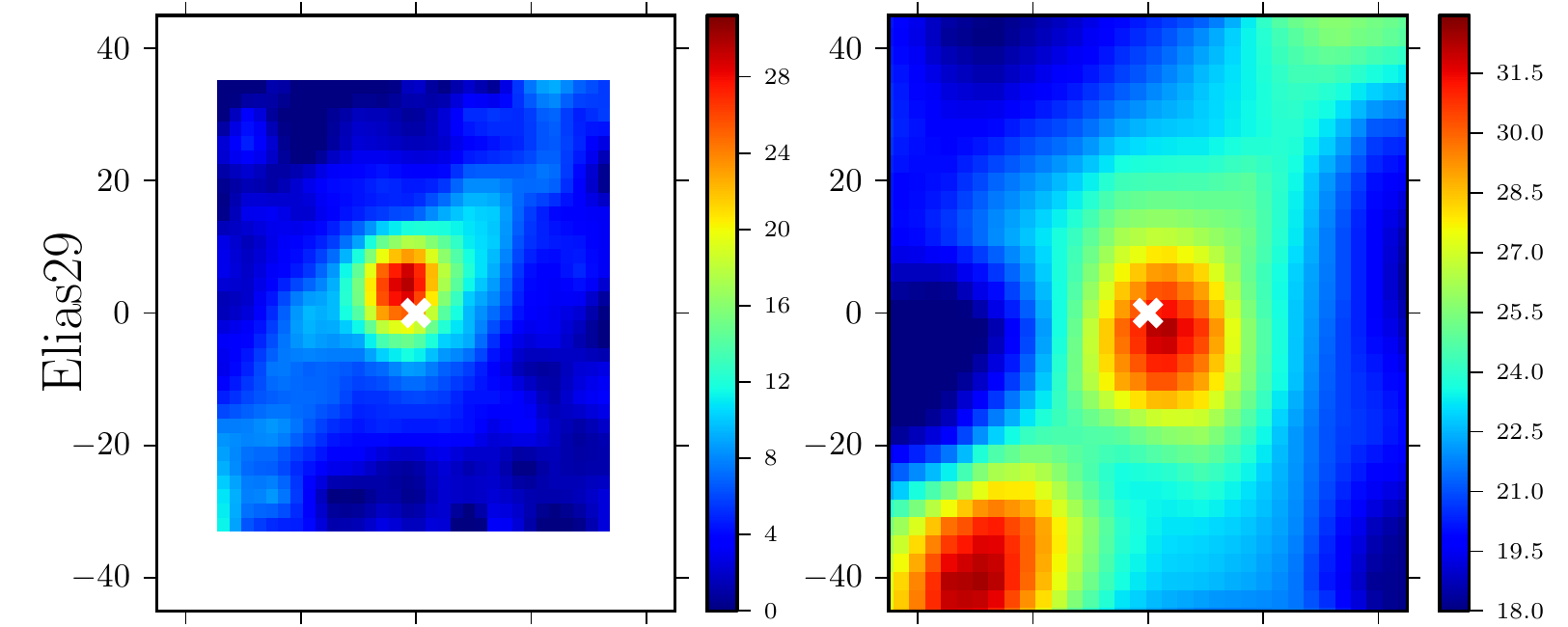}   
    \includegraphics[scale=0.51]{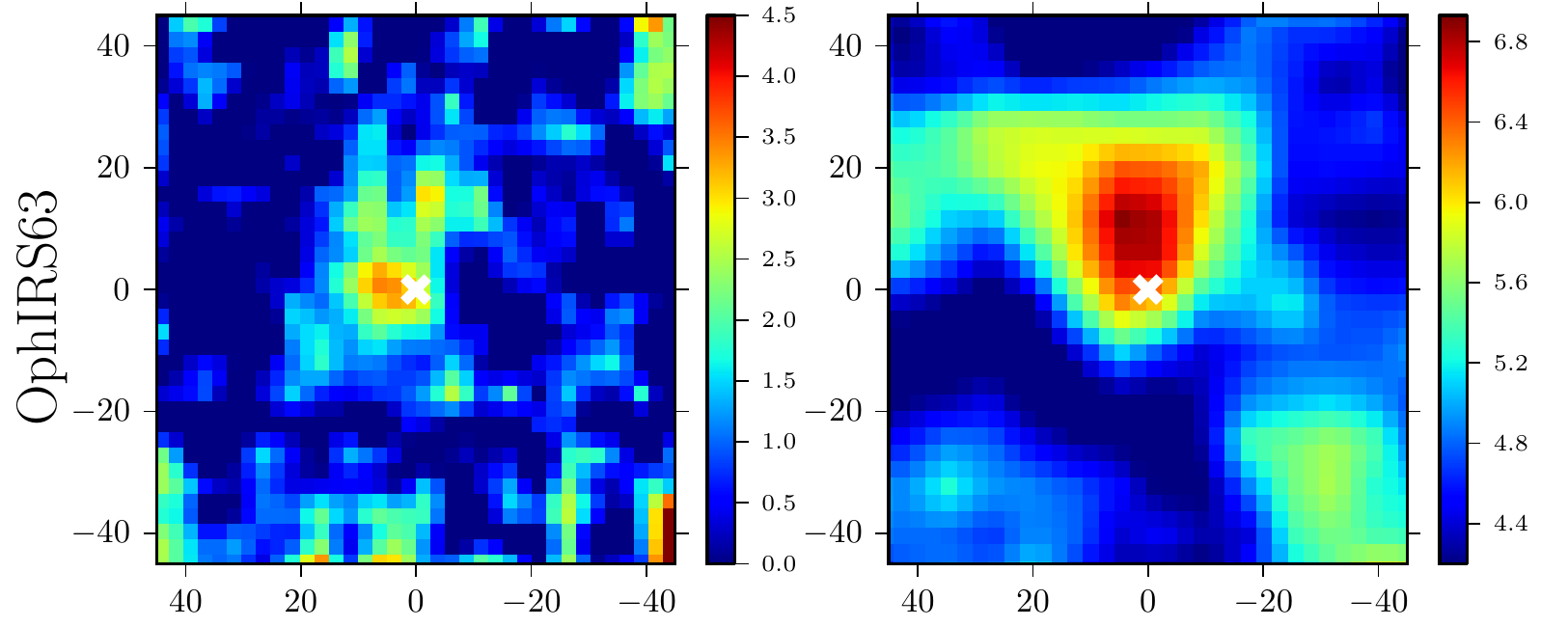} \\
    \includegraphics[scale=0.51]{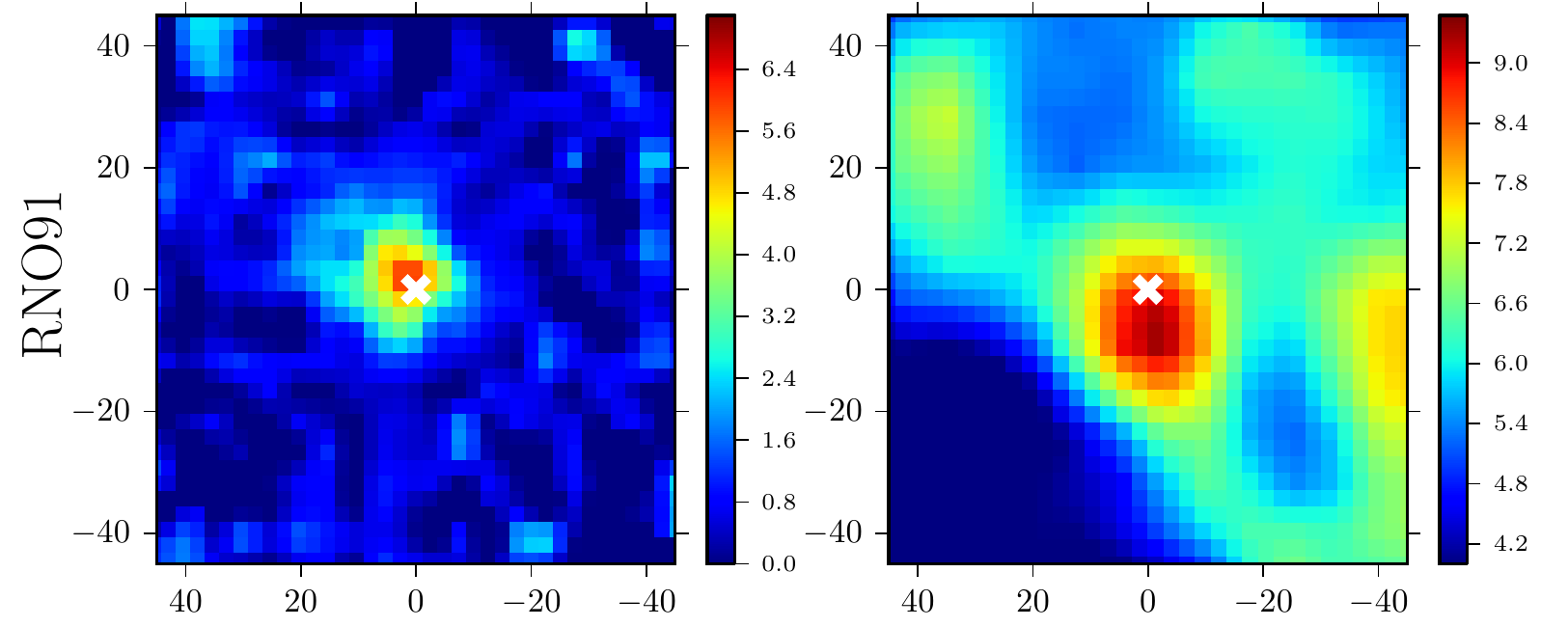}   
    \includegraphics[scale=0.47]{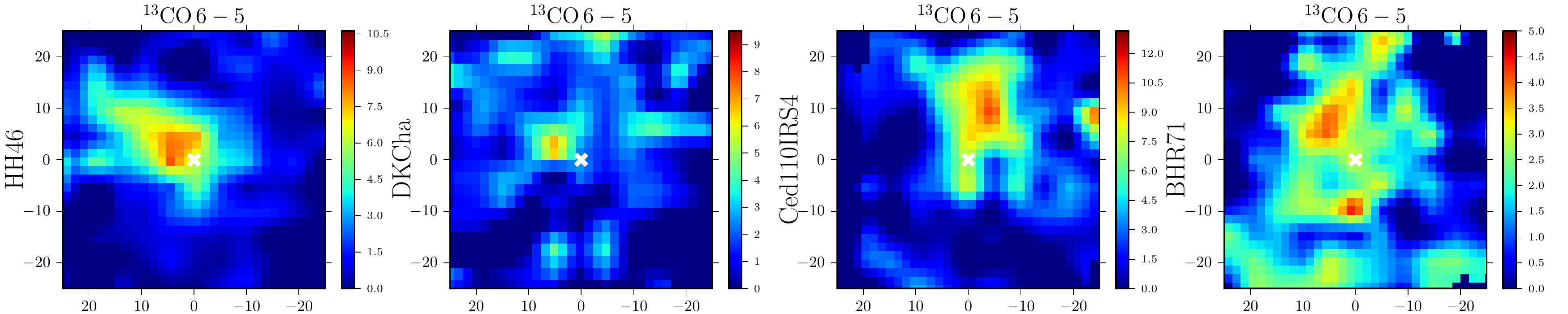}   
    \includegraphics[scale=0.51]{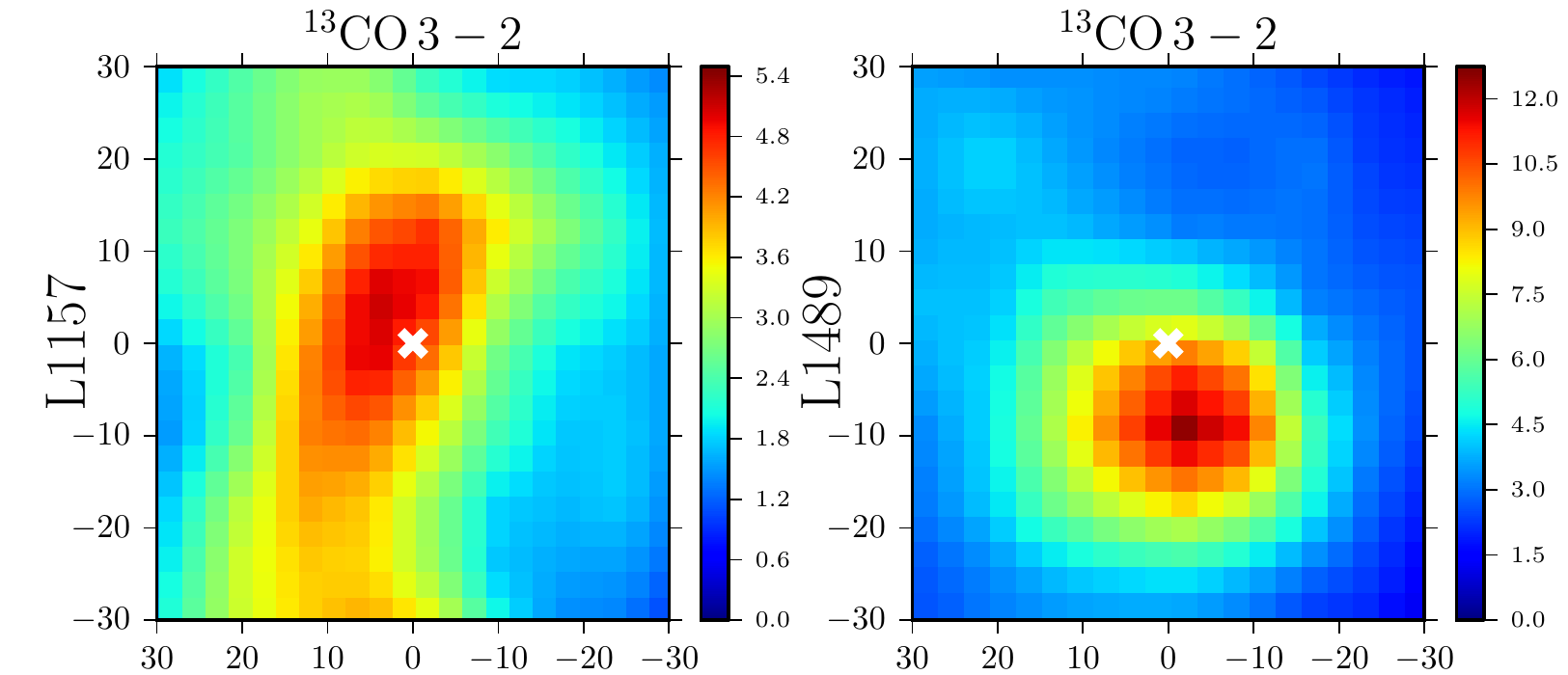}   
\end{flushleft}
    \caption{\small Caption is same as for Fig.~\ref{fig:All13COs1}.}
    \label{fig:All13COs2}
\end{figure*}

\begin{figure*}[!ht]
    \centering
    \includegraphics[scale=0.50]{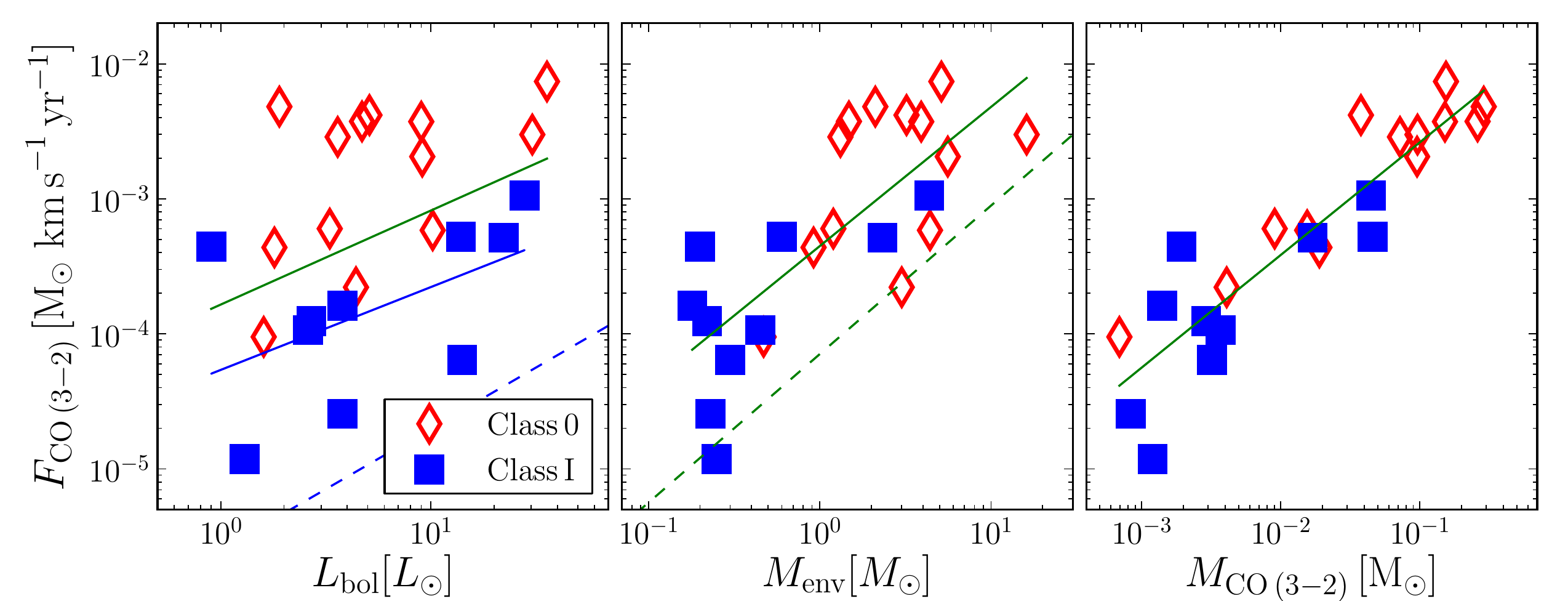}
    \caption{\small Correlations between \FCO\ as measured from CO 3--2 and 
    bolometric luminosity, envelope mass and outflow mass as determined from CO 3--2.}
    \label{fig:corrFCO32}
\end{figure*}

\begin{figure*}[htb]
    \centering
    \includegraphics[scale=0.20]{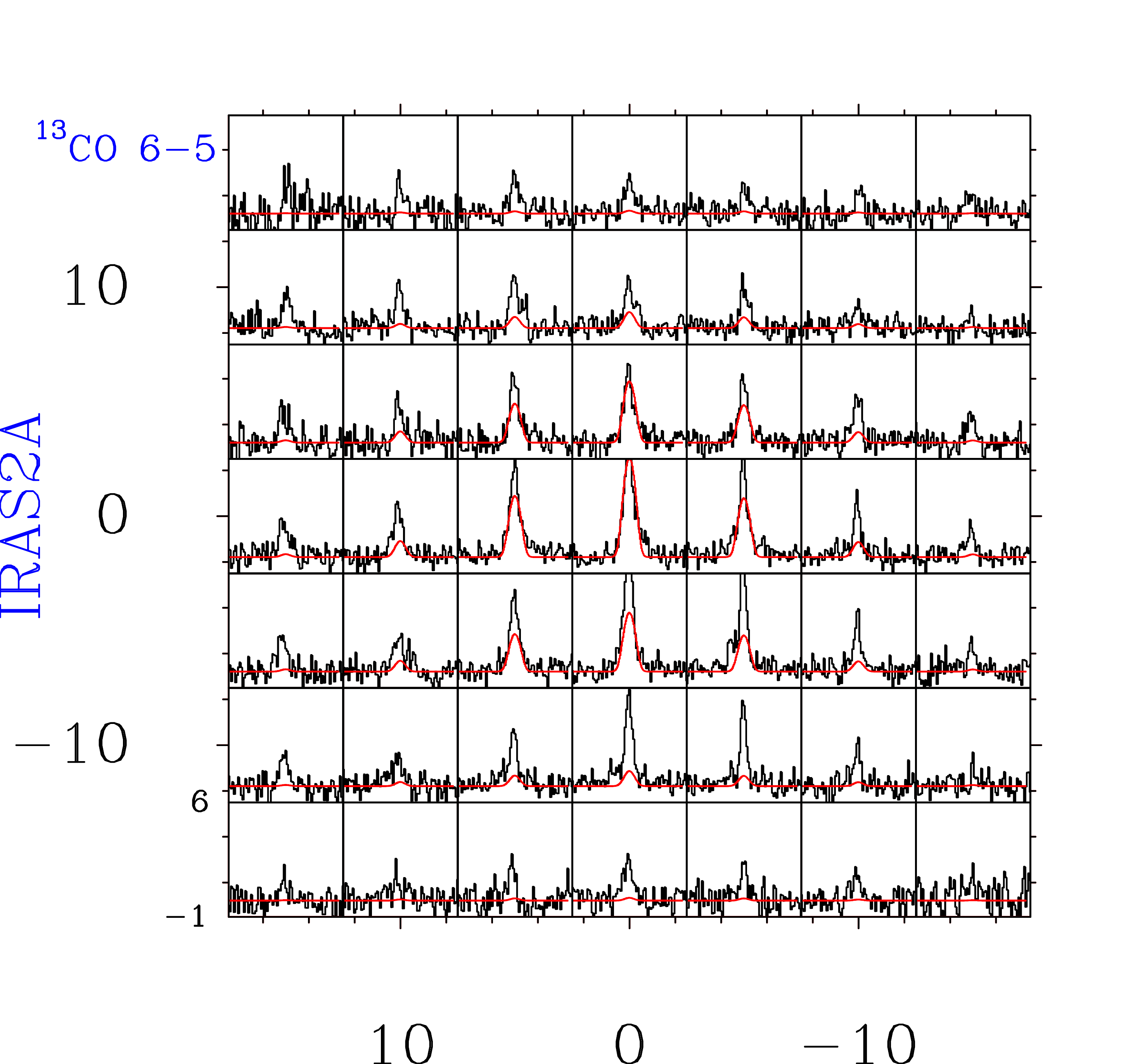}   
    \includegraphics[scale=0.20]{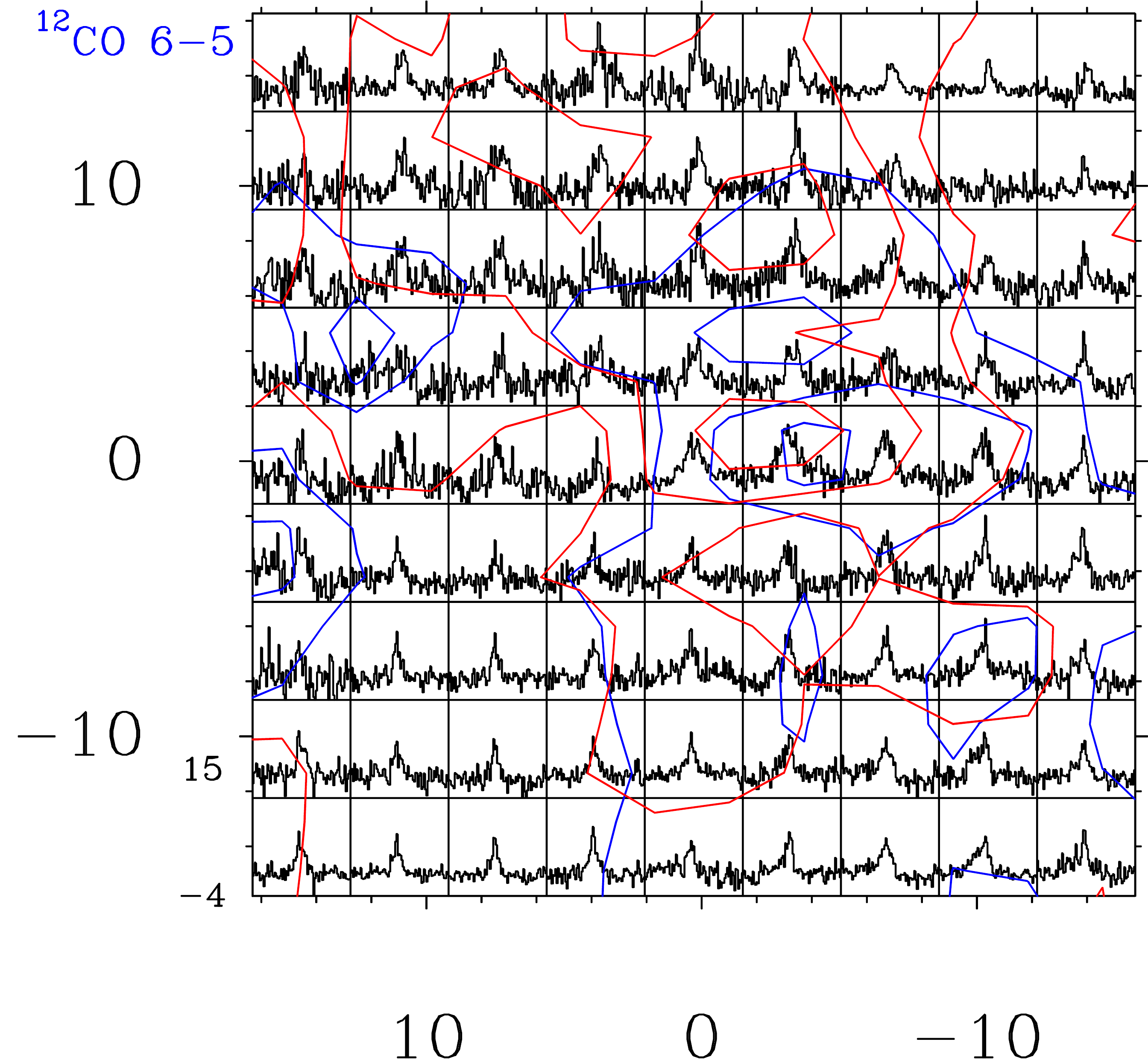}   
    \includegraphics[scale=0.20]{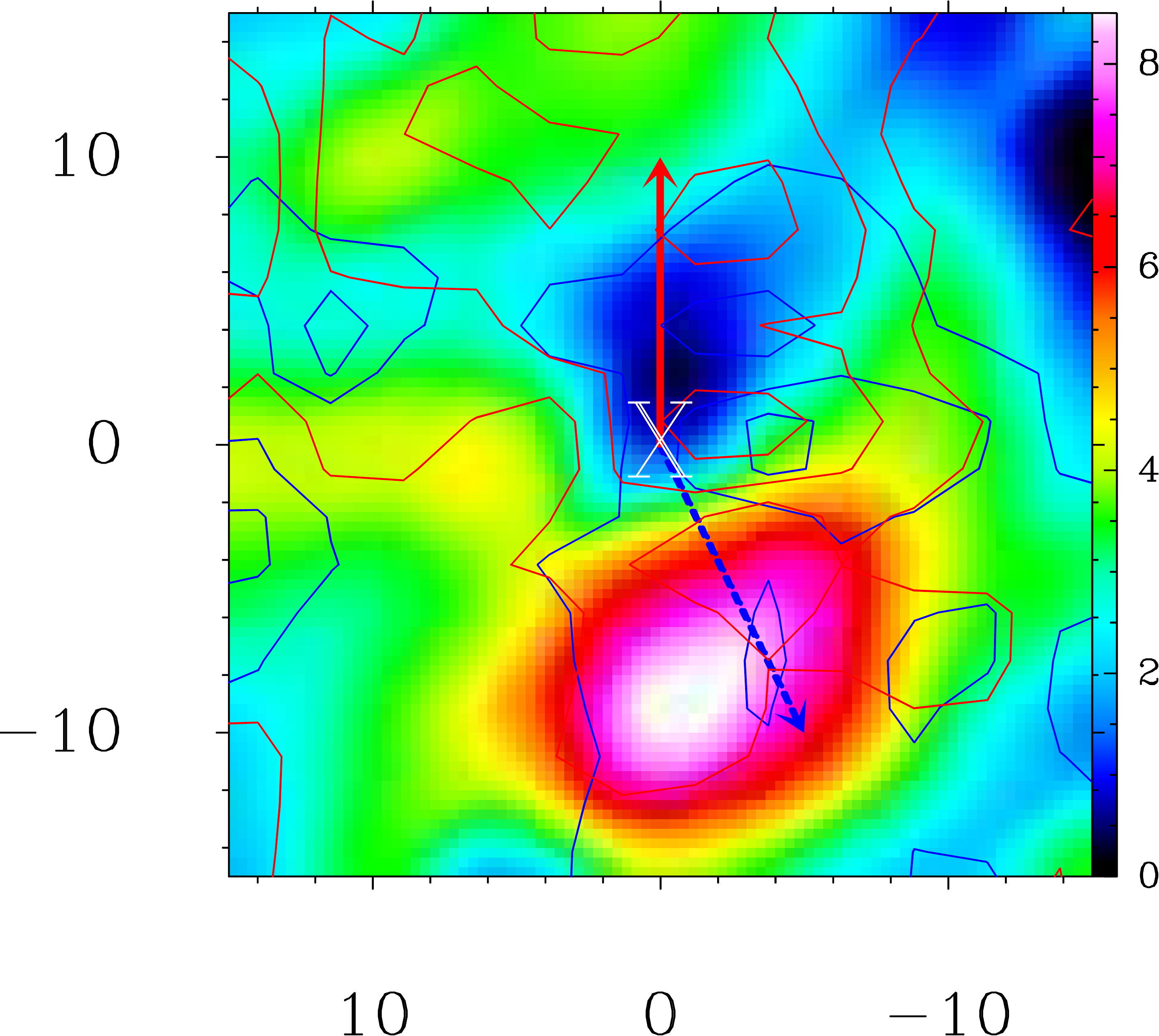} \\
    \includegraphics[scale=0.20]{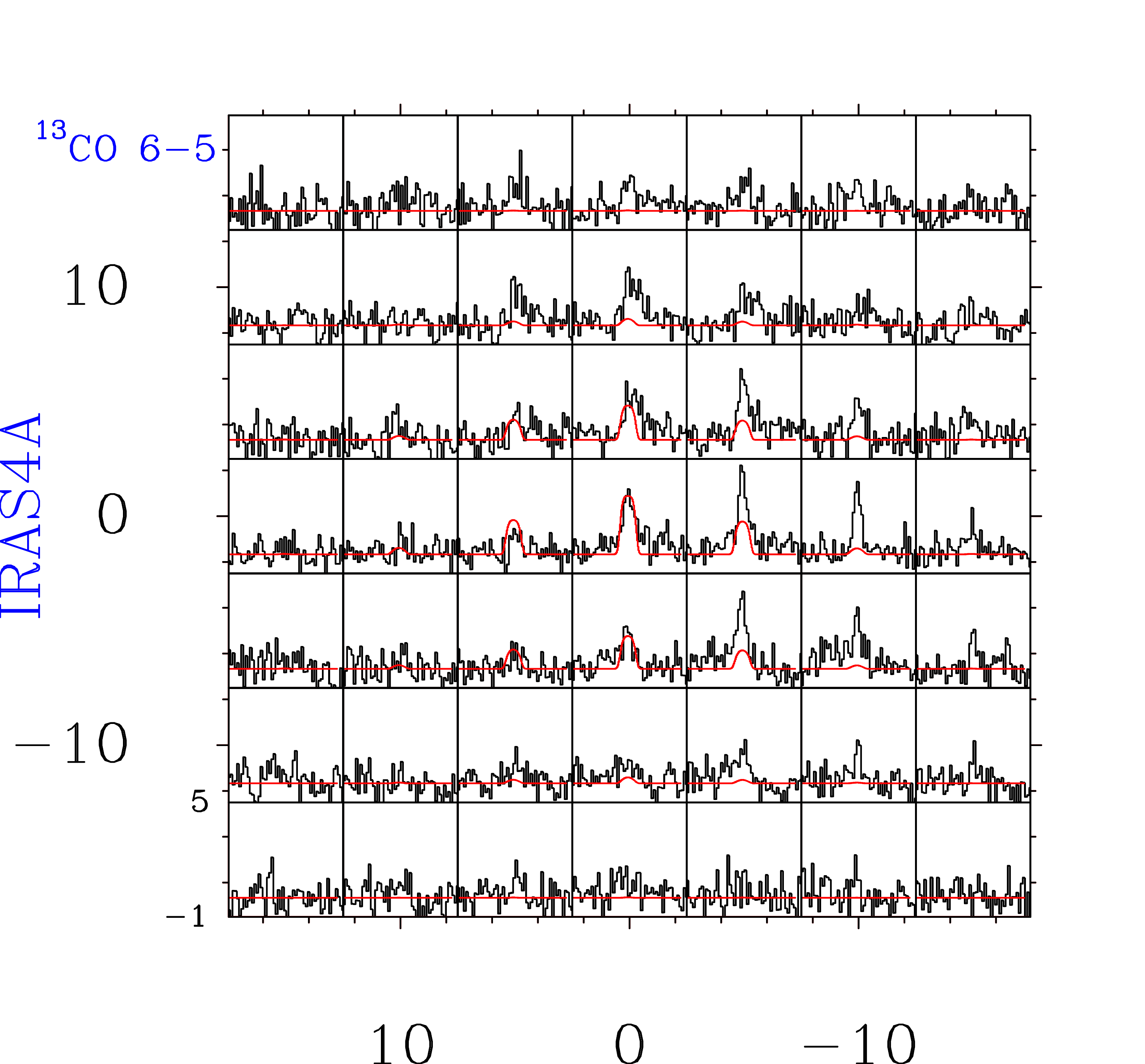}   
    \includegraphics[scale=0.20]{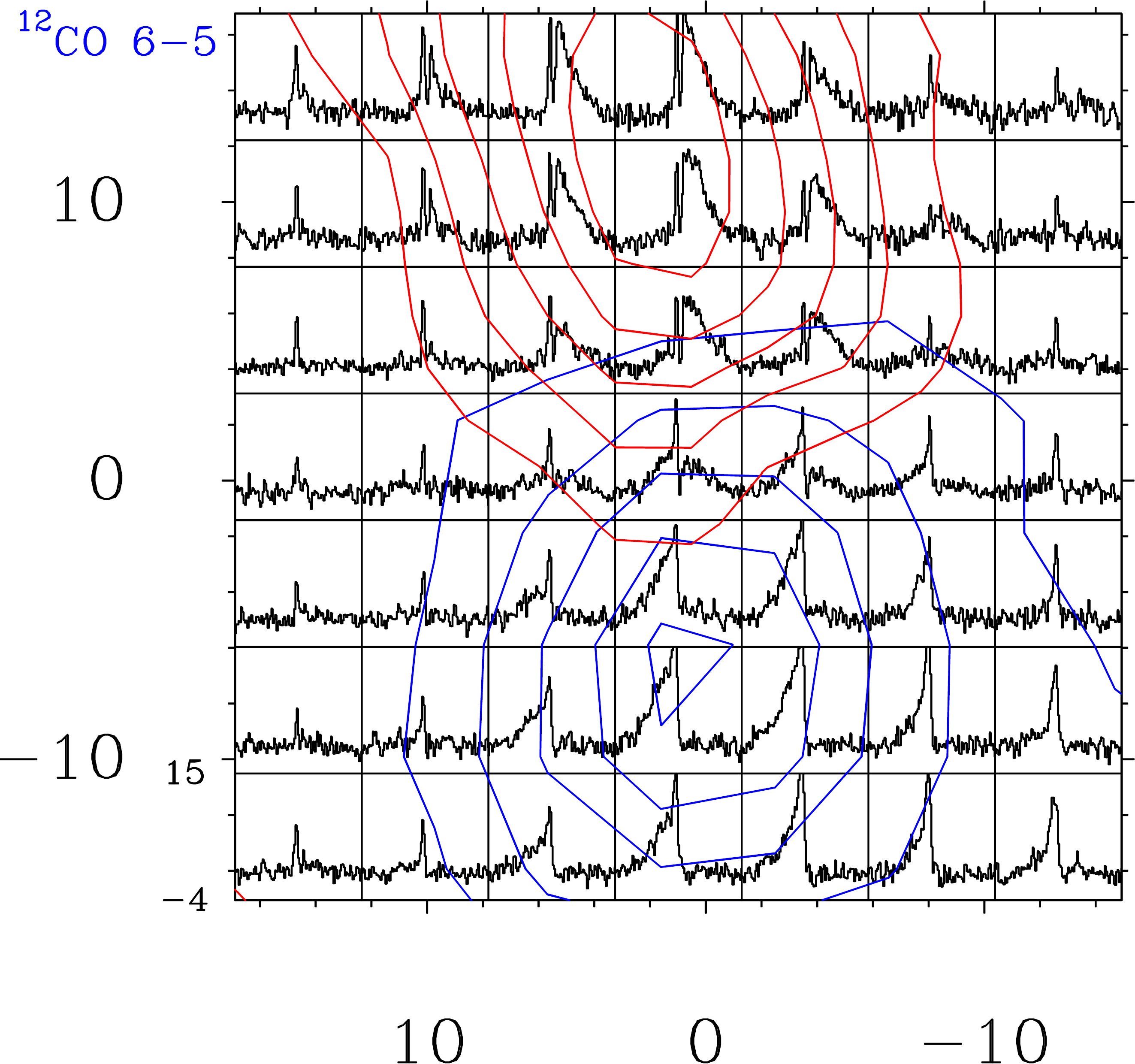}   
    \includegraphics[scale=0.20]{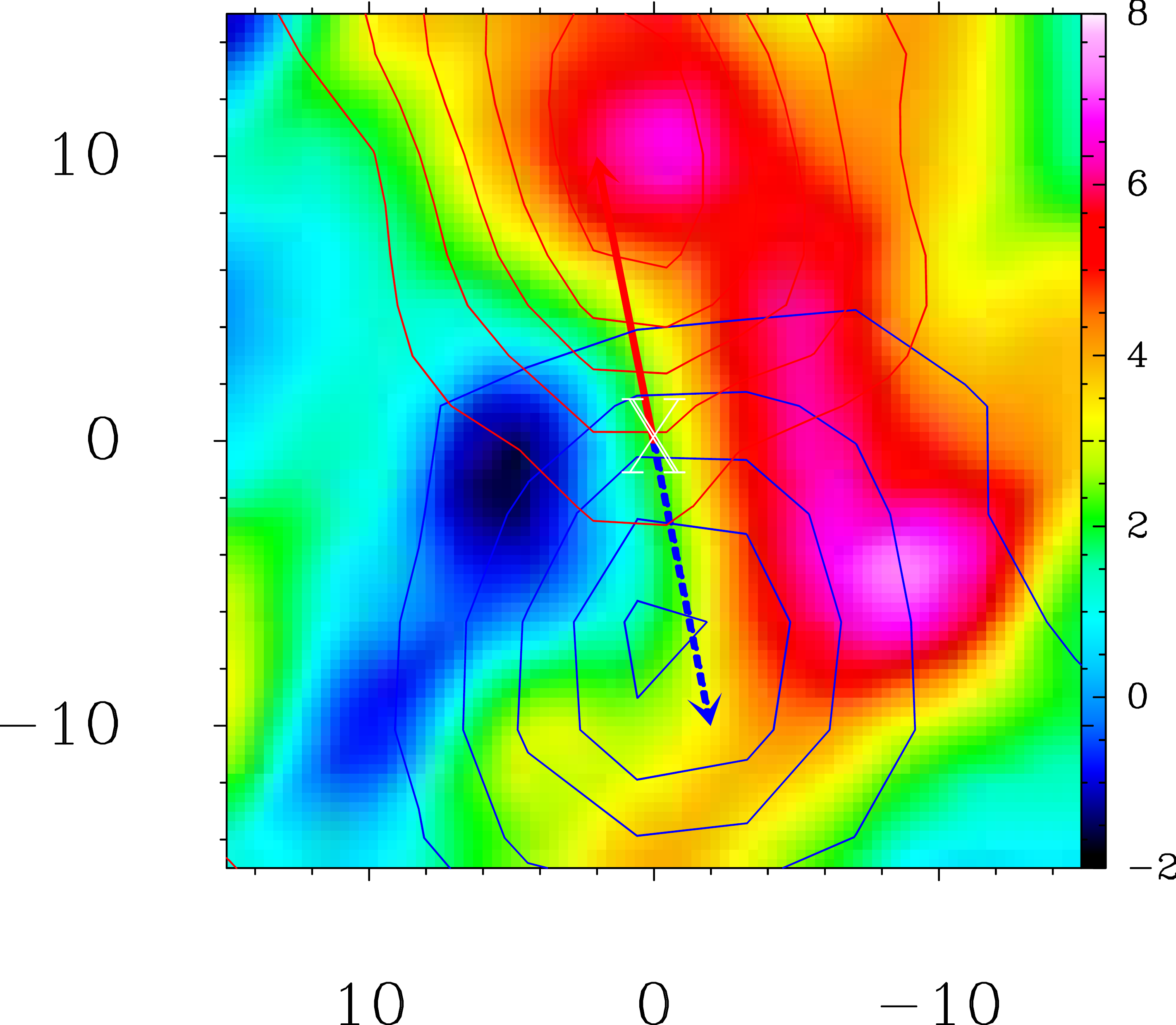} \\
    \includegraphics[scale=0.20]{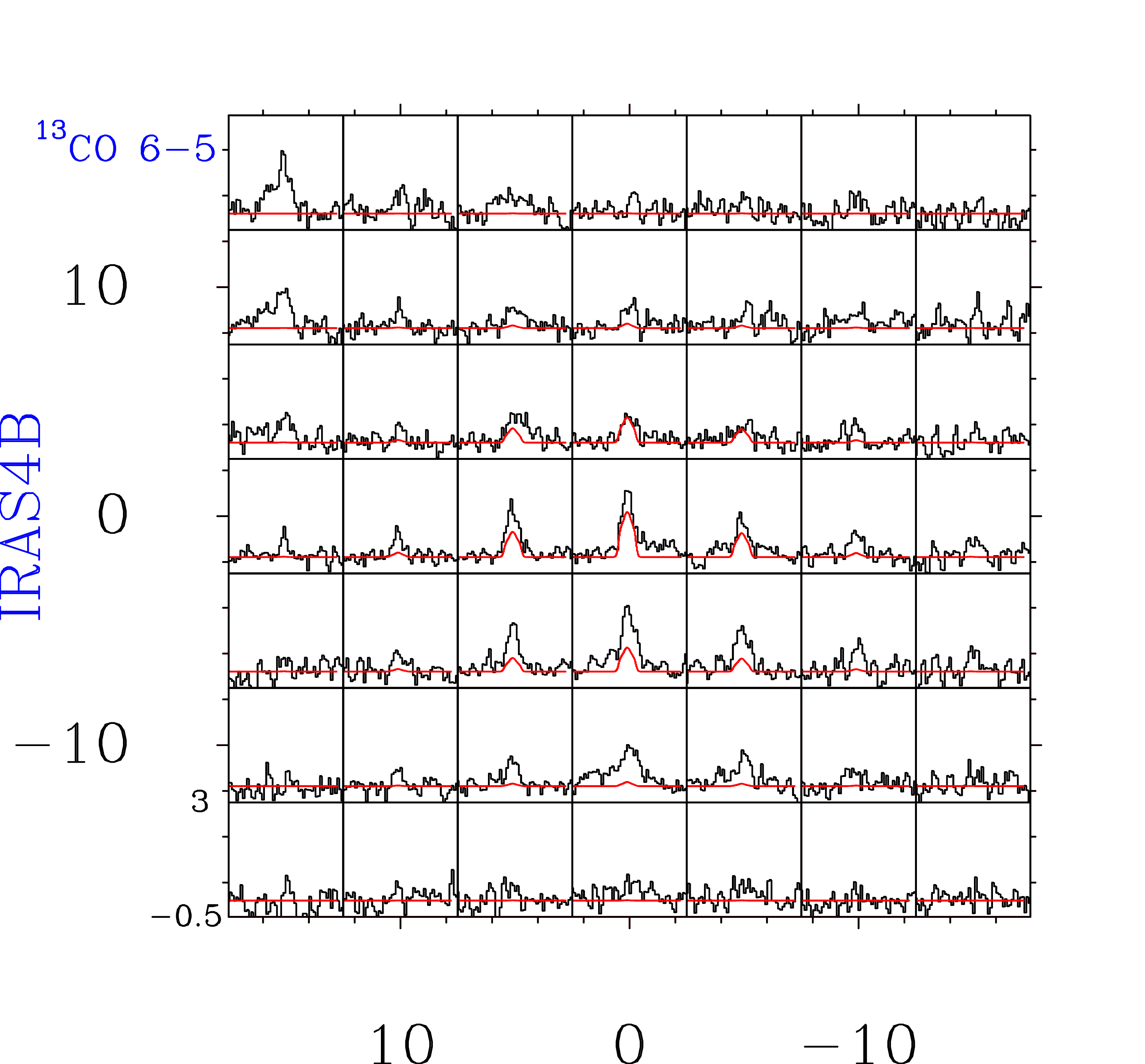}   
    \includegraphics[scale=0.20]{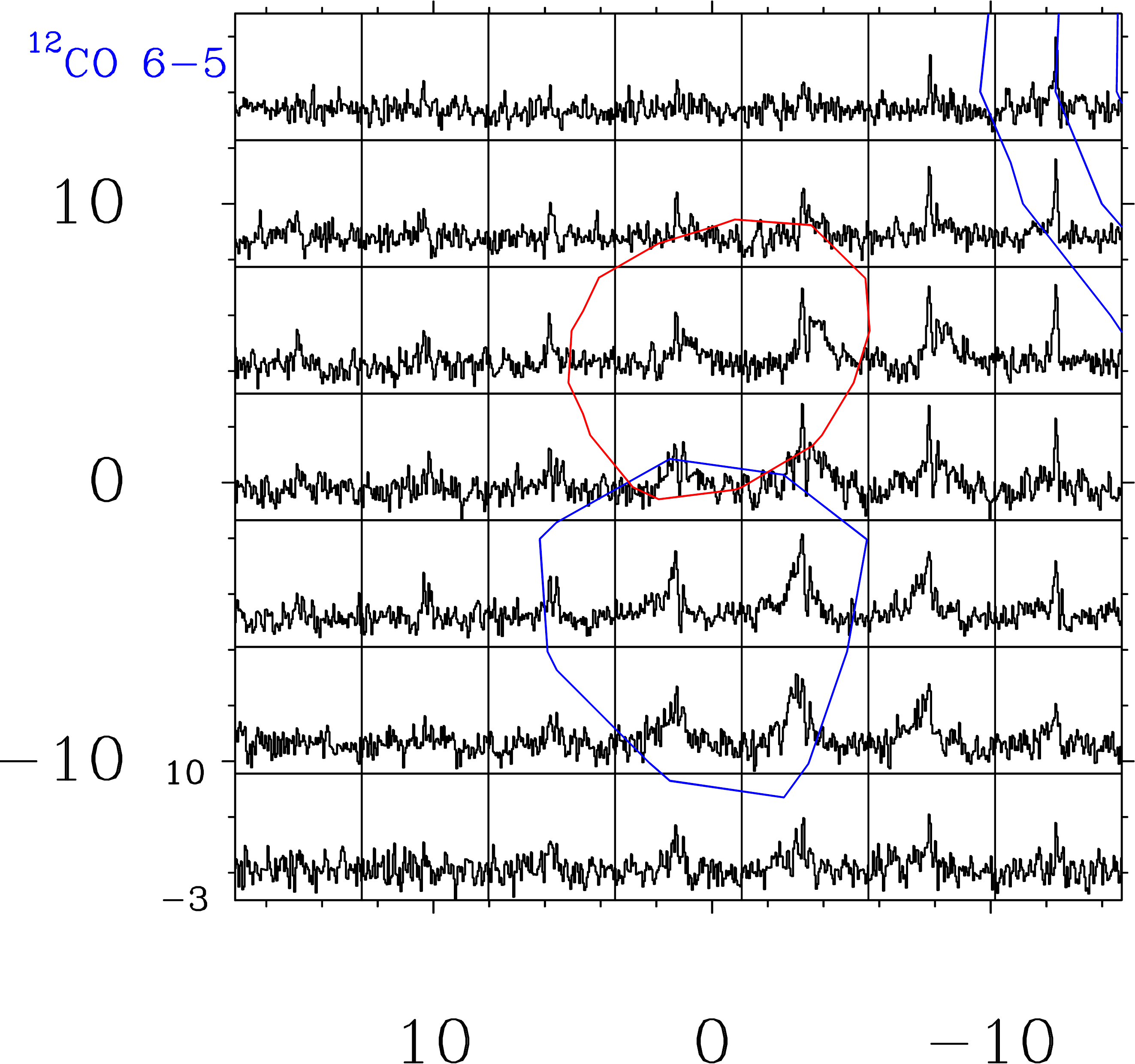}   
    \includegraphics[scale=0.20]{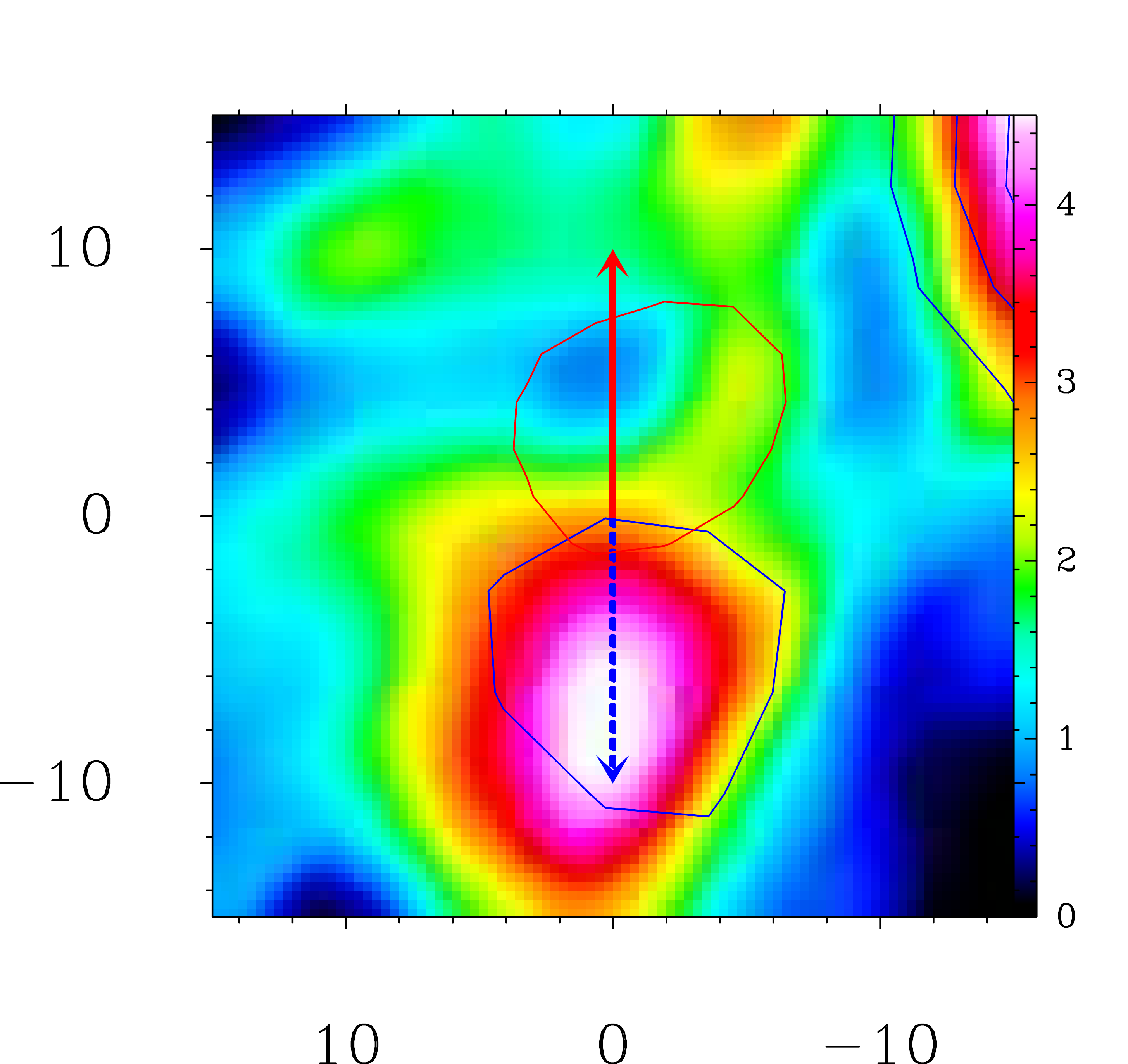} \\
    \includegraphics[scale=0.20]{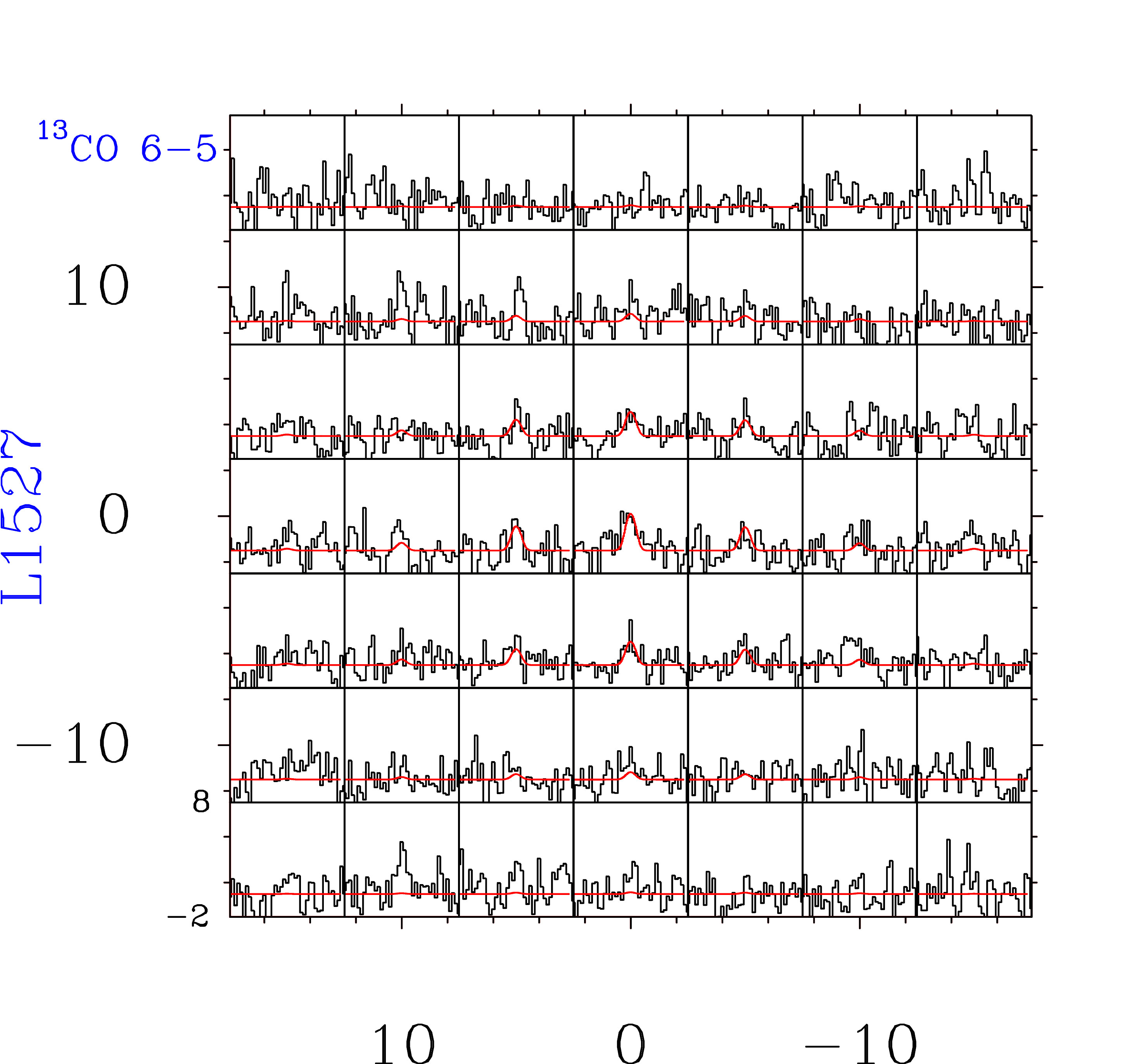}   
    \includegraphics[scale=0.20]{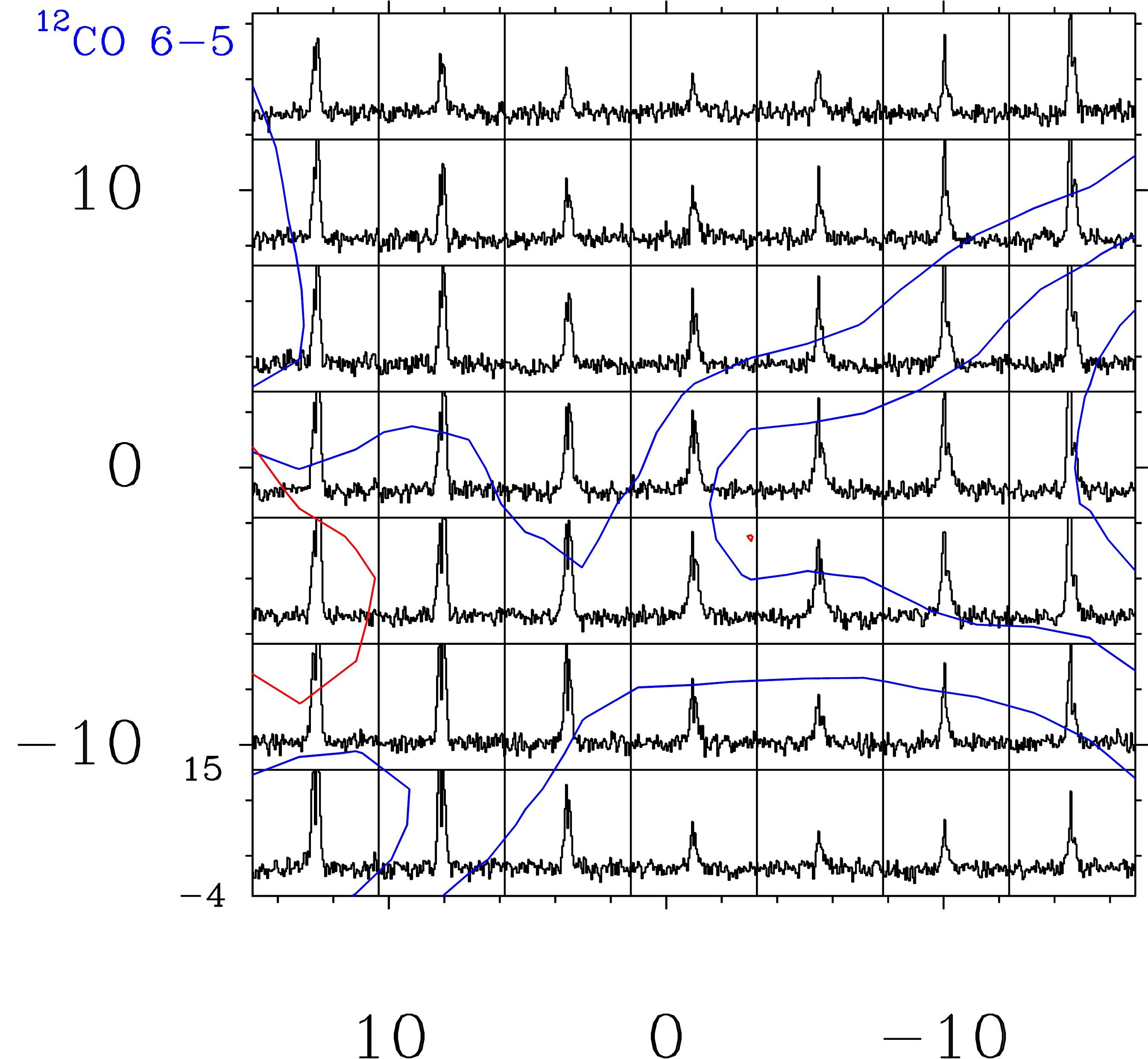}   
    \includegraphics[scale=0.20]{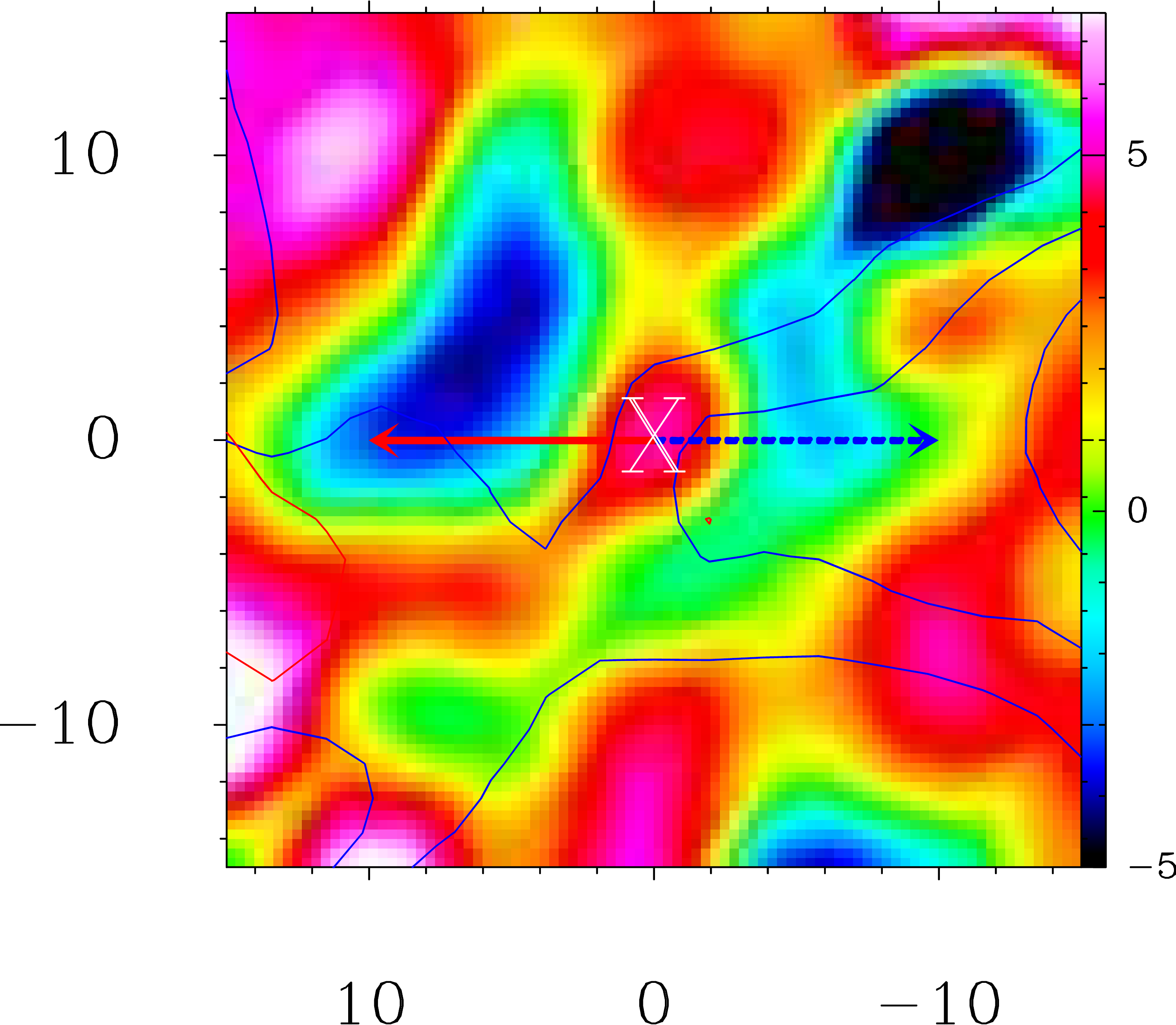}   
    \caption{\small 7$\times$7 pixel fully sampled maps are extracted
      toward the central positions of the sources in \thco\ 6--5 ({\it
        left}) and in \twco\ 6--5 ({\it middle}) transitions. The axes
      represent the equatorial offsets ($\Delta$$\alpha$,
      $\Delta$$\delta$) in arcsec. The main beam temperature intensity
      scale of each box are shown in the y-axes of the bottom-left
      box in Kelvins. The velocity range in each box is $\pm$8 \kms\ for the
      \thco\ spectra, and $\pm$25 \kms\ for the \twco\ spectra. The red lines in the 
      left-hand panels are the $^{13}$CO 6--5 model line intensities for the passively 
      heated envelope. The excess emission in the observations compared with 
      these model profiles corresponds to the UV-heated gas and is shown
      as an image in the {\it right} panel with the intensity scale in K km s$^{-1}$. 
      The middle and right panels contain the red and blue outflow lobes with 
      the contour levels given in Table~\ref{tbl:contourlevels}. The blue and red
      arrows in the right-hand panels show the direction of the outflow lobes.
    }
    \label{fig:specmap13CO65_1}
\end{figure*}

\begin{figure*}[htb]
    \centering
    \includegraphics[scale=0.20]{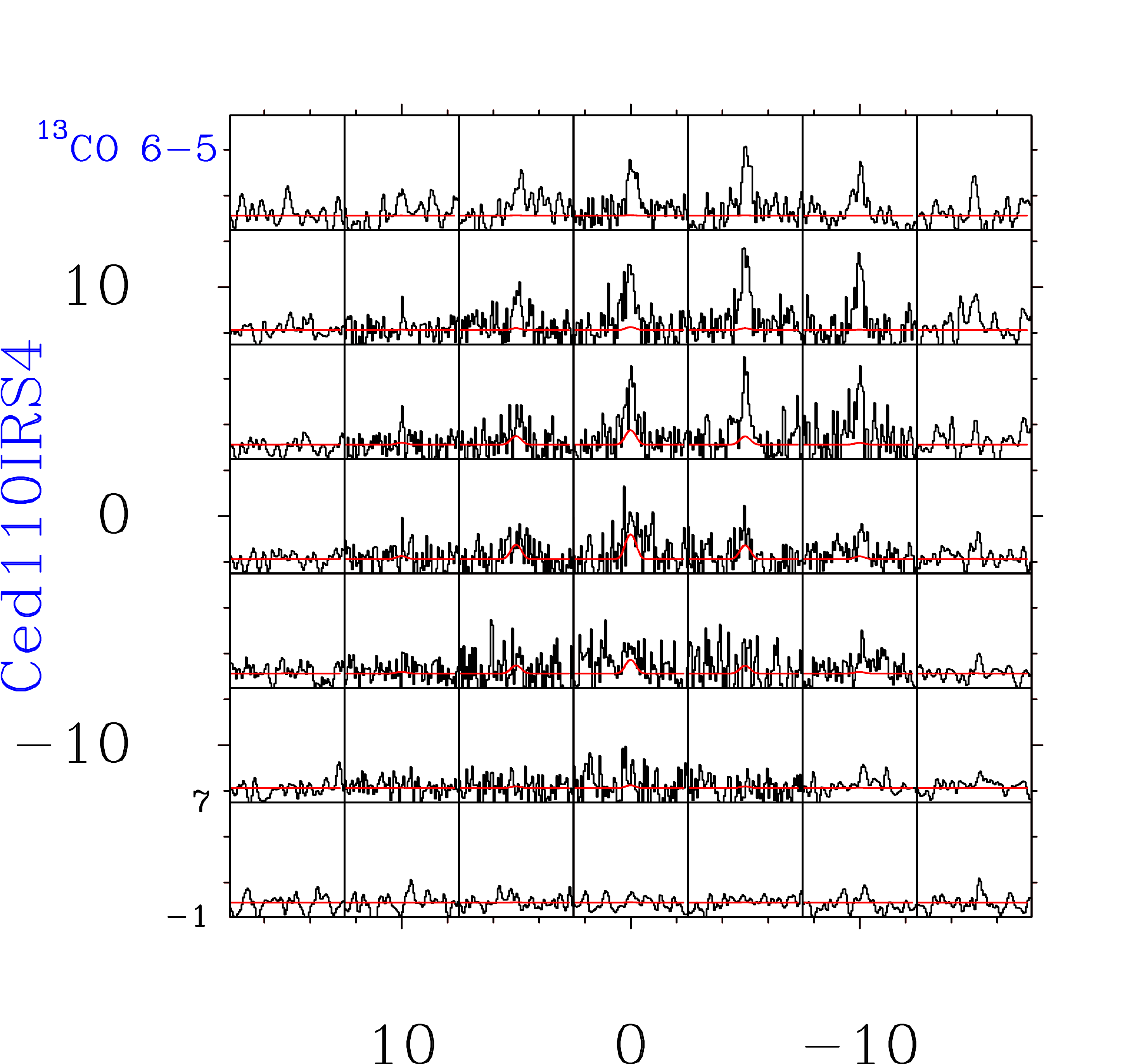}   
    \includegraphics[scale=0.20]{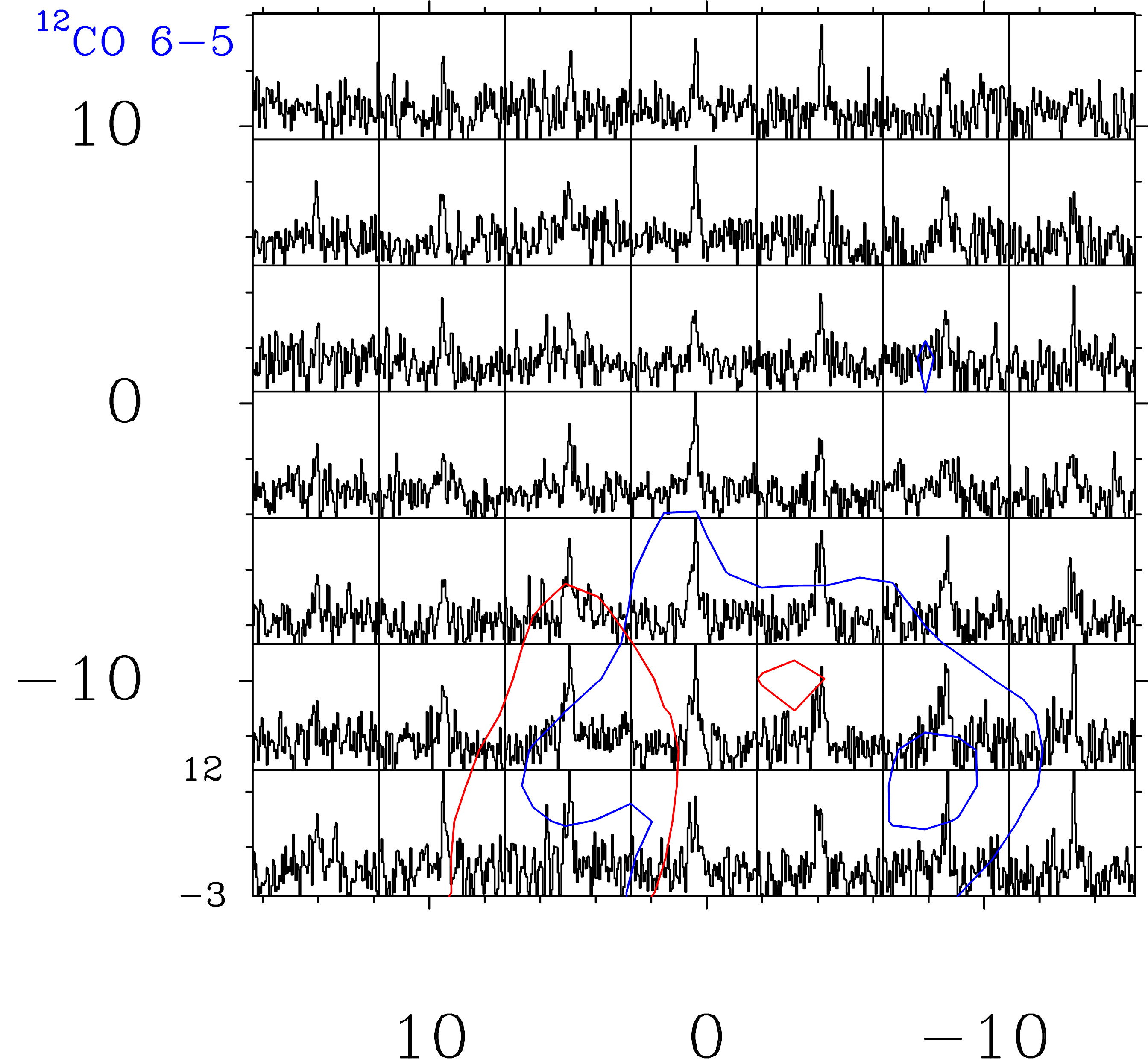}   
    \includegraphics[scale=0.20]{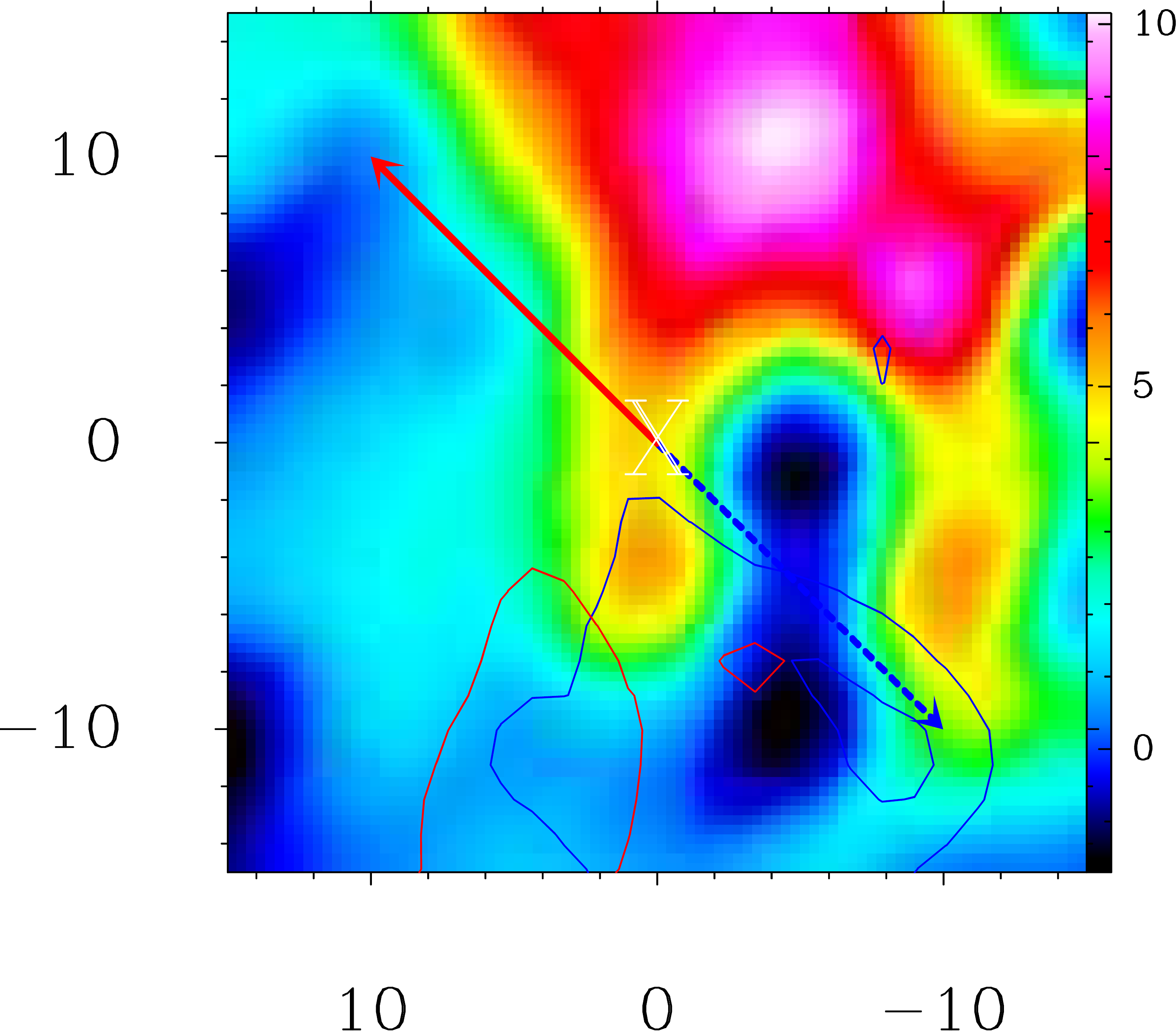} \\
    \includegraphics[scale=0.20]{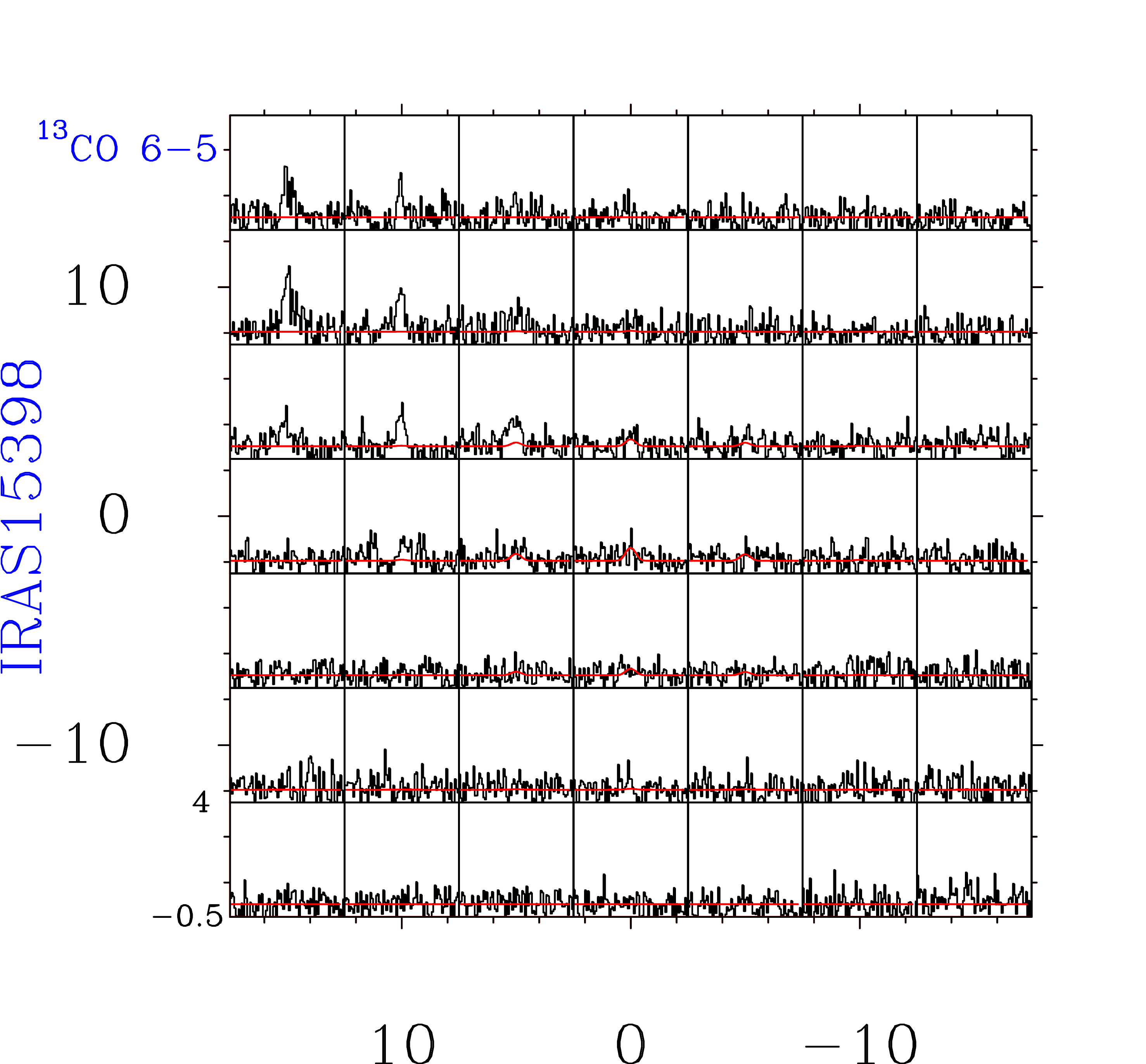}   
    \includegraphics[scale=0.20]{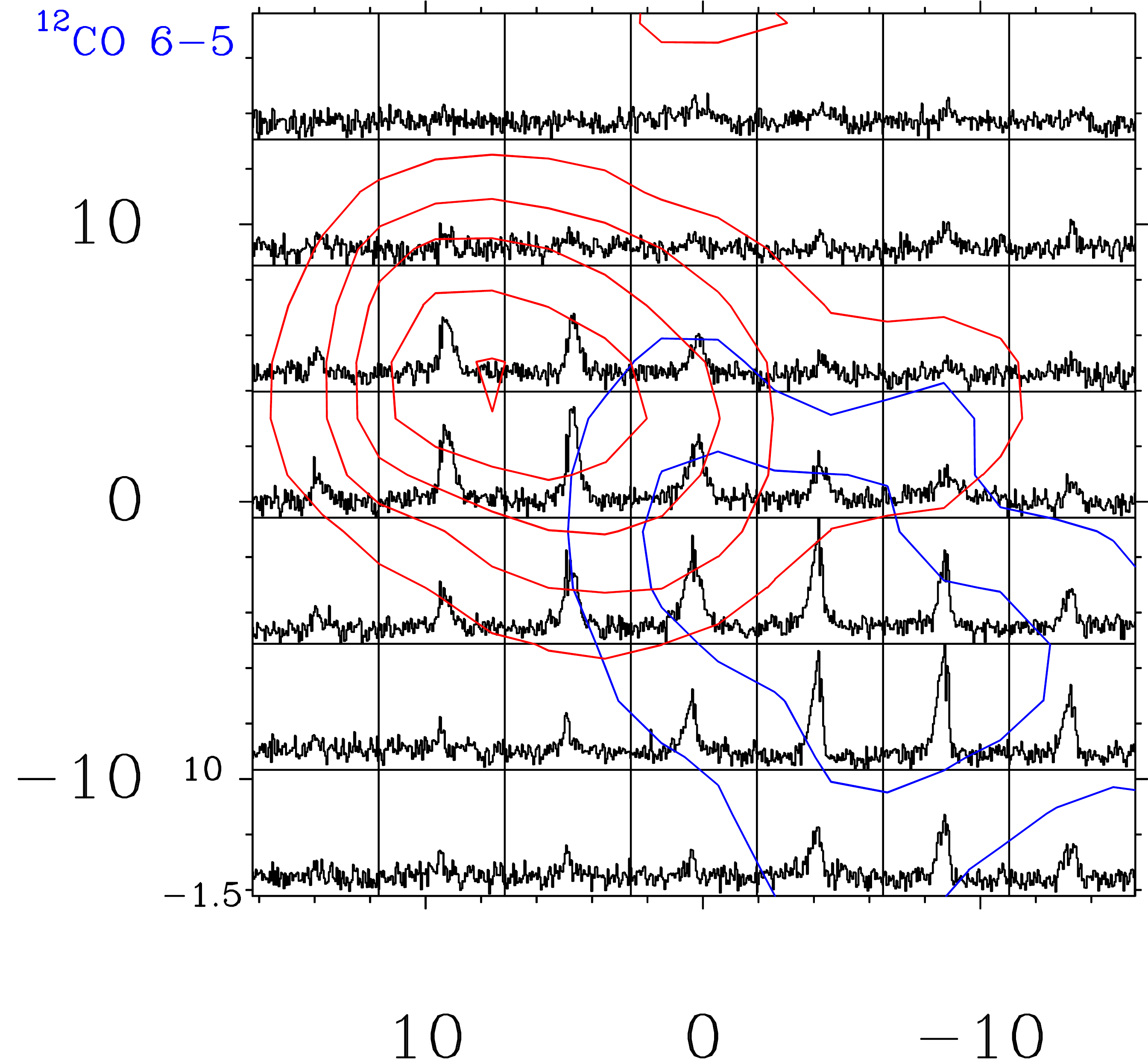}   
    \includegraphics[scale=0.20]{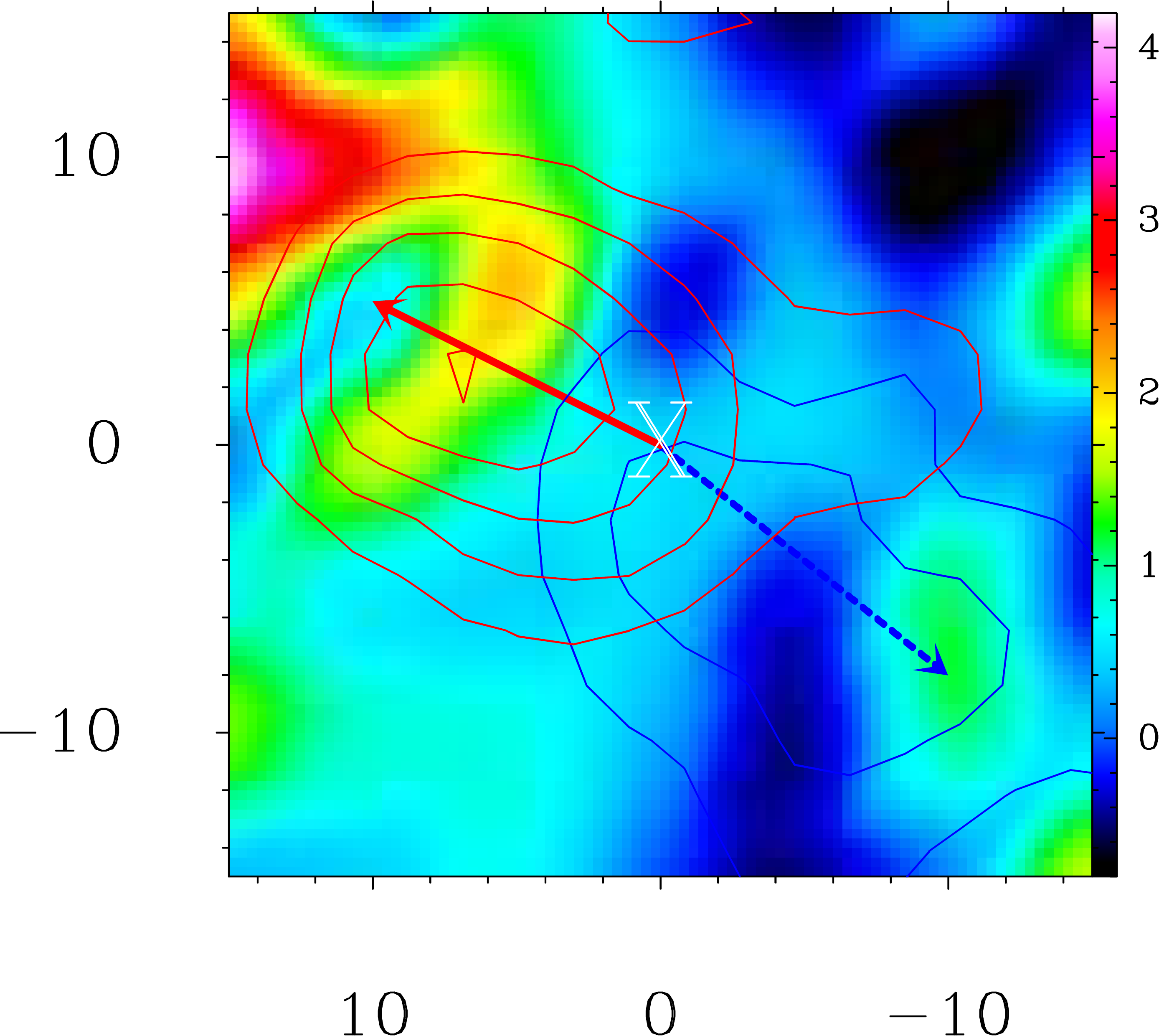} \\
    \includegraphics[scale=0.20]{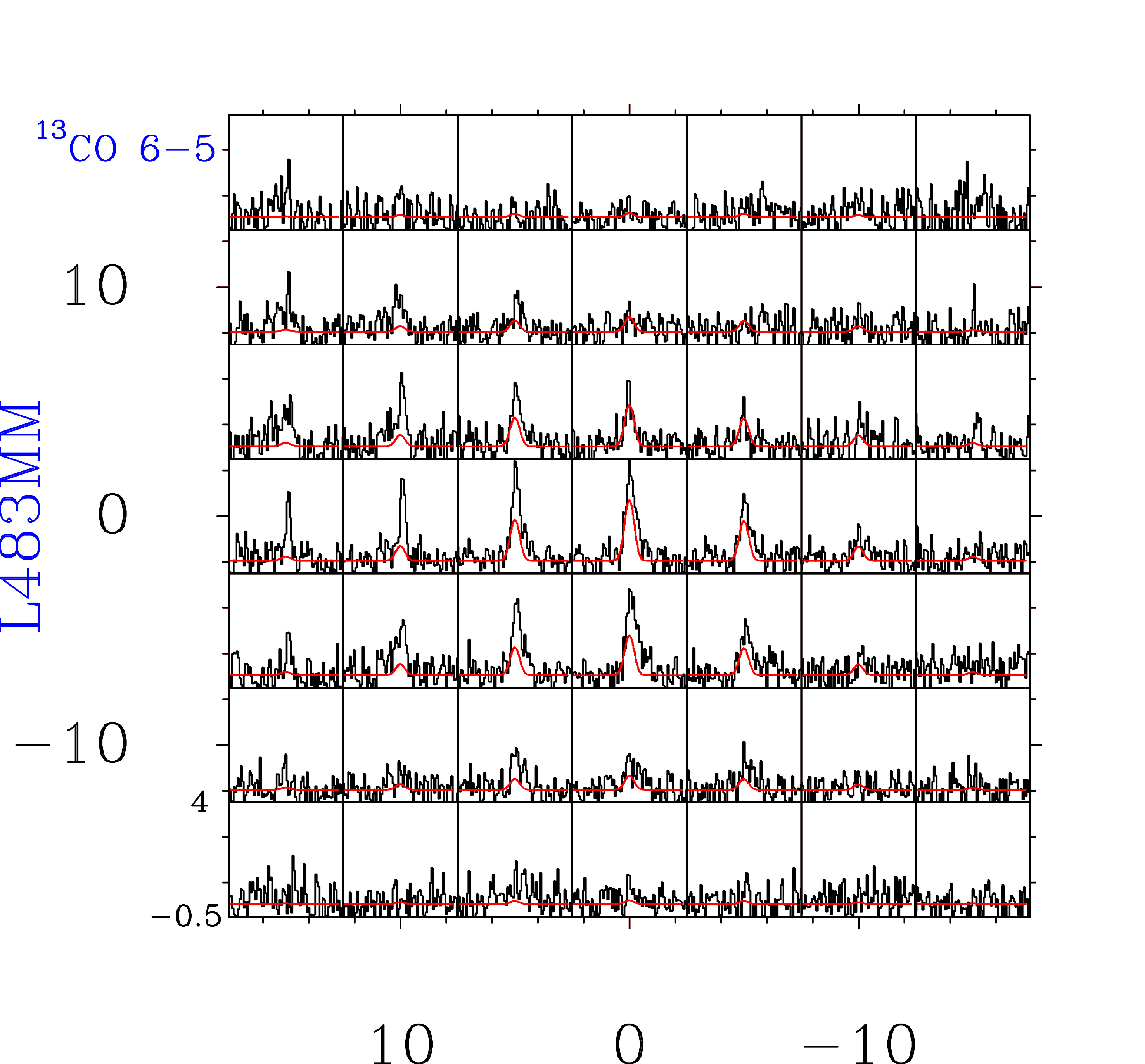}   
    \includegraphics[scale=0.20]{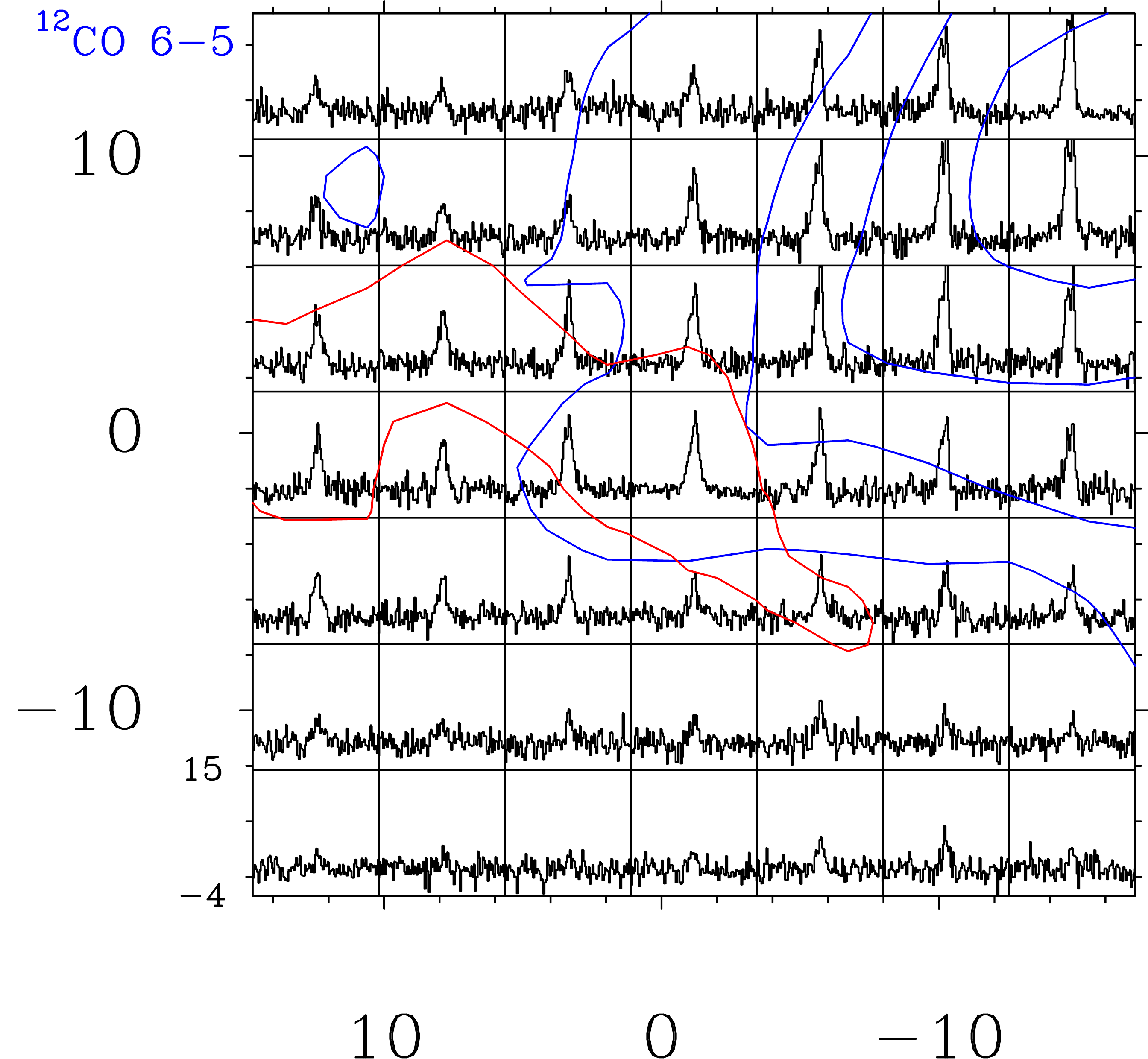}   
    \includegraphics[scale=0.20]{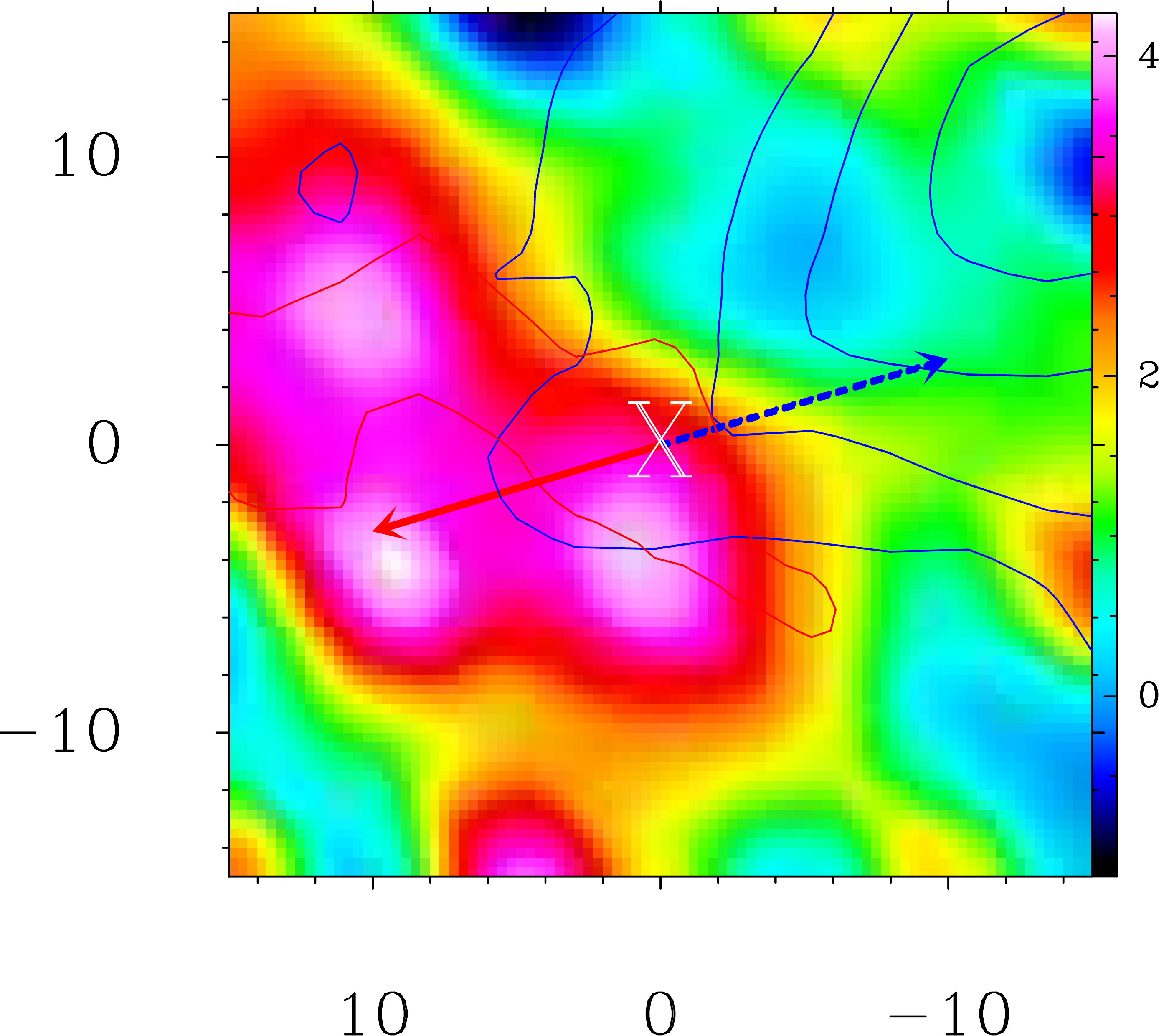} \\
    \includegraphics[scale=0.20]{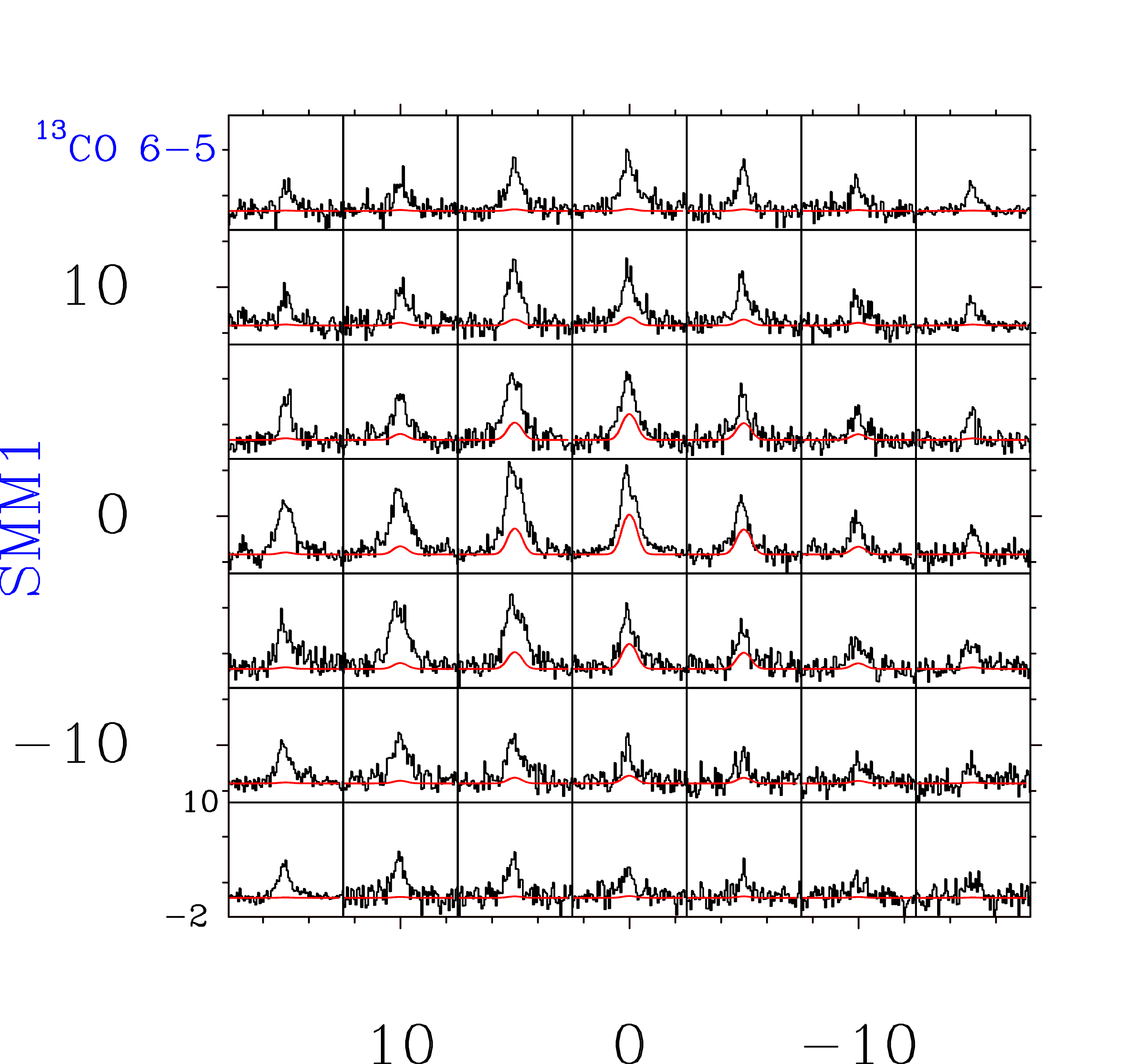}   
    \includegraphics[scale=0.20]{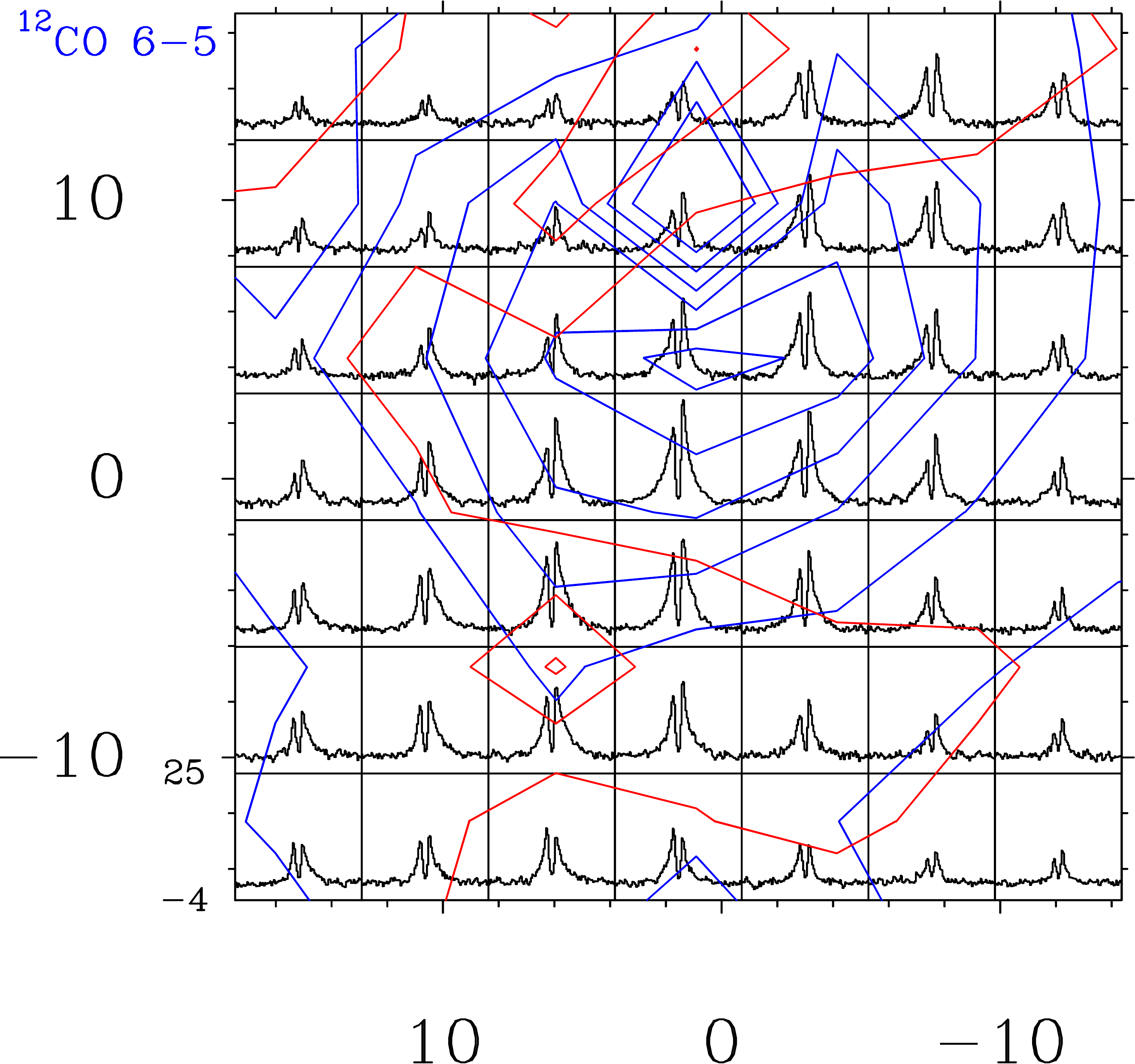}   
    \includegraphics[scale=0.20]{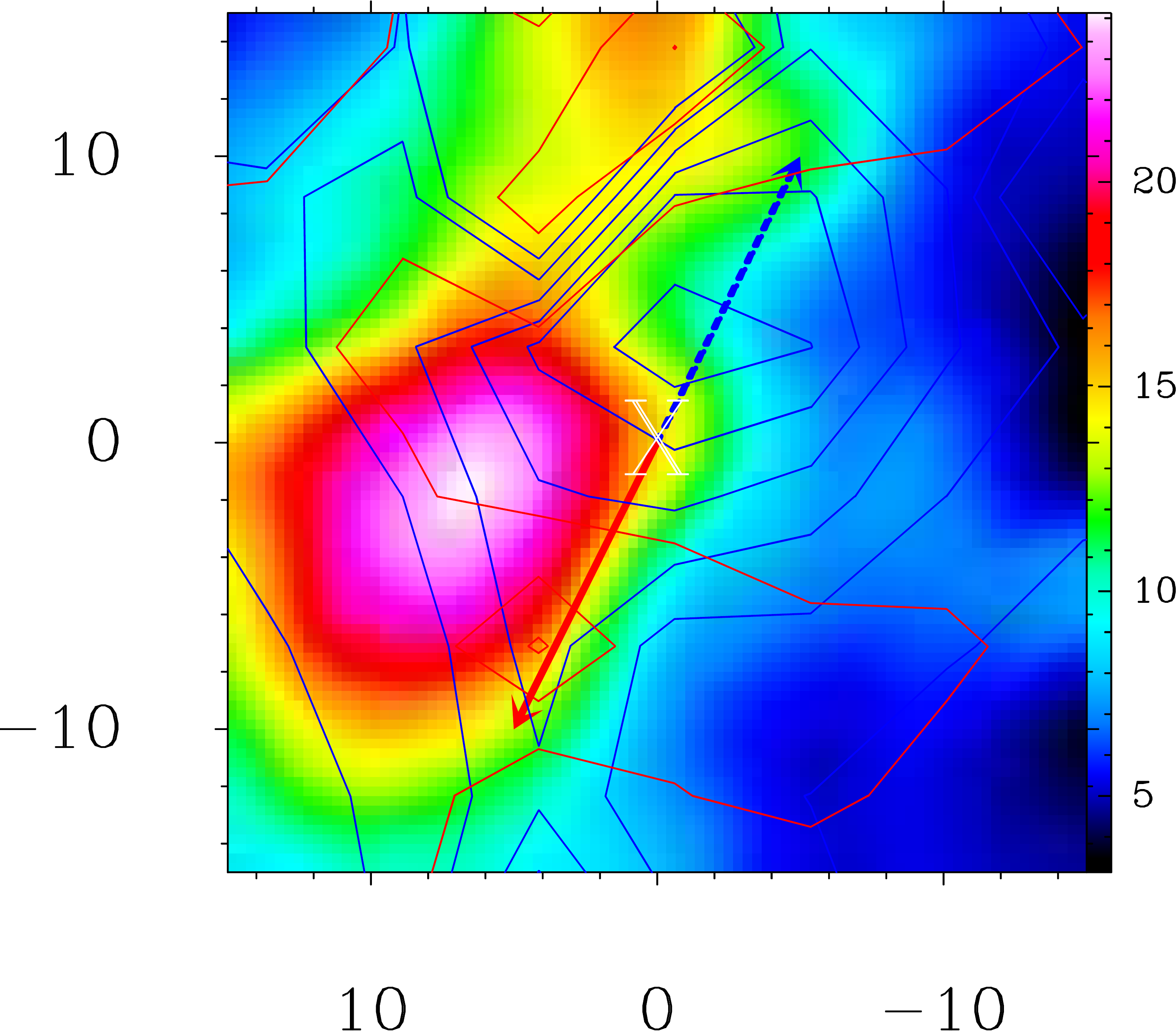}   
    \caption{\small Same as Fig. \ref{fig:specmap13CO65_1}.}
    \label{fig:specmap13CO65_2}
\end{figure*}

\begin{figure*}[htb]
    \centering
    \includegraphics[scale=0.20]{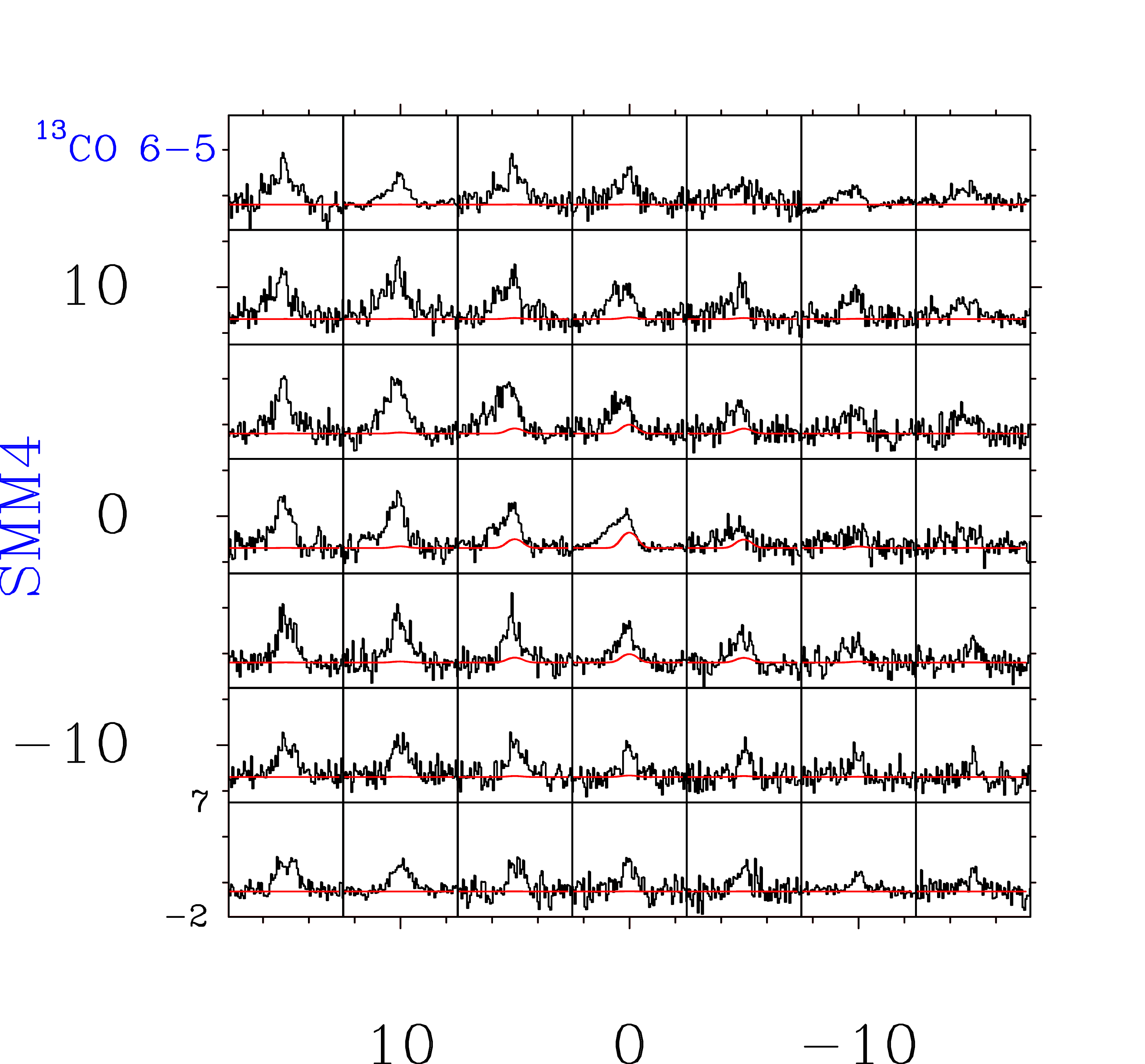}   
    \includegraphics[scale=0.20]{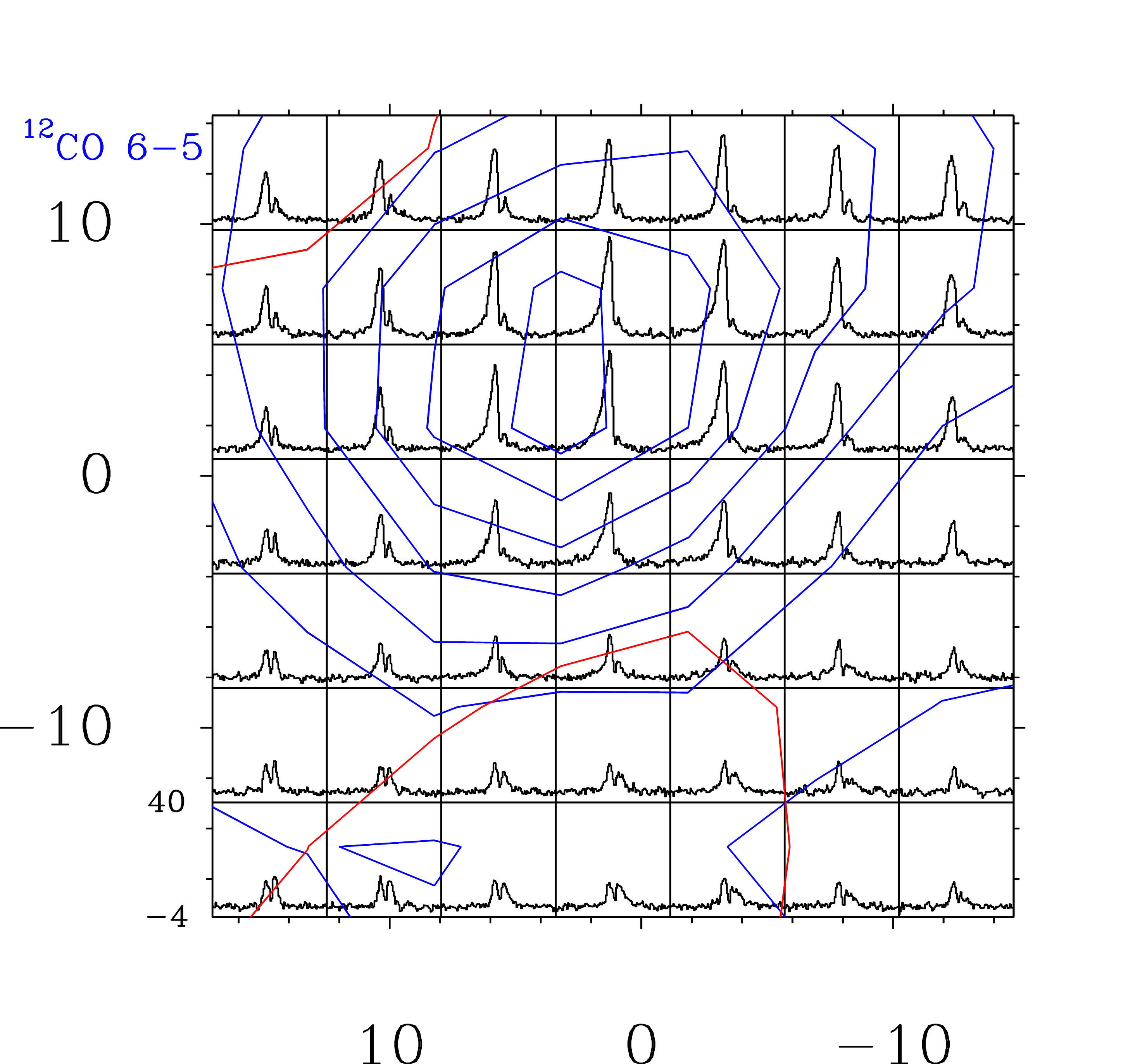}   
    \includegraphics[scale=0.20]{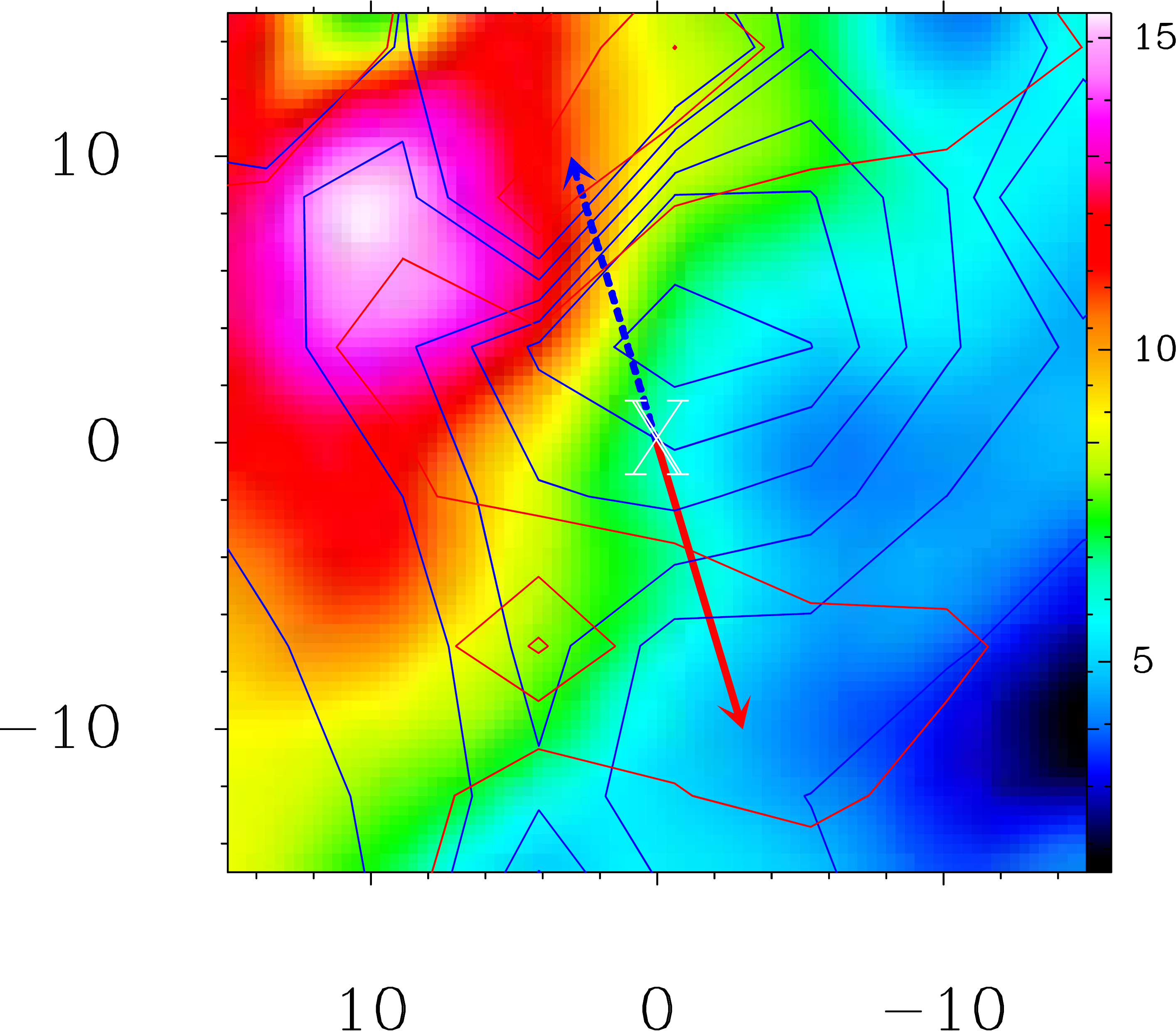} \\
    \includegraphics[scale=0.20]{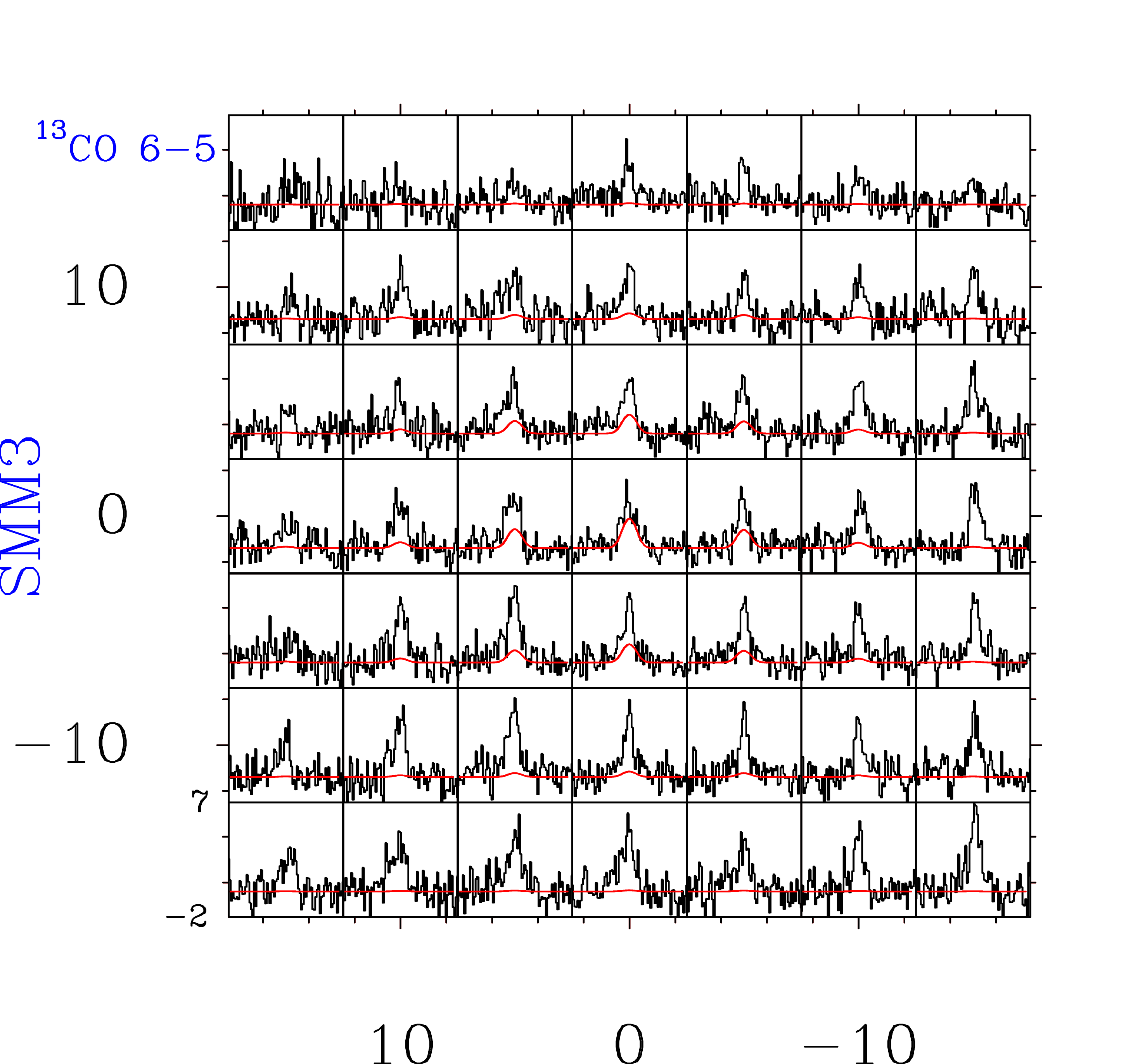}   
    \includegraphics[scale=0.20]{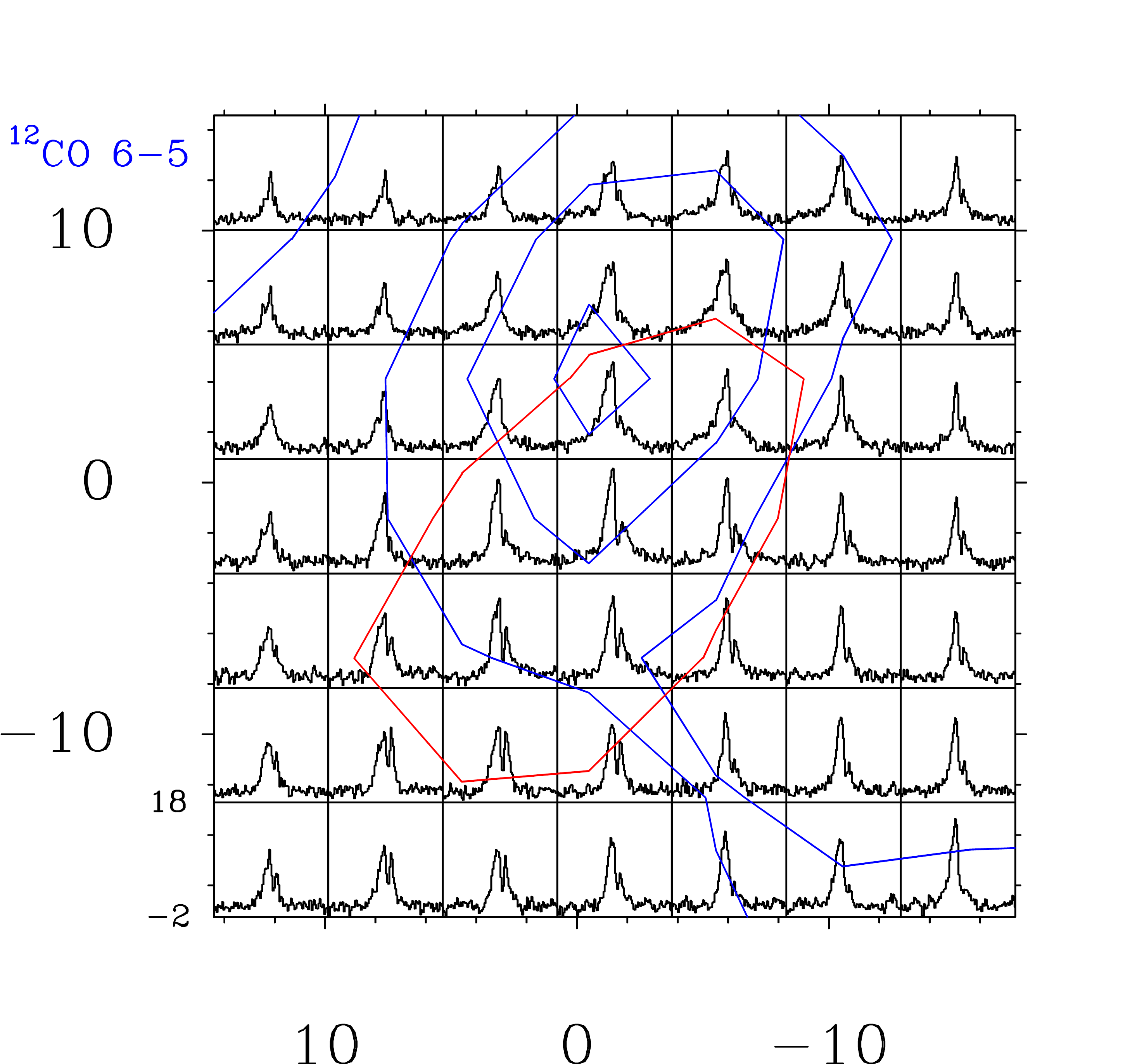}   
    \includegraphics[scale=0.20]{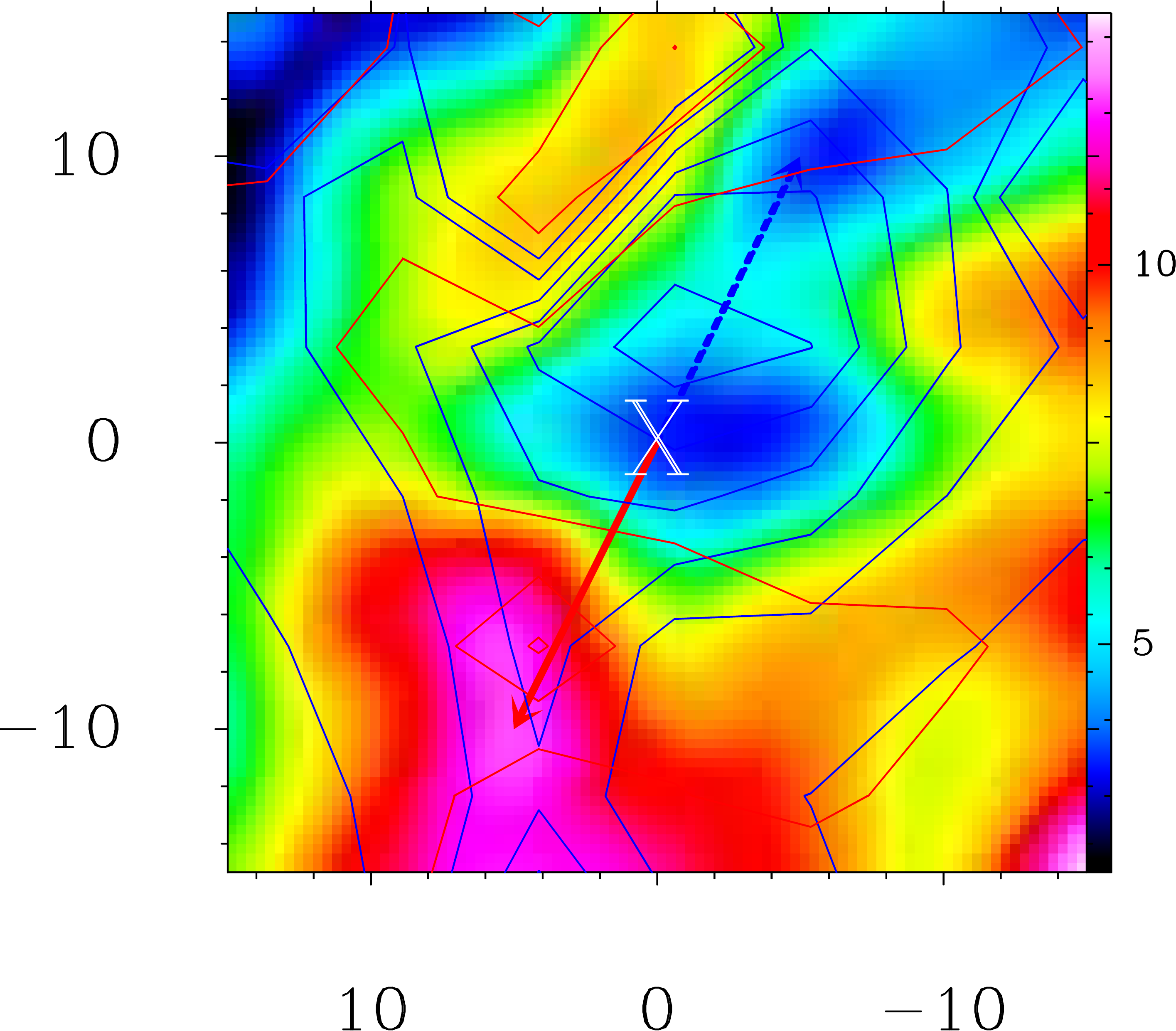} \\
    \includegraphics[scale=0.20]{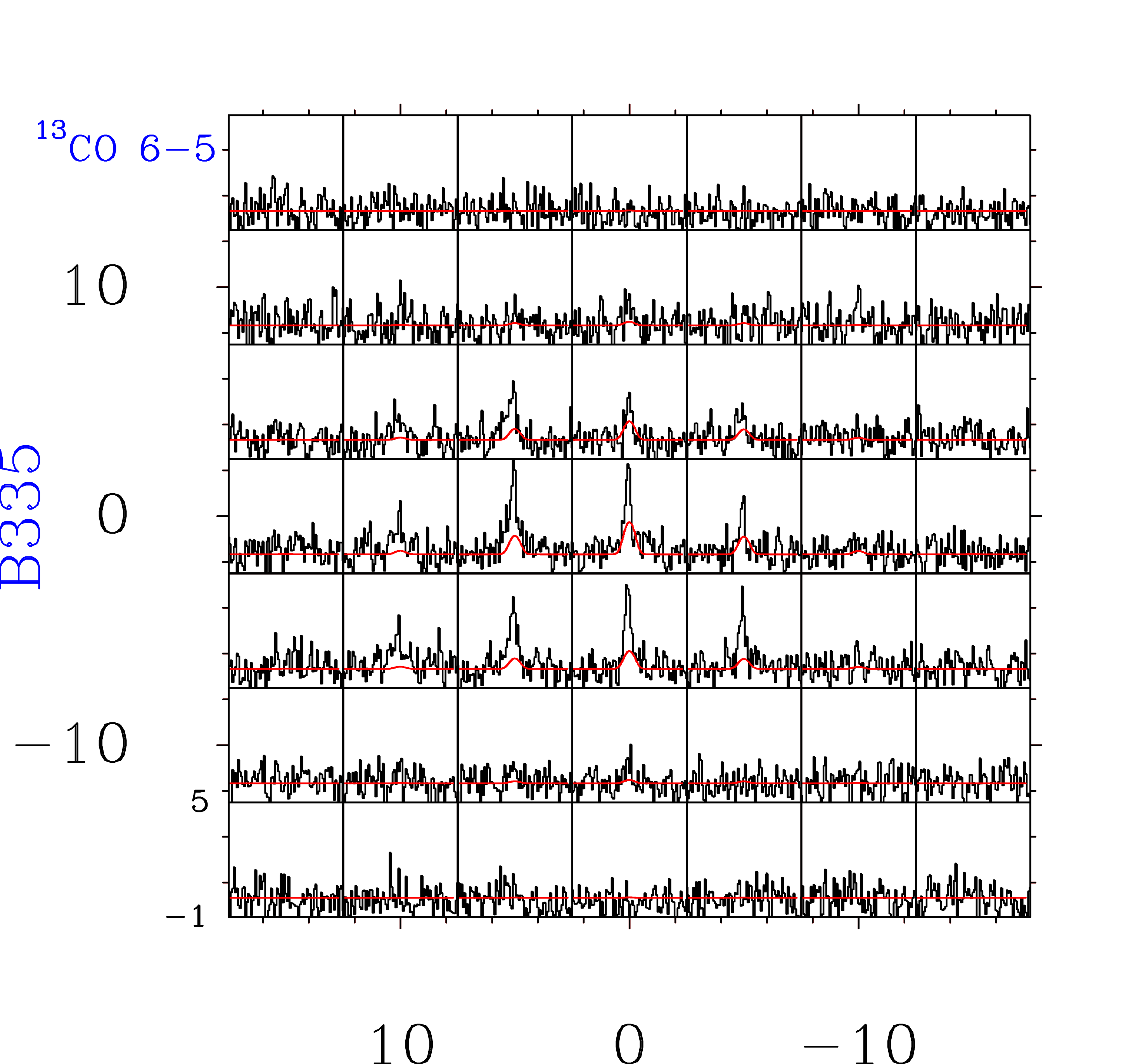}   
    \includegraphics[scale=0.20]{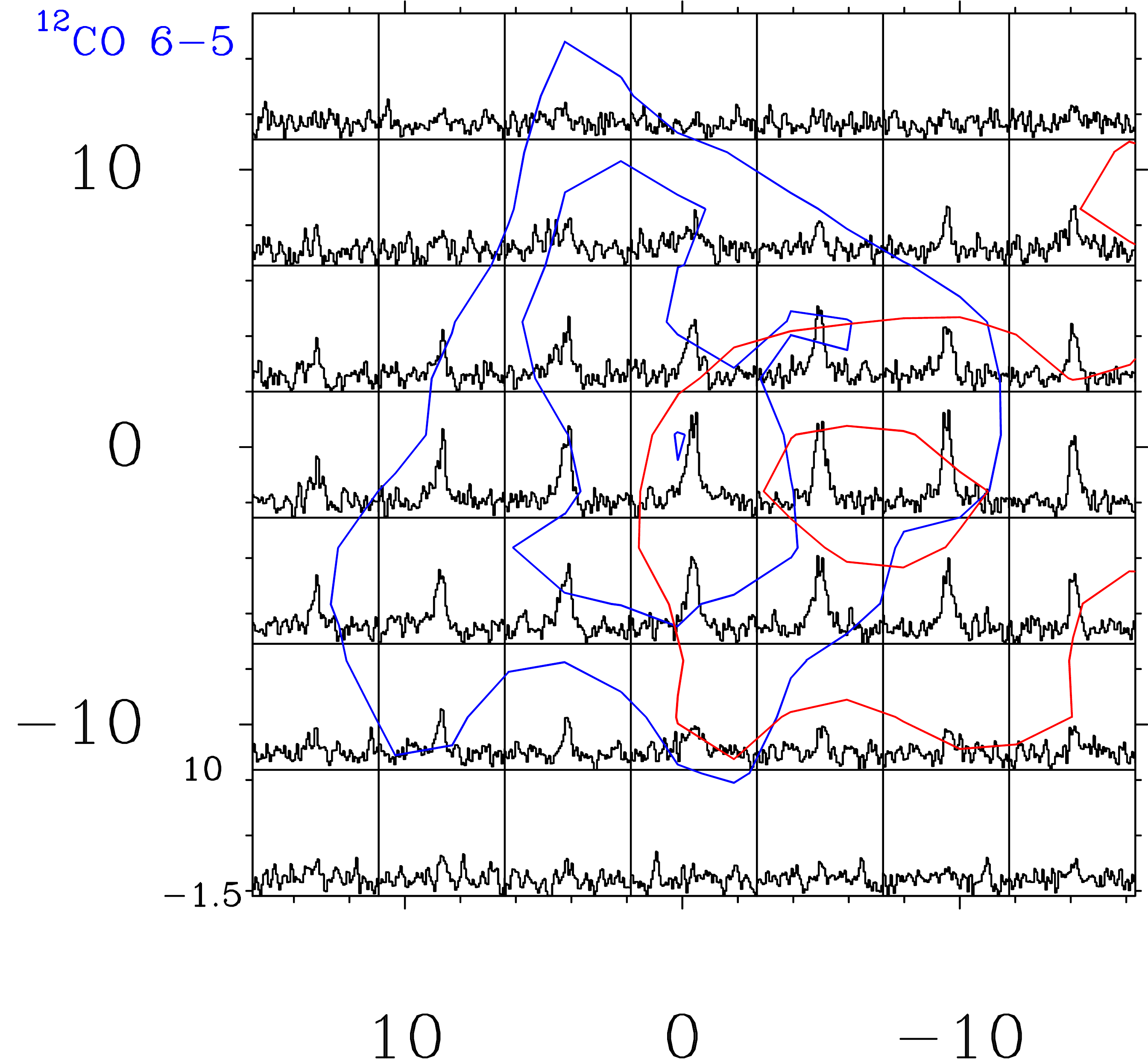}   
    \includegraphics[scale=0.20]{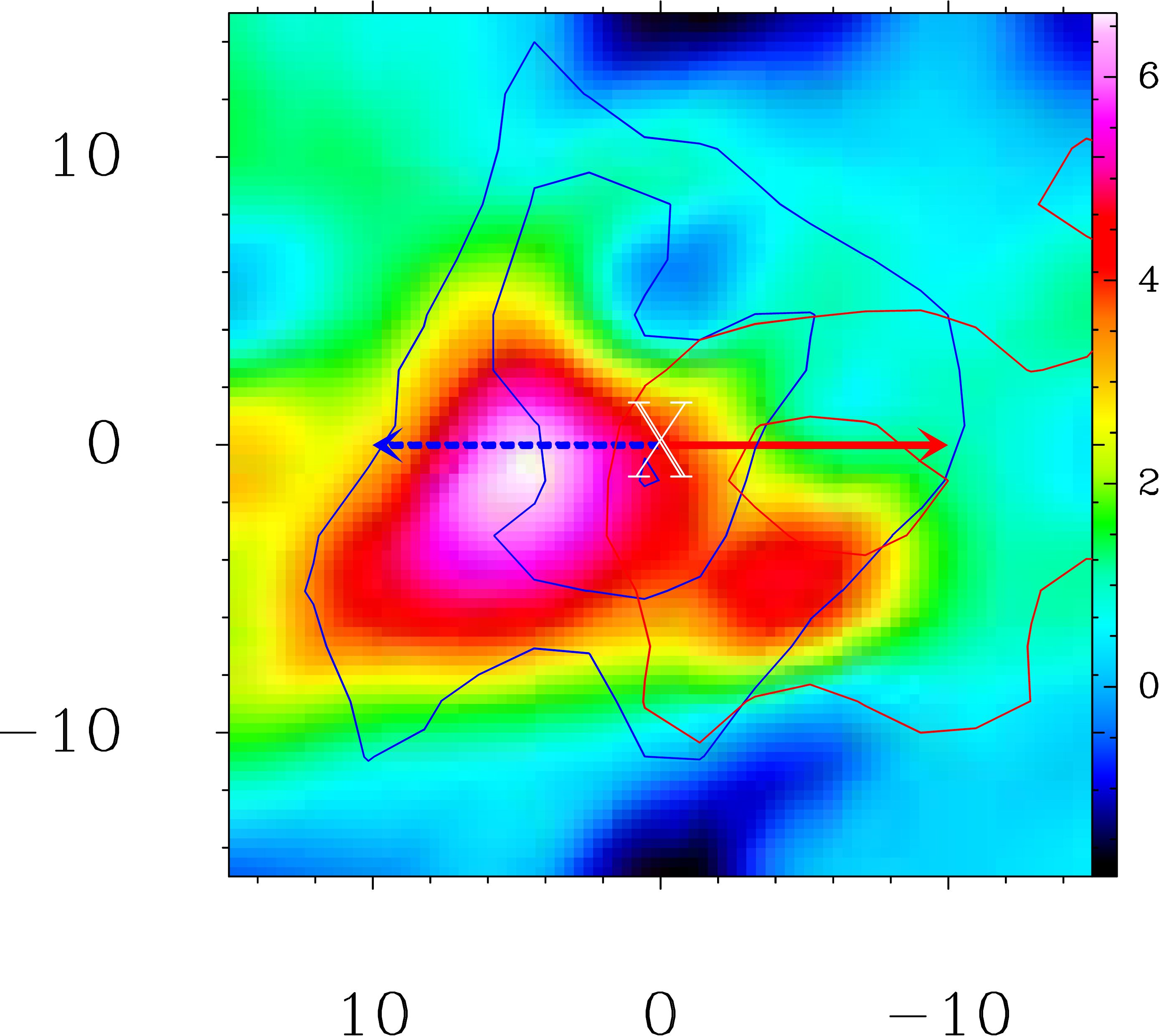} \\
    \includegraphics[scale=0.20]{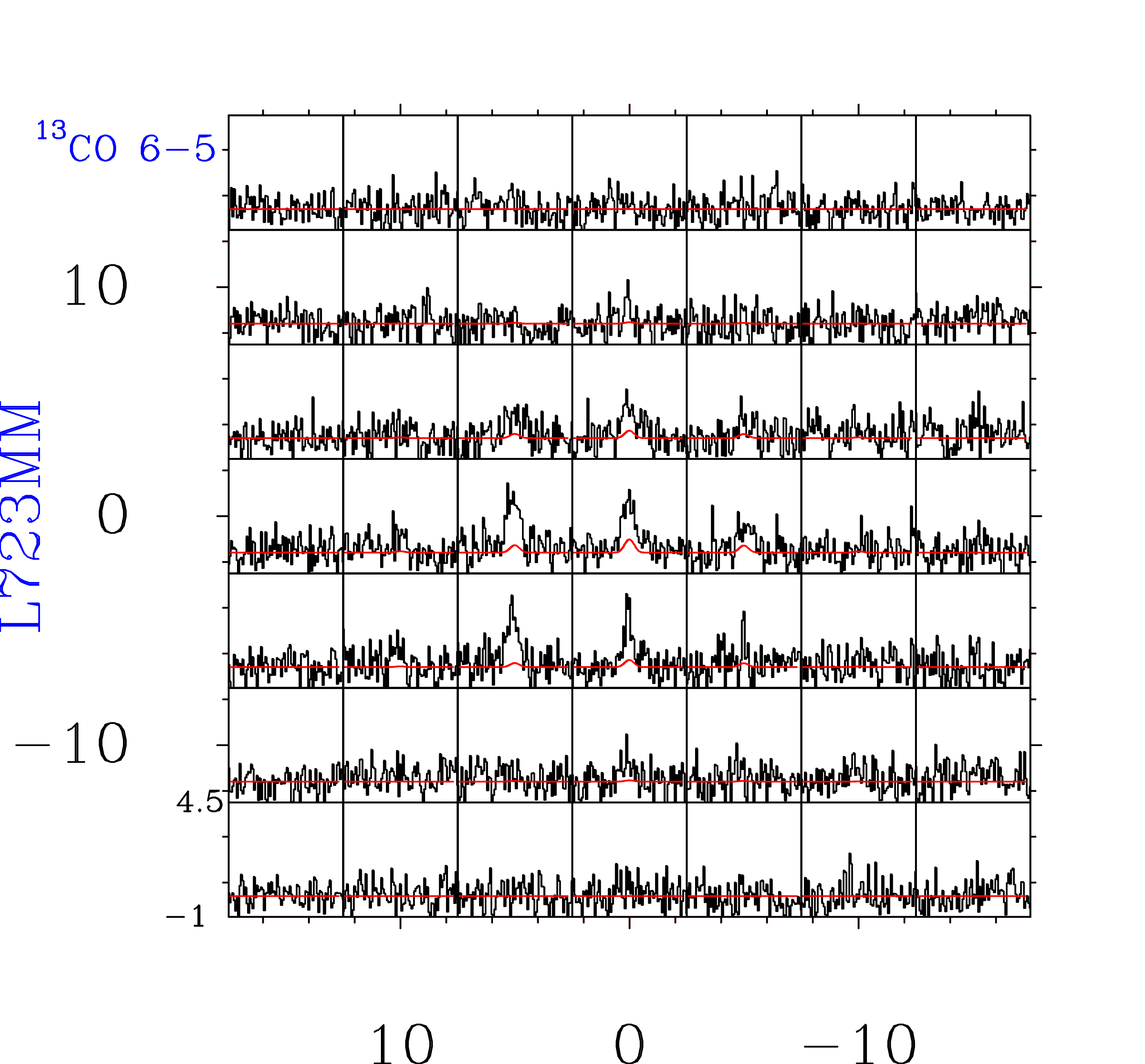}   
    \includegraphics[scale=0.20]{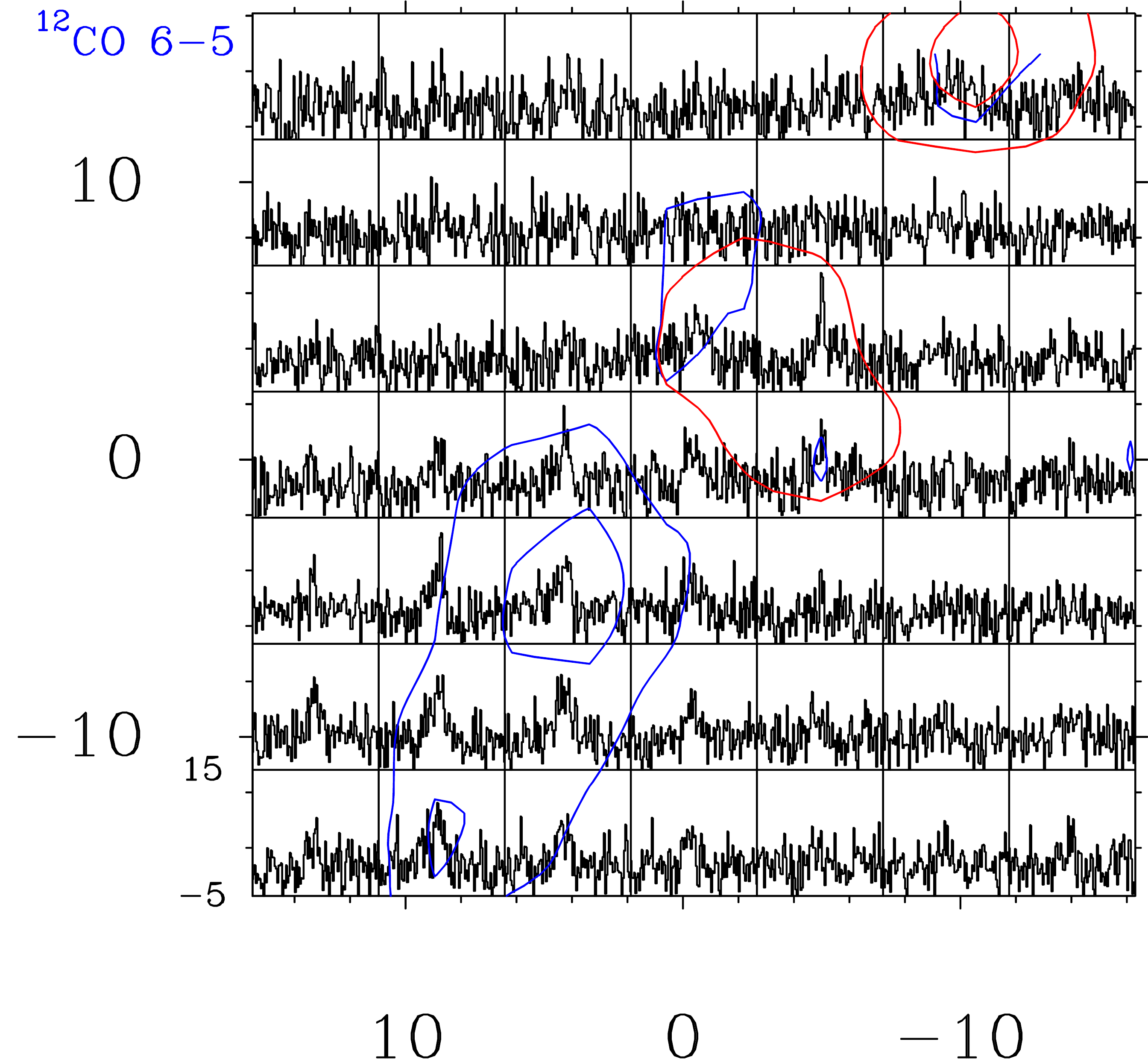}   
    \includegraphics[scale=0.20]{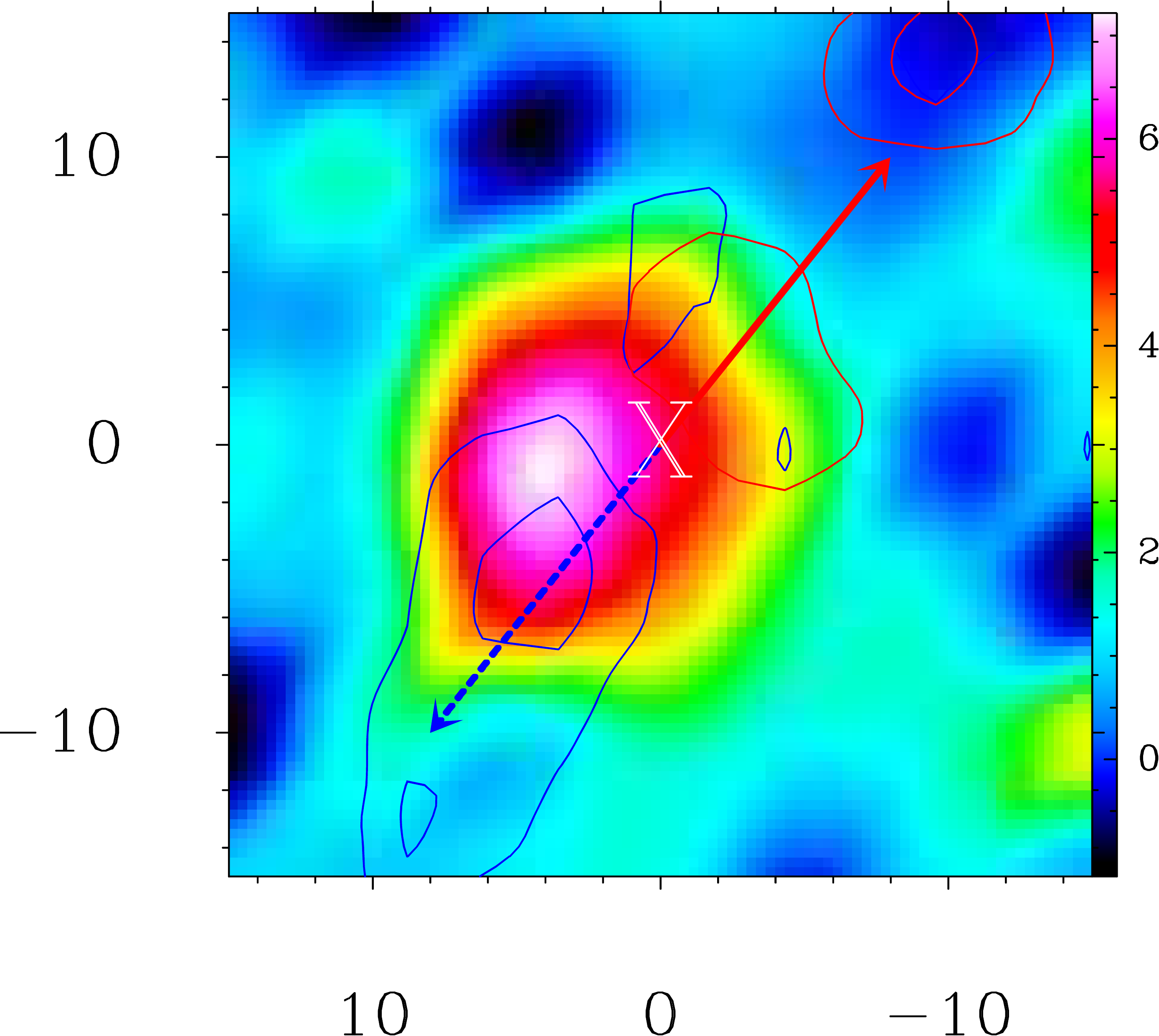}   
    \caption{\small Same as Fig. \ref{fig:specmap13CO65_1}.}
    \label{fig:specmap13CO65_3}
\end{figure*}

\begin{figure*}[htb]
    \centering
    \includegraphics[scale=0.20]{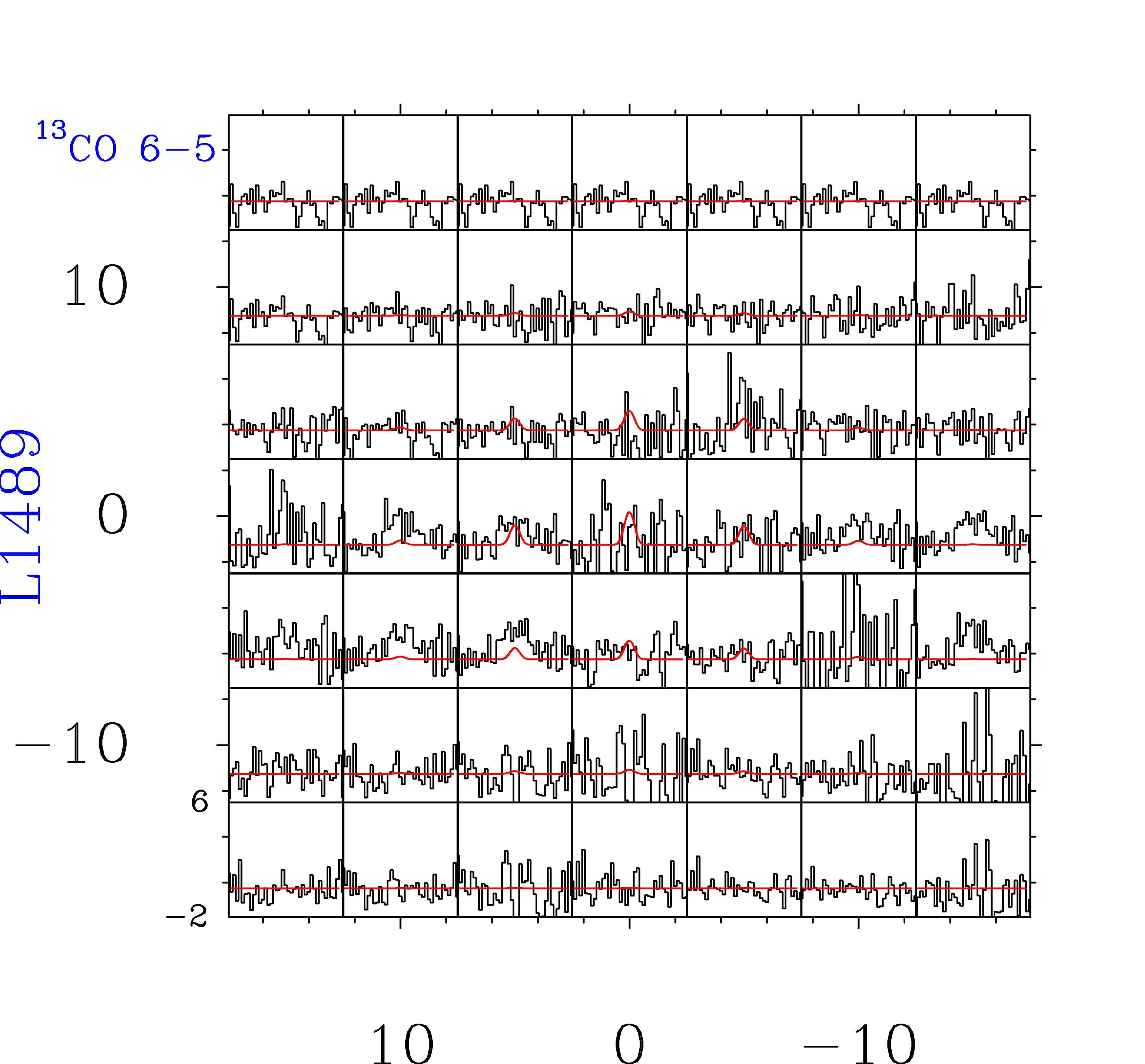}   
    \includegraphics[scale=0.20]{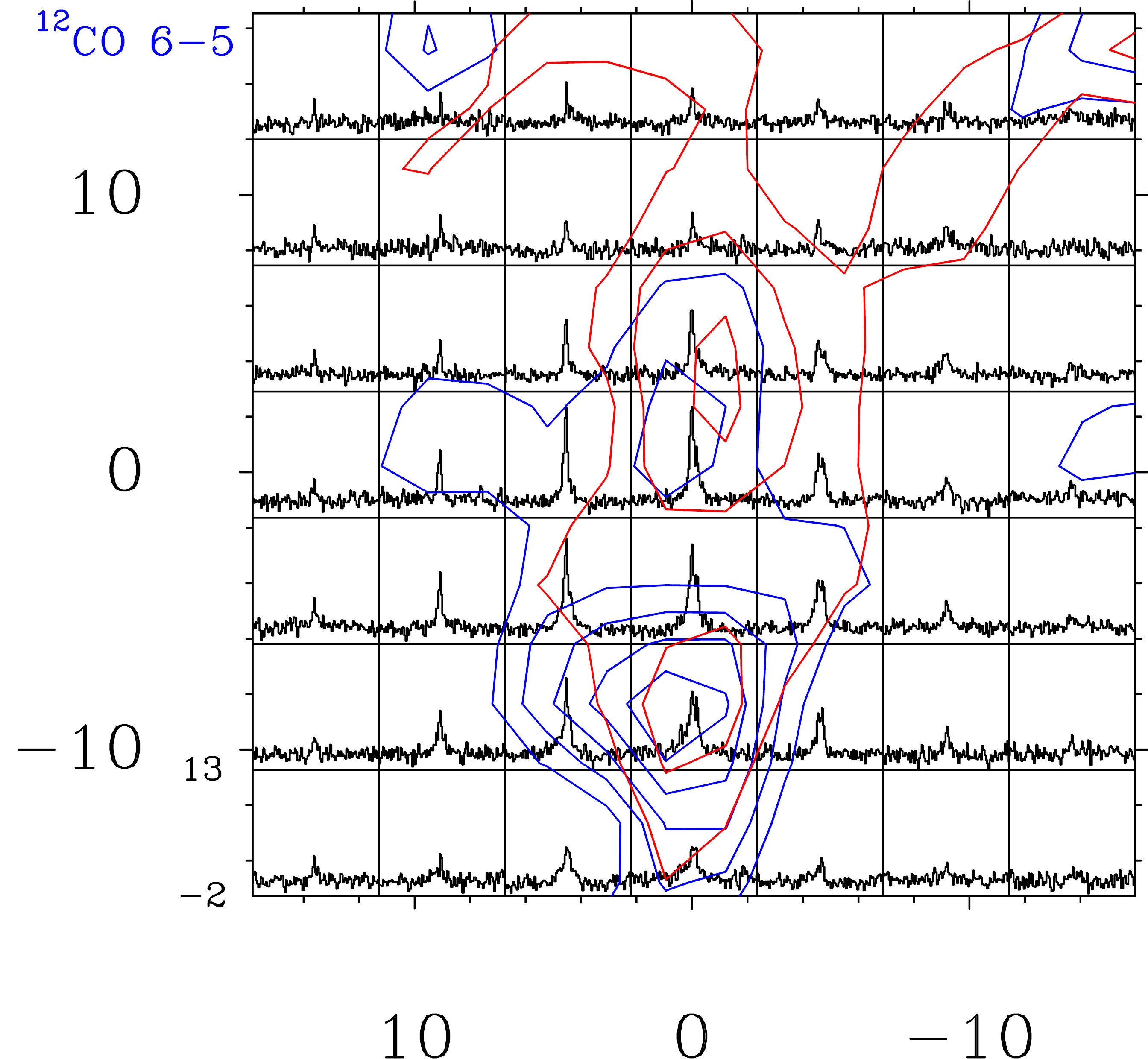}   
    \includegraphics[scale=0.20]{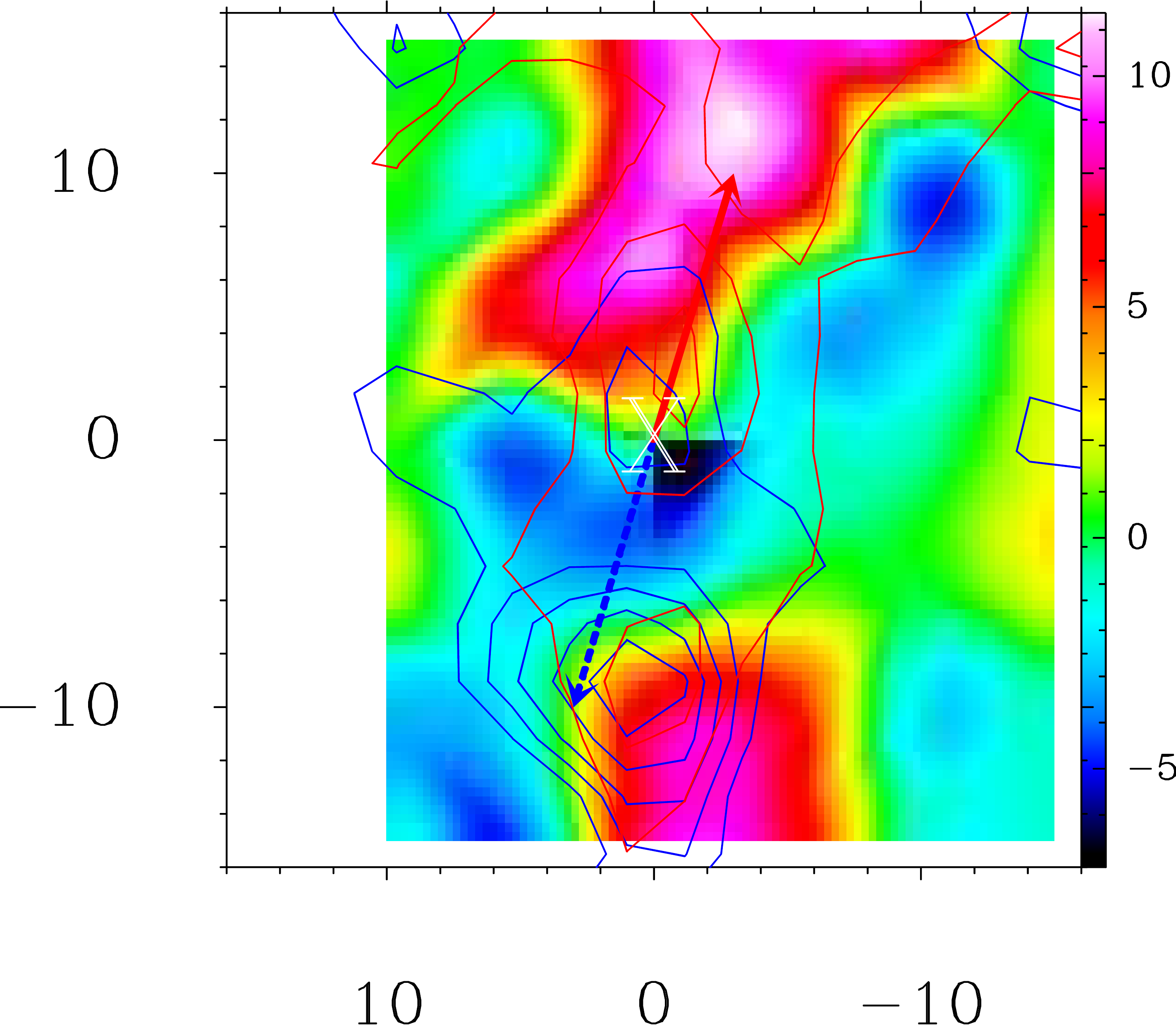} \\
    \includegraphics[scale=0.20]{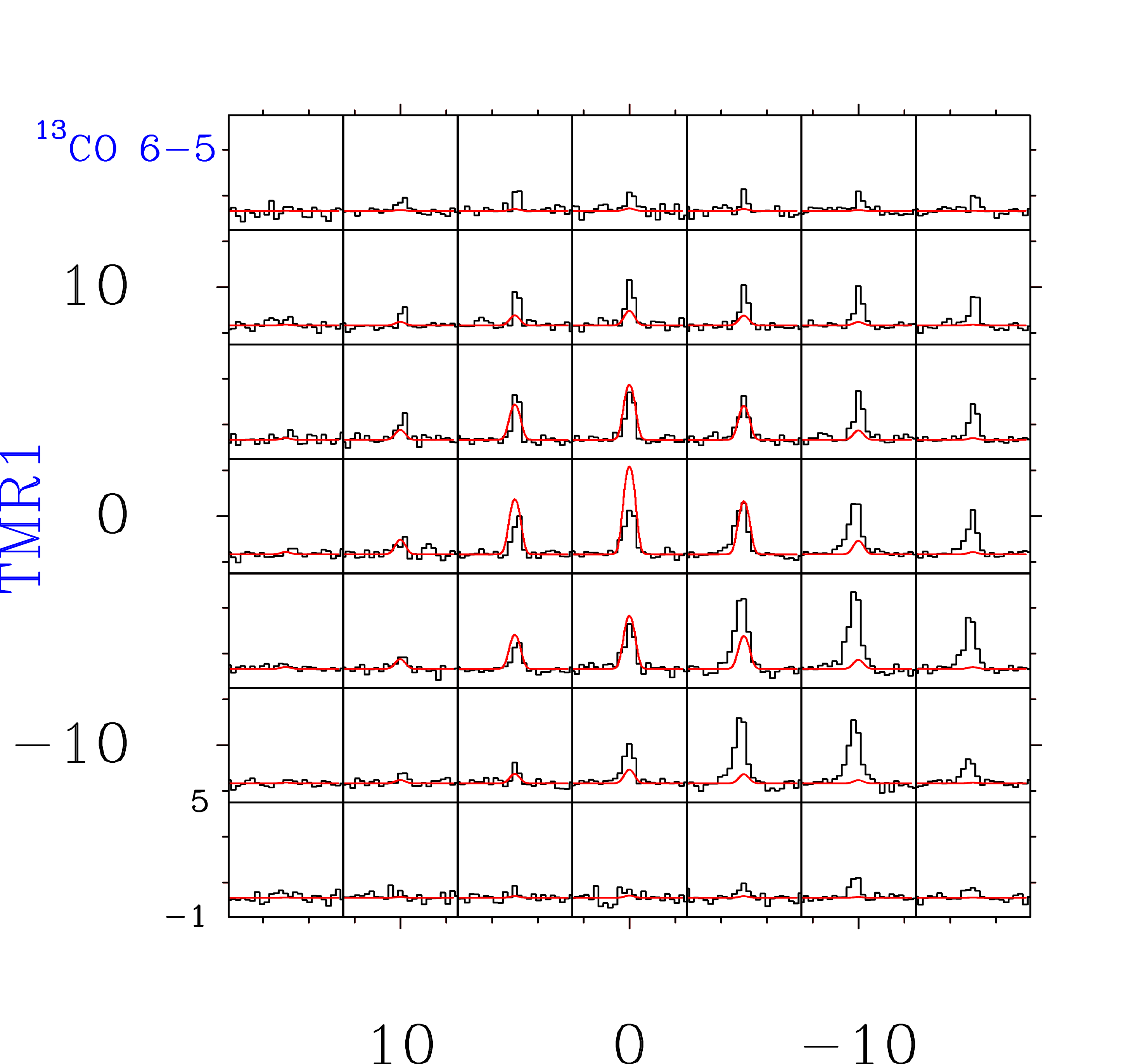}   
    \includegraphics[scale=0.20]{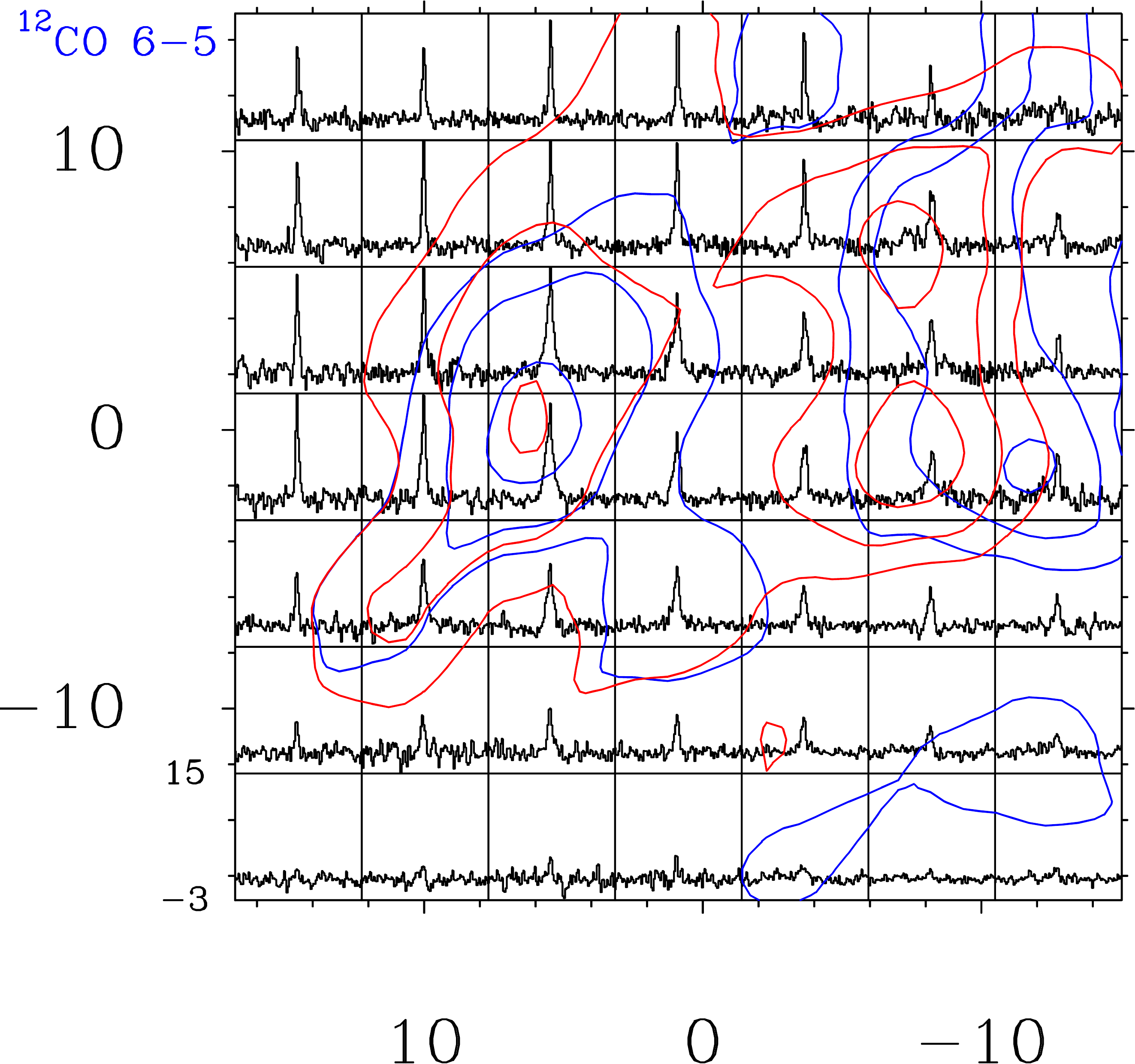}   
    \includegraphics[scale=0.20]{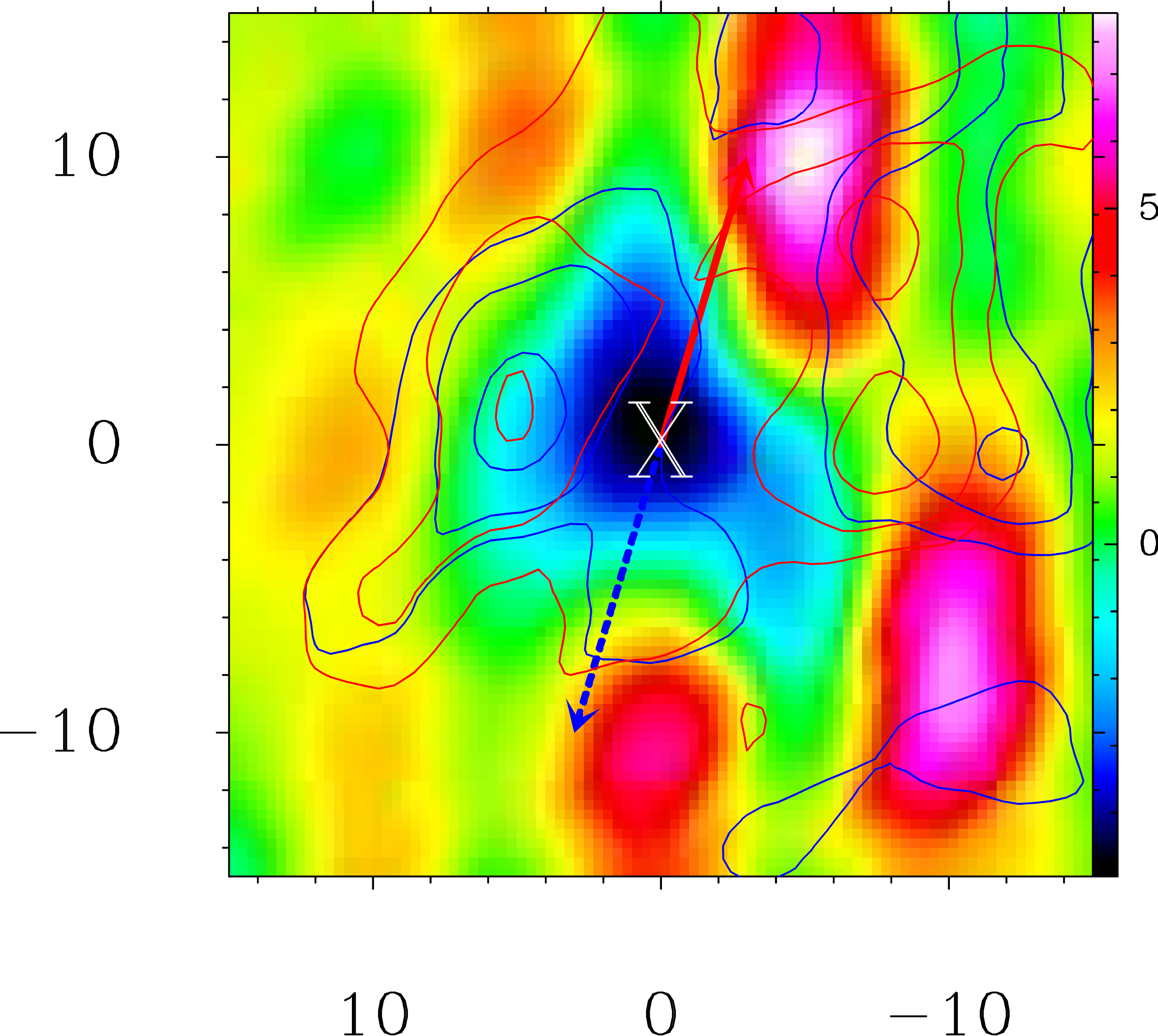} \\
    \includegraphics[scale=0.20]{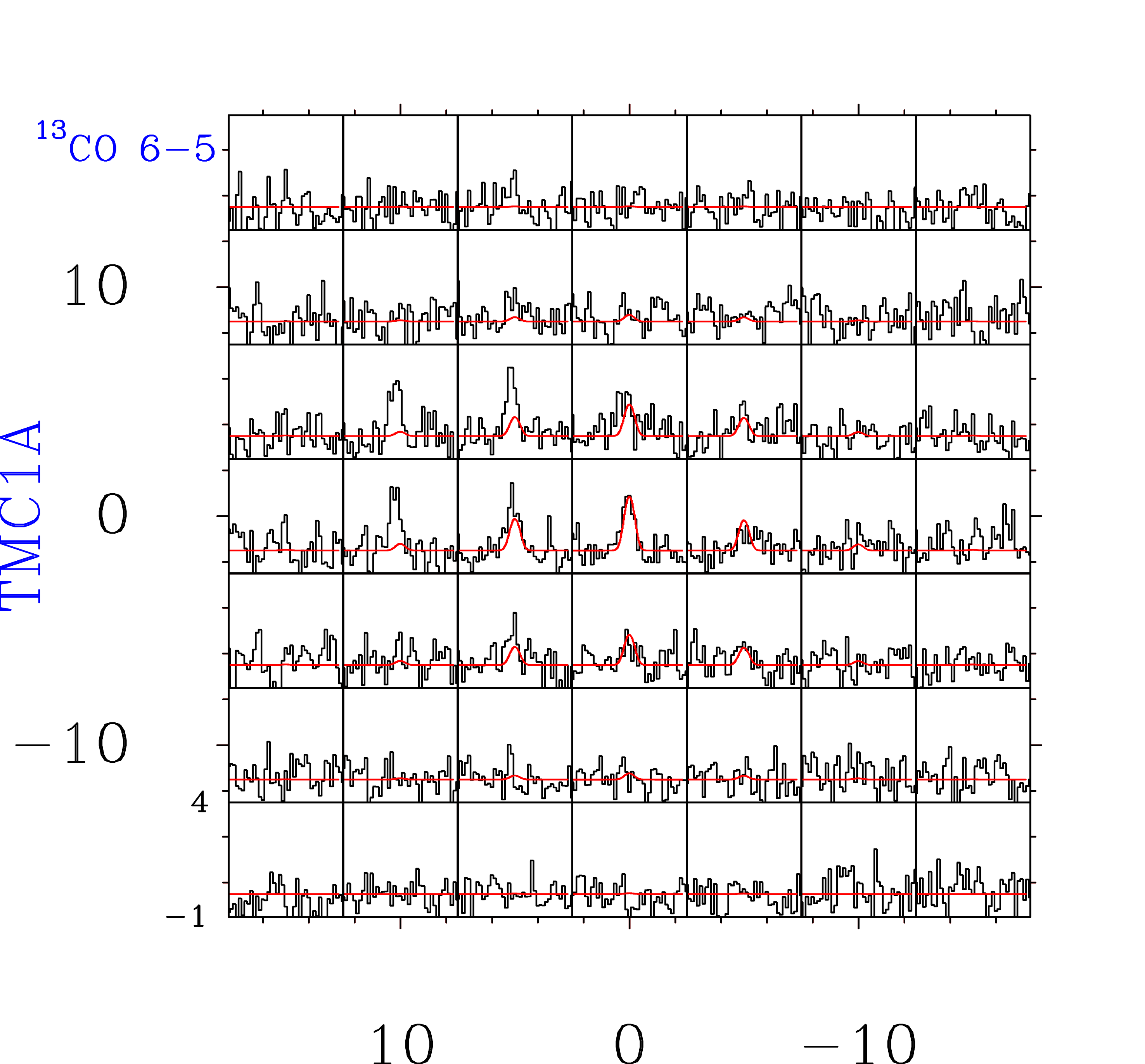}   
    \includegraphics[scale=0.20]{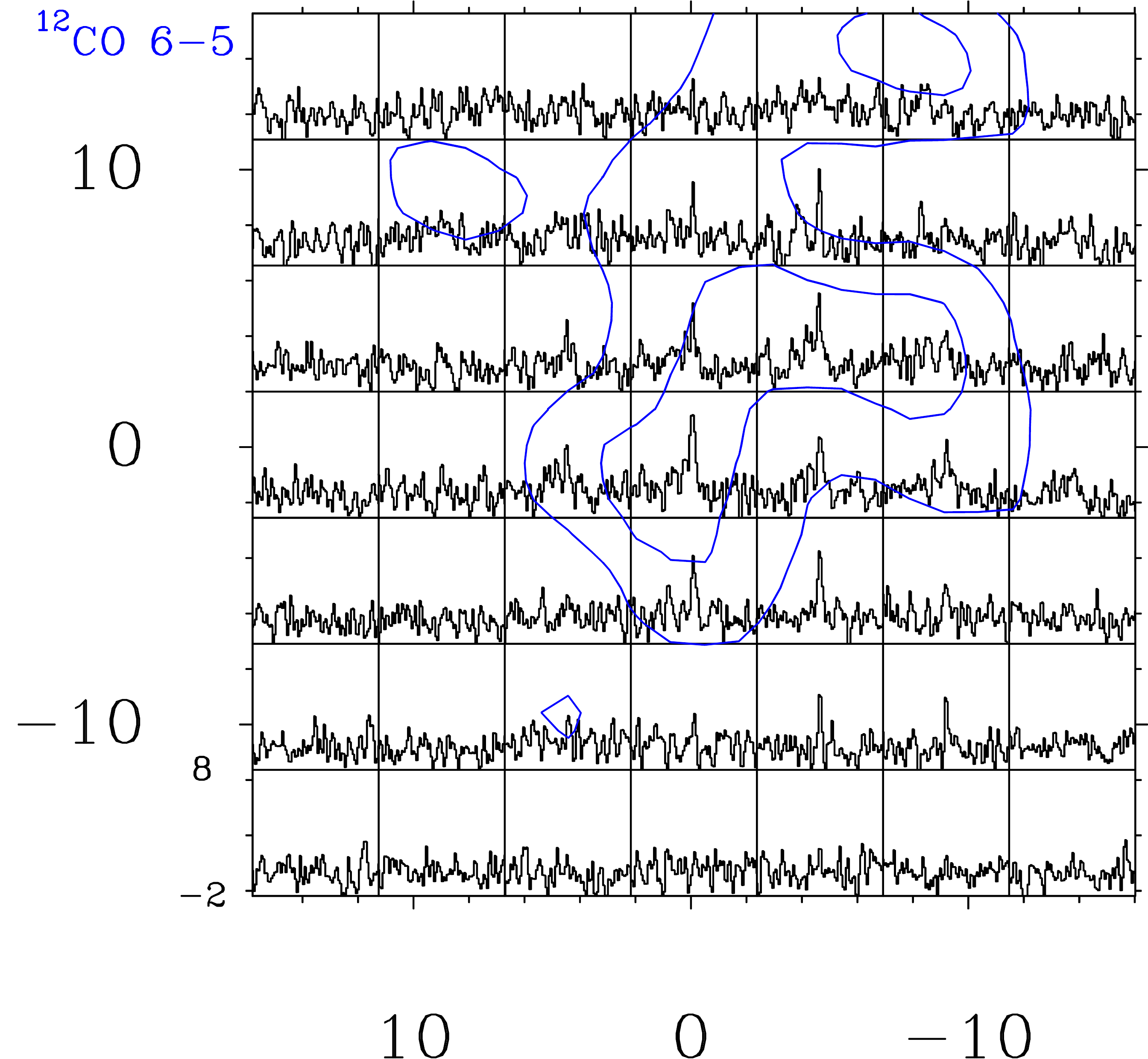}   
    \includegraphics[scale=0.20]{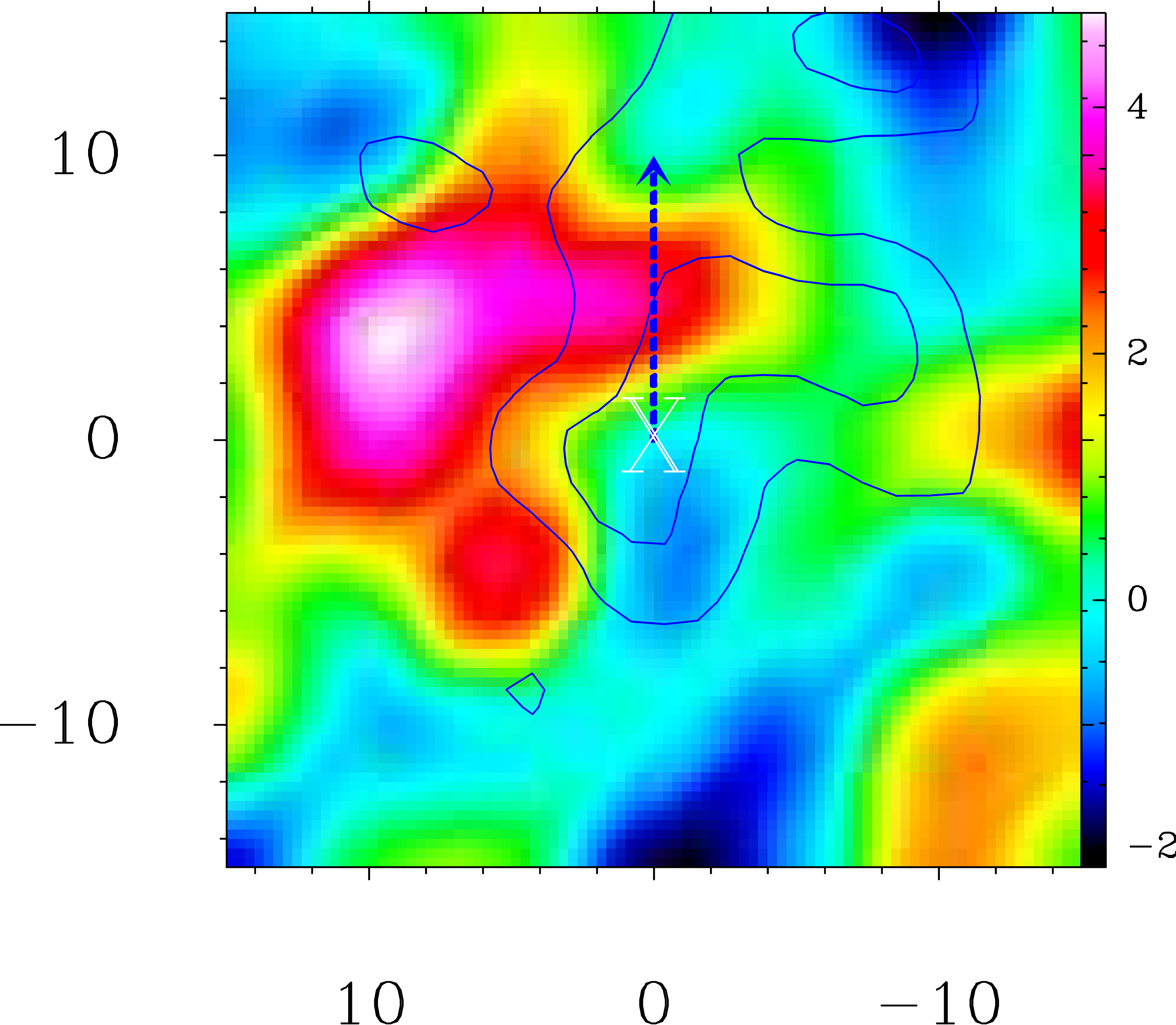} \\
    \includegraphics[scale=0.20]{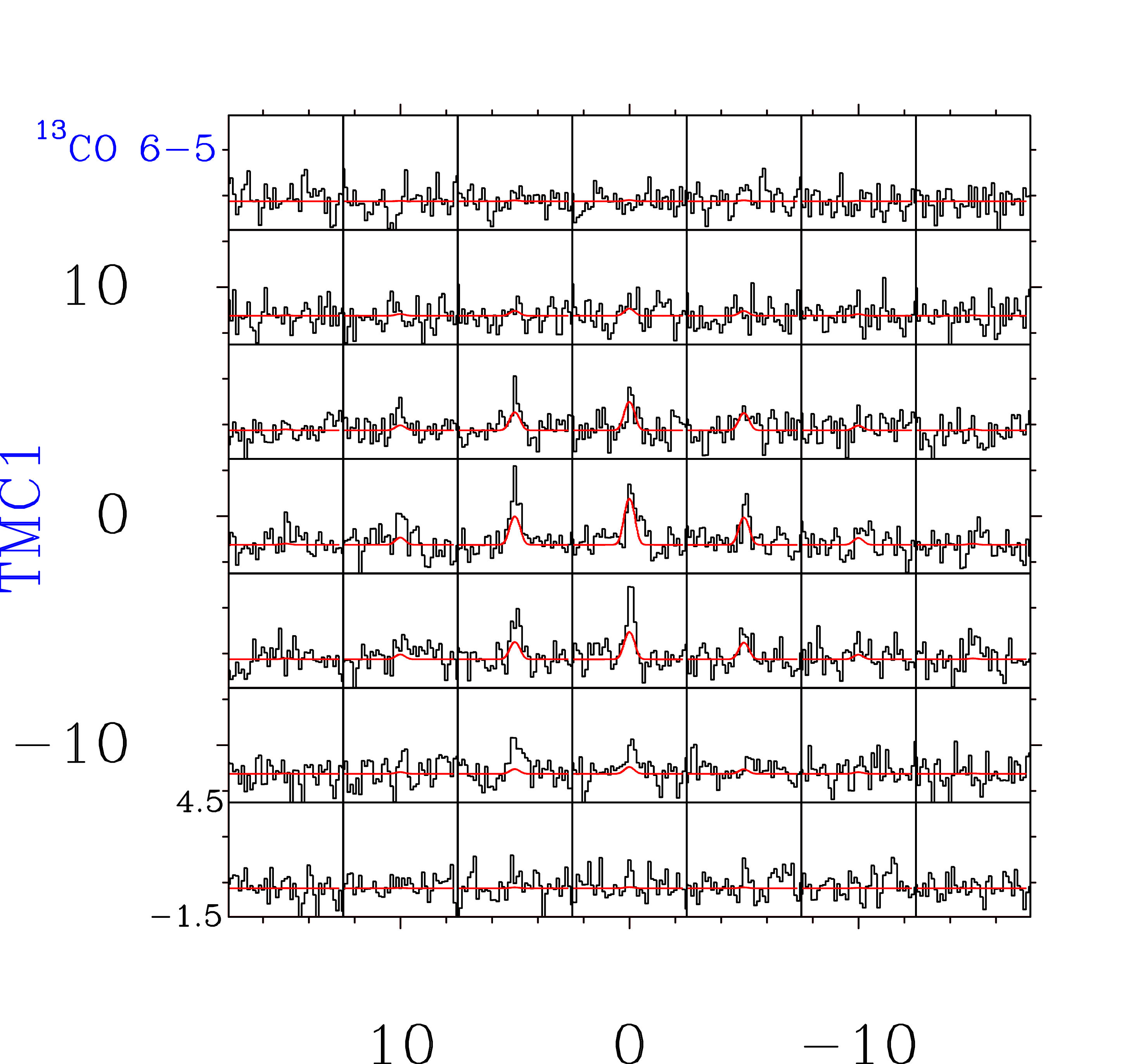}   
    \includegraphics[scale=0.20]{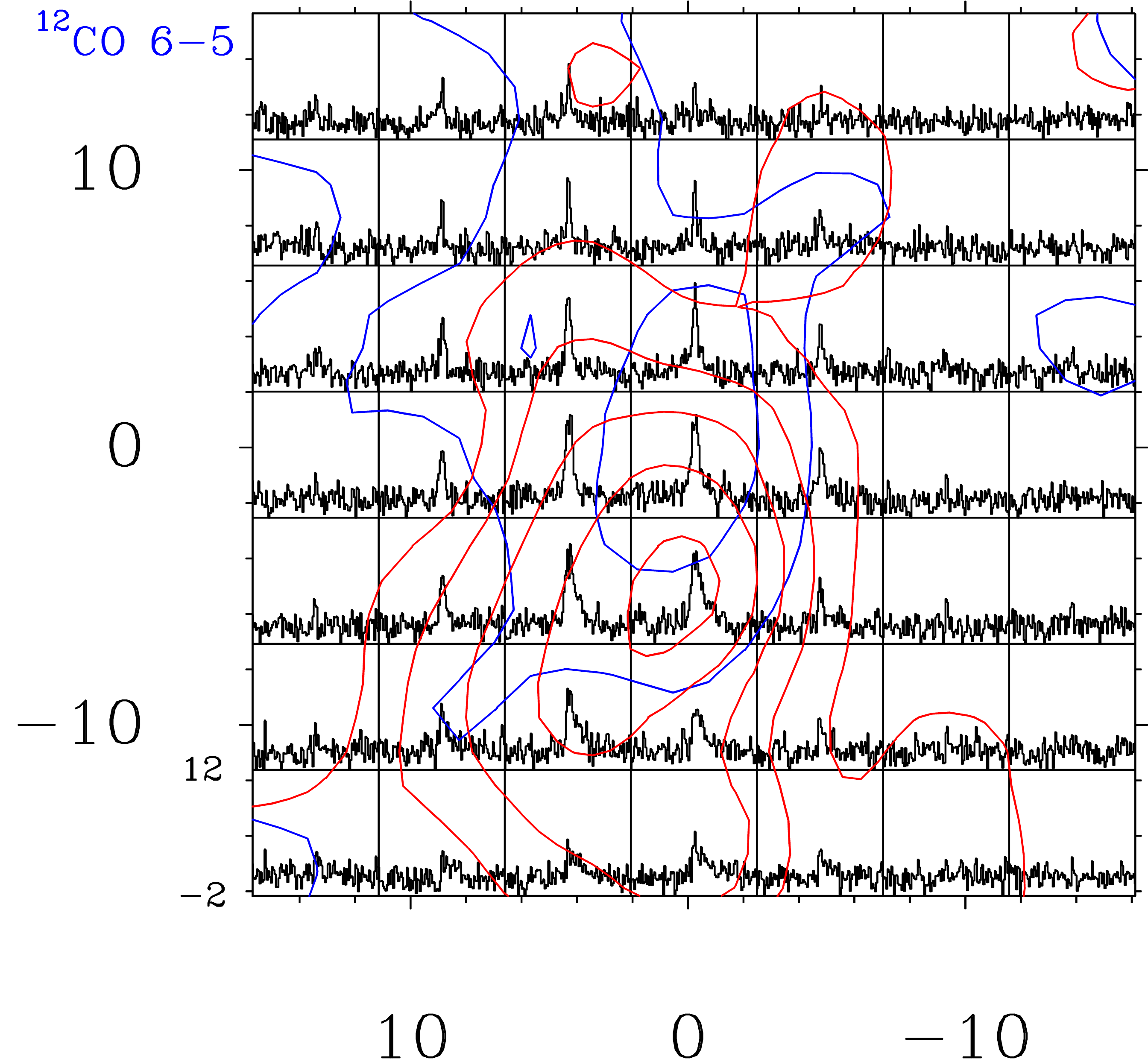}   
    \includegraphics[scale=0.20]{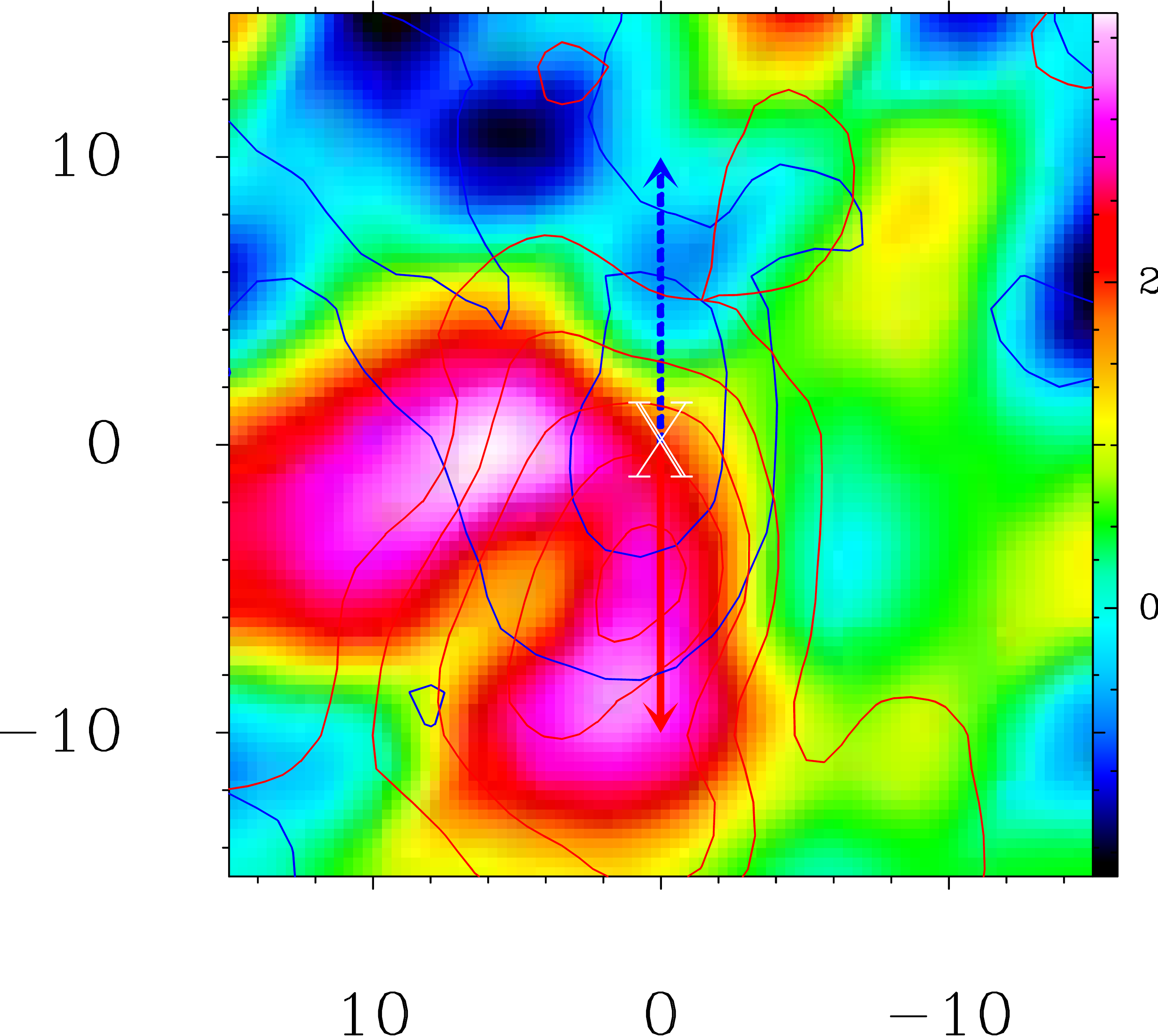}   
    \caption{\small Same as Fig. \ref{fig:specmap13CO65_1}.}
    \label{fig:specmap13CO65_4}
\end{figure*}

\begin{figure*}[htb]
    \centering
    \includegraphics[scale=0.20]{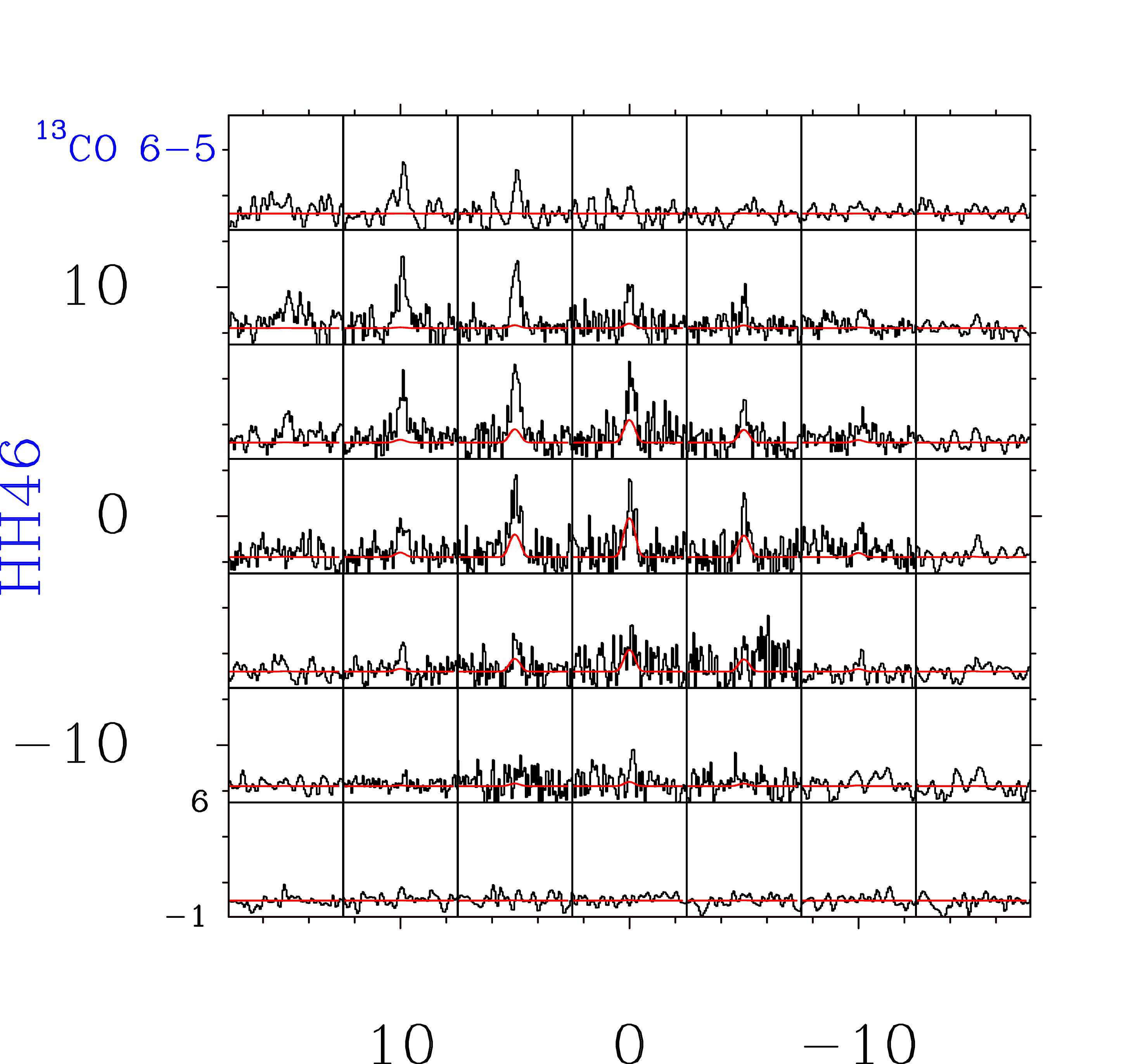}   
    \includegraphics[scale=0.20]{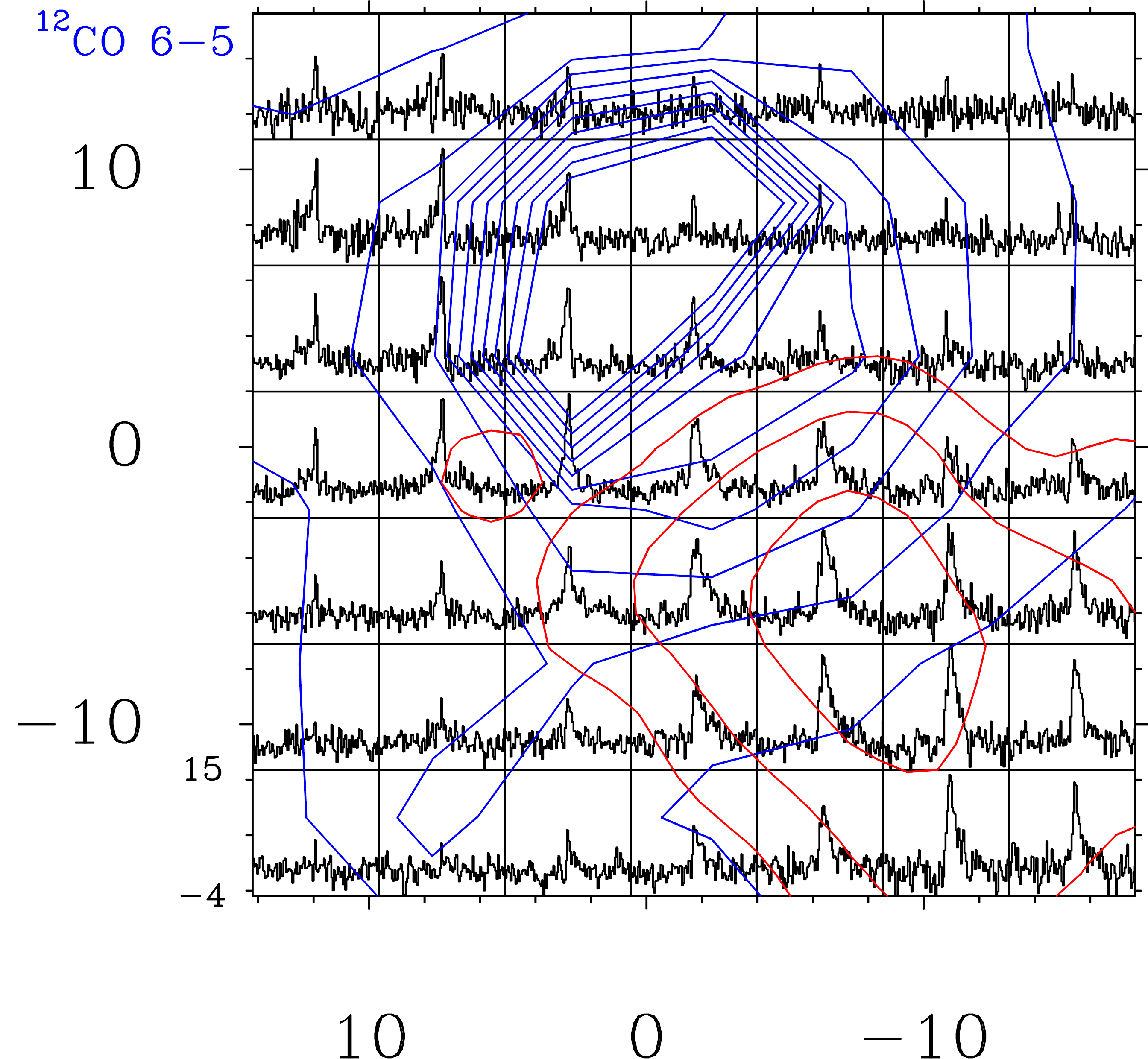}   
    \includegraphics[scale=0.20]{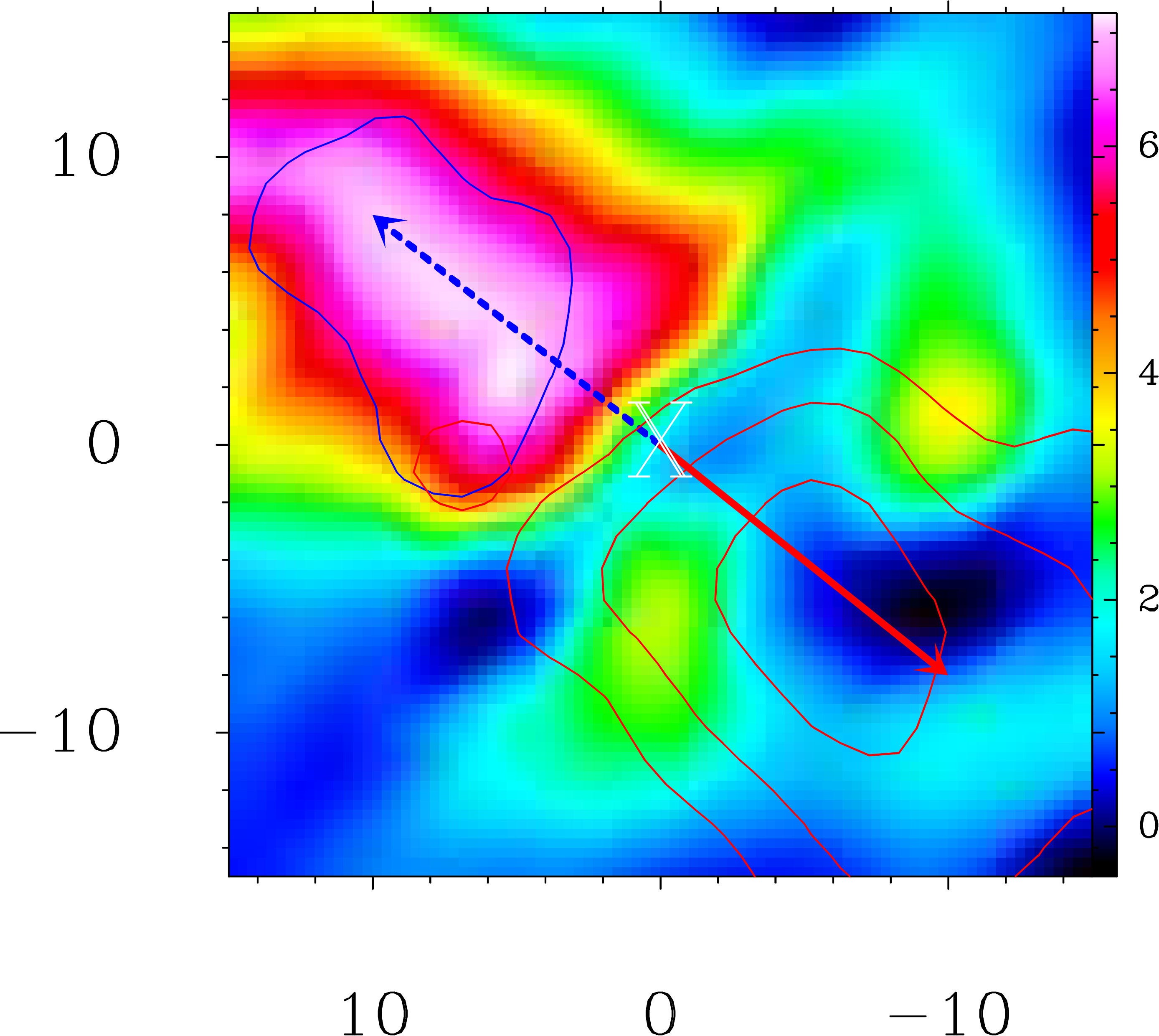} \\
    \includegraphics[scale=0.20]{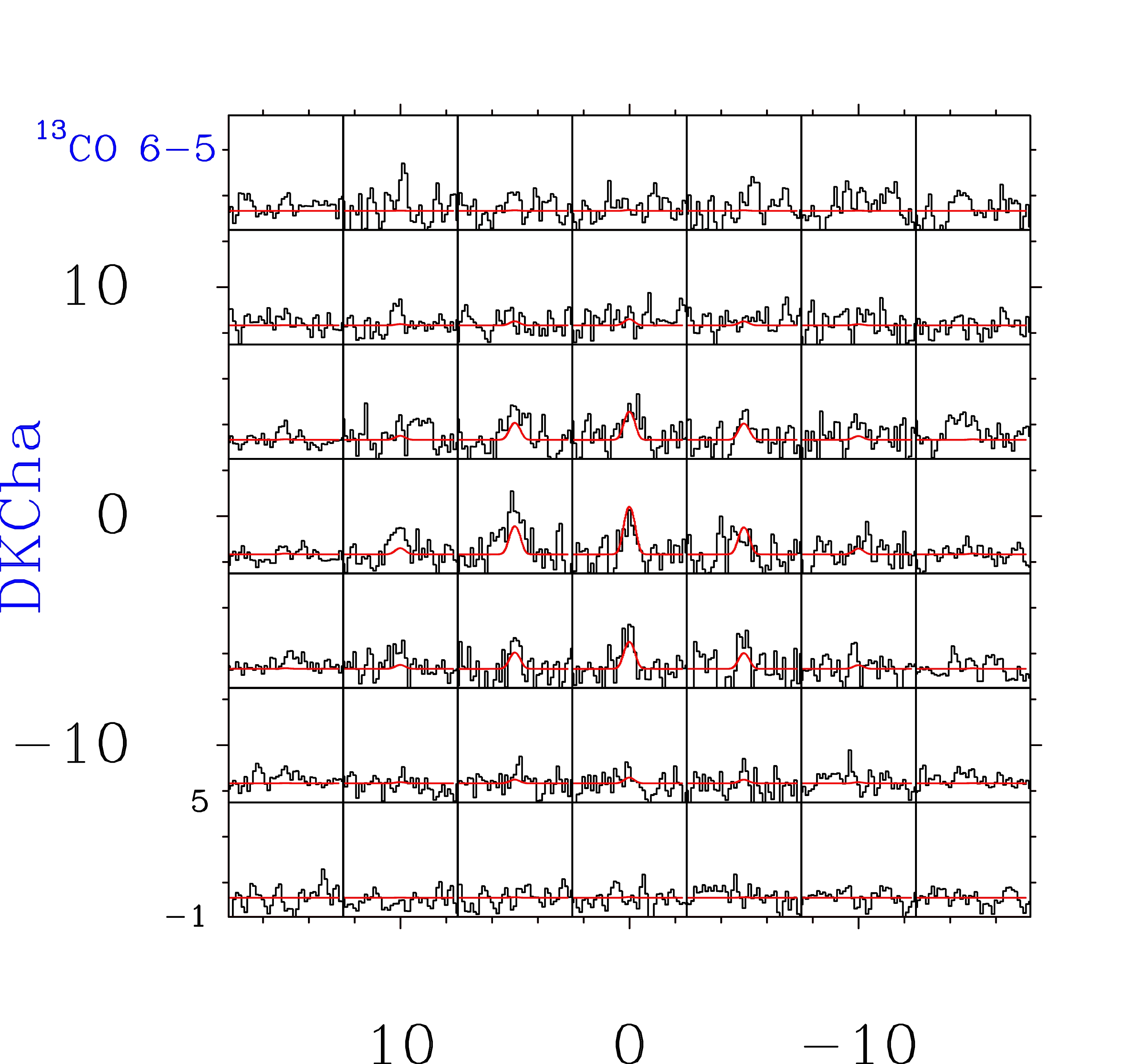}   
    \includegraphics[scale=0.20]{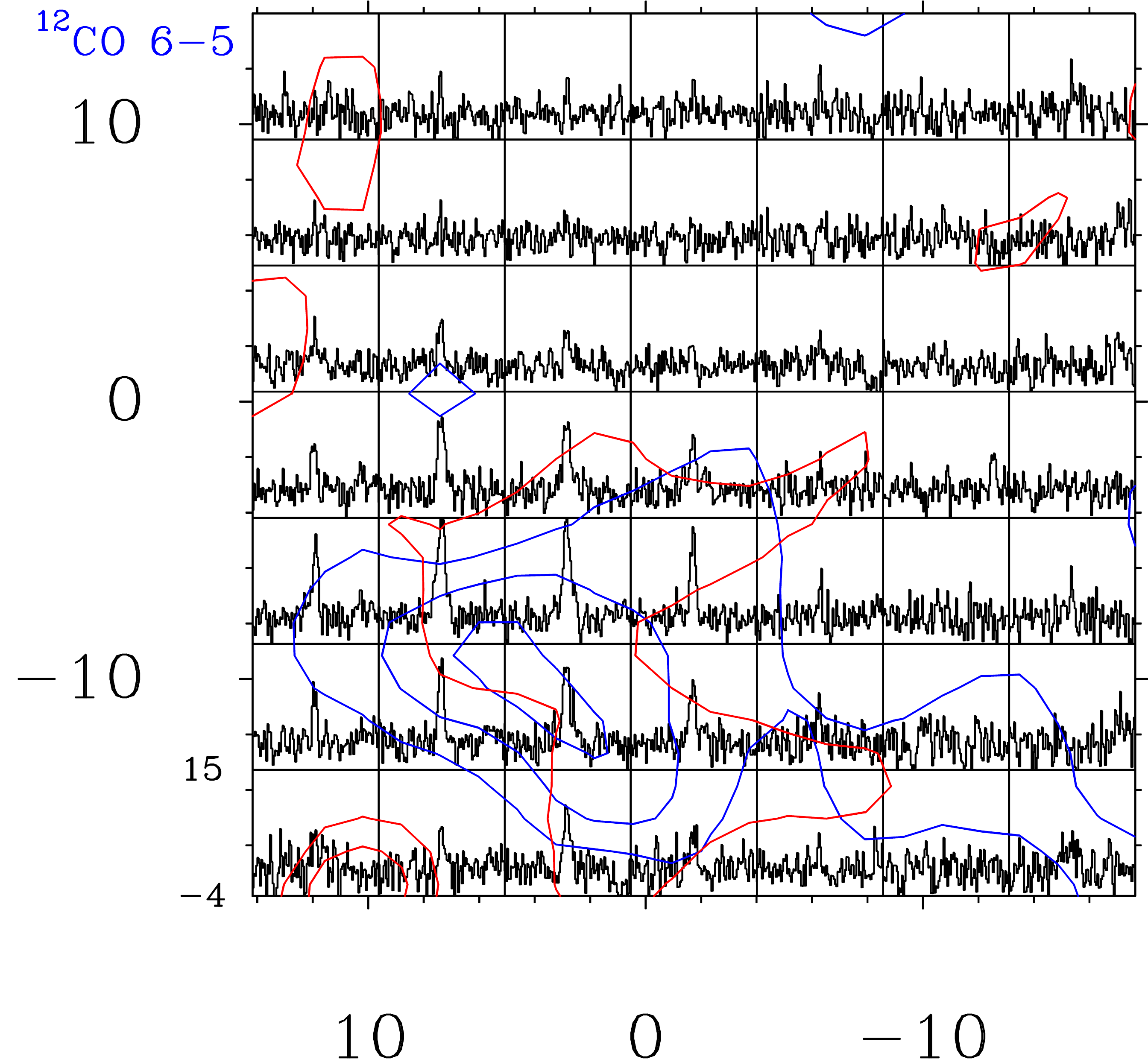}   
    \includegraphics[scale=0.20]{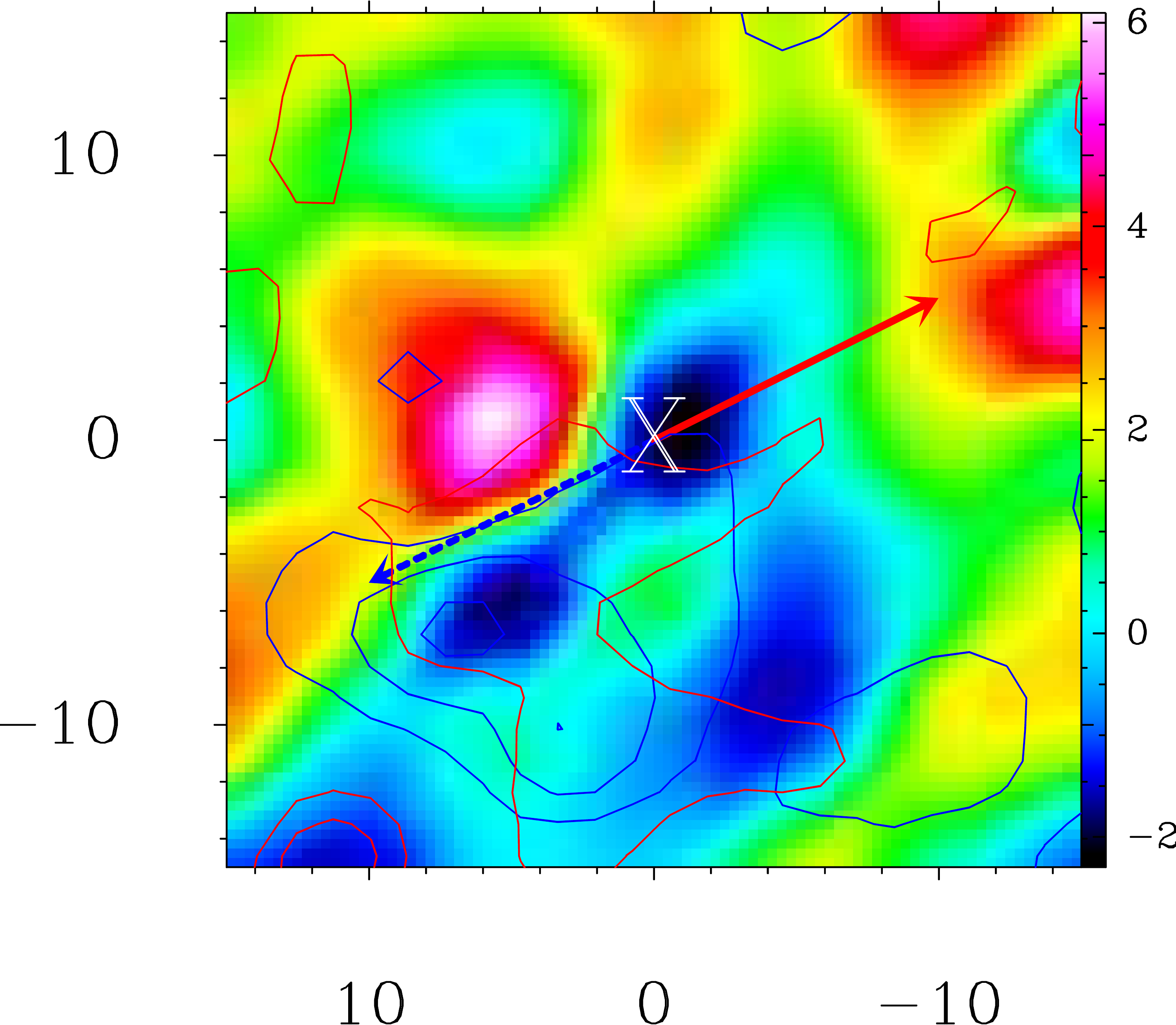} \\
    \includegraphics[scale=0.20]{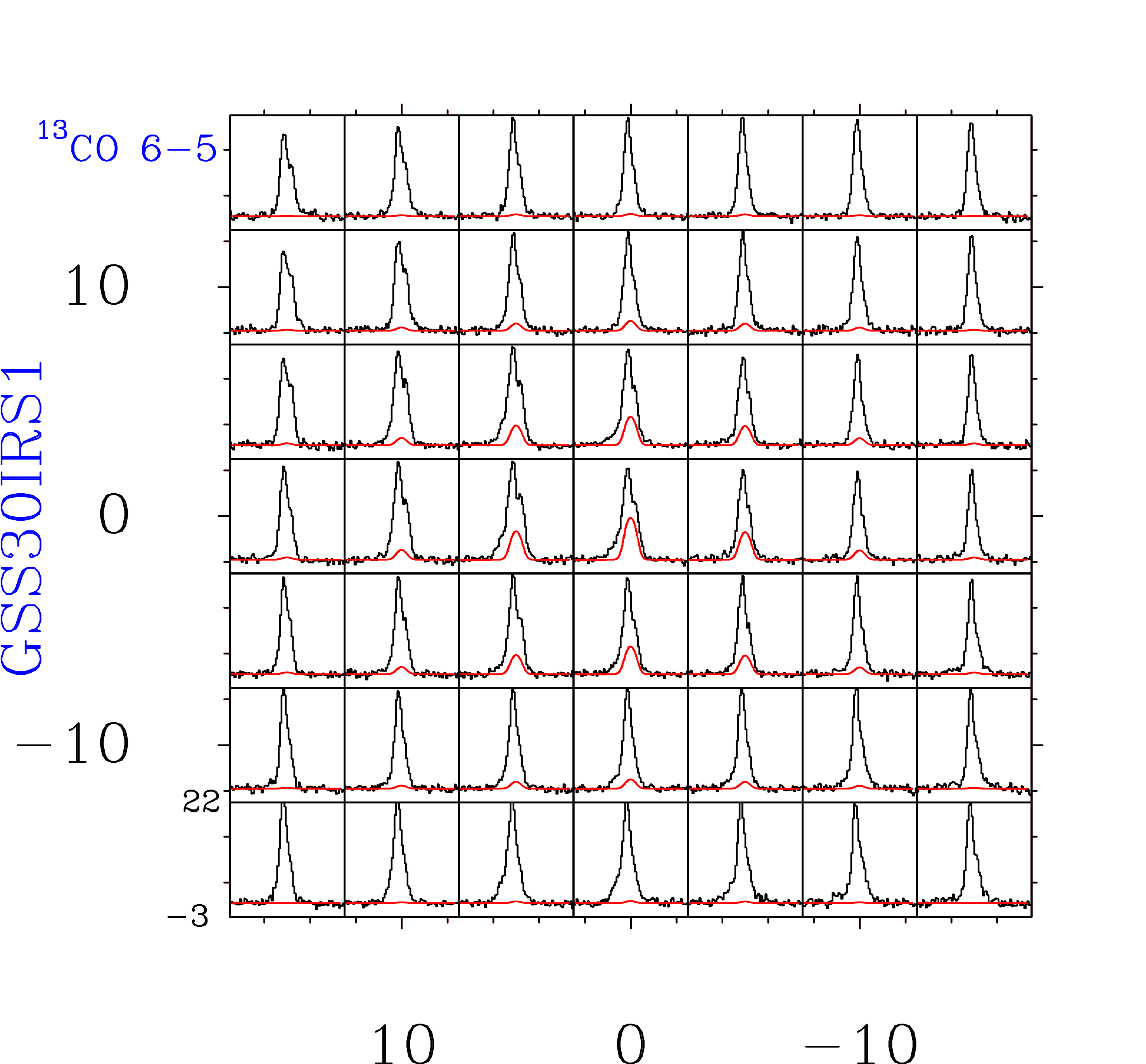}   
    \includegraphics[scale=0.20]{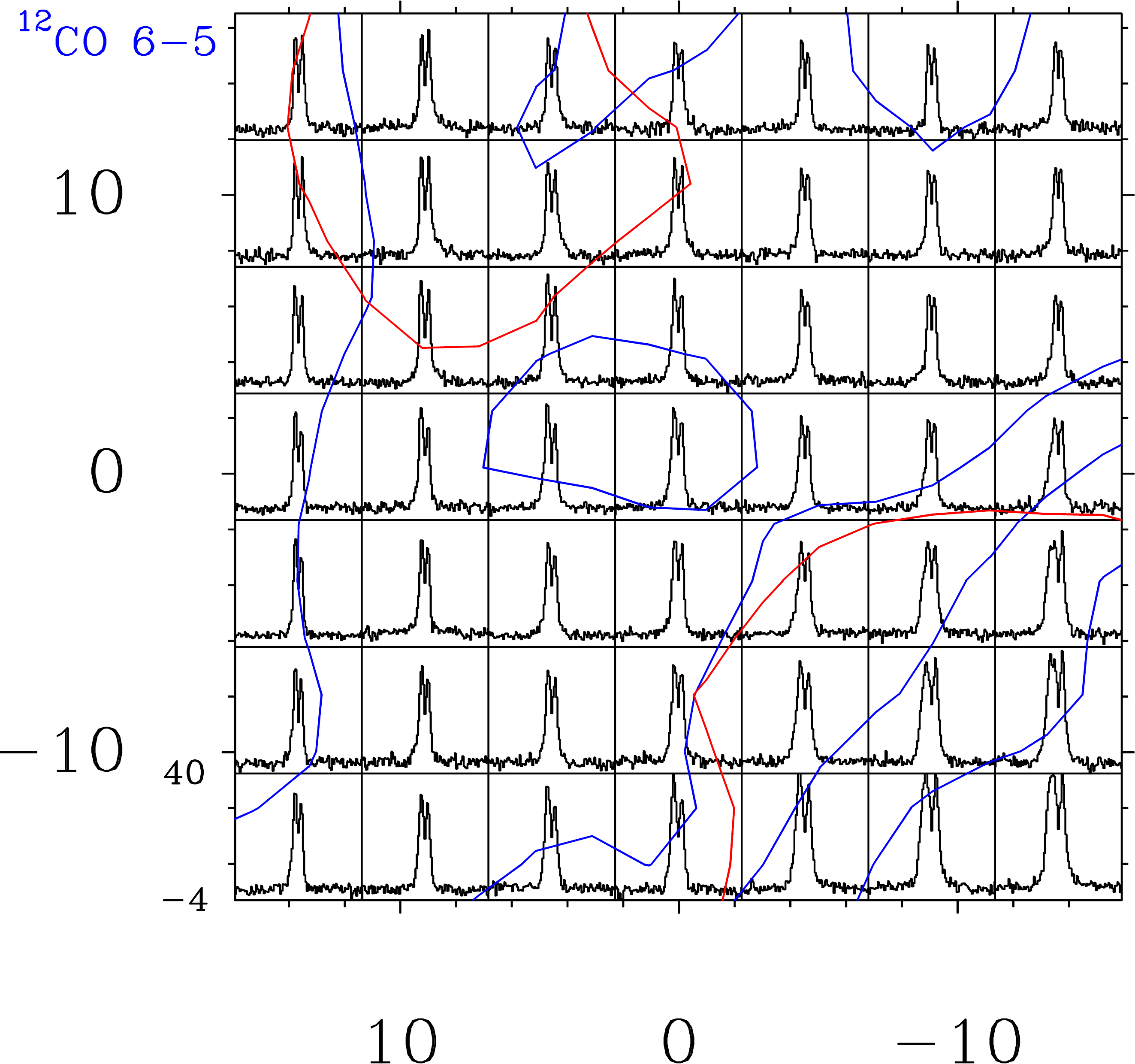}   
    \includegraphics[scale=0.20]{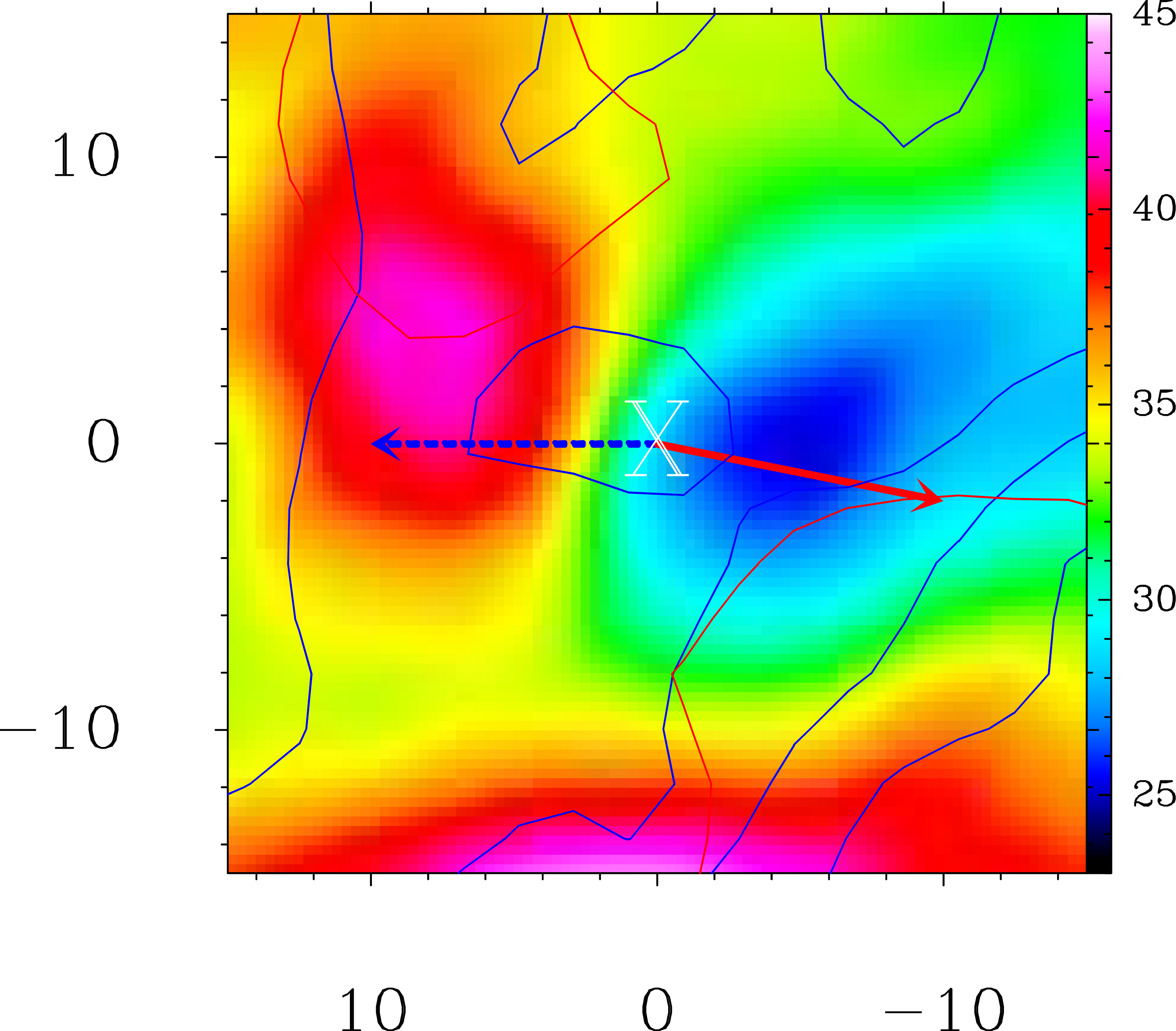} \\
    \includegraphics[scale=0.20]{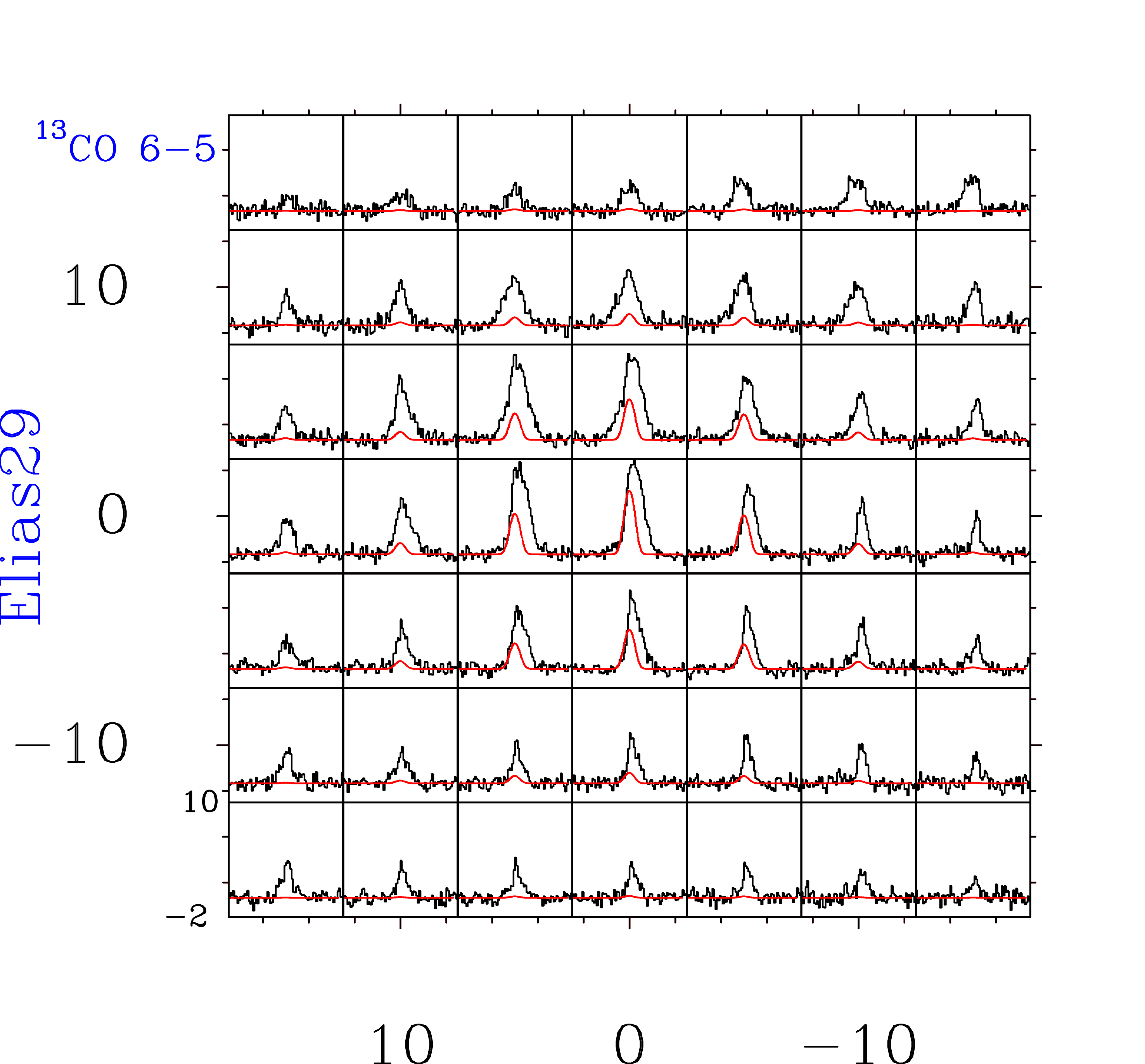}   
    \includegraphics[scale=0.20]{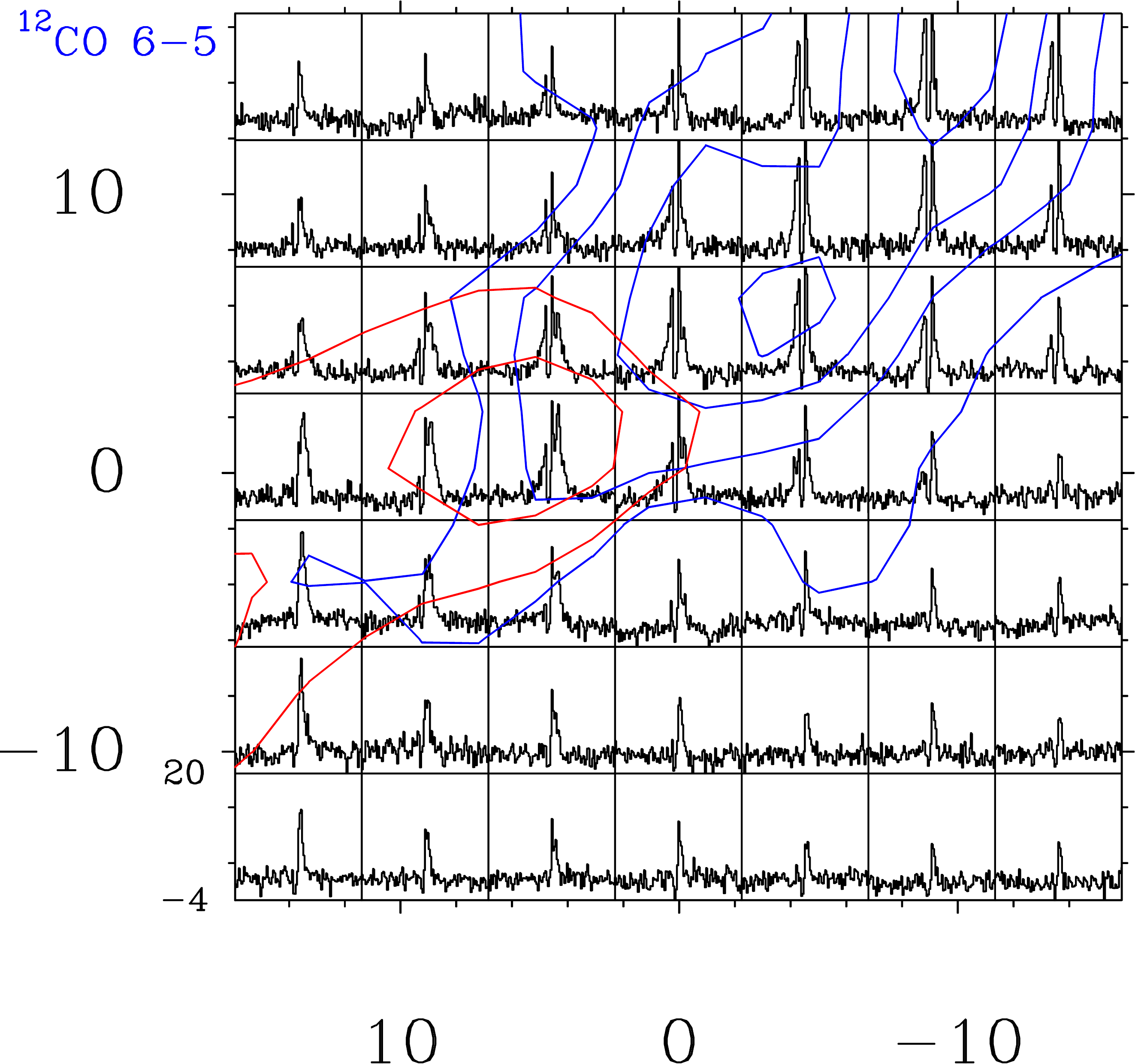}   
    \includegraphics[scale=0.20]{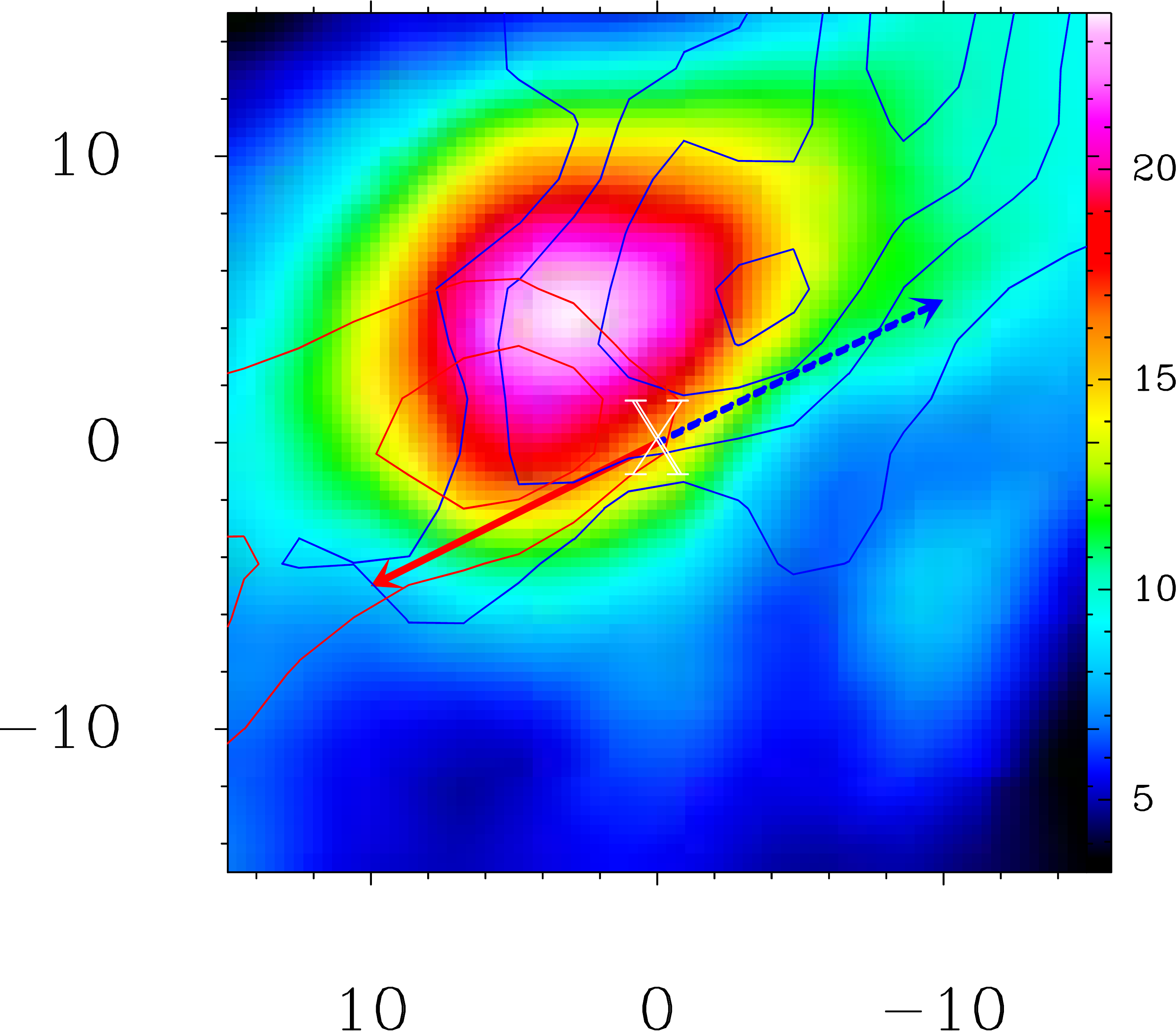}   
    \caption{\small Same as Fig. \ref{fig:specmap13CO65_1}.}
    \label{fig:specmap13CO65_5}
\end{figure*}

\begin{figure*}[htb]
    \centering
    \includegraphics[scale=0.20]{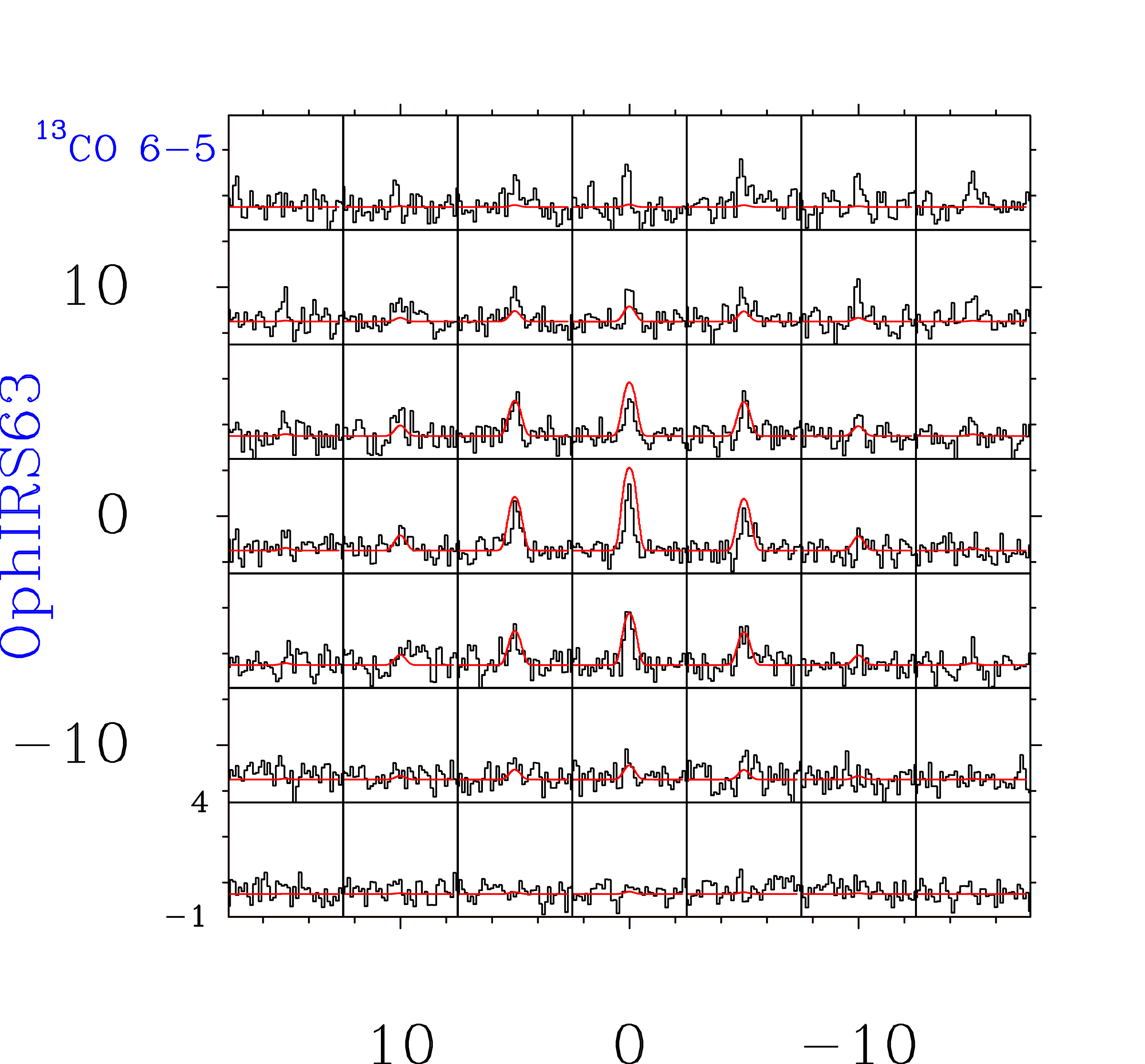}   
    \includegraphics[scale=0.20]{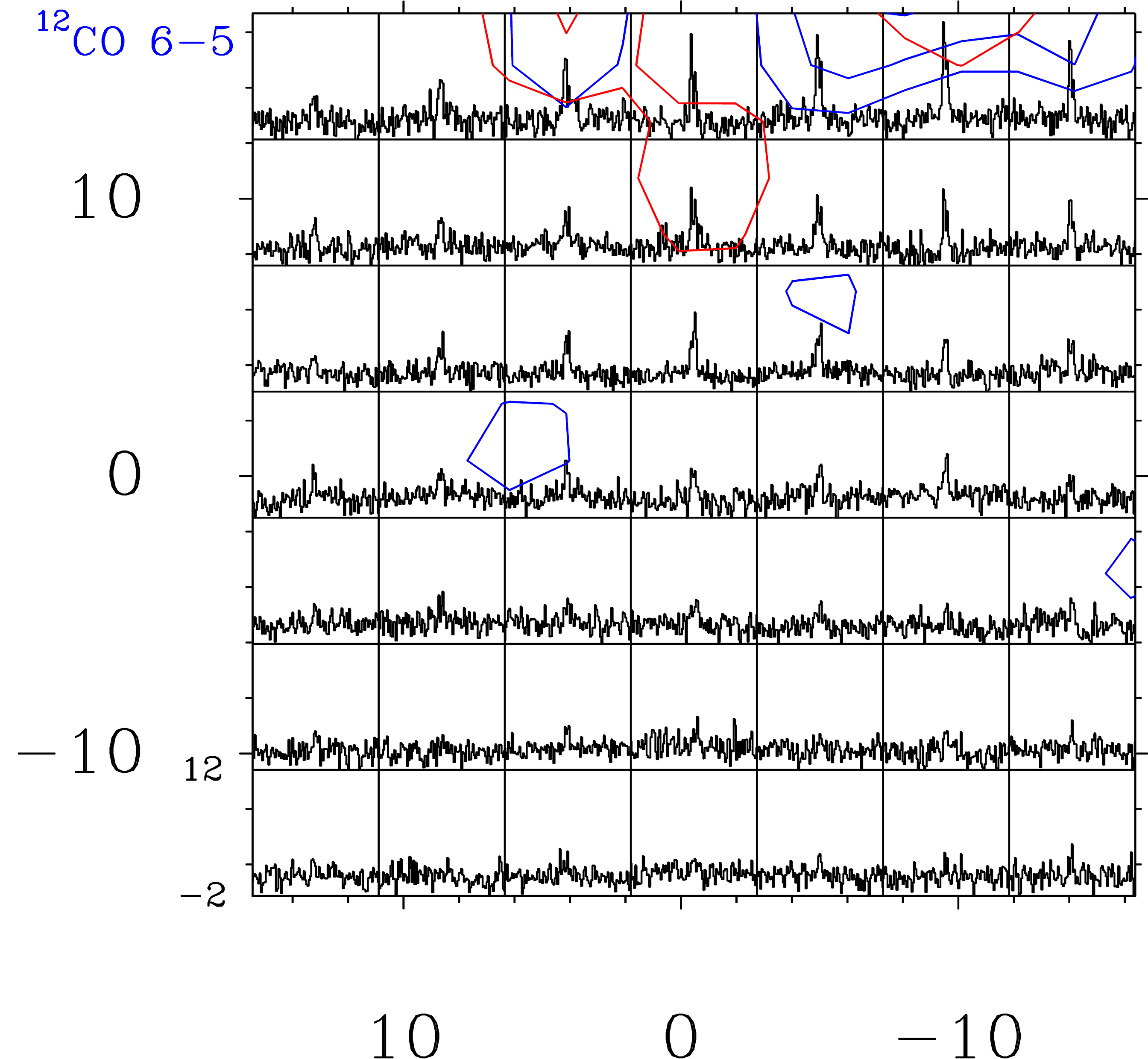}   
    \includegraphics[scale=0.20]{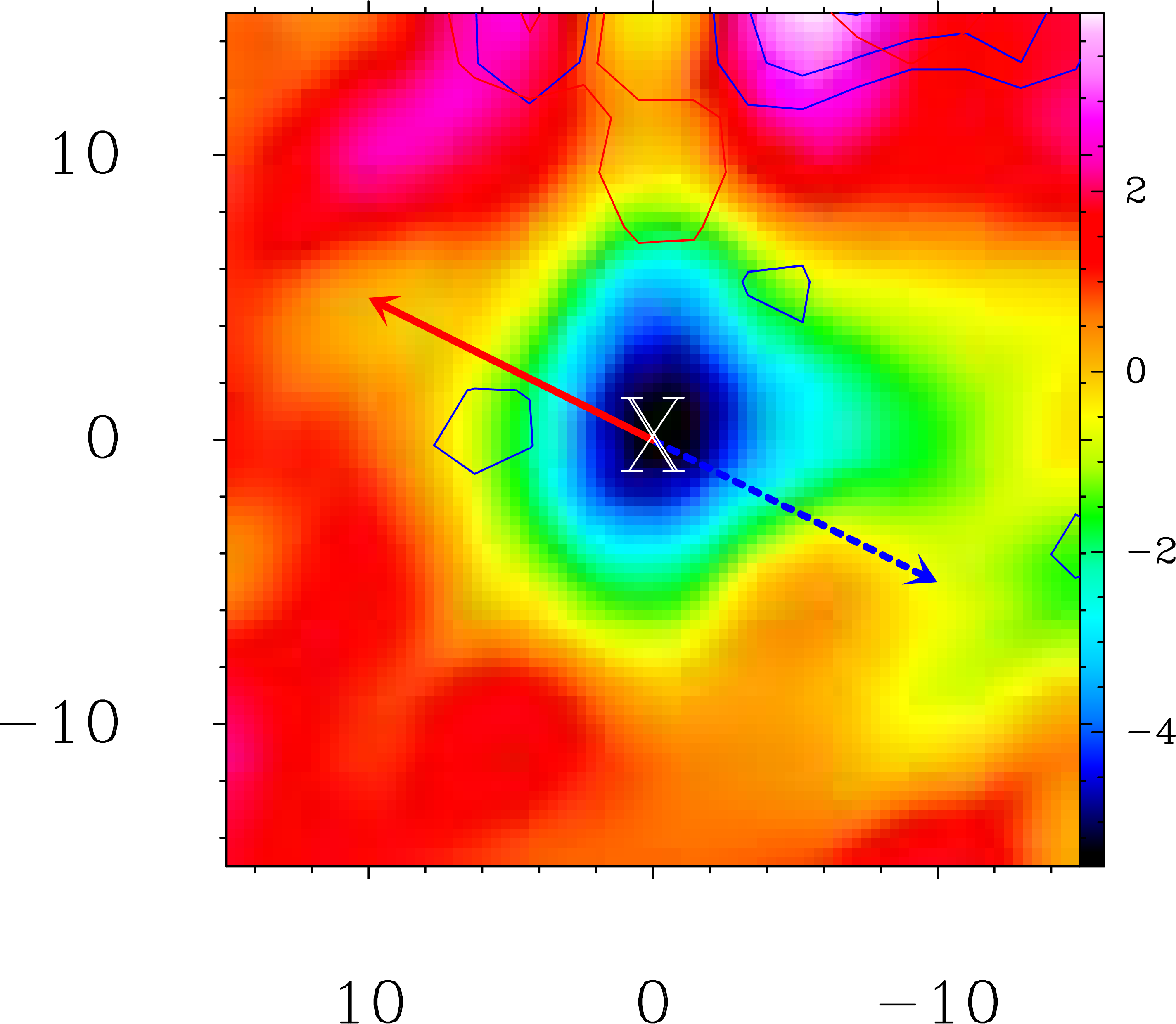} \\
    \includegraphics[scale=0.20]{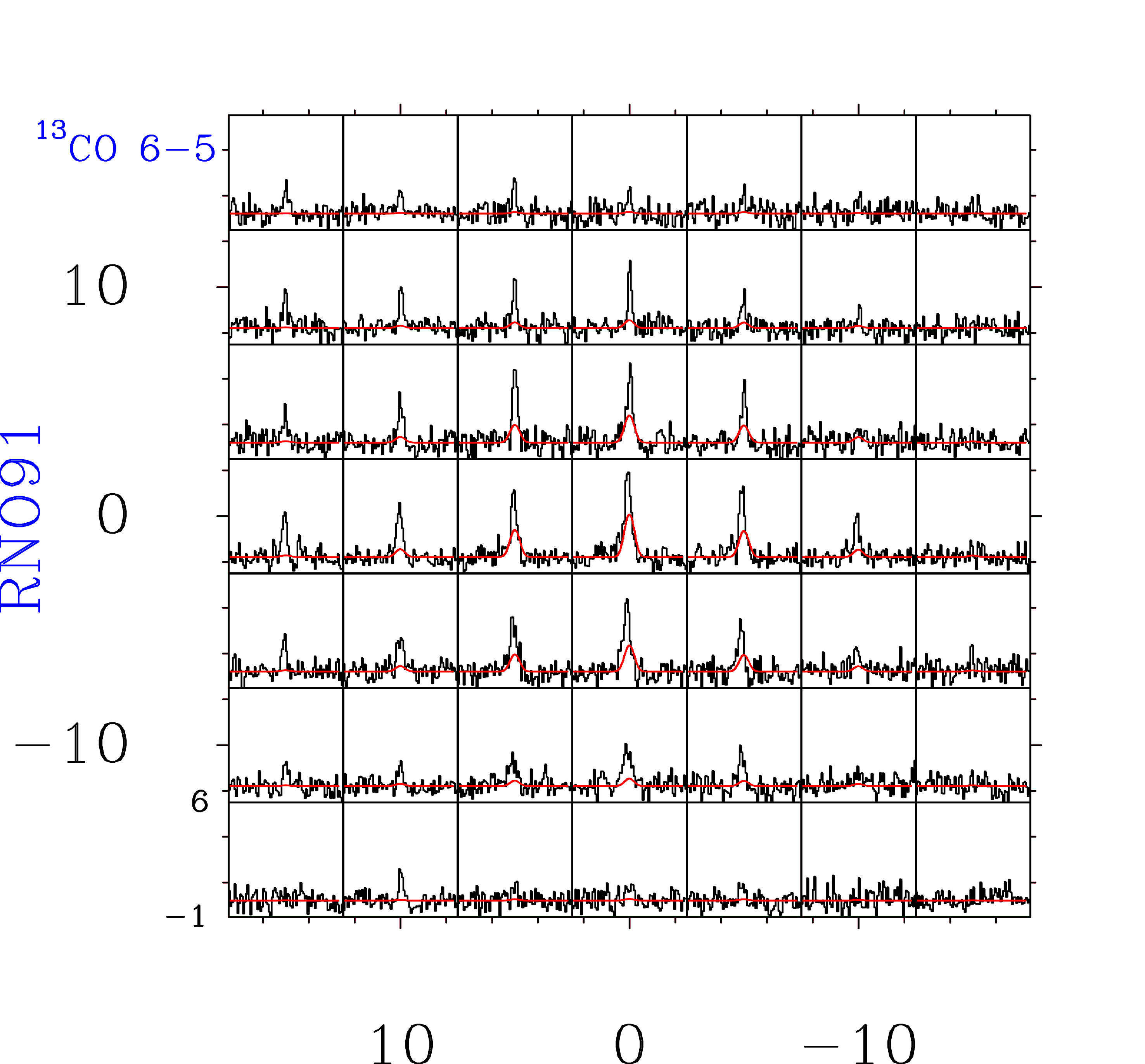}   
    \includegraphics[scale=0.20]{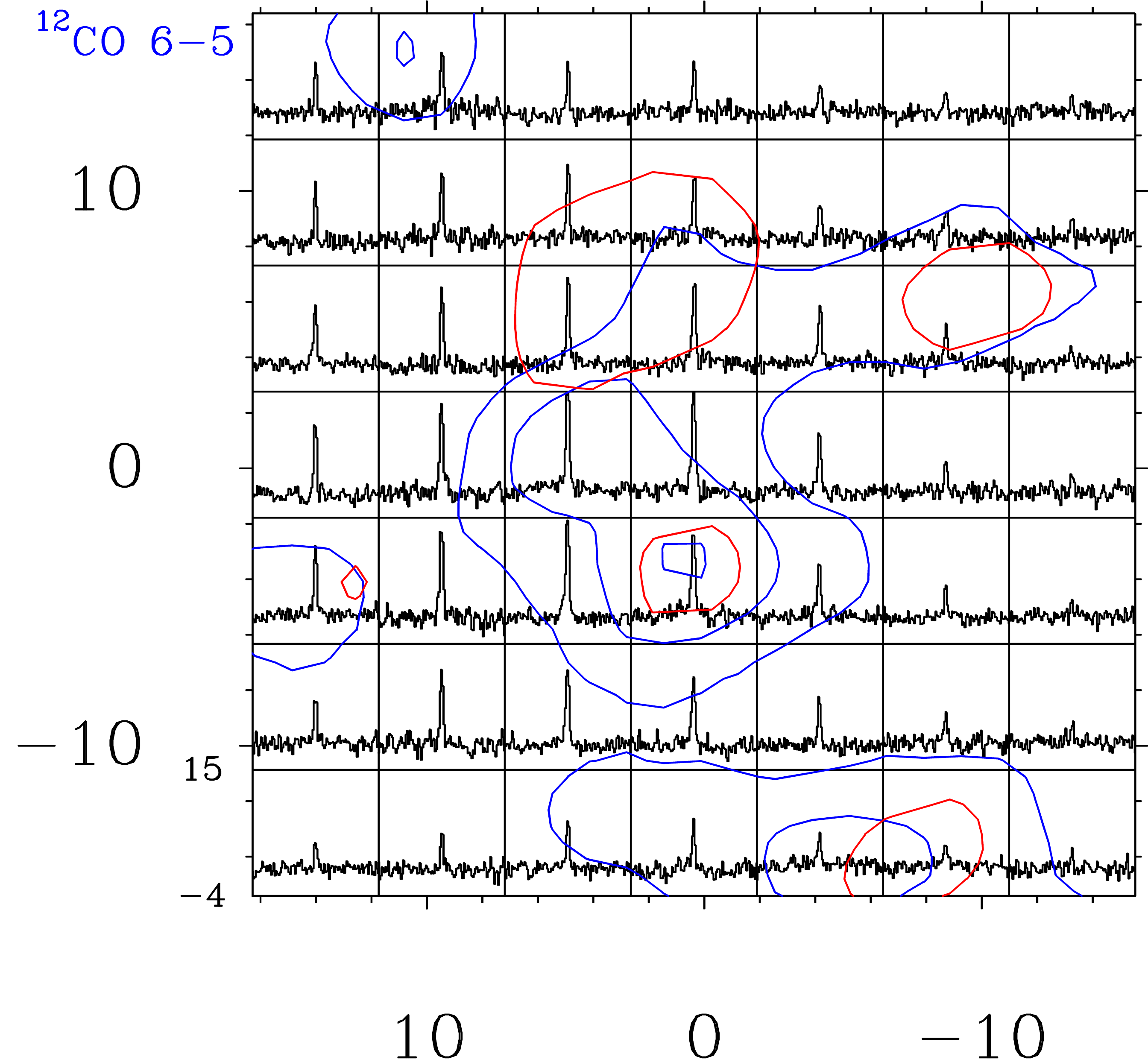}   
    \includegraphics[scale=0.20]{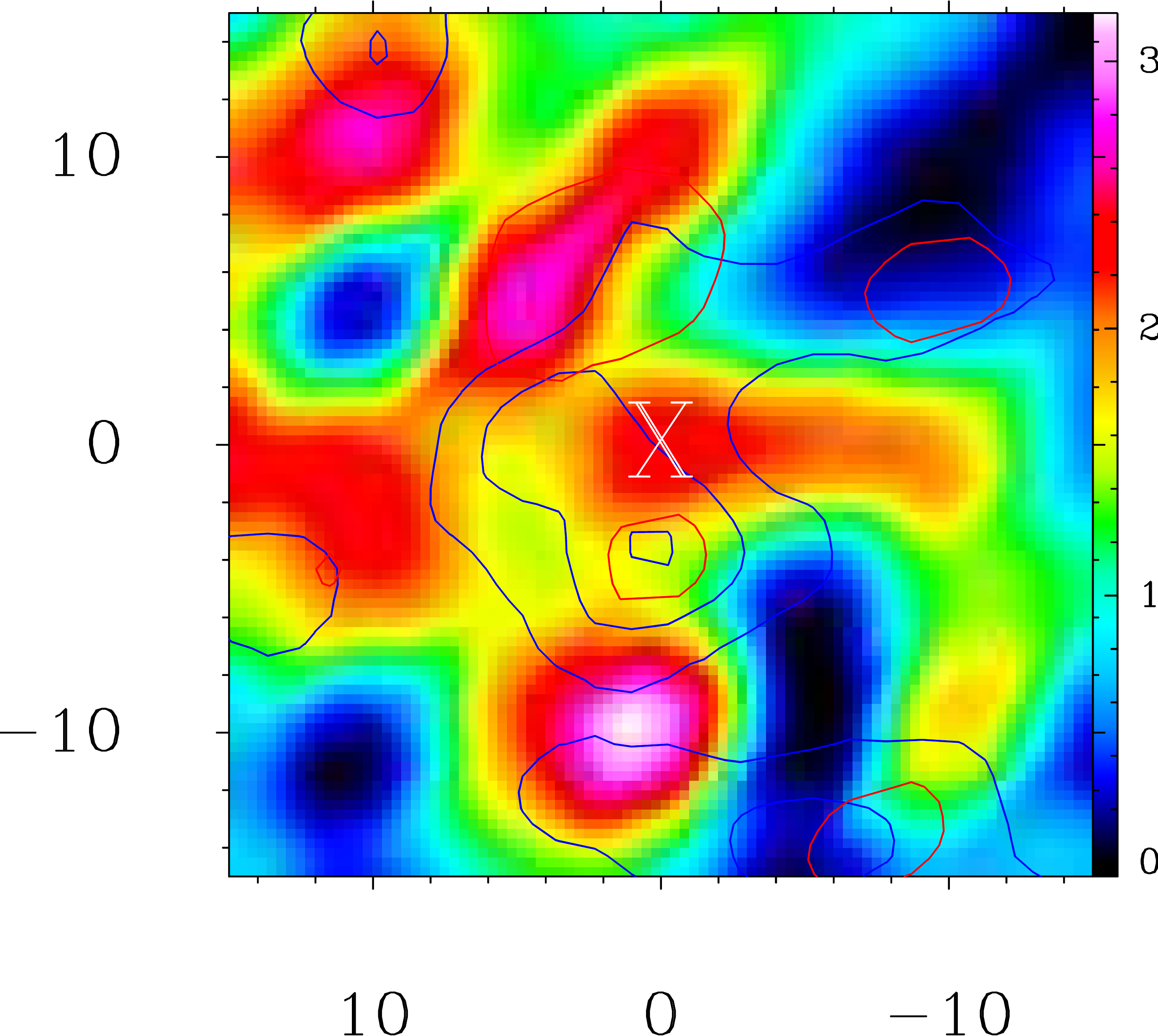} \\
    \includegraphics[scale=0.20]{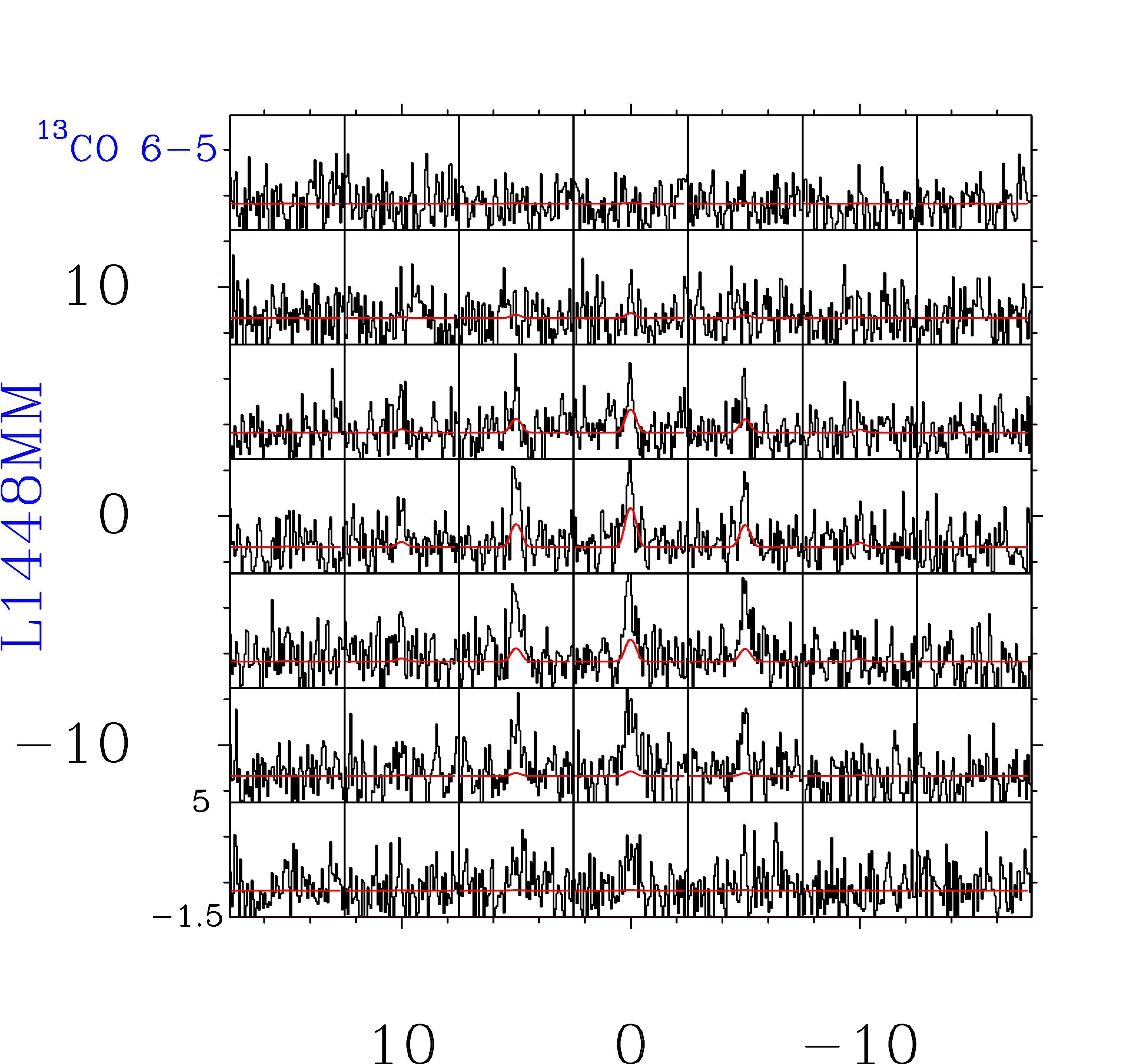}   
    \includegraphics[scale=0.20]{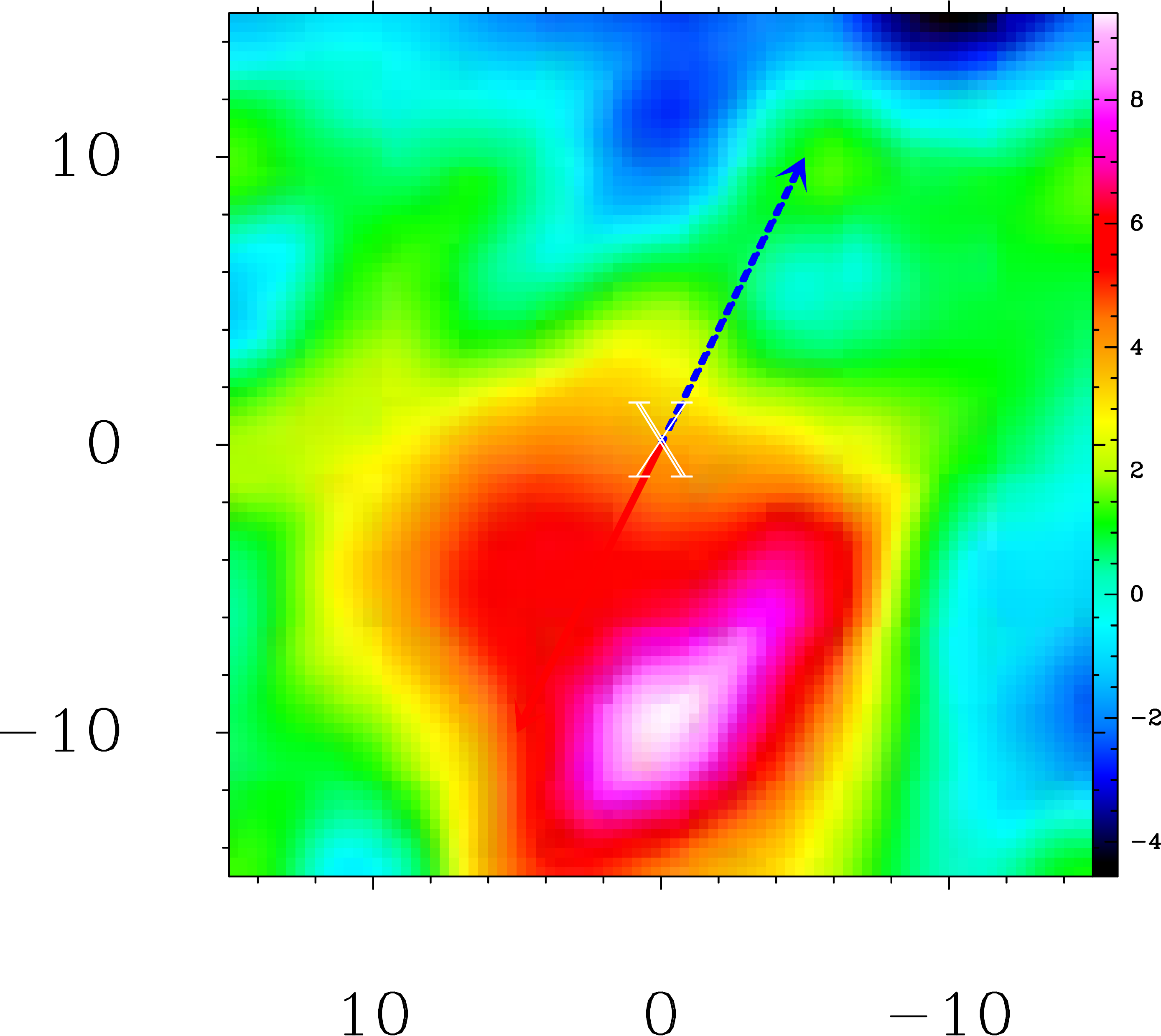} \\
    \includegraphics[scale=0.20]{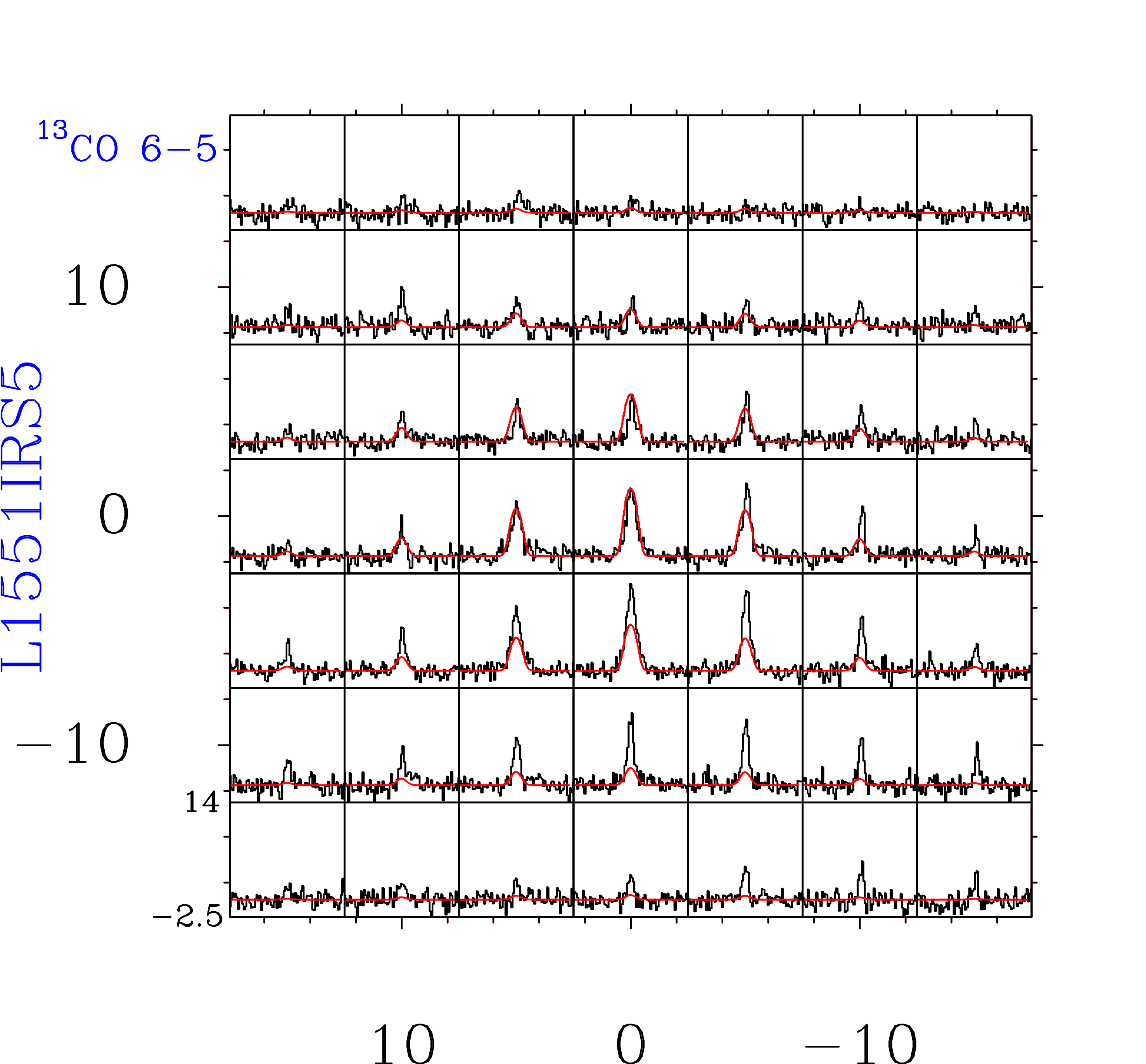}   
    \includegraphics[scale=0.20]{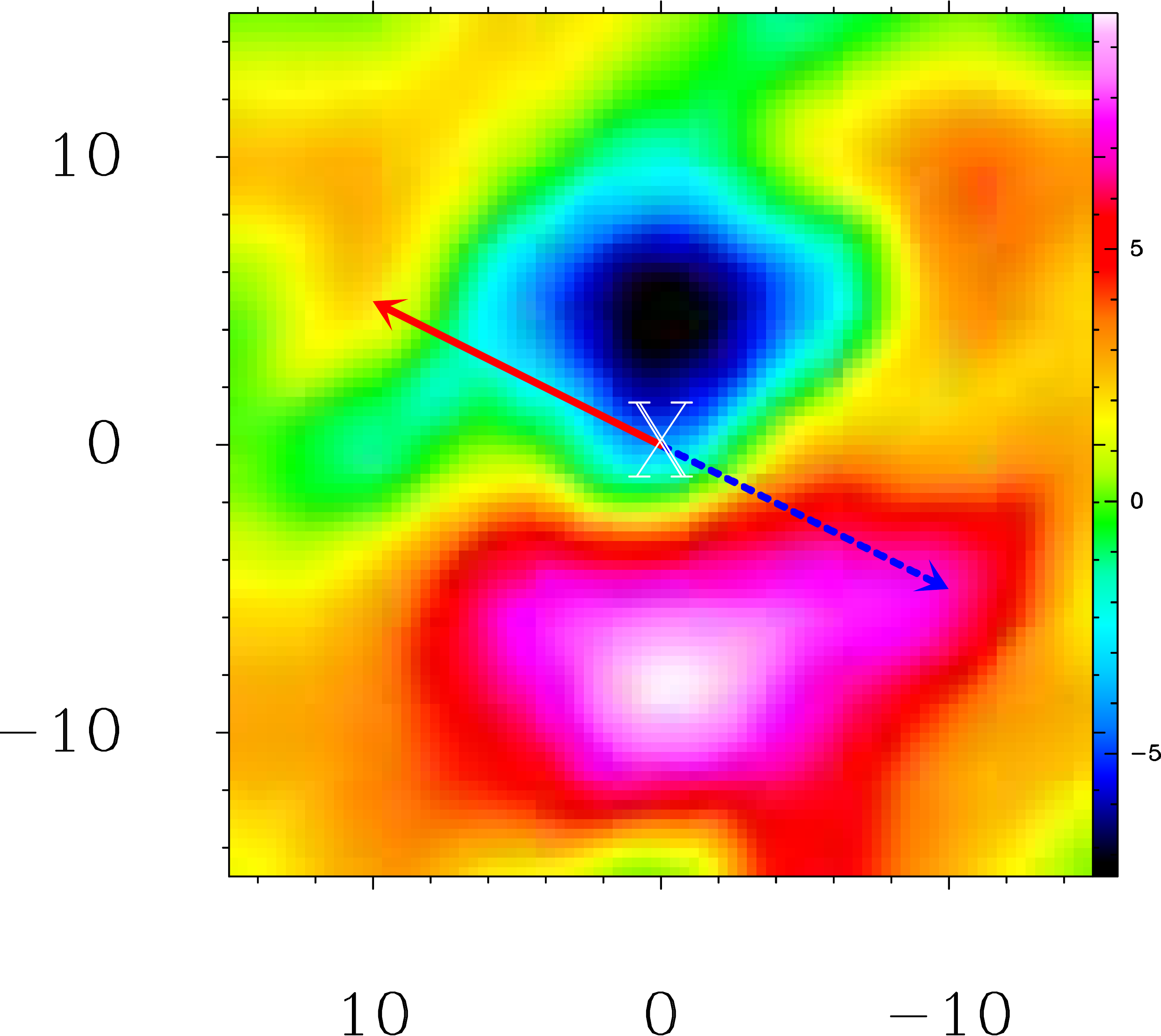}   
    \caption{\small Same as Fig. \ref{fig:specmap13CO65_1}. \twco\ 6--5 transitions were not observed for L1448MM and L1551\,IRS5 in our observing campaign, therefore left blank.}
    \label{fig:specmap13CO65_6}
\end{figure*}

\begin{figure}[tb]
    \centering
    \includegraphics[scale=0.5]{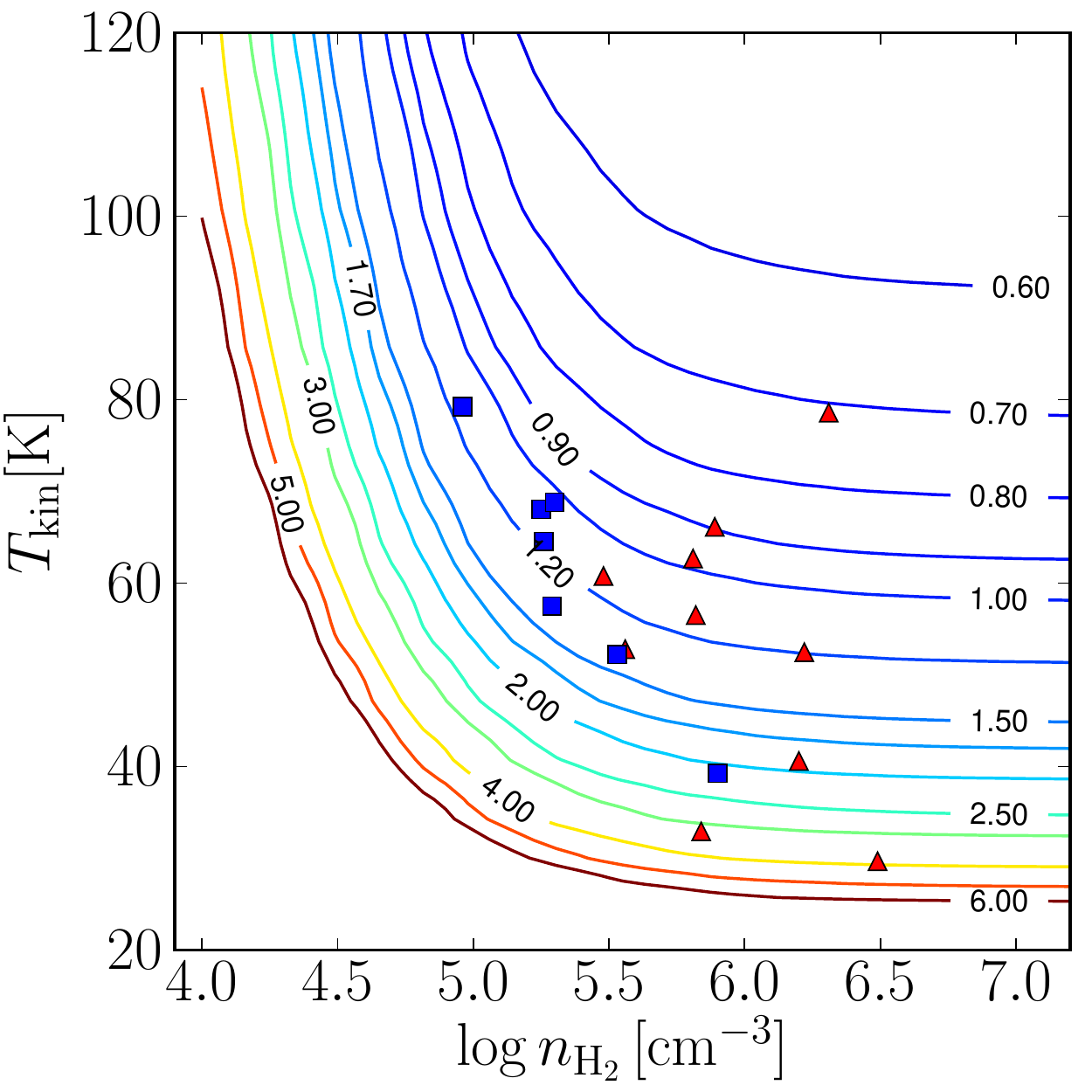}
    \caption{\small \thco\ 3--2/6--5 intensity ratio as a function of
      density and gas temperature calculated via RADEX for
      $N(^{13}{\rm CO})$=1.5$\times$10$^{14}$ \cmtwo. Red markers
      indicate the observed intensity ratios for the central pixels
      for Class~0 sources whereas blue markers are for Class~I
      sources. Both pixels are taken to be 15$\arcsec$ diameter. The
      corresponding densities are the values at the 7.5$\arcsec$
      radius found in the power-law envelope models of
      \citet{Kristensen12}.}
    \label{fig:radex13coratio}
\end{figure}

\end{document}